\documentclass[3p]{elsarticle}   

\usepackage[T1]{fontenc}
\usepackage{url}
 \usepackage{lscape}

\usepackage[pagewise,displaymath]{lineno}

\usepackage{aas_macros}
\usepackage{multirow}            
\usepackage{array}
\usepackage{hyperref}
 \usepackage[utf8]{inputenc}
 \usepackage{slashed}
  \usepackage{amsmath}
  \usepackage{amssymb}
\usepackage[dvipsnames]{xcolor}
\usepackage{xcolor}
\usepackage{graphicx}
\usepackage{hyperref}
  \usepackage{amsfonts}
\usepackage{multicol}
 \usepackage{tikz}
\usepackage{cancel}
 \usepackage{mathtools}

\biboptions{sort&compress}

\usepackage[normalem]{ulem}    
\modulolinenumbers[1]                

\definecolor{nicered}{rgb}{0.7,0.1,0.1}
\definecolor{nicegreen}{rgb}{0.1,0.5,0.1}
\definecolor{red}{rgb}{1.0, 0, 0}
\definecolor{darkblue}{rgb}{.0,.0,.8}
\definecolor{niceblue}{rgb}{0,0,1}
\definecolor{niceviolet}{rgb}{0.5,0,1.0}
\definecolor{blue}{rgb}{0,0,1}
\hypersetup{colorlinks=true,citecolor=nicegreen,linkcolor=nicered,urlcolor=nicered}

\usepackage[normalem]{ulem}    
\modulolinenumbers[1]                

\journal{Physics Reports}

\newcommand{\beq}{\begin{equation}}
	\newcommand{\eeq}{\end{equation}}
\newcommand{\bea}{\begin{eqnarray}}
	\newcommand{\eea}{\end{eqnarray}}

\begin{document}
	
\begin{frontmatter}
		
\title{Axion Astrophysics}

\author{Pierluca Carenza\corref{corr1}}
\ead{pierluca.carenza@fysik.su.se}
\address{The Oskar Klein Centre, Department of Physics, Stockholm University, Stockholm 106 91, Sweden}
\author{Maurizio Giannotti\corref{corrauth}}
\cortext[corrauth]{Corresponding author}
\ead{mgiannotti@unizar.es}
\address{Centro de Astropart{\'i}culas y F{\'i}sica de Altas Energ{\'i}as (CAPA), Universidad de Zaragoza, Zaragoza, 50009, Spain and Department of Chemistry and Physics, Barry University, 11300 NE 2nd Ave., Miami Shores, FL 33161, USA}
\author{Jordi Isern\corref{corr3}\,}
\ead{isern@ieec.cat} 
\address{(Institut de Ciències de l'Espai (ICE, CSIC), Campus UAB, C/ de Can Magrans s/n, E-08193 Cerdanyola del Vallès, Spain \\ 
    Institut d'Estudis Espacials de Catalunya (IEEC), C/ Gran Capità 2-4, E-08034 Barcelona, Spain\\ Fabra Observatory, Royal Academy of Sciences and Arts of Barcelona (RACAB),La Rambla 115, E-08002 Barcelona, Spain}
\author{Alessandro Mirizzi\corref{corr2}}
\ead{alessandro.mirizzi@ba.infn.it}
\address{Dipartimento Interateneo di Fisica “Michelangelo Merlin”, Via Amendola 173, 70126 Bari, Italy,\\
    Istituto Nazionale di Fisica Nucleare - Sezione di Bari, Via Orabona 4, 70126 Bari, Italy}
\author{Oscar Straniero\corref{corr4}\,}
\ead{oscar.straniero@inaf.it} 
\address{INAF, Osservatorio Astronomico d’Abruzzo, 64100 Teramo, Italy,\\
    Istituto Nazionale di Fisica Nucleare - Sezione di Roma, Piazzale Aldo Moro 2, 00185 Roma, Italy}

\begin{abstract}
Stars have been recognized as optimal laboratories to probe axion properties. In the last decades there have been significant advances in this field due to a better modelling of stellar systems and accurate observational data. In this work we review the current status of constraints on axions from stellar physics.  We focus in particular on the Sun, globular cluster stars, white dwarfs and (proto)-neutron stars.	
\end{abstract}



\end{frontmatter}
\newpage
\tableofcontents
\newpage

\section{Introduction}
\label{sec:Intro}

Stellar interiors represent unique environments to probe very rare processes in particle physics. For example, the plasmon decay process $\gamma^* \rightarrow \nu \bar{\nu}$~\cite{Adams:1963zzb} is the most efficient energy drain mechanism during the Red Giant (RG) phase of stars, even though the probability of such a decay to occur between successive interactions of the plasmon is just $\sim 10^{-26}$~\cite{Friedland:2012hj}. Signatures of this process, predicted by the theory of weak interactions, are also easily discernible in the luminosity function of White Dwarf (WD) stars, which provides arguably its most clear \emph{experimental} test. Another significant example is the neutrino pair production $e^+ e^-\to  \nu  \bar \nu$ process, driven by the weak interactions. To occur, this process must compete with the much more efficient $e^+e^-\to \gamma\gamma$, driven by the electromagnetic interactions, with a relative branching ratio of
\begin{align}
\label{eq:}
\frac{e^+e^-\to  \nu  \bar \nu}{e^+e^-\to \gamma \gamma}\approx 10^{-19}\,.
\end{align}
This extremely small branching ratio makes observations of the process $e^+ e^-\to  \nu \bar \nu$ in a terrestrial laboratory prohibitive. Yet, this process governs the evolution of stars in late evolutionary stages and its validity can be tested in a variety of astronomical observations. Thus, stars are of extreme importance for particle physics, providing in some cases the only method to test very rare processes. In fact, as we shall see in this Report, throughout the years they have provided some of the strongest bounds on various theories of exotic particles with feeble interactions. 

The importance of astrophysics and, in particular, of stellar observations to study subatomic and fundamental physics became evident soon after the fundamental role of nuclear reactions in stellar evolution was recognized.  Sixty years ago, in a seminal paper published in 1963, J. Bernstein and collaborators studied the effects of neutrino electromagnetic form factors in the Sun and derived a strong bound on the neutrino magnetic moment~\cite{Bernstein:1963qh}. The bound was, at the time, more stringent than the existing constraints from dedicated laboratory experiments. In the following years, a series of seminal papers analyzed neutrino properties, including lepton charge~\cite{Gribov:1968kq} or radiative neutrino decays~\cite{Cowsik:1977vz,Falk:1978kf} from stellar evolution arguments. The potential of stars to constrain the properties of new exotic particles was recognized in the following years, particularly after the introduction of the \emph{axion}, in 1977~\cite{Peccei:1977hh,Peccei:1977ur,Wilczek:1977pj,Weinberg:1977ma}. Within a year from the axion hypothesis, a series of powerful arguments were proposed to constrain this particle through astrophysical observations (see, e.g., Refs.~\cite{Vysotsky:1978dc,Dicus:1978fp,Mikaelian:1978jg,Sato:1978vy}).  These arguments set the basis for the modern approach to the field of stars as laboratories for fundamental physics, as lucidly exposed in Ref.~\cite{Raffelt:1996wa}. 

In more recent years, the use of astrophysical methods to constrain new physics has received an additional boost thanks to the increasing interest in the physics of \emph{Feebly Interacting Particles (FIPs)}. FIPs are, in general, fields with specific properties (spin, parity, etc...), coupled to the Standard Model (SM) through effective interactions, suppressed by very high energy scales (see Refs.~\cite{Agrawal:2021dbo,Antel:2023hkf} for recent reviews). From a modern point of view, the study of these particles is interesting since they are the most likely low-energy realizations of unknown UV-complete theories. Furthermore, the existence of these fields is frequently associated with unsolved issues in particle physics, astrophysics and cosmology. In particular, light FIPs, also known as  \emph{WISPs (weakly interacting slim particles)}~\cite{Jaeckel:2010ni} are often invoked as dark matter candidates. The QCD axion is, arguably, the most motivated and best studied WISP candidate. Other well studied members of the WISP family include axion-like particles, dark photons, sterile neutrinos, and chameleons. Such WISPs can be produced in stellar media and, owing to their feeble couplings, provide an efficient energy loss/transport mechanism. Depending on their properties, WISPs might compete with standard energy loss and transport processes—such as those involving photons and neutrinos—thereby influencing stellar observables in measurable ways. Notably, the same weak couplings that allow WISPs produced in stars to escape freely—making stars exceptional WISP factories—also make their direct detection in colliders particularly challenging, as they pass through detectors with ease. This is why stars have provided some of the most stringent bounds on WISPs, often surpassing the potential of terrestrial experiments. Clearly, studying WISPs demands a paradigm shift from the \emph{energy frontier} to the \emph{intensity frontier}~\cite{Proceedings:2012ulb}, and stars serve as excellent laboratories for probing this frontier.

More intriguingly, if new light and weakly interacting particles do exist in nature, one might expect their signatures to be discerned in stellar observations \emph{before} a direct laboratory detection~\cite{Giannotti:2015kwo,Giannotti:2017hny,DiLuzio:2021ysg}. Stars have already demonstrated their complementary role in guiding laboratory searches for new physics. A notable example is the famous claim made by the PVLAS experiment in 2005, which reported an optical rotation in vacuum induced by a magnetic field, potentially attributed to an axion-like particle (ALP)~\cite{PVLAS:2005sku}. However, it was quickly pointed out that such a particle would be incompatible with the very existence of our Sun~\cite{Raffelt:2006cw}. Thus, based on well-established astrophysical arguments, the axion interpretation was immediately rejected, as confirmed also by the PVLAS collaboration~\cite{PVLAS:2007wzd}, which shortly afterwords refuted the claim.

\begin{figure}[t!]
\centering
\includegraphics[width=0.85\columnwidth]{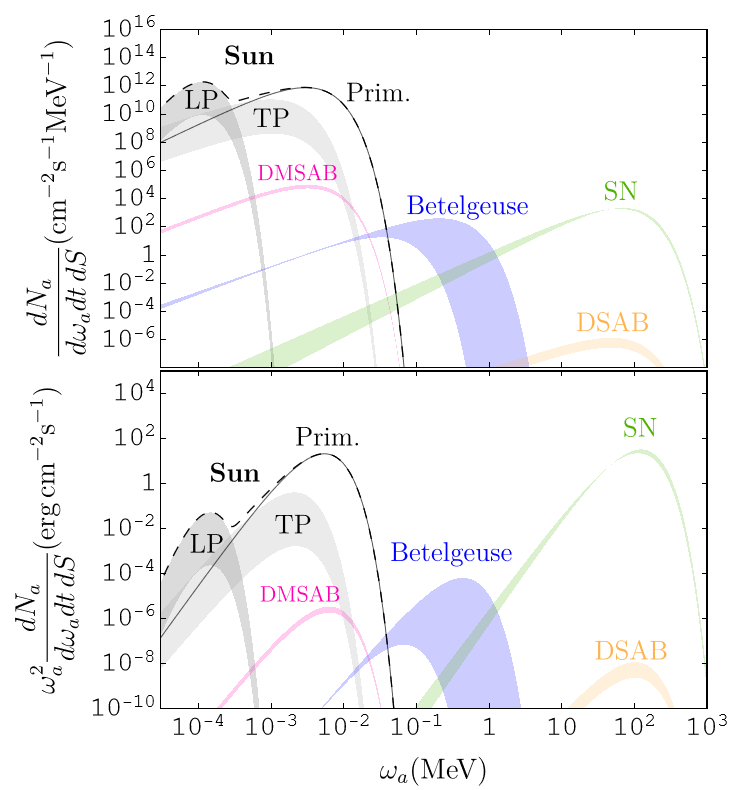}
\caption{ 
Stellar axion spectra at Earth (number flux in the upper panel, and energy flux in the lower panel) vs axion energy $\omega_a$, when considering axions coupled to photons and produced by different sources: the Sun (total flux in dashed line) via Primakoff~\cite{Raffelt:1985nk,Raffelt:1987np} -- black line, and in case of a magnetic field longitudinal plasmon (LP) conversion~\cite{Caputo:2020quz}  and transverse plasmon (TP) conversion~\cite{Guarini:2020hps} -- gray bands; a close-by red supergiant like Betelgeuse~\cite{Xiao:2020pra} (blue band) at a distance $d=197$~pc, and a Galactic SN at a distance of 10~kpc with a progenitor of $18~M_{\odot}$~\cite{Lucente:2020whw} (green band). There are represented also the diffuse fluxes from extra-galactic main-sequence stars~\cite{Nguyen:2023czp} [Diffuse Main-Sequence Axion Background (DMSAB) -- magenta band] and from extra-galactic SNe, the Diffuse Supernova Axion Background (DSAB)~\cite{Calore:2020tjw,Calore:2021hhn} (orange band). It is assumed an axion-photon coupling $g_{a\gamma}=10^{-11}~{\rm GeV}^{-1}$ and an axion mass $m_{a}\ll {\rm eV}$.}
\label{fig:stellaraxions}
\end{figure}

The types of arguments used to probe axions\footnote{In this review, we will often consider axions and ALPs on the same footing, since in many cases the differences between the two are irrelevant for our discussion. In this case, we talk about axions but include ALPs as well. We will clarify explicitly whenever this distinction is relevant.}  from stars fall in two main categories. On one side, one might aim at a \emph{direct} detection of stellar axions. Fig.~\ref{fig:stellaraxions} shows the expected axion flux from several stellar systems, assuming an axion-photon interaction at the current experimental bound. The Figure includes also the diffuse fluxes from galactic Main Sequence (MS) stars and from extra-galactic Supernovae (SNe). The main strategy, in this case, is to make visible these fluxes exploiting their conversions into observable photons in celestial or artificial magnetic fields. These photon fluxes can be observed via existing X- or gamma-ray telescopes~\cite{Xiao:2020pra,Meyer:2016wrm,Calore:2021hhn} or via specific axion experiments (e.g. helioscopes~\cite{CAST:2004gzq,Irastorza:2011gs,IAXO:2019mpb}). The other possibility is to \emph{indirectly} probe axions by monitoring the discrepancy induced by their production on different stellar observables. These probes include, e.g., the reduction of helium burning stars in Globular Clusters (GCs)~\cite{Raffelt:1987yb,Ayala:2014pea}, the impact on WD luminosity function~\cite{MillerBertolami:2014rka} or on Neutron Star (NS) light-curves~\cite{Beznogov:2018fda,Buschmann:2021juv}.


So far, stellar observations have provided immensely powerful tools to study the properties of various WISPs, such as neutrinos, axions and dark photons. In fact, several of the most stringent constraints on these particles were derived from astrophysical and, in particular, stellar observations. Will this trend continue in the near future? We think that there is enough evidence to support this belief, considering the pace at which the astrophysical instrumentation is evolving. New telescopes and space experiments will undoubtedly expand substantially our understanding of stellar evolution in the next decade or so. 

In this Report, we focus on axions and ALPs, and summarize strategies and practical methodologies used to constraint their properties from astronomical observations. There are excellent reviews on axion astrophysics in the literature and the interested reader should consider, at least, some of them to fully understand the evolution of this field. In particular, Georg Raffelt's classic manual \emph{``Stars as Laboratories for Fundamental Physics''}~\cite{Raffelt:1996wa}, though almost thirty years old, is a gem and has provided the main guide for us and for many young researchers in this field. The older Physics Report reviews by Michael Turner~\cite{Turner:1989vc} and Georg Raffelt~\cite{Raffelt:1990yz} are also still very valuable and should be read by anyone interested in this topic. Some updates can be found in the more recent reviews~\cite{Raffelt:2006cw,Caputo:2024oqc}, which have also a high pedagogical value. 

So, why another review? The objective of this review is not to replace but to update and, in some cases, complete the previous monoghraps. The field of axion astrophysics has experienced a substantial development in the recent years. Practically all axion bounds from stellar evolution have been updated. Furthermore, the methodology itself has changed in recent years. Several new astrophysical mechanisms and observables, previously largely ignored, are now considered in the attempt to extract axion properties in the most reliable way. Finally, the increasingly precise astrophysical observations ask for more accurate numerical simulations and for a careful analysis of the uncertainties that affect observations and simulations. Thus, we consider it timely to provide a comprehensive and updated review on the state-of-the-art axion astrophysics, which includes the modern developments and might provide a good reference for both theorists and experimentalists interested in axion physics. We hope this will complete the list of recently published excellent reviews on axion cosmology~\cite{Marsh:2015xka}, axion experiments~\cite{Irastorza:2018dyq,Sikivie:1983ip}, and axion models~\cite{DiLuzio:2020wdo}. Finally, since astrophysical arguments have for decades guided experimental axion searches -- a fact that is getting more and more relevant nowadays, since new axion experiments are finally reaching the potential to probe parameter regions of astrophysical interest~\cite{Giannotti:2017law,DiLuzio:2021ysg} -- we hope that our review could provide also a good reference for the upcoming experimental searches.

We realize that summarizing all the astrophysical axion probes would be an enormous task. Therefore, this work will focus mostly on \emph{stars and stellar observations}. The plan of the paper is the following. In Sec.~\ref{sec:stellarev}, we revise the basic equations leading the stellar evolution and we outline the different phases of stellar life. We also discuss the impact of FIPs on energy-loss and energy-transfer in stars. This Section is more pedagogical than the rest of the paper since we aim at offering studying material to the young researchers who are interested in getting involved in this field. In Sec.~\ref{sec:axionproduction}, we introduce axions and their models and we  present the main production processes in stellar environments, associated to couplings with photons, electrons and nucleons. In Sec.~\ref{sec:Sun}, we consider axion production,  detection strategies and bounds from the Sun and from other nearby sources (MS stars, supergiants and pre-supernovae).  Sec.~\ref{sec:GC}  is devoted to axion bounds from energy-loss in GC stars. In particular the focus is on the tip of Red Giant Branch (RGB), on the $R$ and $R_2$ parameters and Horizontal Branch (HB) stars. Sec.~\ref{sec:white_dwarfs} focuses on constraints from WD, notably from secular drift of the pulsation period and from luminosity function. Sec.~\ref{sec:SN_NS} is devoted to axion constraints from core-collapse SNe and from NS. A detailed review is presented on the bounds from SN 1987A neutrino observation. Perspectives for future SN observations are also discussed. The complementary bounds from the cooling of old and young NS are also presented. Finally, in Sec.~\ref{sec:conclusions} we summarize our results, we present the future perspectives of this topic and we conclude. Two appendices follow: in one we fix our unit conventions and in the other notation and acronyms.

Other interesting aspects of axion astrophysics, including very recent results from X-ray observations~\cite{Conlon:2017qcw,Marsh:2017yvc,Reynolds:2019uqt}, the quite studied topic of very-high-energy astrophysics~\cite{Mirizzi:2007hr,Hooper:2007bq,DeAngelis:2007dqd,DeAngelis:2007wiw,DeAngelis:2011id,Fermi-LAT:2016nkz,Galanti:2022chk}, the conversions of dark matter axions in the magnetic field of astrophysical objects like NS~\cite{Hook:2018iia,Safdi:2018oeu,Edwards:2019tzf,Foster:2020pgt}, (primordial) black holes~\cite{Ferreira:2024ktd} and the Sun~\cite{An:2023wij,Todarello:2023ptf}, or axion constraints from  the modified equations of state in compact stars~\cite{Balkin:2022qer,Gomez-Banon:2024oux} or black-hole superradiance~\cite{Cardoso:2018tly,Day:2019bbh} have not been included in our review.  

\section{Stellar evolution and feebly-interacting particles}
\label{sec:stellarev}

In this Section we will revise the physics principle ruling the stellar evolution and how to account for the role of FIPs. In Sec.~\ref{sec_basic} we present the differential equations describing the stellar evolution. In Sec.~\ref{sec_energy} we describe the main sources and sinks of energy in stellar interiors. In Sec.~\ref{sec_transport} we characterize the radiative energy transport, while in Sec.~\ref{sec_mixing} we consider the effect of convection. In Sec.~\ref{sec_theralaxions} we describe the impact of FIPs on the equations of stellar evolution. Finally, in Sec.~\ref{sec:stelev} we present a brief overview of the stellar evolution, introducing different types of stars we will refer to in the next Sections. We remark that, for pedagogical reasons, in this section we employ CGS units while in the rest of the report we adopt natural units. A discussion on the different unit systems is provided in Appendix A.

\subsection{Stellar structures: a minimal set of equations}\label{sec_basic}
The following $4+n$ differential equations describe the structure of a self-gravitating, spherically symmetric stellar model~\cite{Kippenhahn:2012qhp}
%
\begin{align}
  & \frac{dP}{dm_{r}}=-\frac{Gm_{r}}{4 \pi r^{4}} \,\ ,
 \,\,\   \,\,\
 \,\,\   \,\,\ \,\  
   \,\,\   \,\,\ \,\ \,\,\ \,\   \,\,\ \,\  \,\,\ \,\  \,\,\ \,\  \,\,\ \,\  \,\,\ \,\  \,\,\ \,\  \,\,\ \,\  \,\  \,\  \,\  
    \textrm{(hydrostatic equilibrium)}\label{eq_hydrostatic}\\
  &\frac{dr}{dm_{r}}=\frac{1}{4 \pi r^{2} \rho}  \,\ ,
  \,\,\   \,\,\ \,\  
   \,\,\   \,\,\ \,\ \,\,\ \,\   \,\,\ \,\  \,\,\ \,\  \,\,\ \,\  \,\,\ \,\  \,\,\ \,\  \,\,\ \,\  \,\,\ \,\  \,\  
    \,\,\ \,\  \,\  \,\,\ \,\    
  \textrm{(mass conservation)} \label{eq_continuity} \\
  &\frac{dL_{r}}{dm_{r}}=\varepsilon_{\rm nucl}-\varepsilon_{\nu}  + \varepsilon_{\rm grav}  \,\ ,
    \,\,\   \,\,\ \,\ \,\,\ \,\   \,\,\ \,\  \,\,\ \,\  \,\,\ \,\  \,\,\ \,\  \,\,\ \,\  \,\,\ \,\  \,\,\ \,\  \,\  \textrm{(energy conservation)} 
  \label{eq_conserv}\\  
  &\frac{dT}{dm_{r}}=-\nabla\frac{Gm_{r}}{4\pi r^{4}}\frac{T}{P}  \,\ ,
   \,\,\   \,\,\ \,\  \,\,\ \,\   \,\,\ \,\  \,\,\ \,\  \,\,\ \,\  \,\,\ \,\  \,\,\ \,\  \,\,\ \,\  \,\,\ \,\  \,\,\ \,\  \,\,\ \,\  \,\,\  \textrm{(energy transport)} 
  \label{eq_transport} \\
    &\frac{dY_{i} }{dt} =\left( \frac{dX_{i} }{dt} \right)_{\rm nucl}+\left( \frac{dX_{i} }{dt} \right)_{\rm mix},  \quad i=1,.....,n  \,\ . \label{eq_varchi}
     \,\,\   \,\,\ \,\  \textrm{(chemical comp. variation)} 
\end{align}
They provide an appropriate description of the internal structure of the majority of the stars, when the effects of rotation and magnetic fields can be neglected. The first equation represents the equilibrium between gravity and pressure. Here, $m_{r}$ is the mass contained within the sphere of radius $r$ and $P$ is the pressure at $r$. The second equation describes the mass conservation. Here, $\rho$ is the density at $r$. 
The third is the energy conservation equation. $L_{r}$ is the luminosity at $r$, i.e., the radiated electromagnetic power through the spherical surface with radius $r$, while the three $\varepsilon$ coefficients in the right hand side of this equation represent the rates of energy generation or sink (per unit mass) as due to the release of gravitational energy ($\varepsilon_{\rm grav}$) and nuclear energy ($\varepsilon_{\rm nucl}$), and the energy loss due to the production of thermal neutrinos ($\varepsilon_{\nu}$). Then, the fourth equation describes the energy transport that is explicitly connected to the temperature gradient. The nabla function ($\nabla=\frac{d\log T}{d\log P}$) depends on the specific transport mechanism, radiation, convection or conduction. Finally, Eq.~\eqref{eq_varchi} describes the variations of the chemical composition due to the occurrence of nuclear reactions or bulk motion of material, such as that induced by thermal convection. Here, $X_i$ are the mass fractions and $n$ is the number of isotopes. When coupled to the equation of state (EoS), $P=f(\rho,T,\mu)$, and appropriate expressions for the energy generation and energy sink rates, $\varepsilon_{\rm grav,nucl,\nu}$, numerical solutions of Eqs.~\eqref{eq_hydrostatic}--\eqref{eq_varchi} provide values of $4+n$ dependent variables (e.g., $r$, $L$, $P$, $T$ and the $n$\quad$X_i$) as a function of the mass coordinate, i.e., the Lagrangian variable $m_r$, and the time $t$. The quantity $\mu$ is the mean molecular weight. In case of a fully ionized plasma, it is
\begin{equation}
    \mu=\left ( \sum_{i}\frac{X_i(1+Z_i)}{A_i} \right )^{-1} \,\ ,
\end{equation}
where $Z_i$ and $A_i$ are the atomic number and the mass number, respectively. The sum should be extended to all the chemical species. 

From the equations described so far, it appears that a star evolves because of energy losses and changes in the internal chemical composition due, for example, to nuclear reactions.
  In particular, time derivatives appear explicitly in Eqs.~\eqref{eq_varchi}, which describe variations of the chemical composition, and implicitly in Eq.~\eqref{eq_conserv}, where the left-hand side represents the outward energy fluence, while the coefficients $\epsilon$ in the right-hand side represent rates of energy production or consumption. Also, mass-loss or mass-accretion processes should be taken into account. In practice, the stellar structure is continuously adjusted in order to maintain thermal and hydrostatic equilibrium. In principle, the $n+4$ differential equations [Eqs.~\eqref{eq_hydrostatic})--\eqref{eq_varchi})] are strictly coupled and should be solved simultaneously. This is a somewhat time consuming, and in most cases the solution of the first four equations and that of the $n$ equations [Eq.~\eqref{eq_varchi}] are obtained in two separate steps. Firstly, the $n$ equations describing the evolution of the chemical composition are solved by keeping the temperature and the density profiles fixed during a time step, and then, the 4 equations describing the physical structure are solved by taking the chemical composition fixed. This is a good approximation when physical quantities, such as $T$ or $L$, change slowly compared to variations in the chemical composition. In other words, the evolution timescale of the physical structure must be much longer than the nuclear burning timescale. When dynamical instability, such as convection, are at work, also the mixing timescale should be short compared to the timescales of hydrostatic relaxation. In general, this condition is quite well fulfilled by core-H and core-He burning stars, but they are easily violated during more advanced phases of stellar evolution. In all the cases the majority of the stellar evolution codes use an implicit method to solve the stellar structure equation, as firstly proposed by Ref.~\cite{1964ApJ...139..306H}. So, differences in the models found in the extant literature are mainly due to differences in the input physics [EoS, radiative opacity, nuclear network and nuclear reaction rates, energy loss rates]. The numerical accuracy (mesh point spacing and time steps) may also produces discrepancies. Modern stellar evolution codes use adaptive mesh and time step algorithms to ensure accurate solutions all along a complete evolutionary sequence.

\subsection{Sources and sinks of energy}\label{sec_energy}
The main energy sources in stellar interiors are thermonuclear reactions. For fusions between $i$ and $j$ nuclei the energy production rate is
\begin{equation}\label{eq_enucl}
    \varepsilon_{k}=Y_iY_j\rho N_A \langle\sigma v \rangle_k Q_k \,\ ,
\end{equation}
where $N_A$ is the Avogadro's number and $\langle\sigma v\rangle$ is the rate of the reaction. In the case of a reaction between charged particles, it is
\begin{equation}
   \langle\sigma v \rangle=\left( \frac{8}{\pi m} \right)^\frac{1}{2}\left( \frac{1}{kT} \right)^\frac{3}{2}\int_{0}^{\infty}\sigma\left( E \right)Ee^{-\frac{E}{kT}}dE \,\ ,
\end{equation}
where $m$ is the reduced mass of the interacting nuclei and 
\begin{equation}
   \sigma(E)=\frac{S(E)}{E}\exp({-b/\sqrt{E}}) \,\ ,
\end{equation}
is the cross section. Here $b=\frac{\sqrt{2m}\pi Z_1Z_2}{\hslash}=0.99\,Z_1Z_2m^{1/2}$~MeV$^{1/2}$ determines the penetrability of the Coulomb barrier, while $S(E)$, the so-called {\it astrophysical factor}, depends on the structure of the compound nucleus and represents the probability to get a successful reaction output, provided that the Coulomb barrier has been penetrated.  With only few exceptions, the $S$ factor is usually derived from laboratory experiments.  Note that typical thermal energies in stellar interiors are quite low, from a few keV to a few MeV. When the compound nucleus does not present low-energy states, which may enhance the cross section, the astrophysical factor is a smooth function of the energy. In any case, low-energy cross-section measurements are often limited by the noise, as due to cosmic ray, beam induced background and natural radioactivity. As a matter of fact, in most case the cross section is measured at high energy and, then, the $S(E)$ is extrapolated to the low-energy. When resonances whose strength cannot be easily determined  affect the $S(E)$ factor, the resulting reaction rate may be rather uncertain. Finally the $Q_k$ factor in Eq.~\eqref{eq_enucl} is the net energy released by one reaction. It is usually given by the difference of the total mass of the reagents and the total mass of the products. In some cases, however, if neutrinos are released by the reaction, their energy should be subtracted. Indeed, neutrinos are weak interactive particles and, once produced, they freely escape from the star, bringing out part of the thermal energy. If the energy subtracted by the neutrinos produced by nuclear reactions is directly accounted into the $Q$ factors, the energy loss due to the production of neutrinos by thermal processes, i.e., $\varepsilon_\nu$, should be explicitly considered in Eq.~\eqref{eq_conserv}. In practice, neutrinos by plasma oscillations (or plasmon decay, $\gamma^* \to \nu  \overline{\nu}$) are efficiently produced in the degenerate core of red-giant and asymptotic-giant stars. The same process is also active in hot WD (see Sec.~\ref{sec:white_dwarfs}). In addition, the production of thermal neutrinos by Compton effect ($\gamma  e^- \to e^- \nu  \overline{\nu}$) and by electron-positron pair annihilation ($e^+ e^- \to \nu  \overline{\nu}$) is the main energy-loss mechanism in the carbon burning phase of massive stars and up to their final core collapse. The stellar energy-loss rates for this neutrino processes are calculated according to the quantum electrodynamics theory~\cite{Itoh:1984xy,Itoh:1984uv,Haft:1993jt,Esposito:2001if,Esposito:2003wv}.

Finally, the third term in Eq.~\eqref{eq_conserv}, i.e., $\varepsilon_{\rm grav}$, represents the heat gain or loss as due to variations of internal energy, volume and/or chemical composition, namely
\begin{equation}
    \varepsilon_{\rm grav}=-T\frac{dS}{dt}=-\frac{dU }{dt}+\frac{P}{\rho^2}\frac{d \rho}{dt}+\sum_{i}\mu_i\frac{dn_i}{dt} \,\ ,
\end{equation}
where $U$ and $S$ are the specific internal energy and entropy, respectively, and $\mu_i=-T\left ( \frac{\partial S}{\partial n_i} \right )_{U,\rho, n_j\neq i}$ is the chemical potential of the chemical species $i$ and electrons. 

\subsection{Radiative energy transport}\label{sec_transport}
Overall, the presence of a negative temperature gradient naturally induces a radiative energy transfer from the stellar core toward the surface. The photon mean free path (mfp) is limited by various electromagnetic interactions. Inelastic scattering on electrons (Thomson or Compton) dominates in the central fully ionized regions, while bound-bound and bound-free absorption are active in the cooler external zones where matter is only partially ionized. In general, the mfp can be written as 
\begin{equation}
    l_f=\frac{1}{\kappa_{f} \rho} \,\ ,
\end{equation}
where $\kappa_{f}$ is the mean radiative cross section per unit mass of photons with frequency $f$, also called monochromatic opacity. In all cases, the photon mfp is quite short in stellar interiors. In the Sun, the average mfp is $\sim 1$~cm. In these conditions, the radiative energy transport may be treated as a diffusive process (see. e.g., Ref.~\cite{Kippenhahn:2012qhp}). In practices, the $\nabla$ term in Eq.~\eqref{eq_transport} is
\begin{equation}\label{eq_radiative_gradient}
    \nabla=\nabla_{\rm rad}=\frac{3}{16acG}\frac{\kappa_\gamma L_rP}{M_rT^4} \,\ ,
\end{equation}
where $\kappa_\gamma$ is the so-called Rosseland mean opacity, namely
\begin{equation}
   \frac{1}{\kappa_{\gamma}}=\frac{\pi}{acT^3}\int_{0}^{\infty}\frac{1}{\kappa_f}\frac{\partial B}{\partial T} df \,\ .
\end{equation}


\subsection{Transport by convection and other dynamical and secular instabilities}\label{sec_mixing}
Several phenomena may induce instabilities causing bulk motion. The most important one is thermal convection, but some stars show anomalous composition of their atmospheres that are believed to be the result of the activation of additional mixing processes. For instance, bright RG stars show low carbon isotopic ratios likely due to the operation of an internal process that mixes the region between the H-burning shell and the bottom of the convective envelope. Processes capable to drive such an extra-mixing are thermohaline circulation~\cite{1972ApJ...172..165U}, gravity waves~\cite{1993A&A...271L..29S,Kumar:1999mc} or magnetic buoyancy~\cite{Parker:1970xv}. Also massive MS stars usually show chemical variations that bring signatures of material processed by the CNO cycle, such as anomalous enhancements of nitrogen~\cite{2019A&A...631A..97F}. In that case, the chemical anomaly is usually ascribed to meridional circulation induced by rotation, even if gravity waves may also contribute to the mixing in the radiative envelope of these massive stars. 

In general, the heat capacity is so large in stellar interior that the energy loss in the bulk motion triggered by convection is negligible. In practice, convection may be treated as an adiabatic process. This is the case, for example, of the convective cores powered by a central thermonuclear burning. In this case, the $\nabla$ term in Eq.~\eqref{eq_transport} can be replaced with
\begin{equation}\label{adiabatic}
 \nabla=\nabla_{\rm ad}=\frac{\gamma-1}{\gamma\,\chi_T} \,\ ,
\rm \end{equation}
where $\gamma=c_p/c_v$ is the ratio of the specific heats at constant pressure and volume, while $\chi_T=\left ( \frac{\partial \ln P}{\partial \ln T} \right )_\rho$ (see, e.g., Ref.~\cite{1987Sci...235..465C}). However, when convection is active in the envelope of a star, the energy transfer by radiation cannot be neglected and the temperature gradient becomes super-adiabatic. In general, in a layer unstable by convection the temperature gradient is intermediate between the adiabatic and the radiative one, namely
\begin{equation}\label{adiabatic}
 \nabla_{\rm ad}\leq\nabla\leq\nabla_{\rm rad} \,\ .
\end{equation}
A characteristic scale height for stellar convection is the pressure scale height
\begin{equation}
    H_P=-\frac{dr}{d\ln P}=\frac{P}{\rho g} \,\ .
\end{equation}
Usually, this scale height is quite large, so that convection occurs under a turbulent regime. Owing to the difficulty to treat non-linear hydrodynamic equations, a phenomenological theory of convection, the mixing-length theory originally  introduced in Ref.~\cite{1958ZA.....46..108B}, is often preferred to estimate the actual temperature gradient. In this case, the $\nabla$ parameter depends on the average mfp of a convective bubble, the mixing length, usually expressed in terms of the pressure scale height, namely
\begin{equation}
    \Lambda=\alpha H_P \,\ ,
\end{equation}
where $\alpha$ is a free parameter, which is usually calibrated by reproducing the measured solar radius.  

\subsection{Impact  of novel feebly-interacting particles on stellar evolution}\label{sec_theralaxions}
Like neutrinos, FIPs can be produced in stellar cores by thermal processes. The dominant production process would depend on the type of the FIP and on the coupling with ordinary matter particles. In particular, depending on the interaction strength, FIPs can be in a \emph{free-streaming regime}, in case of weak coupling, or in a \emph{trapping regime}, in case of stronger coupling, where weak or strong coupling are defined based on how the mfp compare with the  characteristic size of the stellar system under consideration. 

\subsubsection{Energy loss}
In case of weak coupling, the FIP mfp may be larger than stellar radii and, once produced in the hottest internal layers, these particles freely escape, bringing out of the star the thermal energy spent for their production. Like neutrinos, this additional energy sink mechanism may be accounted by adding a negative term ($\varepsilon_X$) in the right hand side of Eq.~\eqref{eq_conserv}. In this case, FIPs contribute a channel of \emph{energy-loss}. When an efficient thermonuclear burning is active within the stellar core, the effect induced by this additional energy sink is to accelerate the fuel consumption. 

A striking example is the possible production of axions in the core of HB stars belonging to a GC (see Sec.~\ref{sec:Rparam}). In these He-burning stars, the evolutionary timescale is shorter in case of axions. The opposite occurs in stars in which there are no active nuclear burning within the core, as it happens in low-mass RGB stars. In this case, degenerate electrons provide a major contribution to the pressure in the helium-rich core. As fresh helium is accumulated on top of the core by the hydrogen shell burning, its mass increases, while its radius decreases. Hence, the core temperature is determined by the balance between the energy gain driven by the core contraction and the energy loss by plasma-neutrinos, coupled to an efficient thermal conduction by degenerate electrons. As a result, the core temperature slowly increases with time, until the conditions for the helium ignition are attained. This occurrence coincides with the tip of the RGB. In case of a more efficient energy-loss, as due to the FIPs production during the RGB phase, the helium ignition is delayed and the Tip of the Red-Giant Branch is brighter (see Sec.~\ref{sec:tip}).

\subsubsection{Radiative energy transfer by (not-too-much) feeble interacting particles}
If the mfp is smaller than the core radius, FIPs may contribute to the energy redistribution within a stellar core. If the FIP mfp is sufficiently small, this energy transport mechanism may be treated as a diffusive process, as already done for the photons. Thus, a sort of Rosseland mean opacity can be defined also for FIPs. For instance, for a bosonic FIP $X$,   the mean opacity is~\cite{Raffelt:1988rx,Raffelt:1990yz} 
\begin{equation}
    (\kappa_X\,\rho)^{-1} = \frac{1}{4\,a\,T^3}\int_{m_X}^{\infty} dE\, \beta_E\, \lambda_E\, \frac{\partial B_E}{\partial T}\,,
\label{eq:kappa}
\end{equation}
where $\beta_E=\sqrt{1-m_X^2/E^2}$ is the FIP velocity, $\lambda_E$ is the FIP decay length and $B_E$ is the FIP thermal spectrum
\begin{equation}
    B_E = \frac{1}{2\pi^2} \frac{E^2 \sqrt{E^2 - m_X^2}}{e^{E/k T}-1}\,\ .
\label{eq:BE}    
\end{equation}
Hence, the opacity in Eq.~\eqref{eq_radiative_gradient} can be replaced by
\begin{equation}
    \kappa=\left ( \frac{1}{\kappa_{\gamma}} + \frac{1}{\kappa_{e}} + \frac{1}{\kappa_{X}} \right )^{-1} \,\ ,
\end{equation}
where $\kappa_{\gamma}$, $\kappa_{e}$  and $\kappa_{X}$ are the opacities of photons, conductive electrons and FIPs, respectively. The largest impact on the stellar evolution occurs when the FIP opacities are comparable or smaller than the standard ones, favoring the energy transfer. For instance, in case of core-collapse SN or NS the major effect is found when the FIP opacities are comparable to those of trapped neutrinos (see Sec. \ref{sec:SN_NS}).

Recently, the energy transfer by bosonic FIPs in stars has been studied in details in Ref.~\cite{Caputo:2022rca}, where have been found explicit volume-integral expressions for the boson luminosity, reaching from the free-streaming to the strong-trapping limit. The latter is seen explicitly to correspond to quasi-thermal emission from a ``FIP sphere'' according to the Stefan-Boltzmann law.

\subsubsection{Constraining the physics nature of FIPs }
Once FIPs effects are included in stellar structure equations, the next question is how to get possible constraints on their existence. The general strategy is straightforward: once an observable stellar property much sensitive to the FIP ingredient is identified, a  comparison between the theoretical prediction and the measured value of this stellar property provides a hint (or a bound) to the proposed new physics hypothesis. To be competitive with laboratory experiments, the error budget should be reduced as much as possible. Actually, the main issue is the correct evaluation of all the sources of errors, those affecting both the theoretical predictions and their observational counterparts. The main risk is to underestimate the global error. A simple example may better illustrate the procedure. According to Eq.~\eqref{eq_conserv}, the luminosity of a star is
\begin{equation}\label{eq_luminosity}
   L=\int_{0}^{R_{\rm star}}\frac{\partial L}{\partial r}dr=\int_{0}^{R_{\rm star}}4\pi r^2 \rho (\varepsilon_{\rm nucl}-\varepsilon_{\nu}  + \varepsilon_{\rm grav})dr\,.
\end{equation}
Photometric measurements give the apparent brightness of the star, from which the intrinsic luminosity can be derived, provided that the distance is known. Now, suppose to find a discrepancy between the theoretical prediction and the measured value. As usual, this tension may be due to either errors affecting the theoretical recipe and/or the observed luminosity, or to some missing physics ingredient. At this regard, already different stellar systems present intriguing hints of extra-cooling that might be attributed to FIP emission~\cite{Giannotti:2015kwo,Giannotti:2017hny,DiLuzio:2021ysg}. Nevertheless, before claiming for an exotic particle physics solution, one should first clarify possible systematics in  the astrophysical input. Examples of observational uncertainties  are those due to the calibration of the photometric measurements, the evaluation of light extinction along the line of sight or the estimate of the distance. Examples of theoretical uncertainties are those affecting the energy generation rates, i.e., the $\varepsilon$ factors in the energy conservation Eq.~\eqref{eq_conserv}, such as unknown low-energy nuclear states that may affect fusion cross sections and, in turn, nuclear reaction rates. Once all the possible sources of error are under control, from the observed discrepancy we can infer possible deviation from standard model predictions that could affect the stellar luminosity.  

The previous argument has been applied to several classes of stars (Sun, RG, He-burning stars, WD, NS, etc...) and several FIP candidates. A general introduction is provided in Ref.~\cite{Raffelt:1996wa}. In this work we will focus on the production and the impact of \emph{axions} on stellar evolution. In the next sections, a brief overview of stellar evolution for different stellar masses is provided, focusing, in particular, on the evolutionary phases in which a larger impact on FIP production is expected.

\subsection{Overview of stellar evolution}\label{sec:stelev}

Stars are self-gravitating objects that maintain thermal and hydrostatic equilibrium for most of their lives. They evolve because they lose energy, mainly from the surface, by radiation, but in some important cases also from their hot interiors, by production of the weakly interacting neutrinos. As stars lose energy, their internal temperature increases. This may appear counter-intuitive, but it is not. Briefly, to replace the energy loss and maintain the equilibrium, the stellar core contracts and, doing so, extracts energy from its gravitational field. However, according to the virial theorem (see next Section), the amount of gravitational energy released in this way exceeds the energy leakage of about a factor of 2, so that the internal energy must increase. Then, when the temperature becomes large enough, thermonuclear reactions enter into the game. If an efficient nuclear fire is active near the stellar center, the core contraction stops and all the energy lost by radiation or thermal neutrino emission is sustained by an equivalent nuclear energy release. In practice, the nuclear reactions rate is regulated by the stellar brightness. 
  
Summarizing, the  energy lost by radiation or neutrinos is replaced by part of the gravitational energy released as a consequence of a core contraction or by the activation of exothermic thermonuclear reactions. In the latter case, the star evolves slowly, adjusting its structure to account for the variation of the chemical composition. On the contrary, when the nuclear power is off, the core must contract and faster variations of the macroscopic stellar properties take place. The alternation of phases controlled by the core contraction and phases controlled by the nuclear power ends when the density is high enough that a high degeneracy of the electron component of the stellar plasma develops. In this case, owing to the peculiar equation of state of fully degenerate electrons, the increase of the density produce an increase of the pressure, but the temperature is practically insensitive to a further contraction. As discussed in Sec.~\ref{sec:degenerate_eos}, this will lead to a dichotomy in the final destiny of stars, which depends on their core mass and composition.

\subsubsection{The virial theorem}\label{sec:virial}
\label{intro}

Making use of Eq.~\eqref{eq_continuity}, let us rewrite the hydrostatic equilibrium equation [Eq.~\eqref{eq_hydrostatic}]
\begin{equation}
\frac{dP(r)}{dr}+\frac{Gm(r)}{r^2}\rho(r)=0 \,\ .
\label{eq_idro}
\end{equation}
Here, we have neglected the centrifugal force, possibly induced by stellar rotation. In most cases, this is a reasonable approximation (e.g., for the Sun). In general, rotation velocities are higher in massive stars. On the other hand, thermal equilibrium implies the well-known thermodynamic relation between pressure $P$ and internal energy per unit volume 
$u$, namely
\begin{equation} 
P=(\gamma -1)u \,\ ,
\label{eq_pu_relation}
\end{equation}
where $\gamma$ depends on the equation of state. In case of a classical perfect gas, it is $\gamma= \frac{c_P}{c_V}$, i.e., the ratio of the specific heats at constant pressure and constant volume, and it only depends on the total number of degrees of freedom. Thus, combining Eqs.~\eqref{eq_idro}-\eqref{eq_pu_relation}, after some algebraic manipulations (see, e.g.,  Ref.~\cite{Kippenhahn:2012qhp} for the details), we obtain a fundamental equation known as the {\it virial theorem} 
\begin{equation} 
3(\gamma -1)E_{i}+\Omega_{\rm grav}=0 \,\ , 
\label{eq_virial}
\end{equation}
where
\begin{equation} 
E_{i}=\int_{0}^{M}\frac{u}{\rho}dm=\int_{0}^{M} \frac{1}{\gamma -1}\frac{P}{\rho}dm \,\ ,
\end{equation}
is the total internal energy, and
\begin{equation} 
\Omega_{\rm grav}=-\int_{0}^{M}\frac{Gm}{r}dm \,\ ,
\end{equation}
is the gravitational potential energy. For a mono-atomic and non-relativistic perfect gas, $\gamma =\frac{5}{3}$ and the coefficient multiplying the internal energy in Eq.~\eqref{eq_virial} is just 2. It implies that 50~\% of the gravitational potential energy released during a contraction of the stellar structure is transformed into internal energy, while the other 50~\% is transferred outside by radiation and eventually lost. In this case, gravity may replace all or part of the energy lost by the core. However, in case the gas develops relativistic conditions (see next Section), the $\gamma$ factor drops down to a critical value, $\sim \frac{4}{3}$, for which all of the released gravitational energy is transformed into internal energy. Obviously, since the energy loss never stops, hydrostatic equilibrium cannot be maintained for long. The same conclusion may be reached looking at the total energy $W$. From Eq.~\eqref{eq_virial} we have 
\begin{equation}
W=E_{i}+\Omega_{\rm grav}=-\frac{\Omega_{\rm grav}}{3(\gamma -1)}+\Omega_{\rm grav}=\frac{3\gamma -4}{3\gamma -3}\Omega_{\rm grav} \,\ .
\label{eq_etot}
\end{equation}
Therefore, a stellar structure is bound ($W<0$) only if $\gamma >\frac{4}{3}$. In practice, stars dominated by relativistic particles are not stable. Furthermore, relativistic stars have not a hydrostatic scale length, so that any injection or removal of energy induces either large expansions (explosions) or contractions (collapse).
 
Stars form in the collapse of molecular clouds. Various phenomena may trigger this phenomenon, such as the compression induced by the shock wave generated by the explosion of a nearby SN. As the density rises up, the electromagnetic interactions of photons with atoms or molecules becomes more effective.  When the density is large enough, so that the gas cloud becomes optically thick, the hydrostatic equilibrium settles on. Later on, the contraction of the proto-star proceeds under the control of the virial theorem. As a consequence, the internal energy eventually increases, until the temperature becomes large enough to ignite the fusion of hydrogen nuclei. When H is almost completely consumed near the center, the core contraction restarts. Once again, the virial theorem controls the increase of temperature and density within the stellar core, until conditions for the next nuclear burning, i.e., the He burning, are attained. In general, the core density progressively increases, until the development of a strong electron degeneracy determines the end of the stellar evolution. 

\subsubsection{The equation of state of degenerate electrons}\label{sec:degenerate_eos}
Among the various components of the stellar plasma, electrons are the lightest fermions. A simple estimate of the degree of degeneracy may be obtained by comparing
the de Broglie wavelength of the electrons and the mean 
inter-particles distance. The former is
 \begin{equation}
 \lambda=\frac{h}{p}\approx \frac{h}{(3m_ekT)^{1/2}} \,\ ,
 \label{eq_deb}
 \end{equation}
where $m_e$ and $p$ are the electron mass and momentum, while $h$ and $k$ are the Planck and the Boltzmann constants, respectively. Here we have approximated the momentum by means of $\frac{p^2}{2m_e}\approx \frac{3}{2}kT$, where $\frac{3}{2}kT$ represents the average (Maxwellian) thermal energy. Hence, the mean inter-particle distance is given by
\begin{equation}
d \approx \frac{1}{n_e^{1/3}} = \left ( \frac{1}{Y_e N_A \rho} \right )^{1/3} \,\ ,
 \label{eq_particle_distance}
\end{equation}
where $n_e$ is the number of electron per unit volume,  $N_A$ is the Avogadro number and $Y_e$ is the average number of electrons per nucleon. In case of full ionization, the latter is given by
\begin{equation}
Y_e=\sum _i\frac{Z_i}{A_i}X_i \,\ ,
\label{eq_ye}
\end{equation} 
where the summation extends to all the chemical species. Therefore, electron degeneracy develops when $\lambda > d$ and, by combining Eq.~\eqref{eq_deb} and Eq.~\eqref{eq_particle_distance}, this condition implies
\begin{equation} 
\rho > \frac{1}{Y_e N_A}\left ( \frac{3m_e kT}{h^2} \right )^{3/2} \,\ ,
\end{equation}
or $\rho > 10^{-9} T^{3/2}$~g/cm$^3$. Independently of the initial mass, the late evolution of any star is marked by the development of core densities well above this threshold. In this case, the pressure by degenerate electrons provides, by far, the largest contribution to the total pressure. This is a consequence of the Pauli exclusion principle for which only two electrons, with opposite spins, may occupy the same elementary cell (state) in the phase space. Let us illustrate this important point in more details.

According to the Heisenberg's principle, the volume of an elementary cell in the phase space is $(2\pi \hbar)^3$, so that a volume $d\tau = dp_xdp_ydp_z\,dxdydz=d^3p\,dV$ may contain
\begin{equation}
dn=g \frac{d^3p\,dV}{(2\pi \hbar)^3} 
\end{equation}
particles. Here, $g$ is the state multiplicity and represents the maximum number of particles that share the same elementary cell. For the electrons, whose spin is 1/2, $g=2$. Therefore, if $f(p)\,d^3p$ is the number of electrons with momentum between $p$ and  $p+d^3p$, the total number of particle in a finite volume $V$ will be
\begin{equation}
n=\int  \frac{2}{\left ( 2\pi \hbar \right )^3}f(p)dVd^3p=\frac{2V}{\left ( 2\pi \hbar \right )^3}\int f(p)d^3p =\frac{2V}{\left ( 2\pi \hbar \right )^3}\int_{0}^{\infty }f(p)4\pi p^2dp \,\ ,
\end{equation}
where we used $d^3p=4\pi p^2\,dp$ due to isotropy. Hence, according to the Fermi-Dirac distribution function
\begin{equation}
f(\varepsilon)=\frac{1}{\exp\left({\frac{\varepsilon -\mu}{k T}}\right)+1} \,\ ,
\end{equation}
where $\varepsilon $ and $\mu$ are the electron energy and chemical potential, respectively. In case of strong degeneracy it reduces to the well-known step distribution:
\begin{equation}
f(\varepsilon)=  \bigg\{\begin{matrix} 1  & \textrm{for}
\quad \varepsilon\in\left[ 0,\varepsilon_F \right]
\\ 0 &\hspace{0.2cm}  \textrm{for}\quad\varepsilon\in \left( \varepsilon_F,\infty \right] \end{matrix}  \,\ .
\nonumber
\end{equation}
Here $\varepsilon_F$ is the Fermi energy, which represents the maximum energy of the electrons at zero temperature (if $T$ is not zero the thermal tail extends beyond the Fermi energy). It also implies a maximum momentum. For non-relativistic electrons, it is $p_F^2=2m_e\varepsilon_F$. Therefore, taking $f(p)=1$ for $p\leq p_F$ and 0 otherwise, we get
\begin{equation}
n=\frac{8\pi V}{\left ( 2\pi \hbar \right )^3}\int_{0}^{p_F }p^2dp=\frac{Vp_F^3}{3\pi^2 \hbar^3} \,\ ,
\label{nele}
\end{equation}
so that
\begin{equation}
p_F=\left ( 3\pi^2 \right )^{1/3}\hbar \left ( \frac{n}{V} \right )^{1/3} \,\ ,
\label{pfermi}
\end{equation}
and
\begin{equation}
\varepsilon _F=\left ( 3\pi^2 \right )^{2/3}\frac{\hbar^2}{2m_e} \left ( \frac{n}{V} 
\right )^{2/3} \,\ .
\label{efermi}
\end{equation}
Eventually, from Eq.~\eqref{nele}, we can derive the internal energy per unit volume $u$
\begin{equation}
u=\frac{8\pi}{\left ( 2\pi \hbar \right )^3}\int_{0}^{p_F }\varepsilon p^2dp=\frac{1}{2m_e\pi^2 \hbar^3}\int_{0}^{p_F } p^4dp=
\frac{p_F^5}{10m_e\pi^2 \hbar^3} \,\ ,
\end{equation}
\noindent
and, recalling the relation in Eq.~\eqref{eq_pu_relation} between pressure and internal energy with $\gamma=\frac{5}{3}$ (non-relativistic electrons)
\begin{equation}
P=\frac{p_F^5}{15m_e\pi^2 \hbar^3}=\frac{\left ( 3\pi^2 \right )^{2/3} \hbar^2}{5m_e}\left ( \frac{n}{V} \right )^{5/3} \,\ ,
\end{equation}
\noindent
where we have used Eq.~\eqref{pfermi}.
Noteworthy, the EoS of a degenerate electron gas has a polytropic form. Indeed, with the substitution 
$\left ( \frac{n}{V} \right )=N_AY_e\rho$, we find
\begin{equation}
P=k\rho^\gamma \,\ ,
\label{eq_polytropic} 
\end{equation}
where
\begin{equation}
    k=\frac{\left ( 3\pi^2 \right )^{2/3} \hbar^2}{5m_e}\left( N_{A}Y_e \right )^{5/3}~~~   
      {\rm and}~~~  
      \gamma=5/3 \,\ ,
\label{eq_pdeg_norel} 
\end{equation}
The EoS here derived is valid for non-relativistic electrons. Nevertheless, one can repeat the calculation also for relativistic electrons, using the appropriate expressions for the energy, namely $\varepsilon =\sqrt{p^2c^2+m_e^2c^4}$. In case of ultra-relativistic particles, $\varepsilon\sim pc$ and the result is still a polytropic EoS with
\begin{equation}
    k=\frac{\left ( 3\pi^2 \right )^{1/3} c\hbar}{4}\left( N_{A}Y_e \right )^{4/3}~~~   
      \textrm{and}~~~  
      \gamma=4/3 \,\ .
\label{eq_pdeg_rel} 
\end{equation}
Note that the larger the density, the larger the Fermi energy [Eq.~\eqref{efermi}]. In the iron core of a massive star, where densities above $10^9$~g/cm$^3$ develop, $\varepsilon_F \gg m c^2$ and, hence, $\gamma\sim 4/3$. As we have seen in the previous Section, this represents a marginal stability condition: the collapse of the core
is just around the corner.

\subsubsection{The Chandrasekhar mass}
Let us consider a degenerate stellar core. As usual, we define the polytropic index as
\begin{equation}
n=\frac{1}{\gamma-1} \,\ ,
\end{equation}
so that $n=\frac{3}{2}$ and 3, for non-relativistic and relativistic electrons, respectively.
By combining Eq.~\eqref{eq_idro} and Eq.~\eqref{eq_polytropic}, one may easily get the Lane-Emden equation
\begin{equation}
\frac{1}{\xi ^{2}}\frac{d}{d\xi }\left ( \xi ^2\frac{d\theta }{d\xi } \right )=-\theta ^n
\label{eq_lane_emden} \,\ ,
\end{equation}
where the two dimensionless variables are
\begin{equation}
\rho=\rho_c\theta ^n \,\quad {\rm and} \quad r=\alpha \xi \,\ ,
\label{theta}
\end{equation}
with 
\begin{equation}
\alpha=\sqrt{\frac{n+1}{4\pi G}k\rho_{c}^{(1-n)/n}} \,\ ,
\end{equation}
(for a detailed derivation of the Lane-Emden equation, see Ref.~\cite{Kippenhahn:2012qhp}). Hence, the core mass $M$ can be written in terms of $\theta$ and $\xi$
\begin{equation}
M=\int_{0}^{R}4\pi r^2 \rho dr= \int_{0}^{\xi_S}4\pi \rho_c \theta^n \alpha^3 \xi^2 d\xi \,\ ,
\label{eq_mcore}
\end{equation}
where $\xi_S=\frac{R}{\alpha}$ is the value of $\xi$ at the core surface. Then, by means of the Lane-Emden equation, we obtain
\begin{equation}
M= -4\pi \alpha^3  \rho_c \int_{0}^{\xi_S}\frac{d}{d \xi}\left ( \xi^2\frac{d\theta}{d\xi} \right ) d\xi
= -4\pi \alpha^3  \rho_c  \xi_S^2\left ( \frac{d\theta}{d\xi} \right )_{\xi_S} \,\ ,
\label{emmep}
\end{equation}
being $\xi=0$ at the center.

The Lane-Emden equation can be easily solved for $n=0$ and $1$. In these cases, the term $\xi_S^2\left ( \frac{d\theta}{d\xi} \right )_{\xi_S}$ in the previous equation is $2\sqrt{6}$ and $\pi$, respectively. For $n=1.5$ and 3 numerical solutions provide approximate values, which are $\sim 2.7$ and $\sim 2.4$, respectively. Then, in case of non-relativistic electrons ($n=1.5$), we make use of $R=\alpha \xi_S$, as derived from the definition in Eq.~\eqref{theta}, to eliminate $\rho_c$, so that Eq.~\eqref{emmep} becomes
\begin{equation}
M=22.4~{\rm M}_\odot \frac{Y_e^5}{R_9^3} \,\ ,
\label{eq_MR}
\end{equation}
where $R_9=\frac{R}{10^9~{\rm cm}}$ and $M_{\odot} = 1.989\times 10^{33}~{\rm g}$. For instance, taking $Y_e=0.5$, as for a C-O composition, and $R=1.05\times 10^9$~cm, the mass of the degenerate structure is $M=0.6$~M$_\odot$. These values are representative of the most common C-O WD, which are the final product of the evolution of low- an intermediate-mass stars. 

According to Eq.~\eqref{eq_MR}, if the mass of a degenerate core increases, its radius becomes smaller. However, the larger the core mass the higher the central density and, in turn, the larger the Fermi energy. In practice, if the core mass is large enough, the electrons becomes relativistic, so that $\gamma$ decreases from 5/3 to 4/3 (or $n$ increases from 1.5 to 3), while the $k$ constant in the polytropic EoS must be replaced with that in Eq.~\eqref{eq_pdeg_rel}. As a result, the product $\rho_c \alpha^3$ in Eq.~\eqref{emmep} becomes a constant and equal to $\left ( \frac{k}{\pi c} \right )^{3/2}$, so that we get a unique solution for the mass
\begin{equation}
M_{\rm Ch} = 5.81~{\rm M}_{\odot} Y_e^2 \,\ .
\label{chandra}
\end{equation}
In other words, for degenerate and relativistic electrons the mass-radius relation vanishes. Practically, it exists a maximum mass for a stable degenerate core, namely, the one given by Eq.~\eqref{chandra}. This is the so-called Chandrasekhar mass limit. When the stellar core approaches this limit, electrons become relativistic, so that $\gamma\sim4/3$, just at the stability border. When a star, in the course of the evolution, develops a degenerate core with mass similar to or exceeding this limit, as it happens at the end of the Si burning in massive stars, a collapse will occur, eventually  triggered by the onset of electron captures. On the contrary, if the mass of the degenerate core is smaller than this limit, a cool and compact WD forms at the end of the evolution, as in the case of low- and intermediate-mass stars.

Note that the Chandrasekhar mass depends on the chemical composition. Indeed, $Y_e$ scales as the mean $Z/A$ ratio [Eq.~\eqref{eq_ye}]. In practice, $Y_e$ is smaller if the gas contains neutron-rich isotopes and, in turn, the corresponding Chandrasekhar mass is smaller. For example, for a C-O core ($Y_e\sim 0.5$) the Chandrasekhar mass is $M_{\rm Ch}=1.44$~M$_\odot$, while for an iron core ($Y_e\sim 0.46$) it will be $M_{\rm Ch}=1.25$~M$_\odot$. The most extreme case would be a degenerate pure-H core, for which $Y_e=1$ and the corresponding Chandrasekhar mass is $5.81$~M$_\odot$. 

\begin{figure}
\centering     
\includegraphics[width=1\columnwidth]{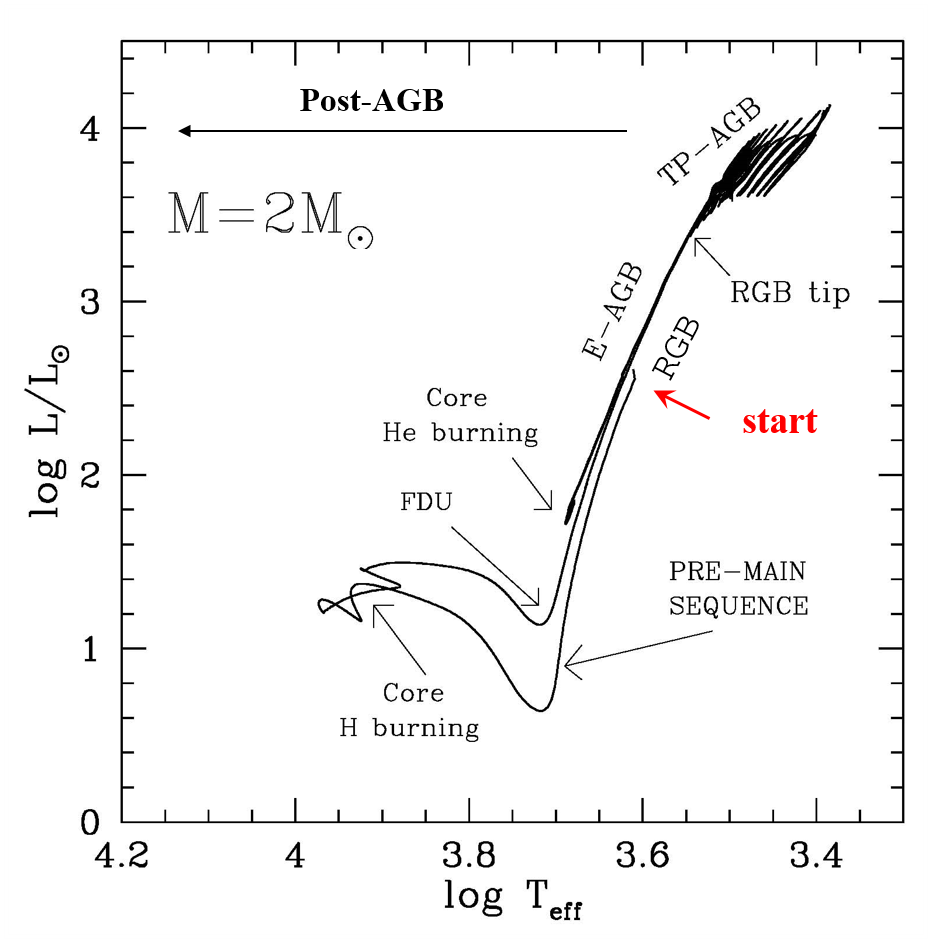}
\caption{The evolutionary track of a $2$~M$_\odot$ model (solar composition) in the Hertzsprung–Russell diagram.  The starting point of the track corresponds to the first homogeneous model in hydrostatic equilibrium. The following evolutionary phases are marked with labels. The calculation has been done by means of the FuNS (Full Network Stellar evolution) code (see Ref.~\cite{Straniero:2019dtm}, for more details).}
\label{fig_hr2}  
\end{figure}

\subsubsection{H and He burning}\label{sec:stelev_H_He}
In Fig.~\ref{fig_hr2}, the evolutionary track in the luminosity/effective temperature plane of a 2~M$_\odot$ model (solar composition) is shown. During the pre-MS, according to the virial theorem, the core contracts until the central temperature is large enough to activate fusions of the lighter isotopes. The Zero-Age Main Sequence (ZAMS) is attained when all the secondary isotopes of the pp chain and the CNO cycle reach a balance between production and consumption. This event marks the beginning of the core-H burning. Low-mass stars, like the Sun, burn H mainly through the pp-chain, while stars with mass of at least 1.3~M$_\odot$ develop larger central temperatures ($>20$~MK) and the H burning proceeds mainly through the CNO cycle. In the first case (pp-chain), the radiative gradient in the core never exceeds the adiabatic one, so that the energy is transported away by radiation (see Sec.~\ref{sec_transport} and \ref{sec_mixing}). On the contrary, the external layers are cool enough that partial recombination of various chemical constituents takes place. Owing to bound-bound and bound-free electron transition in atoms and molecules, the radiative opacity is quite large and convection develops in these external layers. An opposite stratification occurs in the more massive stars, those in which the CNO cycle provides most of the nuclear energy during the core-H burning. In this case, the temperature profile is steeper near the center, so that a convective core develops, while in their envelope, which is hotter than that of stars with lower mass, a radiative energy transport sets on.

Upon the central H exhaustion, stars leave the MS and move toward the RGB. In this phase the He-rich (or H-exhausted) core contracts and its temperature and its density progressively increase. Meanwhile, an efficient shell-H burning sets at the base of the H-rich envelope and the outer layers, inflated by this nuclear fire, expand and cool down, so that the external convection penetrates inward. As a consequence of this deep mixing process, the surface composition is enriched with the yields on the internal nucleosynthesis (first dredge up).

\begin{figure}[t!]
\centering     
\includegraphics[width=0.5\textwidth]{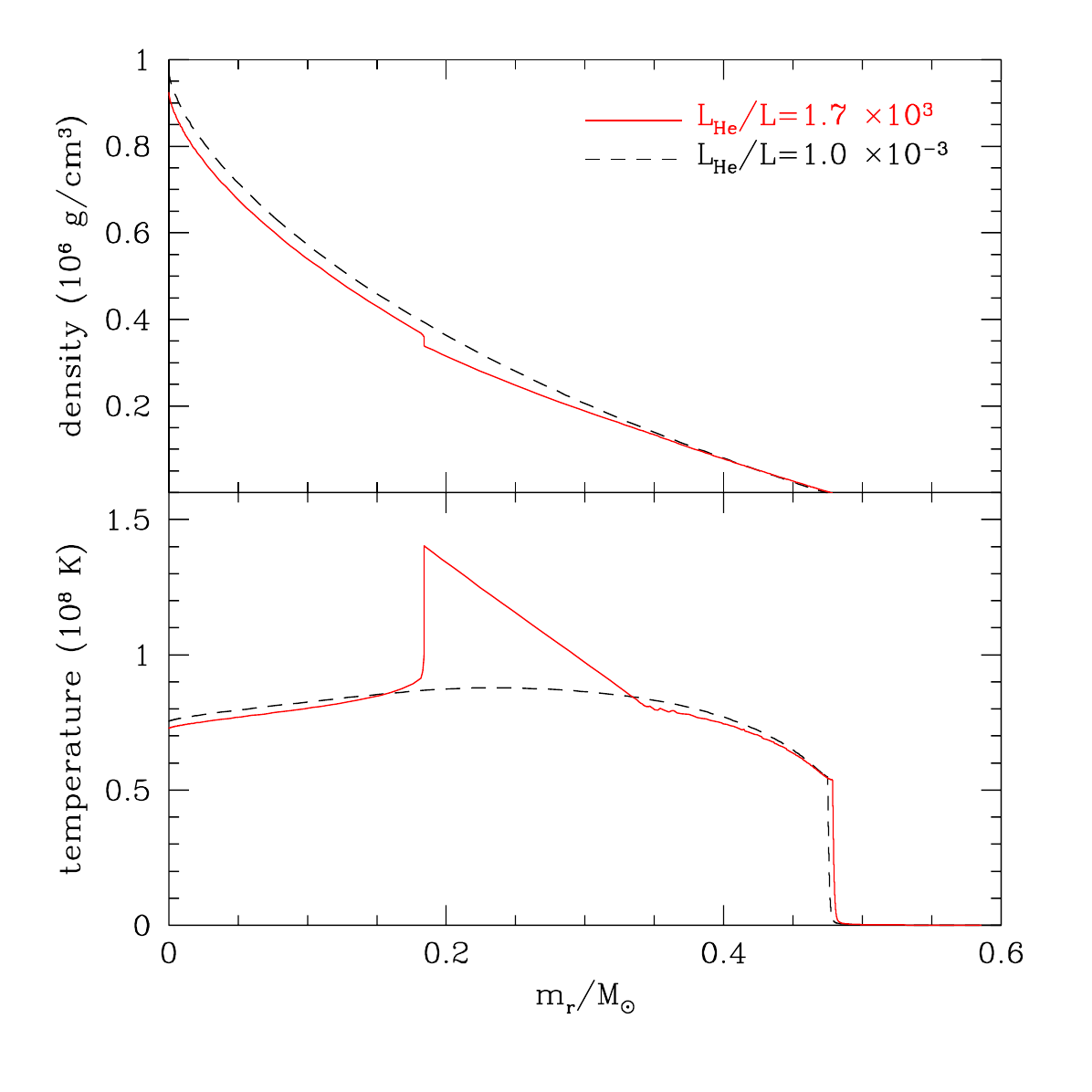}
\caption{Upper panel: core density profile within the core of a RGB star (initial mass $M=0.82~{\rm M}_\odot$, $Z=2.3\times 10^{-3}$, $Y=0.25$). The black-dashed line refers to a model just before the He ignition, when the nuclear power generated by the triple-$\alpha$ reactions are yet a small fraction of the stellar luminosity $L$, while the red-solid line refers to a model at the first and more powerful He-flash, when the luminosity of the He burning is more than $10^3 L$. Lower panel: corresponding temperature profiles. Note the sharp rise of $T$ in the layer where most of the nuclear energy is released.}
\label{fig_RGB_1}  

\centering     
\includegraphics[width=0.5\textwidth]{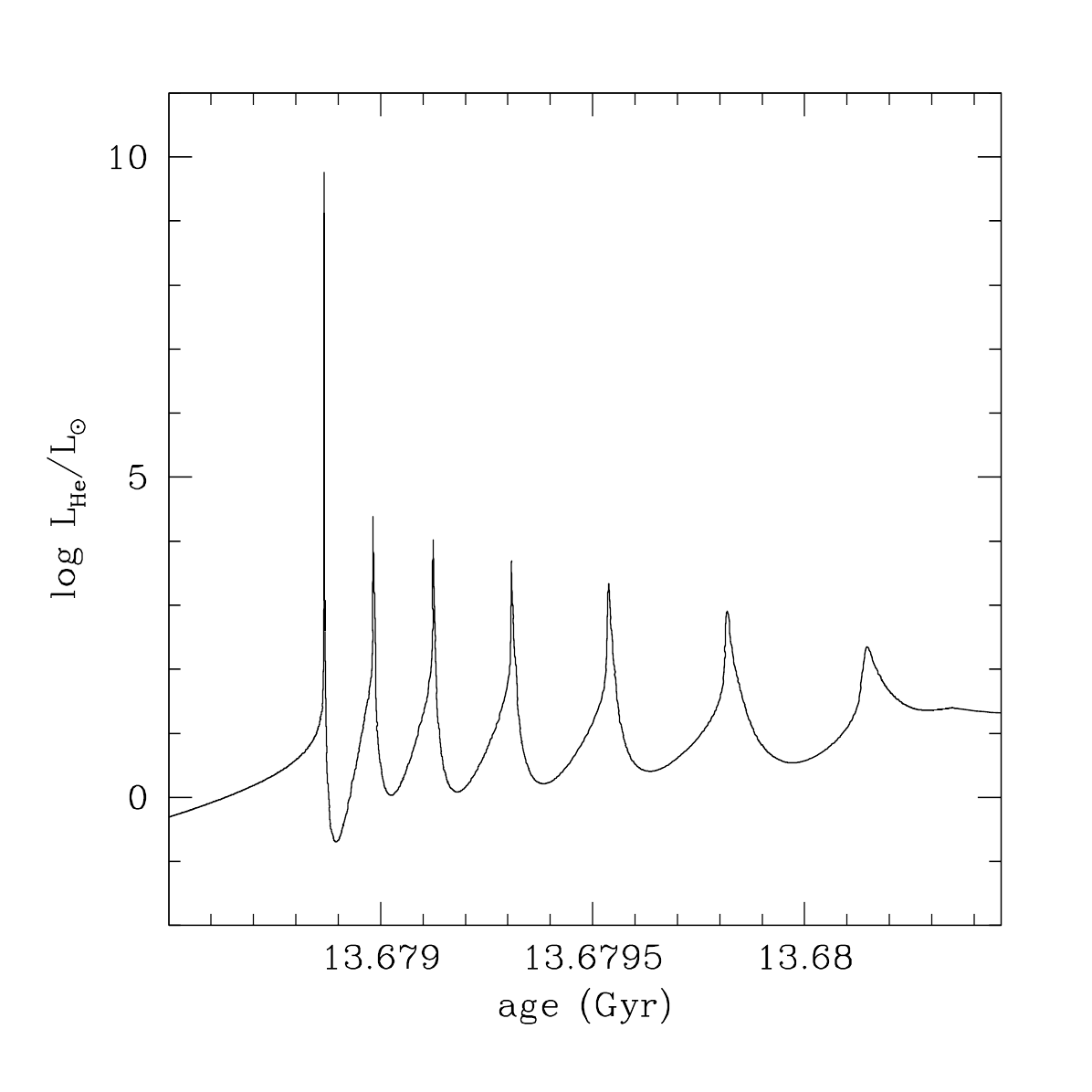}
\caption{The series of thermonuclear runaways (He flashes) starting when the star reaches the tip of the RGB. The model input parameters are: mass $M=0.82~{\rm M}_\odot$, $Z=2.3\times 10^{-3}$, $Y=0.25$. The first and more-powerful flash occurs where the temperature profile attains a maximum value (see Fig.~\ref{fig_RGB_1}). It is followed by 6 less-powerful flashes, which take place closer to the center. Each flash generates a convective shell. Eventually, the He burning arrives at the center, becomes quiescent and an extended convective core develops. This occurrence coincides with the so-called ZAHB.
}
\label{fig_RGB_flash}  
\end{figure}

In low-mass stars ($M < 2~{\rm M}_\odot$) electron degeneracy develops in the He-rich core. This occurrence has two major effects. Firstly the electron conductivity increases, contributing to an efficient energy redistribution within the core. As a result, an almost isothermal temperature profile develops within the core. On top of that, a plasma neutrino emission is more effective in the innermost layers, where the density is higher. As a consequence of this neutrino cooling, the maximum temperature moves out of the center (see Fig.~\ref{fig_RGB_1}). Then, an off-center He ignition occurs, which causes a thermonuclear runaway known as {\it He-flash}. In practice, the release of nuclear energy by the triple-$\alpha$ reactions causes an increase of the temperature not counterbalanced by an increase of the pressure. This is a consequence of the degenerate equation of state, for which the electron pressure mainly depends on their density, while it weakly depends on $T$. As shown in Fig.~\ref{fig_RGB_flash}, up to $10^{10}$~$L_\odot$ of nuclear power are attained by the first and most powerful flash, which is followed by a series of weaker thermonuclear runaways, until the electron degeneracy is removed in the whole core. This occurrence marks the beginning of the quiescent He burning and it is called the Zero Age Horizontal Branch (ZAHB). 

\begin{figure}[t!]
\centering
\includegraphics[width=0.6\textwidth]{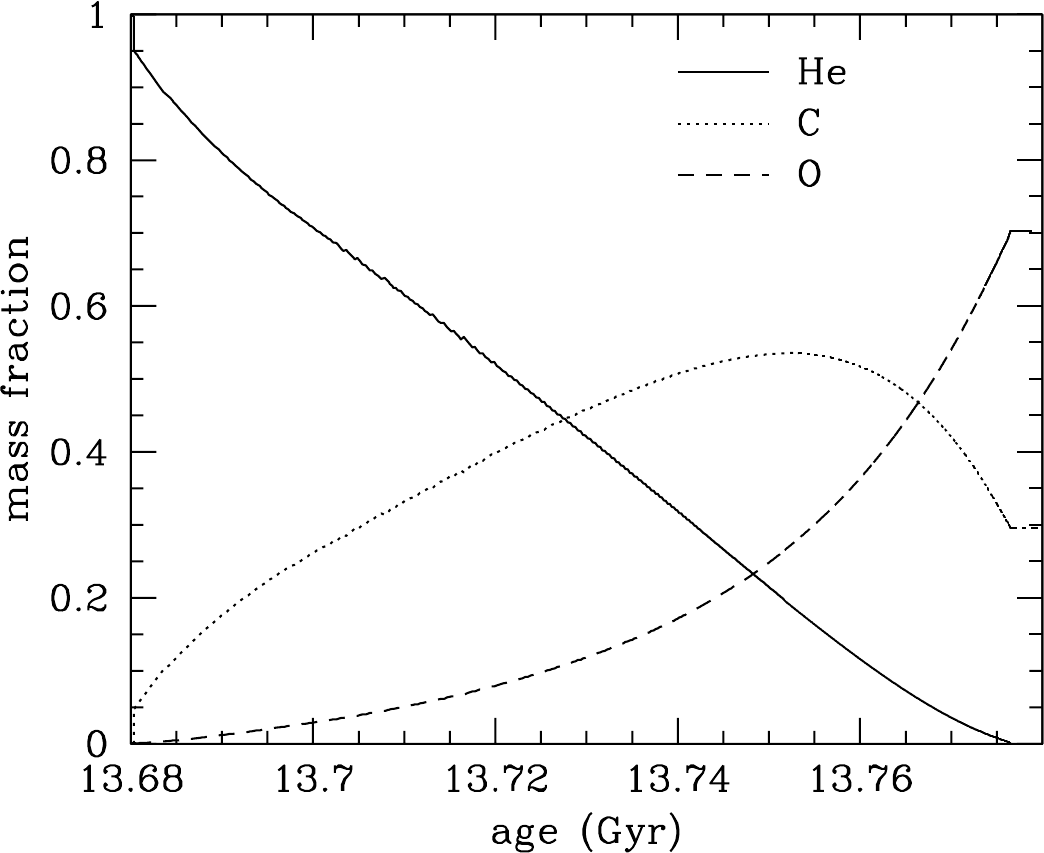}
\caption{Evolution of the central abundances of He, C and O, during the core-He burning phase.
}
\label{fig:gc_heco}

\centering     
\includegraphics[width=0.6\textwidth]{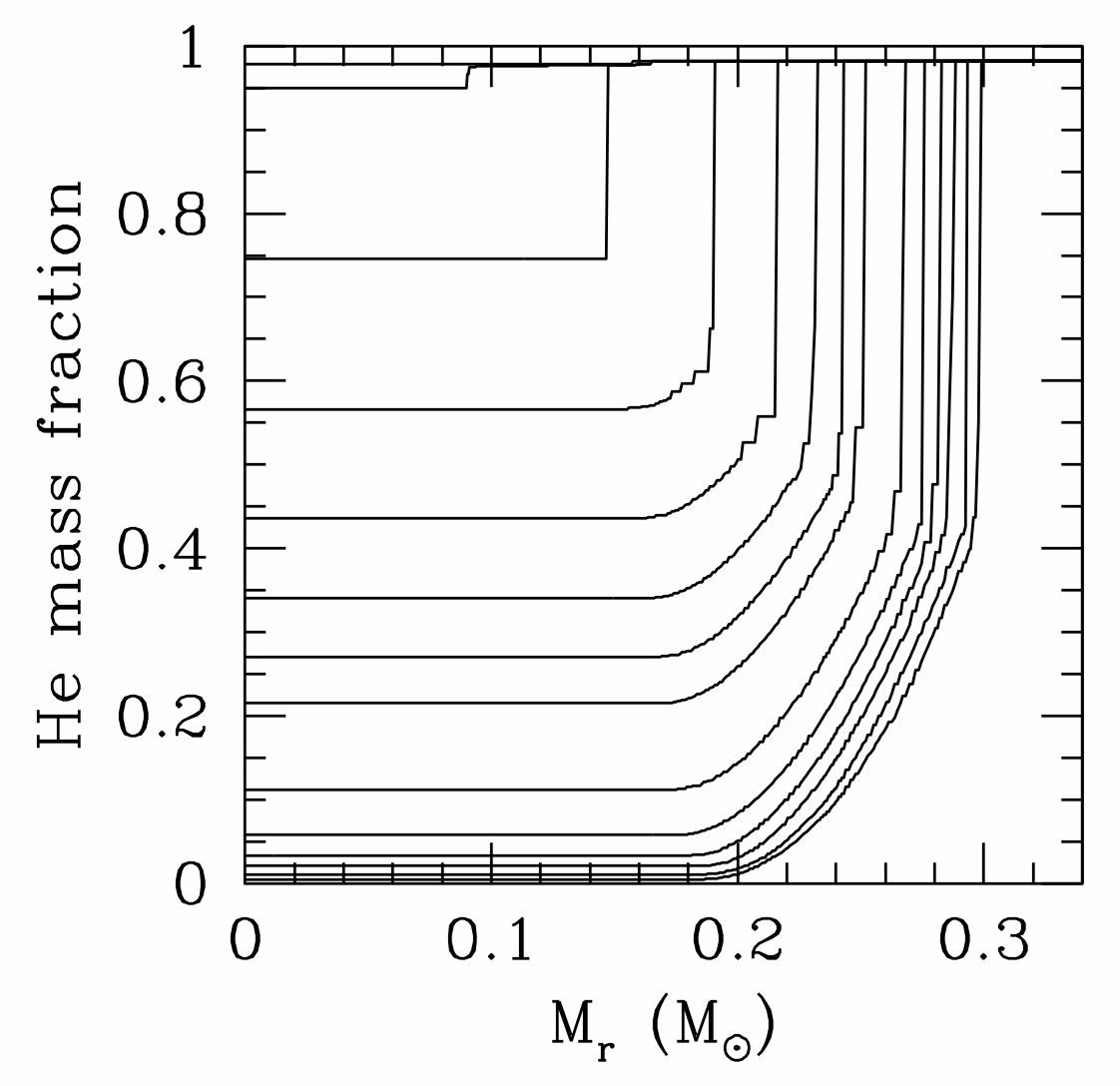}
\caption{Evolution during the He-burning phase of the He mass fraction profile within the core of a 2~${\rm M}_\odot$ model. Each curve represents a different time step. The sequence starts with a fully homogeneous core with $X_{\rm He}\sim 0.98$. Then, a convective core develops that homogenizes a central He depleted regions. As He is converted into C and O, the mass of the convective core increases. Later on, a semiconvective region forms beyond the convective core. Three zones can be clearly distinguished: 
    {\it i)} a more internal homogeneous zone, which is the region modeled by the fully convective core; 
    {\it ii)} an intermediate zone, where the He mass fraction increases linearly outward, as modeled by the semiconvection;
    {\it iii)} a more external zone, where He abundance has not changed.
}
\label{fig:semico}  
\end{figure}

In stars with $M > 2.5~{\rm M}_\odot$, the He ignition occurs in a non-degenerate core. In all cases, an extended convective core develops, as powered by the central nuclear burning. Two are the leading nuclear process, namely: the  $^{4}$He$(2\,^{4}$He$,\gamma)^{12}$C and the  $^{12}$C$(^{4}$He$,\gamma)^{16}$O. Fig.~\ref{fig:gc_heco} shows the evolution of the He, C and O central abundances for a star of $M=5$~M$_\odot$. After a first phase during which the main nuclear fusion yield is carbon, in the late part of the core-He burning the oxygen production, through $^{12}$C$(^{4}$He$,\gamma)^{16}$O, becomes the fastest. This occurrence has an important impact on the duration of the He-burning phase. Indeed, although the two leading nuclear reactions release a similar amount of energy, the $^{12}$C$(^{4}$He$,\gamma)^{16}$O consumes just one $\alpha$ particle, compared to the three consumed by the $^{4}$He$(2\,^{4}$He$,\gamma)^{12}$C.

Also the extension and the mixing efficiency in the zones that are unstable against convection affect the He-burning lifetime. As He is converted into C and O, the radiative opacity increases in the whole convective core. As a result, the radiative temperature gradient ($\nabla_{\rm rad}$), which linearly depends on the opacity, increases and a discontinuity takes place at the external border of the convective core that becomes unstable. It is like a ball on top of a hill: even a small perturbation moves the ball out of its unstable equilibrium. Similarly, even a minimal overshoot of C and O enriched material from the convective core into the radiative layer above it, coupled to a symmetric backward flux of fresh He from outside, causes an increase of the $\nabla_{\rm rad}$ just outside the convective core, while it decreases inside. In practice, the convective instability is pushed forward and the mass of the convective core progressively increases, until the external convective border becomes stable against a further overshoot~\cite{1971Ap&SS..10..355C}. On top of that, a minimum in the radiative temperature gradient forms, so that the overshoot mechanism described above can proceed until the marginal stability condition, i.e., $\nabla_{\rm rad}=\nabla_{\rm ad}$, is attained at the $\nabla_{\rm rad}$ minimum. Then, the layer above the minimum remains detached from the underlying fully convective core. Nevertheless, since $\nabla_{\rm rad} > \nabla_{\rm ad}$ in this layer, a mixing of He-rich material from outside occurs, causing a decrease of $\nabla_{\rm rad}$, so that the convective instability is progressively inhibited. In practice, this mixing can continue until the marginal stability condition is established in a quite extended zone above the fully convective core. This phenomenon is called semiconvection. Summarizing, convection, overshoot and semiconvection determine the chemical profiles within the  core (see Fig.~\ref{fig:semico}): an homogeneous central region, where a full convection is active ($\nabla_{\rm rad} > \nabla_{\rm ad}$), is surrounded by a semi-convective shell ($\nabla_{\rm rad} = \nabla_{\rm ad}$), where only a partial mixing takes place. 

How to describe this complex interplay of convection and semiconvection taking place in the core of He burning stars is the matter of a longstanding debate. Different treatment schemes have been developed and adopted by different authors. Owing to the non linearity of the hydrodynamical equations describing turbulent convection, all the approaches adopted in the extant stellar model calculations are based on phenomenological descriptions of the dynamical instabilities occurring in stellar interiors. For a discussion about the different mixing schemes adopted in the calculations of He-burning stellar models see Ref.~\cite{Straniero:2002pf} and some examples are illustrated in Sec.~\ref{sec:Rparam}. Let us here stress that the adoption of different treatments of convection may imply non-negligible differences in the computed duration of the core-He burning phase and in the chemical profile left after this evolutionary phase. The latter have also important consequences for the more advanced phases of stellar evolution, namely, the Asymptotic Giant Branch (AGB) phase and the WD cooling sequence. We will be back to this problem in Sec.~\ref{sec:Rparam}. 

A last issue often encountered in the calculations of models of He-burning stars concerns some numerical instabilities occurring during the late part of this evolutionary phase, when the mass fraction of He within the convective core drops below  $\sim 0.1$ (see Ref.~\cite{1973ASSL...36..221S}). It occurs that any small fluctuation of the location of the external border of the convective core may substantially enhance the amount of fuel available in the center for the He-burning, causing a sudden increase of the nuclear energy flux that push the convective border more outside, so that more He is transported toward the center. After Ref.~\cite{1985ApJ...296..204C}, the occurrence of this phenomenon was called {\it breathing pulse}.

\subsubsection{The final destiny of massive stars}
\label{sec2}
\begin{figure}[t!]
\centering      
\includegraphics[width=0.5\columnwidth]{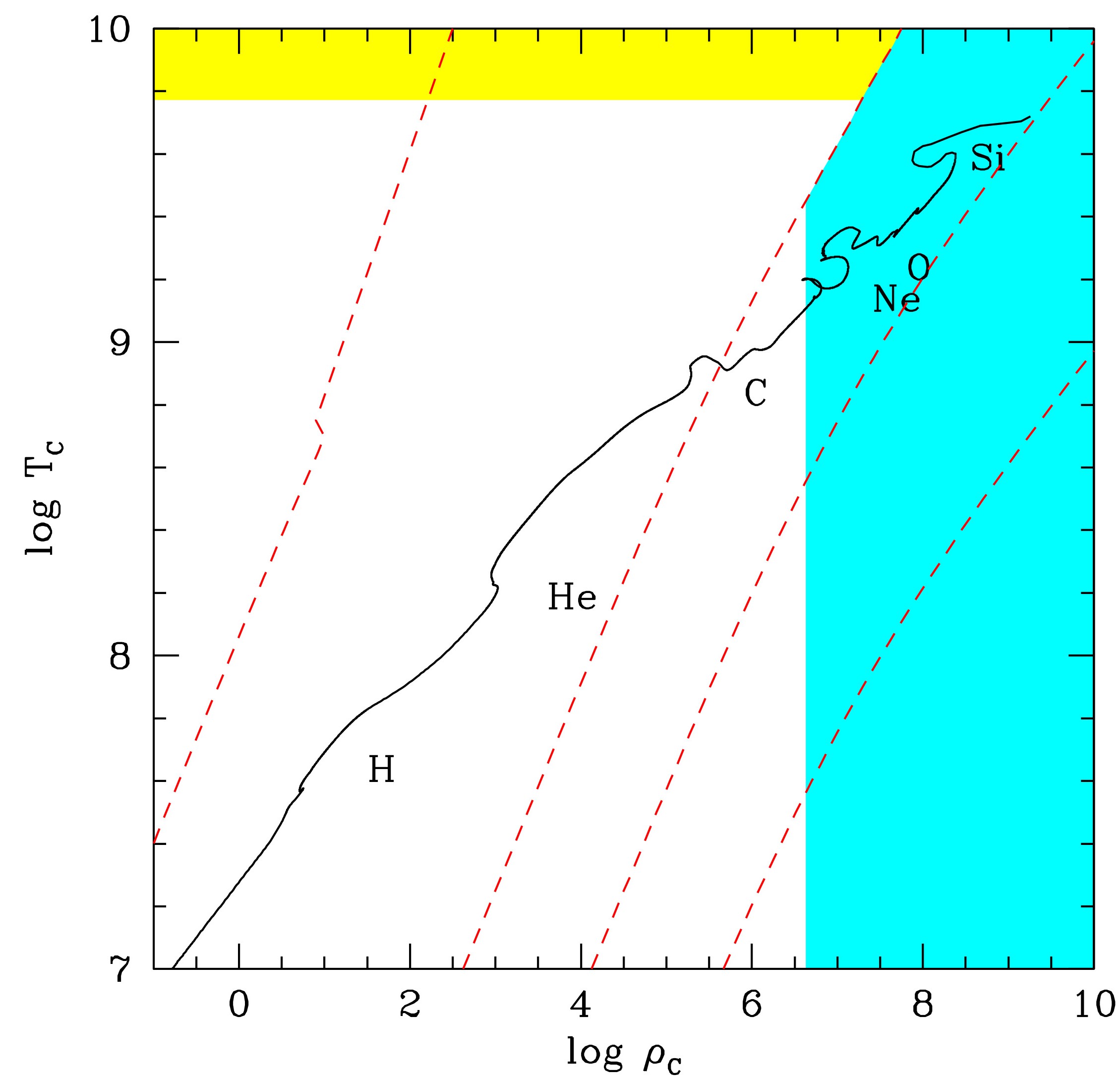}
\caption{Central temperature versus central density during 
the evolution of a massive star with solar composition and $M=20~{\rm M}_\odot$. Labels mark the epochs of the various burning phases. The point with the lowest $T_c-\rho_c$ corresponds to the first model in hydrostatic and thermal equilibrium, as it is established after the collapse of the original molecular cloud, while the last point of the evolutionary track, that with the highest $T_c-\rho_c$, corresponds to  the onset of the final core collapse (see Ref.~\cite{Straniero:2019dtm} for details about the model calculation). The red-dashed curves represent constant degeneracy parameter lines, namely: $\phi=\frac{\mu}{kT}=-10, 0, 10, 100$, left to right. Here we have assumed $Y_e=0.464$, which is appropriate for an iron-rich mixture. Note that electrons develop partial degeneracy after the C burning. The yellow area represents the region where electrons are relativistic but non-degenerate ($T>mc^2/kT$), while the cyan area shows the region occupied by relativistic and degenerate electrons. Extreme relativistic-degenerate conditions are attained during the late evolution (convective Si burning).  
}
\label{fig1}   
\end{figure}

The evolution of the central properties of a typical massive stars ($20$~M$_\odot$) is shown in Fig.~\ref{fig1}. Labels identify the various burning phases, namely: H, He, C, Ne, O and Si burning. Note how the central temperature and density increase in between two successive burning phases. As previously discussed, these are the periods of stellar evolution controlled by the virial theorem.

As a result of the Si burning, an iron core forms, whose density grows up to a few $10^9$~g/cm$^3$ and the temperature up to $\sim 5$~GK. The iron core extends up to $\sim 1.3~{\rm M}_\odot$. Within this core, the major contribution to the pressure comes from degenerate electrons. Because of the quite low electron fraction  of the iron-rich matter ($Y_e\sim0.45$), the core mass exceeds the Chandrasekhar limit ($M_{\rm Ch}\sim 1.25~{\rm M}_{\odot}$, see previous Section). Accordingly, the ratio $\varepsilon_F/mc^2 \sim 100$, so that the electrons are relativistic.

\begin{figure}[t!]
\centering     
\includegraphics[width=0.3\textwidth]{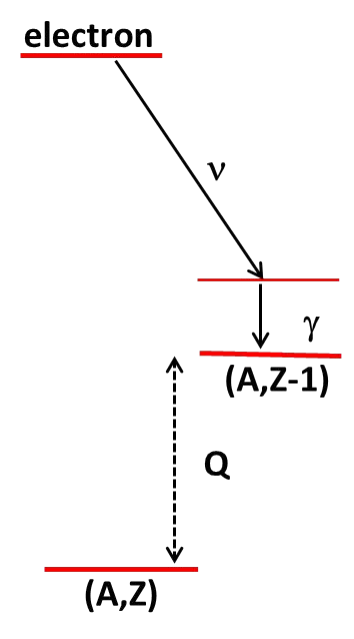}
\caption{Energy level scheme of an electron capture. When the electron is captured by a ($A$, $Z$) nucleus, 1 neutrino plus 1 ($A$, $Z$-1) nucleus in an excited state are released. Then a photon is emitted when the ($A$, $Z$-1) nucleus decays to its ground-state.
}
\label{fig_ecapture}  
\centering     
\resizebox{0.85\textwidth}{!}{\includegraphics{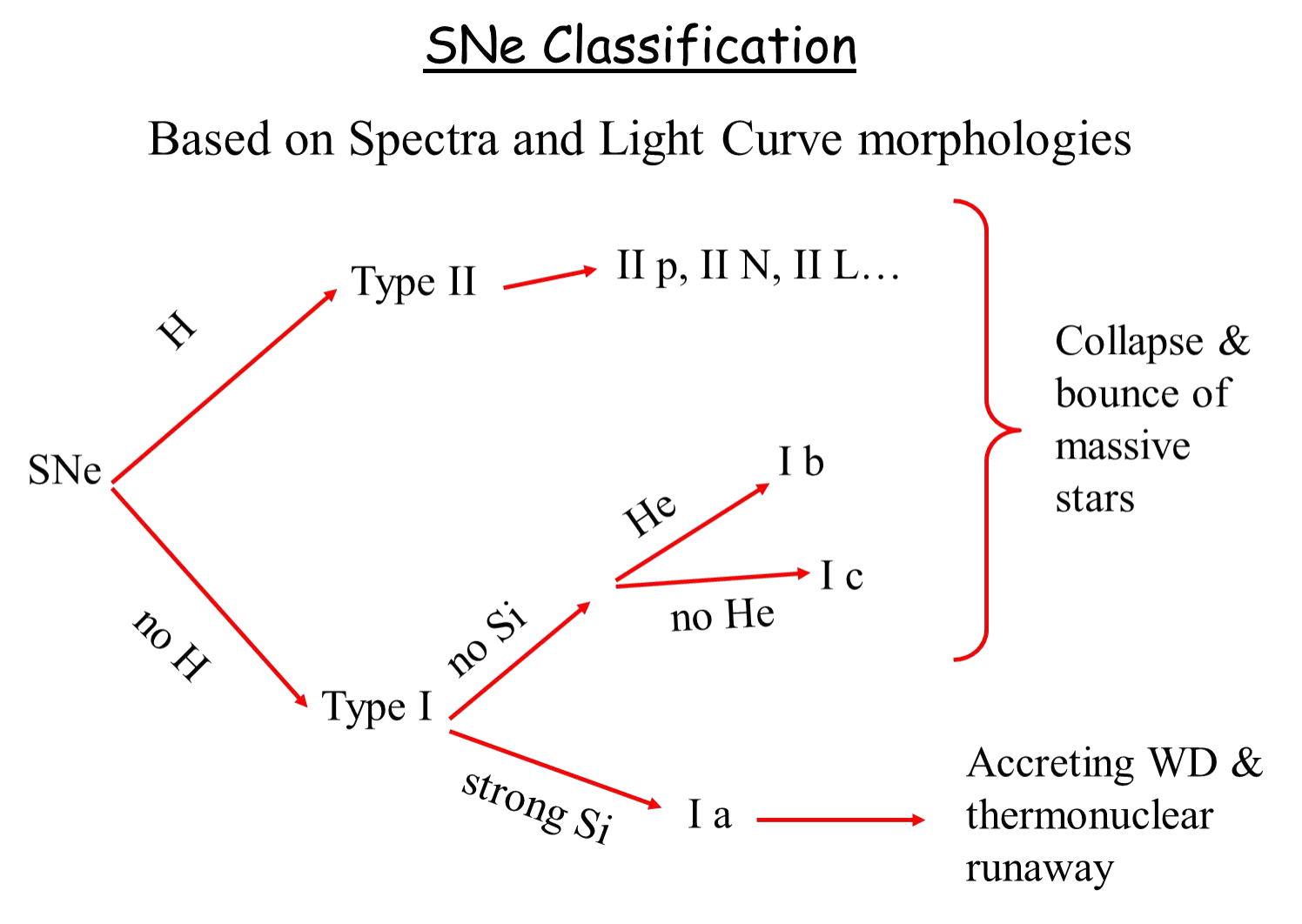}}
\caption{This scheme summarizes the main features of the most common SN types. The supposed SN engine, core collapse or thermonuclear runaway, is also reported.
}
\label{fig4}   
\end{figure}

As the star approaches the core collapse, inverse nuclear reactions (photo-dissociations) counterbalance direct nuclear reactions, a condition known as nuclear statistical equilibrium. The following evolution is driven by the occurrence of electron captures. Indeed, a large fraction of the degenerate electrons has energy large enough to compensate the negative threshold energy ($Q$) for ground-state to ground-state transitions between stable nuclei ($A$, $Z$) and unstable nuclei ($A$, $Z-1$). The electron capture energy diagram is schematically illustrated in Fig.~\ref{fig_ecapture}. 

For each electron capture, a neutrino is emitted and an excited ($A$, $Z$-1) nucleus is produced; then, the latter suddenly decays to its ground-state emitting a photon. In contrast, the $\beta$-decays of ($A$, $Z$-1) nuclei produced in this way are hampered, because the release of (low-energy) electrons, according to the Pauli exclusion principle, are suppressed. Therefore, in case of strong degeneracy, electron captures are favored with respect to $\beta$-decays. 

The consequences of the electron captures are dramatic: the pressure by degenerate electrons suddenly drops and the core, already at the stability borderline, collapses. Once the core density approaches the nuclear density ($\rho \sim 10^{14}$~g/cm$^3$), matter in the core is almost completely converted into degenerate neutrons, whose pressure is large enough to stop the collapse. Then, the endothermic photo-disintegration of iron causes a further energy-loss. As a result of the weak interactions, about $10^{53}$~erg of gravitational potential energy released during the collapse is almost totally converted into neutrinos. During this phase, the entropy is quite low, ${\mathcal O}(1)$~$k$ (Boltzmann constant) per nucleon, so that nuclei remain bound until the nuclear density is reached and the repulsive strong interaction makes matter very hard. Then, a stable neutron-rich core of about $0.8$~M$_\odot$ forms. Meanwhile, the still falling external layers bounce on the hard surface of the Proto-Neutron Star (PNS). The identification of the actual physical conditions for which this bounce gives rise to a SN explosion or, instead, it is followed by a partial or total fall-back, is a rather debated question.  Extant hydrodynamical models (see Refs.~\cite{Janka:2016fox,Burrows:2020qrp,Muller:2016izw} and references therein) show that a shock wave is initiated, which propagates outward. However, as the shock front moves outward, the temperature rises up, so that the in-falling material passing through the shock is efficiently photo-dissociated. As a consequence, the shock loses energy and stalls. Later on, the deposition of energy from  streaming out neutrinos may revitalize the shock. Thus, a neutrino wind from the inner PNS is needed to trigger a SN explosion.  

Among the various core-collapse SNe, the most frequent are of type IIP. Here II indicates the presence of H in the observed spectra, while the ``P'' stays for {\it plateau}. It refers to the rather long constant luminosity phase that follows the sudden maximum brightness of the SN. In the classical type IIP, this plateau may lasts for more than $100$ days. Typically, a few $10^{51}$~erg are converted into radiation and kinetic energy of the ejected debris. This energy is only a small fraction (about $1$~\%) of the total gravitational energy released during the collapse.
 
At variance with type II, type I (a, b or c) SNe do not show any evidence of H in their spectra. However, the Ib and Ic are still the product of the core collapse of a massive star. Their progenitors are stars more massive than the progenitors of the type II or stars whose H-rich envelope has been stripped by a companion in a close binary system.  In contrast, type Ia SNe are thermonuclear explosions of accreting or merging WD. Fig.~\ref{fig4} schematically illustrates the SN classification.

In case of a core collapse SN, the explosion leaves an extended nebula in rapid expansion surrounded by a compact central object, either a NS or a black hole. In particular, a black hole forms when the mass of the central object exceeds the Tolman-Oppenheimer-Volkoff limit. This mass limit is the equivalent for NS of the Chandrasekhar mass for WD: no stable NS can exist with mass larger than this limit. It may happen that this limit may be exceeded in case of an unsuccessful bounce, when the PNS is accreted with the matter that falls back after a failed explosion. The precise value of this mass limit is rather uncertain. It depends on the complex equation of state of nuclear matter that is controlled by nuclear forces. At variance with core-collapse SNe, the thermonuclear engine of SNe Ia does not leave a compact remnant. In practice, the whole progenitor is incinerated after the explosion.

\subsubsection{The final destiny of low-mass and intermediate-mass stars}
\begin{figure}[t!]
\centering      
\resizebox{0.7\textwidth}{!}{\includegraphics{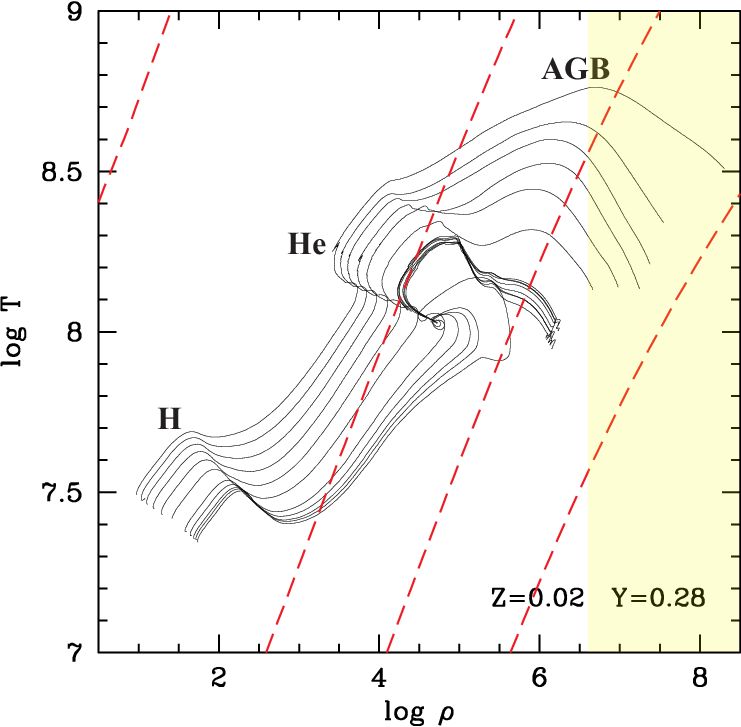}}
\caption{Evolution of the central temperature and  density for stars having masses between 1 and 9 ${\rm M}_\odot$ and nearly solar composition~\cite{Dominguez:1999qr}. The three principal evolutionary phases are labeled, namely: core-H burning (H), core-He burning (He) and the AGB. Note, in particular, the expansion occurring after the He ignition, as due to the release of nuclear energy, and the cooling of the core during the final part of the evolutionary tracks, when stars climb the AGB. This occurrence is a consequence of the production of plasma-neutrinos, which extracts energy from the stellar core. The red-dashed curves represent constant degeneracy parameter lines, namely: $\phi=\frac{\mu}{kT}=-10, 0, 10, 100$, left to right. Here $\mu$ is the electron chemical potential that, in the $T=0$ limit, coincides with the Fermi energy. 
These curves have been calculated assuming an electron fraction $Y_e=0.5$, which is appropriate for the He-rich core of RGB stars as well as for the C-O core of AGB stars. Low-mass stars ($M<2.5~{\rm M}_\odot$) develop a partial degeneracy of the electron ($\phi>0$) during the first ascent of the RGB. Later on, during the AGB phase, the electron degeneracy develops again, in all stars with mass lower than $\sim9~{\rm M}_\odot$. Eventually, the yellow area delimits the region where electrons are degenerate and relativistic. Note that the more massive AGB stars ($M>4~{\rm M}_\odot$) enter in this region when the plasma-neutrino cooling is more efficient.
}
\label{lowmass}   
\end{figure}

The evolution of the central properties of stars with initial mass 1--9~M$_\odot$ is shown in Fig.~\ref{lowmass}. Except for the more massive ($M\geq 8~{\rm M}_\odot$), these stars only experience two burning phases, namely, the H and the He burning. Indeed, stellar evolution calculations have early shown that only stars that at the end of the He-burning have a C-O core massive enough can ignite C. Based on current stellar models, the minimum core mass for the C burning is $\sim 1.1~ {\rm M}_\odot$\footnote{More precisely, after the core-He burning, intermediate-mass stars experience a (second) dredge up, during which the H-rich convective envelope penetrates the H-exhausted core, thus reducing the core mass. Therefore, the C-O core mass limit for the C ignition, $1.1$~M$_\odot$, refers to the core mass after the second dredge up.}~\cite{Dominguez:1999qr}. According to Ref.~\cite{1980ApJ...237..111B}, $M_{\rm up}$ is the minimum initial mass a star should have to develop such a C-O core and, in turn, to burn carbon. In addition, we will call $M_{\rm up}^{\ast}$ the minimum initial mass a star should have to undergo a non-degenerate C ignition. On the base of the current stellar evolution models, $M_{\rm up} \sim 7-8~{\rm M}_\odot$, while $M_{\rm up}^{\ast} \sim 10-11~ {\rm M}_\odot$\footnote{Note that these values depend on  the composition.} (see Refs.~\cite{2019nuco.conf....7S,Limongi:2023bcg} and references therein).

Stars less massive than $M_{\rm up}$  develop a degenerate C-O core whose mass is smaller than $\sim 1.1$~M$_\odot$. Since this core is also less massive than $M_{\rm Ch}$ (about $1.44$~M$_\odot$ for a C-O mixture), it is stable. Indeed, its density is much smaller and the electrons are non-relativistic (see Fig.~\ref{lowmass}). Nevertheless their pressure is high enough to stop the contraction and, in turn, any further release of gravitational energy. In addition, the core is efficiently cooled by the production of plasma neutrinos. As a result, the central temperature begins to decrease, before the conditions for the carbon ignition are reached (see Fig.~\ref{lowmass}). 

During this phase, the star climbs the so-called AGB. The compact C-O core is surrounded by a thin He-rich mantle and a more external and very expanded H-rich envelope. The stellar radius becomes very large, i.e., a few hundreds R$_\odot$. Two burning shells are alternatively actives. For most of the time the shell-H burning provides enough energy to replace the energy lost by radiation. Meanwhile the He-rich mantel is accreted with the ashes of the H-burning, until it reaches a critical mass. Then, recursive thermal instability powered by the periodic activation of the He-burning takes place (thermal pulses). The discovery of these thermal instabilities~\cite{1965ApJ...142..855S} had a great relevance in our understanding of the chemical evolution of the Universe. About half of the elements heavier than iron are produced by low-mass AGB stars through a slow-neutron-capture nucleosynthesis process or s process~\cite{1995ApJ...440L..85S}. Moreover, AGB stars are among the major sources of dust in a galaxy like the Milky Way. The AGB mass loss rate may be as large as $10^{-5}$~M$_\odot$/yr, and the wind from these stars may be as fast as $10^6$~cm/s. Owing to this huge mass loss, the H-rich envelope is rapidly eroded until a bare core remains. These stars end their life as C-O WD.     

\begin{figure}
\centering      
\resizebox{0.6\textwidth}{!}{\includegraphics{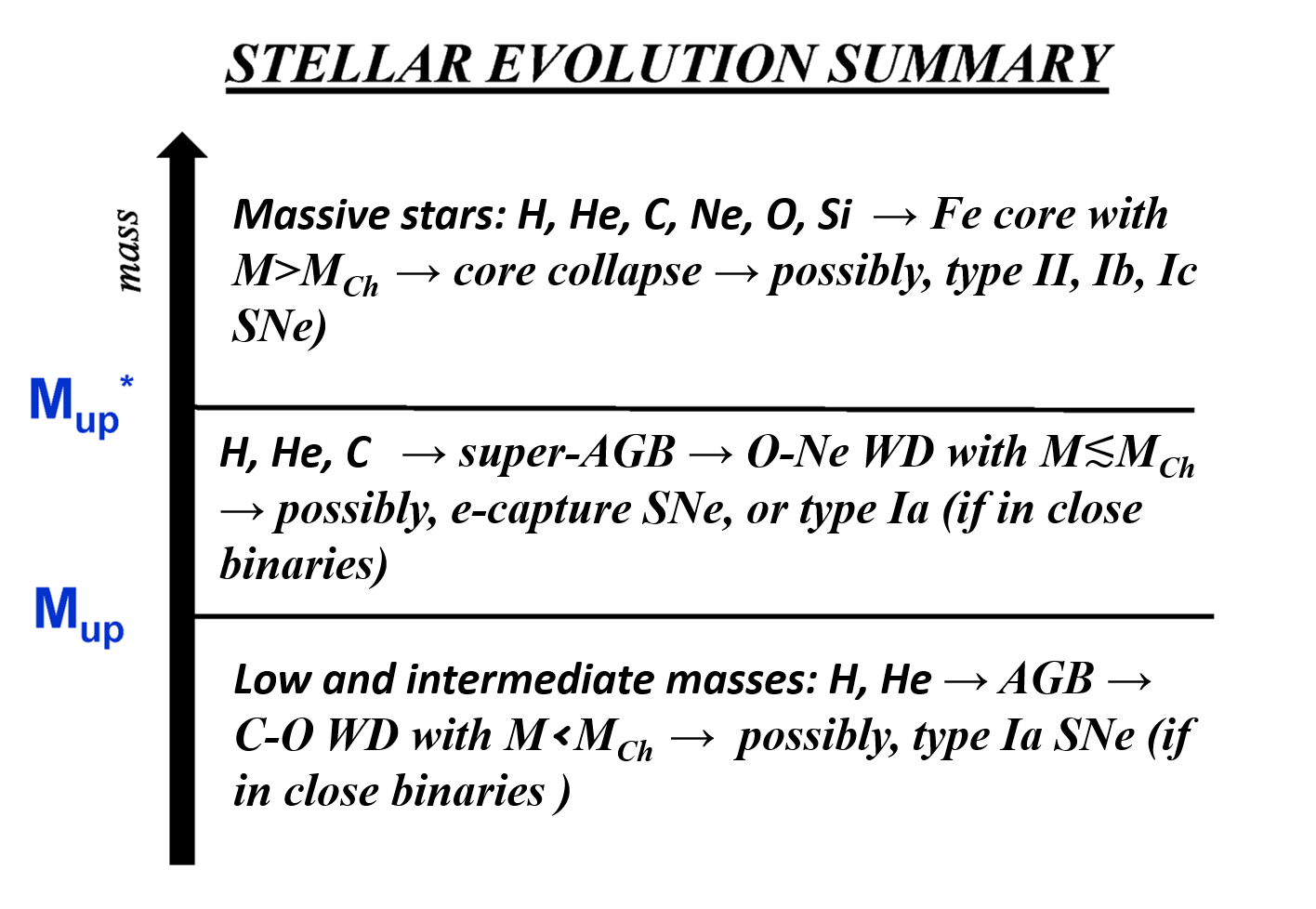}}
\caption{A schematic summary of stellar evolution and its final destiny, as a function of the initial stellar mass. The two critical masses $M_{\rm up}$ and  $M_{\rm up}^\ast$, represent the minimum mass for the C ignition and the minimum mass for the Ne ignition, respectively. 
 }
\label{stelev}   
\end{figure}

Stars with mass $M_{\rm up}<M<M_{\rm up}^{\ast}$, after the H-burning develop core masses in between $1.1$ and $1.3$~M$_\odot$ and may ignite carbon in degenerate conditions. The core mass is still smaller than the Chandrasekhar mass limit and, for this reason, it is stable. As a result of the C burning, these stars form a degenerate O-Ne core. Later on, they enter in the super-AGB phase. As for the less-massive AGB stars, a huge stellar wind erodes the envelope of super-AGB stars. Their final fate may be either the formation of a massive O-Ne WD or an electron capture SN. Indeed, a core collapse may occurs when during the super-AGB phase the core mass grows up to the Chandrasekhar limit. For a O-Ne mixture, this limit is $\sim 1.4~{\rm M}_\odot$. Note that the mass of the O-Ne core  increases because of the shell-C burning, which is active just outside the external border of the core. If this limit is approached, the core cannot maintains the hydrostatic equilibrium and undergoes a rapid contraction. As the density increases, the electron captures start triggering the collapse of the core. However, at variance with normal core-collapse SNe, in this case the collapse could induce a violent thermonuclear runaway, due to the oxygen ignition in the highly degenerate core.

Eventually, stars with initial mass $M>M_{\rm up}^{\ast}$ ignite C in a non-degenerate core later on, they will pass trough all the advanced burning stages (C, Ne, O and Si burnings) up to the formation of a degenerate iron core. As described in the previous Section, these stars may be the progenitors of core-collapse SNe. 
 
So far, we have described the evolution of single stars, assuming that it is independent of possible interactions with the environment. Very interesting deviations from this paradigm may occur when a WD (C-O or O-Ne), as left by the evolution of a low/intermediate mass star, belongs to a close binary system. In this case, the WD may accrete matter from the companion, through Roche-lobe overflow or coalescence. Accreting WDs in close binary systems give rise to very important phenomena, such as, cataclysmic variable stars or Novae. When a stationary accretion may be maintained for enough time, so that the WD mass may approach the Chandrasekhar limit, a C ignition takes place close to the center and a thermonuclear explosion follows. Likely, these objects are the progenitors of type Ia SNe.
 
The different evolutionary scenarios are schematically summarized in Fig.~\ref{stelev}. In all cases, the carbon burning plays a pivotal role. As we have seen, in particular, the range of masses of the various SN progenitors are determined by the occurrence (or not) of the C burning and by the physical conditions at the C ignition.

\pagebreak{}
\section{Axions and their production in stars}
\label{sec:axionproduction}
In this Section we present an  overview on axions and ALPs and on their production in astrophysical environments. We start in Sec.~\ref{sec:axionintro}  with a brief description of the theoretical framework in which these particles emerge, and describe their interactions with Standard Model particles. In Sec.~\ref{sec:Prodution_mechanisms} we describe the main axion production channels in stars, associated with their coupling with photons, electrons, nucleons and neutron electric dipole moment.

\subsection{Axions and axion-like particles}
\label{sec:axionintro}
Several extensions of the SM of particle physics predict a hidden sector of new particles interacting very weakly with SM particles. Often, such scenarios include very light particles, for example as Nambu-Goldstone bosons of global symmetries~\cite{Weinberg:1977ma,Wilczek:1977pj,Chikashige:1980ui,Gelmini:1980re,Georgi:1981pg,Wilczek:1982rv}, or as modes associated with compactified extra-dimensions~\cite{Svrcek:2006yi,Arvanitaki:2009fg,Cicoli:2012sz}. The study of these additional light modes goes under the name of the low-energy frontier of particle physics~\cite{Jaeckel:2010ni} and has become one of the most studied branches of modern particle physics. Indeed, a general interest in the FIPs has emerged in recent years, particularly since they offer explanations to several important cosmological and astrophysical problems (see Refs.~\cite{Antel:2023hkf,Agrawal:2021dbo} for recent reviews).

Axions, and more general ALPs, are among the best motivated and widely studied members of the FIP family. The QCD axion~\cite{Weinberg:1977ma,Wilczek:1977pj}, often known simply as axion, is a prediction of the Peccei and Quinn (PQ) solution of the strong CP problem~\cite{Peccei:1977hh,Peccei:1977ur}. At the origin of this problem, still one of the most puzzling in modern particle physics, is the CP-odd $\theta$-term in the QCD Lagrangian 
\begin{align}
	\label{eq:theta_term}
	L_{\theta} =\frac{g_{s}^{2}}{32\pi^2}
 \theta\,G^a_{\mu\nu}\tilde G^{a\mu\nu} \,\ ,
\end{align}
with $G_{\mu\nu}^a$ the gluon field and ${\tilde G}_{\mu\nu}^a=\frac12\epsilon_{\mu \nu \alpha \beta} G^{a\alpha \beta}$ is its dual, while the index $a$ refers to the color index of $SU(3)_c$ and $g_{s}$ is the strong coupling. The parameter $\theta$ has several phenomenological implications.  For example, it contributes to the nucleon mass difference $m_n-m_p \simeq\left(1.29+0.21 \theta^2+\mathcal{O}\left(\theta^4\right)\right) \mathrm{MeV}$, with consequences for the deuteron binding energy and the triple-alpha process in stars~\cite{Ubaldi:2008nf} and in the primordial nucleosynthesis of light nuclei~\cite{Lee:2020tmi}.

However, by far the most striking impact of the $\theta$ term is its contribution to the neutron electric dipole moment (nEDM)~\cite{Baluni:1978rf,Crewther:1979pi,Pospelov:1999mv}.\footnote{Intriguingly, initial measurements of the neutron electric dipole moment go back to the 1950s~\cite{Smith:1957ht}, long before the experimental confirmation of CP-violation. An excellent account of both theory and measurements can be found in Ref.~\cite{Khriplovich:1997ga}.} The most recent result (at the time of writing), published in February 2020 by the nEDM experiment at PSI, reported the upper bound $|d^{\rm exp}_n |< 1.8 \times 10^{-26}  {\rm e \, cm}$  at $90 \%$ C.L.~\cite{Abel:2020gbr}. This bound requires $\theta \lesssim 10^{-10}$, an extremely low value for a parameter which has no fundamental reason to be zero.\footnote{The contribution to $\theta$ from the weak sector is negligible, of order $ 10^{-17}$~\cite{Georgi:1986kr,Ellis:1978hq}.}

From a modern perspective, the PQ mechanism solves the strong CP problem by promoting $\theta$ from a constant parameter to a dynamical field, the axion $a$:
\begin{equation}
	\label{eq:gluon}
	\frac{g_s^2}{32\pi^2}\, \theta \,G^a_{\mu\nu}\tilde G^{a\mu\nu} 
	\to
	\frac12 (\partial_\mu a)^2 + \frac{g_s^2}{32\pi^2}
	\left( \theta+\frac{a}{f_a} \right) 
	\, G^a_{\mu\nu}\tilde G^{a\mu\nu} \,,
\end{equation}
where $f_a$ is a model dependent energy scale known as the PQ or axion constant. It is possible to show that the energy of vector-like gauge theories such as QCD is minimized at the CP conserving point~\cite{Vafa:1984xg}. In this sense, the PQ mechanism offers a \textit{dynamical solution} of the strong CP problem: since the axion is a dynamical degree of freedom, it will settle in the CP conserving minimum of the potential.

Notice that the introduction of the new energy constant $f_a$ is required by simple dimensional analysis, to compensate for the dimension of the new scalar field. In this way, the original $\theta$-term is transformed into a dimension 5 operator, implying that the theory needs a UV completion. The practical realization of the PQ mechanism proceeds through the introduction of a new abelian axial quasi-symmetry (the PQ symmetry), broken by the color anomaly and spontaneously broken at the PQ scale $f_a$. The axion is interpreted as the pseudo-Goldstone boson of this symmetry. Depending on the UV completion, there are several possible theories, with very specific features and often very different phenomenological predictions. The most studied models, often described as \emph{benchmark QCD axion models}, are the Kim-Shifman-Vainshtein-Zakharov (KSVZ)~\cite{Kim:1979if,Shifman:1979if} axions, representative of the so-called ``hadronic'' axion models since they do not contain an axion-electron coupling at tree level; and the Dine-Fischler-Srednicki-Zhitnitsky (DFSZ)~\cite{Zhitnitsky:1980tq,Dine:1981rt} axions, coupled also to electrons at tree level. The DFSZ axion model requires the introduction of an additional Higgs field (one Higgs gives mass to the up and the other to the down quarks), together with a singlet scalar field. The ratio of the vacuum expectation values of the two Higgs fields defines a fundamental free parameter of the DFSZ axion models: $\tan\beta= v_u/v_d$.\footnote{Perturbative unitarity demands that $\tan\beta \in [0.25, 170]$ (see Ref.~\cite{DiLuzio:2020wdo}).} There exist two DFSZ models, DFSZ~I and DFSZ~II, distinguished according to which Higgs gives mass to the leptons. The two models have slightly different couplings to photons and to electrons (Cf. Tab.~\ref{tab:QCD_axion_couplings}). Furthermore, there are several possible KSVZ models, which can be classified according to their coupling to photons (more exactly, the value of $E/N$ in Tab.~\ref{tab:QCD_axion_couplings},  being $E$ and $N$ the color and electromagnetic anomaly coefficients, respectively). Here, when referring to the KSVZ model, we assume the case with $E/N=0$, which is often considered the reference KSVZ model.  A comprehensive review of the QCD axion model landscape can be found in Ref.~\cite{DiLuzio:2020wdo}. 

\begin{table}[]
\centering
\caption{Values of the coupling constants in the benchmark QCD axion models, defined in Eqs.~\eqref{eq:Laint1} and \eqref{eq:Laint2}.
 }
\label{tab:QCD_axion_couplings}
\renewcommand{\arraystretch}{2.1}
\begin{tabular}{|llll|}
\hline
\multicolumn{2}{|c|}{\textbf{Coupling}} &
  \multicolumn{2}{c|}{\textbf{Expression}} \\ \hline
\multicolumn{1}{|l|}{$g_{a \gamma}=\frac{\alpha}{2 \pi} \frac{C_{a \gamma}}{f_a}$} &
  \multicolumn{1}{l|}{$C_{a\gamma}$} &
  \multicolumn{2}{l|}{$\frac{E}{N} - 1.92(4)$} \\ \hline
\multicolumn{1}{|l|}{\multirow{3}{*}{$g_{a f}=C_{a f} \frac{m_f}{f_a}$}} &
  \multicolumn{1}{l|}{$C_{a e}$} &
  \multicolumn{2}{l|}{$c_e^0+\frac{3 \alpha^2}{4 \pi^2}\left[\frac{E}{N} \log \left(\frac{f_a}{m_e}\right)-1.92(4) \log \left(\frac{\mathrm{GeV}}{m_e}\right)\right]$} \\ \cline{2-4} 
\multicolumn{1}{|l|}{} &
  \multicolumn{1}{l|}{$C_{ap}$} &
  \multicolumn{2}{l|}{$-0.47(3) + 0.88(3) \, c^0_u - 0.39(2) \, c^0_d - C_{a,\, \text{sea}}$} \\ \cline{2-4} 
\multicolumn{1}{|l|}{} &
  \multicolumn{1}{l|}{$C_{an}$} &
  \multicolumn{2}{l|}{$-0.02(3) + 0.88(3) \, c^0_d - 0.39(2) \, c^0_u - C_{a,\, \text{sea}}$} \\ \hline
\multicolumn{1}{|l|}{$g_d=\frac{C_{a n \gamma}}{m_n f_a}$} &
  \multicolumn{1}{l|}{$C_{a n \gamma}$} &
  \multicolumn{2}{l|}{$0.011(5) e$} \\ \hline
\multicolumn{1}{|l|}{} &
  \multicolumn{1}{l|}{$C_{a \pi}$} &
  \multicolumn{2}{l|}{$0.12(1)+\frac{1}{3}\left(c_d^0-c_u^0\right)$} \\ \hline
\multicolumn{4}{|c|}{\textbf{Model-dependent tree-level couplings}} \\ \hline
\multicolumn{2}{|c|}{\textbf{Parameter}} &
  \multicolumn{1}{c|}{\textbf{Value}} &
  \multicolumn{1}{c|}{\textbf{Model}} \\ \hline
\multicolumn{2}{|l|}{\multirow{3}{*}{$E/N$}} &
  \multicolumn{1}{l|}{$8/3$} &
  DFSZ-I \\ \cline{3-4} 
\multicolumn{2}{|l|}{} &
  \multicolumn{1}{l|}{$2/3$} &
  DFSZ-II \\ \cline{3-4} 
\multicolumn{2}{|l|}{} &
  \multicolumn{1}{l|}{$0$} &
  Reference KSVZ model \\ \hline
\multicolumn{2}{|l|}{\multirow{3}{*}{$c_{e_i}^0$}} &
  \multicolumn{1}{l|}{$\frac{1}{3} \sin ^2 \beta$} &
  DFSZ-I \\ \cline{3-4} 
\multicolumn{2}{|l|}{} &
  \multicolumn{1}{l|}{$-\frac{1}{3} \cos ^2 \beta$} &
  DFSZ-II \\ \cline{3-4} 
\multicolumn{2}{|l|}{} &
  \multicolumn{1}{l|}{$0$} &
  KSVZ \\ \hline
\multicolumn{2}{|l|}{\multirow{2}{*}{$c_{u_i}^0$}} &
  \multicolumn{1}{l|}{$\frac{1}{3} \cos ^2 \beta$} &
  DFSZ-I and DFSZ-II \\ \cline{3-4} 
\multicolumn{2}{|l|}{} &
  \multicolumn{1}{l|}{$0$} &
  KSVZ \\ \hline
\multicolumn{2}{|l|}{\multirow{2}{*}{$c_{d_i}^0$}} &
  \multicolumn{1}{l|}{$\frac{1}{3} \sin ^2 \beta$} &
  DFSZ-I and DFSZ-II \\ \cline{3-4} 
\multicolumn{2}{|l|}{} &
  \multicolumn{1}{l|}{$0$} &
  KSVZ \\ \hline
\multicolumn{2}{|l|}{$C_{a, \text { sea }}$} &
  \multicolumn{1}{l|}{$0.038(5) c_s^0+0.012(5) c_c^0+0.009(2) c_b^0+0.0035(4) c_t^0$} &
  KSVZ, DFSZ-I and DFSZ-II \\ \hline
\end{tabular}
\end{table}


Although different models can provide very different phenomenological results, in general, axions are highly constrained by the requirement to solve the strong CP problem (see Ref.~\cite{Giannotti:2022euq} for more details). Here we just summarize a few aspects~\footnote{The interested reader may find more information in Ref.~\cite{DiLuzio:2020wdo}.}:

\begin{enumerate}
	\item  the axion potential can be calculated in QCD~\cite{GrillidiCortona:2015jxo,Borsanyi:2016ksw}. In particular, the axion mass is fixed by QCD parameters once $f_a$ is assigned~\cite{GrillidiCortona:2015jxo}
	\begin{equation}
		\label{eq:ma_fa}
		m_a=\frac{(5.7\pm 0.07)\times 10^6}{f_a/{\rm GeV}}\,{\rm eV} \,.
	\end{equation}
This relation can be relaxed in specific axion models but doing so requires significant changes to the standard QCD, for example the addition of a new QCD sector~\cite{Rubakov:1997vp,Berezhiani:2000gh,Gianfagna:2004je,Giannotti:2005eb}. A more comprehensive discussion and extended bibliography on this aspect can be found in Sec. 6.7 of Ref.~\cite{DiLuzio:2020wdo}.
\item The axion interaction Lagrangian with the SM fields can be written as~\cite{DiLuzio:2020wdo}
\begin{equation} 
\label{eq:Laint1}
\mathcal{L}^{\rm int}_a \supset \frac{\alpha}{8 \pi} \frac{C_{a\gamma}}{f_a} a F_{\mu\nu} \tilde{F}^{\mu\nu}
+ C_{af} \frac{\partial_\mu a}{2 f_a} \bar f \gamma^\mu \gamma_5 f 
+ \frac{C_{a\pi}}{f_a f_\pi} 
\partial_\mu a [\partial \pi\pi\pi]^\mu  
- \frac{i}{2} \frac{C_{an\gamma}}{m_n} \frac{a}{f_a}  \bar n \sigma_{\mu\nu} \gamma_5 n F^{\mu\nu} 
\, ,
\end{equation}
where $F_{\mu\nu}$ is the electromagnetic field, $\widetilde F^{\mu\nu} = (1/2) \, \varepsilon^{\mu \nu \alpha\beta} F_{\alpha \beta}$ its dual, and  $[\partial \pi\pi\pi]^\mu \equiv 2 \partial^\mu \pi^0 \pi^+ \pi^- - \pi_0 \partial^\mu \pi^+ \pi^- - \pi_0 \pi^+ \partial^\mu \pi^-$. The last term represents the axion interaction with the electric-dipole moment (nEDM) operator. This is the most fundamental and model-independent coupling for QCD axions, derived from the axion-gluon interaction in Eq.~\eqref{eq:gluon}. This interaction is required for the solution of the strong CP problem. The value of the constants in Eq.~\eqref{eq:Laint1} are reported in Tab.~\ref{tab:QCD_axion_couplings}~\cite{diCortona:2015ldu,DiLuzio:2020wdo}. Here, $c^0_x$ are model-dependent coefficients.
\end{enumerate}

The axion-pion coupling in Eq.~\eqref{eq:Laint1} is relevant for axion thermalization in the early universe (see, e.g. Ref.~\cite{DiLuzio:2021vjd}), but plays no role in astrophysical context. Therefore, we will neglect it in the following discussion. Instead, the axion coupling to photons and matter field [first two terms in Eq.~\eqref{eq:Laint1}] is typically written  
as 
\begin{equation}
\label{eq:Laint2}
\mathcal{L}^{\rm int}_a \supset 
 \frac{1}{4} g_{a\gamma} a F_{\mu\nu} \tilde F^{\mu\nu} 
+ g_{af} \frac{\partial_\mu a}{2 m_f} \bar f \gamma^\mu \gamma_5 f 
- \frac{i}{2} g_d \, a \, \bar n \sigma_{\mu\nu} \gamma_5 n F^{\mu\nu} \,, 
\end{equation} 
with the relevant couplings defined in Tab.~\ref{tab:QCD_axion_couplings}. By virtue of Eq.~\eqref{eq:ma_fa}, the axion couplings are expected to be proportional to the axion mass. Thus, in this context, light is equivalent to weakly coupled.

As we shall see, the values of the various couplings in Tab.~\ref{tab:QCD_axion_couplings} are strongly constrained by astrophysical arguments, in particular from stellar evolution~\cite{Raffelt:1990yz,Giannotti:2015kwo,Giannotti:2017law,DiVecchia:2019ejf,DiLuzio:2021ysg}. Theoretical arguments also lead to the selection of privileged regions in the axion parameter space. Most notably, a few minimal requirements show that the axion coupling to photons $C_{a\gamma}$, is expected to lay in a relatively narrow band, $0.25\leq C_{a\gamma}\leq 12.7$, which defines the well-known QCD axion band~\cite{DiLuzio:2016sbl,DiLuzio:2017pfr,DiLuzio:2021ysg} (see also discussion in Sec.~3 of Ref.~\cite{Giannotti:2022euq}). It is quite possible to design axion models that live outside the axion window~\cite{DiLuzio:2021pxd,Sokolov:2021ydn}.	However, in general going below that window requires some tuning between independent contributions to the axion-photon coupling~\cite{Kaplan:1985dv} while increasing the coupling above the preferred window may require the addition of several other fields (see, e.g., Ref.~\cite{Darme:2020gyx}).

Light pseudoscalar particles, with properties very similar to QCD axions but with a dynamics unrelated to the strong CP problem, emerge in many extensions of the SM. For example, they are a common prediction of string theory~\cite{Svrcek:2006yi,Arvanitaki:2009fg,Cicoli:2012sz}. These ALPs share a lot of the characteristics of the QCD axion but are, in general, much less constrained. For example, they do not require specific relations between mass and couplings. Since from an astrophysical perspective, one can treat the axion mass and the couplings with ordinary matter as unrelated quantities, in the following the will loosely refer to axions without distinguishing between QCD axions and ALPs. The experimental search for axions has enjoyed an intense revival in the last few years [see Refs.~\cite{Irastorza:2018dyq,Sikivie:2020zpn,Irastorza:2021tdu} for recent reviews]. In particular, new experimental techniques promise the exploration of large, currently unconstrained portions of the axion parameter space. A groundbreaking discovery in the coming decade or so is not such a remote possibility.

\begin{figure}[t]
	\centering
	\includegraphics[width=0.6\linewidth]{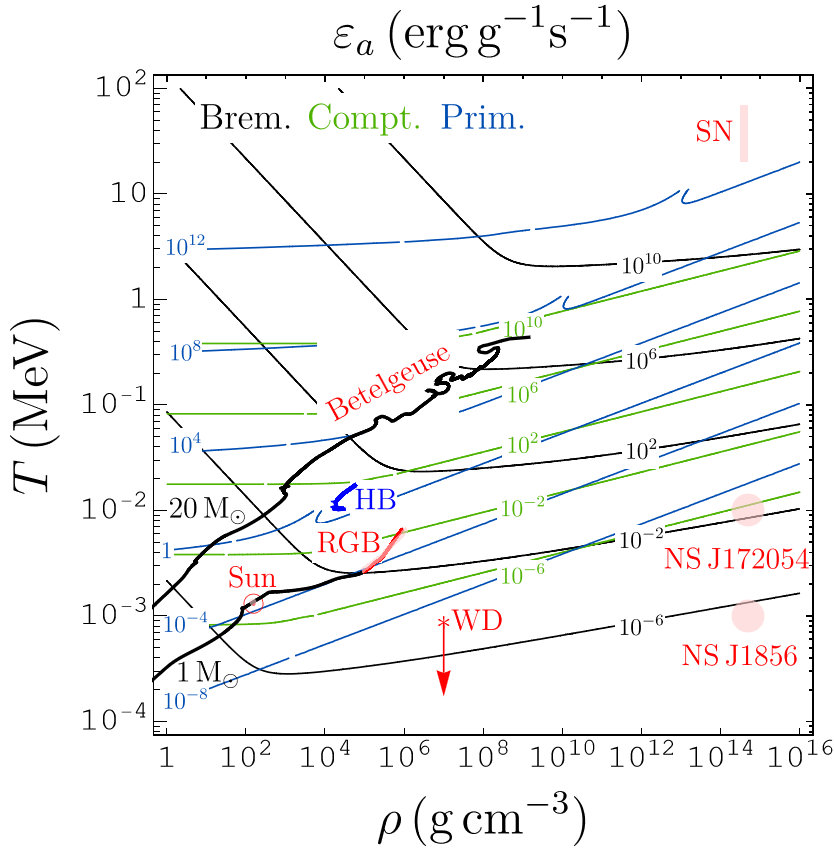}
	\caption{Contours of the axion energy-loss rates per unit mass, $\varepsilon_a$, in erg$\,$g$^{-1}$s$^{-1}$, for a pure He plasma for ordinary stars, while only nucleons and 
 electron fraction $Y_{e}=0.3$ for supernovae (SNe)  and neutron stars (NSs). Different lines correspond to different channels, as shown in the legend. 
		The Primakoff process (blue lines) is calculated for $ g_{a\gamma}=10^{-11} $GeV$ ^{-1} $.
		The Bremsstrahlung (black lines) and Compton (green lines) processes are calculated for $ g_{ae}=10^{-13} $.
The onset of the degeneracy region is visible in the bending of the bremsstrahlung contours. 
		The {shaded regions are indicative of the} central temperature and density of the Sun~\cite{Vinyoles:2016djt}, Red Giant Branch (RGB) stars, horizontal branch (HB) stars and  SNe are also shown.  
		In the case of HB and RGB, these are the results of a numerical simulation of a 0.8 $ M_{\odot} $ model as obtained with the FuNS code~\cite{Straniero:2019dtm}. With the same code also the Betelgeuse model is obtained~\cite{Xiao:2020pra}.
		Regarding degenerate stars, we show the properties of WD L19-2 and two NSs, representing the case of a young NS, J172054 with an age of $\sim 650$~yr, and an old one, J1856 with an age of $4\times10^{5}$~yr~\cite{Buschmann:2021juv}. 
		The thickness of the lines has no significance, except for SNe, where roughly represents the plausible range of density and temperature. 
	}
	\label{fig_bremsstrahlung_compton}
\end{figure}

\vspace{1cm}
\begin{figure}[h!]
	\centering
	\caption{Most relevant axion production mechanisms in stars. 
	}\includegraphics[width=0.87\linewidth]{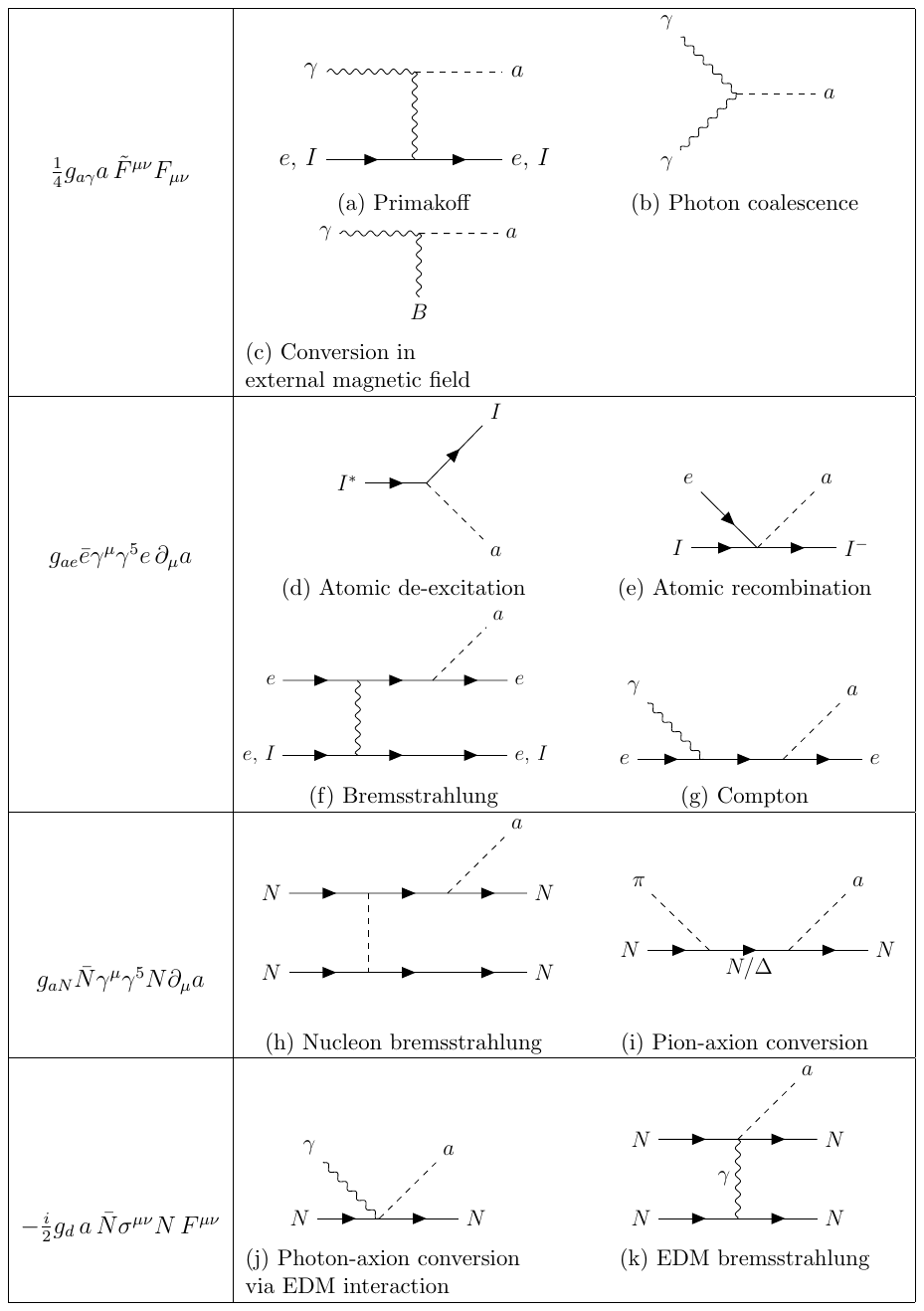}
	
	\label{fig:processes}
\end{figure}
\subsection{Axion production mechanisms in stars}
\label{sec:Prodution_mechanisms}

In this Section, we  present an overview of the principal axion production mechanisms in a stellar plasma, associated with the different couplings with SM particles, described by the Lagrangian of Eq.~\eqref{eq:Laint2}. More details about specific processes will be given in later sections. The typical temperature and density conditions of the stellar environments of interest for this work are represented in Fig.~\ref{fig_bremsstrahlung_compton}  for a number of representative stars including the Sun, HB and RG stars in GCs, and WD stars. The contours represent energy-loss $\varepsilon_a$ (in erg$\,$g$^{-1}$s$^{-1}$) assuming a pure He plasma, associated with the different processes we discuss below. In the case of SNe and NS, we consider a plasma composed of only nucleons and electrons, with an electron fraction $Y_{e}=0.3$.

\subsubsection{Axion-photon coupling}
\label{sec:photon}

\emph{Photon dispersion relation:}
Electromagnetic waves in an isotropic plasma admit three polarizations~\cite{Jackson:1998nia}, two transverse, with the electric field oscillating in the direction orthogonal to the direction of propagation, and one longitudinal. In general, the dispersion relation has the form
\begin{equation}
\omega^2 - k^2 = \pi_{T,L}(\omega, k)\; ,
\end{equation}
where $\omega$ and $k$ are respectively the photon frequency and wave-number and $ \pi_{T,L}(\omega, k)$ are the projections of the photon polarization tensor for the transverse (T) and longitudinal (L) modes, respectively. For the typical energies in stellar cores, the dispersion relation for Transverse Photons (TP) becomes~\cite{Jackson:1998nia}
\begin{equation}
\omega^2 - k^2 \approx \omega_{\rm pl}^2 \; ,
\end{equation}
where
\begin{equation}
\omega_{\rm pl}=\sqrt{ \frac{4 \pi \alpha n_e}{m_e}}
{=28.7\, \mathrm{eV} \left( \frac{Y_e\rho}{{\rm g}\,{\rm cm}^{-3}}\right)^{1 / 2}}  \,\ ,
\label{eq:plasma}
\end{equation}
is the plasma frequency, with $n_e$ the electron density.

The longitudinal mode, also known as Longitudinal Plasmon (LP),  has a  dispersion relation~\cite{Jackson:1998nia}
\begin{equation}
\label{eq:omega_pl_Complete}
\biggl( \frac{\omega}{\omega_{\rm pl}} \biggr)^2 =1+ \frac{3 P}{mn_e} \biggl( \frac{k}{\omega_{\rm pl}}\biggr)^2 \; ;
\end{equation}
where $P$ is the equilibrium pressure. 
In the typical conditions of the stellar plasma, the second term on the right hand side is negligible. Therefore, for most practical purposes the dispersion relation for LP reduces to 
\begin{equation}
\omega^2 \approx \omega^2_{\rm pl} \,\ .
\end{equation}

\emph{Primakoff process:}
At low masses, $m_{a} \lesssim T$, the most efficient production mechanism driven by the axion-photon coupling is the Primakoff process, which involves axion-photon conversions in static electric or magnetic fields.\footnote{This process was originally discussed by Henri Primakoff in 1951 in the case of the neutral pion $\pi^0$~\cite{Primakoff:1951iae}.} In the static limit, this process can be interpreted as the conversion of thermal photons in the electrostatic field of electrons and nuclei~\cite{Raffelt:1985nk} [see Fig.~\ref{fig:processes}-(a)]
\begin{align}
\gamma ~ Ze\to  Ze ~a \,.
\end{align}
In this approach, the collective behavior of the electron motion is reflected in the screening of the Coulomb potential.~\footnote{We remark that in a dynamical treatment, including electrons collective motion, the Primakoff emission can be interpreted as transverse to longitudinal plasmon decay $\gamma_T\to \gamma_L ~a$ or coalescence $\gamma_T$ $\gamma_L \to a $~\cite{Raffelt:1987np}.} Assuming static electric field from electrons and ions the Primakoff transition rate is given by~\cite{DiLella:2000dn,Cadamuro:2010cz,Carenza:2020zil,Lucente:2020whw,Hoof:2021mld}
\begin{equation}
\begin{aligned}
\Gamma_{\gamma\rightarrow a} &=g_{a\gamma}^2\dfrac{T\kappa_s^2}{32\pi} \dfrac{p}{E}\bigg\{\dfrac{\left[\left(k+p\right)^2+\kappa_s^2\right]\left[\left(k-p\right)^2+\kappa_s^2\right]}{4kp\kappa_s^2}\ln\left[\dfrac{(k+p)^2+\kappa_s^2}{(k-p)^2+\kappa_s^2}\right]-\\
&-\dfrac{\left(k^2-p^2\right)^2}{4kp\kappa_s^2}\ln\left[\dfrac{(k+p)^2}{(k-p)^2}\right]-1\bigg\}\,,
\end{aligned}
\label{generalrate}
\end{equation}
where $p=\sqrt{E^2-m_a^2}$ and $k=\sqrt{\omega^2-{\omega_{\rm pl}}^2}$ are the axion and photon momentum respectively, while  the plasma frequency $\omega_{\rm pl}$ plays the role of an ``effective photon mass''~\cite{Raffelt:1996wa}. Notice that only the transverse photon contributes to the production of axions in an external electric field.\footnote{This is evident from the fact that the relevant term in the interaction Lagrangian in this case is $\propto B_{\rm rad}\cdot E_{\rm ext}$, and the fact that the longitudinal mode does not have a magnetic field. }
We take $E=\omega$ since we are ignoring the small energy contribution from the electrons~\cite{Raffelt:1987np,Liang:2023jlz}. Finally, $\kappa_s$ is an appropriate screening scale which accounts for the finite range of the electric field of the charged particles in the stellar medium~\cite{Raffelt:1985nk}. In the case of  non-degenerate plasmas, the charge screening can be well approximated by the Debye screening scale, given by~\cite{Raffelt:1985nk}
\begin{equation}
	\kappa_s^2 = \frac{4\pi \alpha_{\rm em} }{T} n_B \left( Y_e + \sum_j Z_j^2Y_j\right) \, ,
	\label{eq:screening_nondeg}
\end{equation}
where $n_B$ is the baryon density,  $Y_{e}$ and $Y_{j}$ are the number fractions per baryon of the electrons and of the nuclear species $j$ respectively, and $ Z_j $ is its charge number. 

In order to evaluate the energy-loss by Primakoff production, one has to calculate the axion emissivity $Q_a$, which represents the energy emitted via axion production per unit volume and time 
\begin{equation}
Q_a=2\int\dfrac{d^3\textbf{k}}{(2\pi)^3}\Gamma_{\gamma\rightarrow a} \omega f(\omega)=\int_{m_a}^{\infty} dE E \dfrac{d^2n_a}{dtdE}\,,
\label{Qa}
\end{equation}
where the factor 2 comes from the photon polarization degrees of freedom and $f(\omega)=(e^{\omega/T}-1)^{-1}$ is the Bose-Einstein distribution function of the thermal photons. Analogously, one can define the energy-loss rate per unit mass as
\begin{equation}
\varepsilon_a = Q_a/\rho \,\ .
\end{equation}
If one neglects degeneracy effects and the plasma frequency (a good assumption in plasma conditions when the Primakoff process is the dominating axion production mechanism), one can provide a semi-analytical expression for the energy-loss rate per unit mass in axions~\cite{Friedland:2012hj,Choplin:2017auq}
\begin{align}
\label{eq:Primakoff_approx}
\varepsilon_P\simeq 2.8\times 10^{-31} Z(\xi^{2}) \left( \frac{g_{a\gamma}}{\rm GeV^{-1}} \right)^{2}
\frac{T^7}{\rho}\, {\rm erg\,g^{-1}\,s^{-1}}\,,
\end{align}
where $T$ and $\rho$ are in K and in g~cm$^{-3}$ respectively.\footnote{Note that Ref.~\cite{Friedland:2012hj} has a typo concerning the numerical value of the constant in Eq.~\eqref{eq:Primakoff_approx}. The correct value of the constant factor, reported above, can be found in Ref.~\cite{Choplin:2017auq}.} 
The coefficient $Z(\xi^{2})$ is a function of $\xi^{2}\equiv(\kappa_{s}/2T)^{2}$, with $\kappa_{s}$ the Debye-Huckel screening wavenumber. It can be explicitly expressed as an integral over the photon distribution (see Eq.~(4.79) in Ref.~\cite{Raffelt:1990yz}). Ref.~\cite{Friedland:2012hj} proposed the analytical parametrization
\begin{equation}
\label{eq:alternativefit}
{\textstyle Z(\xi^{2})\simeq 
\left(\frac{1.037\xi^{2}}{1.01+\xi^{2}/5.4}+\frac{1.037\xi^{2}}{44+0.628\xi^{2}}\right)
\ln\left(3.85+\frac{3.99}{\xi^{2}}\right)}\,,
\end{equation}
which is better than 2\% over the entire range of $\xi$. In general, $Z(\xi^{2})$ is ${\cal O}(1)$ for relevant stellar conditions. For example, in the core of the Sun, $\xi^{2}\sim12$ and $Z\sim6$ and in the core of a low-mass He burning star, $\xi^{2}\sim2.5$ and $Z\sim3$~\cite{Raffelt:1990yz}, while in a $10 ~{\rm M}_{\odot}$  He burning star, $\xi^{2}\sim0.1$ and $Z\sim0.4$~\cite{Friedland:2012hj}.

As shown in Fig.~\ref{fig_bremsstrahlung_compton}, the Primakoff process has a steep dependence on the stellar temperature, which controls the number of thermal photons, but is suppressed at high density because of the effects of a large plasma frequency and of the reduction of electron targets~\cite{Raffelt:1987yu} (in such conditions,Eq.~\eqref{eq:Primakoff_approx} ceases to be valid). Hence, this process is strongly suppressed in the degenerate core of WDs and RGB stars. Indeed, the strongest bounds on the axion-photon coupling are derived from the analysis of stars with a low density and high temperature in the core.

\begin{figure}[t!]
\centering
\vspace{0.cm}
\includegraphics[width=0.6\columnwidth]{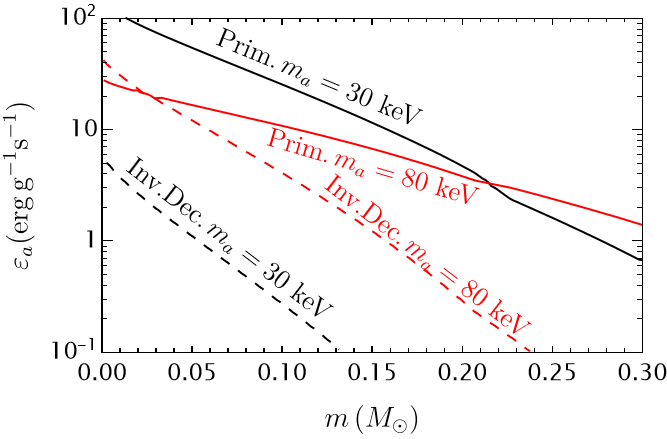}
\caption{Energy-loss rates (in units of erg$\,$g$ ^{-1} \,$s$ ^{-1} $ and normalized for $g_{a\gamma}=10^{-10}\,\ \textrm{GeV}^{-1}$)	as a function of the enclosed mass of the star for Primakoff  ($\gamma+Ze\to a+Ze$) and photon coalescence (or \textit{inverse decay}, $\gamma\gamma\to a$) within the core of a typical HB star. Results for two different axion mass, $m_a=30$~keV and $m_a=80$~keV, are shown.  (Data taken from Ref.~\cite{Lucente:2022wai}).
}
\label{fig:energyloss}
\end{figure}

\emph{Axion coalescence production:}
Another axion production mechanism in stellar plasmas is the axion coalescence (or inverse decay) process, $\gamma\gamma\to a$ [see Fig.~\ref{fig:processes}-(b)], which is particularly relevant at high mass, $m_a\gg T$. However, this process has a kinematical thresholds, and it vanishes for $m_a\leq 2\omega_{\rm pl}$~\cite{DiLella:2000dn}. Above this threshold the production rate is a steep function of the mass and eventually the coalescence dominates over the Primakoff production. The axion coalescence rate in a thermal medium can be easily estimated in the classical limit, replacing the Bose-Einstein photon distribution with a Maxwell-Boltzmann, $f(E)\to e^{-E/T}$, for the photon occupation number~\cite{DiLella:2000dn}. This approximation is justified since we are interested only in axion masses (and thus axion energies) of the order of the temperature or larger (for $m_a\lesssim T$ the coalescence process is practically negligible). The production rate per unit volume of axions of energy between $E$ and
$E+dE$ is~\cite{DiLella:2000dn}
\begin{equation}
d \dot N_a  = \frac{g_{a \gamma}^2 m_a^4}{128 \pi^3}p\left(1- \frac{4 \omega_{\rm pl}^2}{m_a^2} \right)^{3/2} e^{-E/T} dE \,\ ,
\end{equation}
and the axion emissivity (per unit mass)  
\begin{equation}
\varepsilon_a = \frac{1}{\rho}\int  E \frac{d \dot N_a}{d E} dE \,\ .
\end{equation}
In Fig.~\ref{fig:energyloss}, we show the axion energy-loss within the core of a typical HB star for Primakoff and coalescence processes for different axion masses. It is evident that axion production from photon coalescence tends to dominate over that from the Primakoff process for sufficiently high axion masses.

\emph{Plasmon-axion conversions.}  
If a macroscopic magnetic field is present in the stellar interior (as speculated in the case of the Sun and WD), a further production mechanism consists in the conversions of electromagnetic excitations into axions in such an external field~\cite{Caputo:2020quz}, [see Fig.~\ref{fig:processes}-(c)]. In this case, not just the TP but also the LP mode participates to the axion production. For this reason, this process is sometimes indicated as LP/TP. 

For typical conditions in stellar plasma, the photon absorption rate $\Gamma$ is not negligible with respect to the others oscillation parameters. Thus, TP/LP-axion oscillations are interrupted by collisions, happening with a rate $\Gamma$, and one has to  treat simultaneously oscillations and collisions. A suitable formalism is provided by the \textit{kinetic approach} developed for  relativistic mixed neutrinos in the presence of collisions~\cite{Sigl:1992fn}.\footnote{Another  approach is thermal field theory~\cite{Caputo:2020quz,Guarini:2020hps}. Both methods lead to the same results.} This formalism was applied to plasmon-axion mixing in Ref.~\cite{Guarini:2020hps} to which we address the interested readers for further details. 

The result for the TP-axion conversion rate is
\begin{equation}
\Gamma_a^{\rm{TP}}= \biggl[\frac{\Gamma {\Delta_{a \gamma}^T}^2}{\bigl(\Delta_{\rm pl}  - \Delta_{\rm a} \bigr)^2 + \Gamma^2/4} \biggr] \frac{1}{e^{\omega/T}-1} \; ,
\label{eq:ratetrasverso}
\end{equation}
where  
\begin{eqnarray}  
	\Delta_{a\gamma}^T &=& \frac{g_{a \gamma} B_{T}}{2}  \simeq    4.94\times10^{-13} 
	\left(\frac{g_{a\gamma}}{10^{-11}\textrm{GeV}^{-1}} \right)
	\left(\frac{B_T}{\rm G}\right) {\rm km}^{-1}
	\nonumber\,,\\  
	\Delta_a  &=& -\frac{m^2_a}{2 \omega} \simeq 
	-2.5 \times 10^{3} \left(\frac{m_a}{
		{\rm eV}}\right)^2 \left(\frac{E}{{\rm MeV}} \right)^{-1} 
	{\rm km}^{-1}
	\nonumber\,,\\  
	\Delta_{\rm pl}  &=& -\frac{\omega^2_{\rm p}}{2 \omega} \simeq 
	-3.4\times10^{-18}\left(\frac{E}{{\rm MeV}}\right)^{-1}
	\left(\frac{n_e}{{\rm cm}^{-3}}\right) 
	{\rm km}^{-1} \,\ .
 \label{eq:Delta}
	\end{eqnarray}
The expression in Eq.~\eqref{eq:ratetrasverso} is valid both on resonance (i.e. $\Delta_{\rm pl}=\Delta_a$) and off-resonance.

In the case of LP-axion conversion rate one finds 
\begin{equation}
\Gamma_a^{\rm{LP}}= \biggl[ \frac{\Gamma {\Delta_{a \gamma}^L}^2}{(\omega_{\rm pl}-\omega_a)^2 + \Gamma^2/4} \biggr] \frac{1}{e^{\omega/T}-1} \; .
\label{eq:ratelongitudinale}
\end{equation}
Contrarily to the photon transverse modes, the expression in Eq.~\eqref{eq:ratelongitudinale} is valid only on resonance, since it is based on the evolution of on-shell LPs, i.e. it is obtained assuming that $\omega \sim \omega_{\rm pl} \sim \omega_a$ and it is not applicable for $\omega$ very far from $\omega_{\rm pl}$. A general result free of these limitations has been recently obtained in Refs.~\cite{Caputo:2020quz,OHare:2020wum} from a thermal field theory approach.

\emph{Photon-axion conversions in cosmic magnetic fields:} The axion-photon Lagrangian [first term at r.h.s. in Eq.~\eqref{eq:Laint2}] can also trigger photon-axion conversions in the presence of an external magnetic field, outside stellar sources, in low-density regions where the photon absorption is negligible. This effect plays a crucial role both in the presence of astrophysical magnetic fields and in axion direct detection in laboratory experiments. The formalism to study this problem is very similar to the ones used to characterize neutrino oscillations~\cite{Raffelt:1987im}. The conversion probability takes a simple form if one considers the case of a photon beam propagating in a single magnetic domain, where the magnetic field ${\bf B}$ is  homogeneous. One can denote by ${\bf B}_T$ the transverse magnetic field. Then the component of the photon initially polarized orthogonally to ${\bf B}_T$ decouples, while the probability that a photon initially polarized parallel to  ${\bf B}_T$   converts into an axion after a distance $d$ then reads~\cite{Raffelt:1987im}
\begin{equation}
	\label{a16}
	P_{a \gamma}  ={\rm sin}^2 2 \theta \  {\rm sin}^2
	\left( \frac{\Delta_{\rm osc} \, d}{2} \right)~,
\end{equation}
where the oscillation wave number is
\begin{equation}
	\label{a17}
	{\Delta}_{\rm osc} = \left[\left( \Delta_a - \Delta_{\parallel} \right)^2 + 4 \Delta_{a \gamma}^2 \right]^{1/2}~.
\end{equation}
where $ \Delta_\parallel \equiv \Delta_{\rm pl} + \Delta_{\parallel}^{\rm CM} + \Delta_{\rm CMB}$ and the oscillation parameters $\Delta_{a\gamma}$,  $\Delta_a$ and  $\Delta_{\rm pl}$  have been defined in Eq.~\eqref{eq:Delta}. The term $\Delta_{\parallel}^{\rm CM}$ represents the Cotton-Mouton  effect, accounting for the birefringence of fluids in the presence of a transverse magnetic field.  A vacuum Cotton-Mouton effect is expected from QED one-loop corrections to the photon polarization in the presence of an external magnetic field $\Delta_\mathrm{QED} \propto|\Delta_{\perp}^{\rm CM}- \Delta_{\parallel}^{\rm CM}|\propto B^2_T$, precisely $\Delta_\parallel=\frac{7}{2}\Delta_\mathrm{QED}$ and $\Delta_\perp=2\Delta_\mathrm{QED}$~\cite{Raffelt:1987im}. Finally, the term $\Delta_{\rm CMB}
\propto \rho_{\rm CMB}$ represents the 
background photon contribution to the photon
polarization~\cite{Dobrynina:2014qba}.  
Numerically, one can fix 
\begin{eqnarray}  
	\Delta_{\rm QED}&\simeq & 
	2\times10^{-16}\left(\frac{E}{ {\rm MeV}}\right)
	\left(\frac{B_T}{\rm G}\right)^2 {\rm km}^{-1} \nonumber\,, \\
	\Delta_{\rm CMB}&\simeq & 
	2.6\times 10^{-27}\left(\frac{E}{{\rm MeV}}\right)
	{\rm km}^{-1} \,.
	\label{eq:Delta0}
\end{eqnarray}
Finally, the mixing angle $\theta$ in Eq. (\ref{a16}) is given by
\begin{equation}
	\label{theta}
	\theta = \frac{1}{2} \arctan \left(\frac{ 2 \Delta_{a \gamma}}{\Delta_{\parallel}-\Delta_a} \right)
 \,\ .
\end{equation}
For more complex magnetic field configurations, the conversion probability is generally not analytical, and different methods have been developed to address this problem (see, e.g., Refs.~\cite{Mirizzi:2009aj,Kartavtsev:2016doq}).

The conversion of LP in a medium is possible, in the presence of a  magnetic field $B_{\rm coll}$, collinear with the direction of the plasmon propagation. In the case of a constant and homogeneous magnetic field, one finds an expression analogous to Eq.~\eqref{a17} with the substitution $B_T \to B_{\rm coll}$ (see Ref.~\cite{Tercas:2018gxv}). A detailed study of LP-axion conversions and their applications in astrophysical systems has been presented in Ref.~\cite{Caputo:2020quz}.

\subsubsection{Axion-electron coupling}
\label{sec: electroncoupl}

\emph{ABC processes:}
The axion-electron coupling $g_{ae}$  of Eq.~\eqref{eq:Laint2} induces several processes relevant for stellar evolution (see Refs.~\cite{Raffelt:1996wa,Redondo:2013wwa,Hoof:2021mld} for a comprehensive presentation. The most important  are  the Atomic de-excitation and recombination  [see Fig.~\ref{fig:processes}-(d/e)], the electron and ion Bremsstrahlung [see Fig.~\ref{fig:processes}-(f)], and the Compton process [see Fig.~\ref{fig:processes}-(g)], collectively known as the ABC processes.\footnote{Another astrophysical process discussed in the literature is the electron-positron annihilation, $ e^{+} e^{-} \to \gamma ~ a $~\cite{Pantziris:1986dc}, which plays, however, a less significant role in stellar evolution.} The atomic recombination and de-excitation processes are an important contribution to the solar axion spectrum~\cite{Redondo:2013wwa} (see  Fig.~\ref{fig:solarabc}) but can be ignored in  numerical simulations of stellar evolution, as their contribution to the solar energy loss is not significant. 

At high densities, particularly in electron degeneracy conditions, the most efficient axion production mechanism is the electron/ion bremsstrahlung process
\begin{align}
e ~Ze\to  e ~ Ze ~ a\, .
\end{align}  
As evident from Fig.~\ref{fig_bremsstrahlung_compton}, at high density and relatively low temperature, when electrons become degenerate, the bremsstrahlung rate has a very mild dependence on the density. On the other hand, in non-degenerate conditions the rate depends linearly on the density. In both cases, there is also a dependence on the stellar chemical composition. Explicit expressions for the energy-loss rates per unit mass in the degenerate (D) and non-degenerate (ND) limits are provided in Ref.~\cite{Raffelt:1994ry}. Approximately, 
\begin{align}
& \varepsilon_{\rm ND}
\simeq 47\, g_{ae}^{2} T^{2.5}\frac{\rho}{\mu_e}
\sum\frac{X_j Z_j}{A_j}\left(Z_j+\frac{1}{\sqrt{2}}\right) {\rm erg\,g^{-1}\,s^{-1}}\,,\\
& \varepsilon_{\rm D}
\simeq 8.6 \times 10^{-7} F\, g_{ae}^{2} T^{4}\left(\sum\frac{X_j Z_j^2}{A_j}\right)  {\rm erg\,g^{-1}\,s^{-1}}\,,
\label{eq_bremsstrahlung_degenerate}
\end{align}
where $T$ and $\rho$ are in K and in g~cm$^{-3}$ respectively, $\mu_e=\left( \sum X_j Z_j/A_j \right)^{-1}$ is the mean molecular weight per electron, $ X_j $ is the relative mass density of the j-th ion, and $ Z_j,~A_j $ its charge and mass number respectively.\footnote{In the typical plasma conditions where the bremsstrahlung is relevant, one finds $Z_j/A_j\approx 1/2$. So, the rate has a dependence on the chemical composition of the plasma and increases in the case of high $Z$. In particular, the rate is larger in a CO WD core than in the core of a RGB star, composed mostly of He.} The mild density dependence of the degenerate rate is accounted for by the dimensionless function $F$. An explicit expression for this function can be found in Ref.~\cite{Raffelt:1994ry} (see also Sec.~3.5 of Ref.~\cite{Raffelt:1996wa} for a pedagogical presentation). Numerically, it is of ${\mathcal O}$(1) for the stellar plasma conditions, $\rho \sim 10^{5}-10^{6}$ and $T\sim 10^{7}-10^{8}$, of interest for our discussion here, when the degenerate bremsstrahlung process dominates. The intermediate regime between degenerate and non-degenerate conditions in Fig.~\ref{fig_bremsstrahlung_compton} is calculated as $\varepsilon_B=( 1/\varepsilon_{\rm B}^{\rm (d)}+ 1/\varepsilon_{\rm B}^{\rm (nd)})^{-1}$, following the prescription in Ref.~\cite{Raffelt:1994ry}.

The Compton process
\begin{align}
\gamma ~ e \to a ~ e \,\ ,
\end{align}  
accounts for the production of axions from the scattering of thermal photons on electrons. The Compton axion emission rate is a steep function of the temperature
\begin{align}
\varepsilon_{\rm C}\simeq 2.7\times 10^{-22} g_{ae}^2 \frac{1}{\mu_e}\left( \frac{n_{e}^{\rm eff}}{n_e} \right)\,T^6 \,{\rm erg\,g^{-1}\,s^{-1}} \,,
\end{align}
where $n_e$ is the number density of electrons while $n_{e}^{\rm eff}$ is the effective number density of electron targets. At high densities, degeneracy effects reduce $n_{e}^{\rm eff}$, suppressing the rate (cf.~Fig.~\ref{fig_bremsstrahlung_compton}). The Compton process can effectively dominate over the bremsstrahlung only at low density and high temperature. 

\subsubsection{Axion-nucleon coupling}
\label{ref:nucleoncoupling}

The nucleon coupling plays an important role in astrophysics, permitting the production of axions from different nuclear reactions and transitions. We will discuss the most relevant cases later on in this Section and in the next one. Furthermore, this coupling plays a particularly relevant role for axion production in nuclear matter, like in a SN core on in a NS (see Sec.~\ref{sec:SN_NS}). The axion emissivity in these environments is currently a topic of intense investigations~\cite{Hanhart:2000ae,Chang:2018rso,Carenza:2019pxu,Carenza:2020cis,Choi:2021ign,Ho:2022oaw}.

\emph{Nucleon bremmstrahlung:} One of the dominant axion production mechanisms in nuclear matter ($\rho \sim 10^{14}$~g~cm$^{-3}$) is the nucleon-nucleon bremmstrahlung $N ~N \to N~N~a$ [see Fig.~\ref{fig:processes}-(h)]. An accurate description of this process is quite difficult, due to the lack of a solid understanding of the nucleon-nucleon interaction in a dense environment. Originally, this interaction was described trough a simple One-Pion Exchange (OPE) potential [see Fig.~\ref{fig:processes}-(h)]~\cite{Turner:1987by,Burrows:1988ah,Burrows:1990pk}. However, the OPE prescription for calculating the axion emissivity was soon criticized as too simplistic. A comparison with the measured phase shifts showed that the OPE result overestimated the bremsstrahlung axion production by about a factor of four~\cite{Hanhart:2000ae}.  
In the perspective of a precise calculation of the axion emissivity, Refs.~\cite{Raffelt:1991pw,Janka:1995ir,Sigl:1995ac} pointed out the important role of nucleon spin fluctuations, caused by nucleon self-interactions, in reducing the axion emissivity via bremsstrahlung. Typically, in a SN core the nucleon spin fluctuation rate is given by~\cite{Raffelt:2006cw}
\begin{equation}
  \Gamma_{\sigma}=450~{\rm MeV}\left(\frac{\rho}{3\times10^{14}~{\rm g}~{\rm cm}^{-3}}\right)\left(\frac{T}{30~{\rm MeV}}\right)^{1/2}\,,
\end{equation}
and it is large compared to the temperature. Therefore the nucleon propagator must be suppressed to account for the nucleon scatterings, a phenomenon known as the Landau-Pomeranchuk-Migdal (LPM) or multiple-scatterings effect. The resulting axion emissivity per unit mass is~\cite{Raffelt:2006cw}
\begin{equation}
\varepsilon_{a}=3\times10^{37}
\textrm{erg} \,\
 \textrm{g}^{-1} \textrm{s}^{-1} 
g_{aN}^{2}\left(\frac{T}{30~{\rm MeV}}\right)^{4}F\,,
\end{equation}
where $F$ is a factor order unity which includes nuclear physics and the LPM effect. 

Following the attempts of a more accurate calculation of the axion emissivity, Ref.~\cite{Chang:2018rso} criticized the OPE approximation and included corrections to the nucleon bremsstrahlung, including a finite pion mass, corrections for more realistic nuclear interactions and the LPM effect. Taking into account self-consistently all these effects, in Ref.~\cite{Carenza:2019pxu} it was found a reduction by one order of magnitude of the emissivity rate with respect to the naive OPE prescription. It is also worth noticing that in a nuclear medium, finite-density effects can modify the axion-nucleon couplings~\cite{Balkin:2020dsr}. Consequently, studying axion production through nuclear coupling in extremely dense environments, such as SNe or NS, is highly  non-trivial. For instance, Ref.~\cite{Balkin:2020dsr} demonstrated that, unlike in vacuum, the KSVZ axion model may develop a significant coupling to neutrons at finite density. Very recently, Ref.~\cite{DiLuzio:2024vzg} showed that nucleophobia--the property of some specific axion models (e.g., \cite{DiLuzio:2017ogq,DiLuzio:2023tqe}) which do not couple to nucleons in vacuum--is preserved also at the very high densities characteristic of SN and NS. We will come back to this issue in Sec.~\ref{sec:SN_NS}, when discussing the axion emission from SNe. In any case, at the time of writing this matter is still under investigation. 

\emph{Pionic Compton processes:}
The role of the pion-induced reaction in a SN core,   $\pi^{-} p \to n a$ [see Fig.~\ref{fig:processes}-(i)] was first discussed in Refs.~\cite{Turner:1991ax,Mayle:1992sq,Raffelt:1993ix}, and in Ref.~\cite{Keil:1996ju}. However, initial estimates suggested that the thermal pion population was too small for the pion reaction to be competitive~\cite{Raffelt:1996wa}. For this reason, pions and reactions involving pions in SNe have been largely ignored. This approach changed after Ref.~\cite{Fore:2019wib}, demonstrating that strong interactions enhance the abundance of negatively charged pions. On light of this result, the axion emissivity via Compton pionic process was re-evaluated in Ref.~\cite{Carenza:2020cis}, finding an energy emission rate per unit volume 
\begin{equation}
Q^{\pi}_{a} \simeq 
\frac{30}{\pi^{2}}
\frac{{\bar g}_{aN}^{2}}{m_N^{2}}
\left(\frac{g_A}{f_{\pi}} \right)^{2} n_p z_\pi T^{6} \,\ ,
\end{equation}
where $g_A=1.27$ is the axial coupling, $f_{\pi} = 92.4$~MeV is the pion decay constant, $n_p$ is the proton density, $z_\pi=e^{(\mu_\pi-m_\pi)/T}$ is the pion fugacity and \footnote{Note that in Ref.~\cite{Keil:1996ju} there was a wrong negative sign between the first and second term on the left-hand-side of Eq.~\eqref{eq:gbaraNSquared}. }
\begin{equation}
\label{eq:gbaraNSquared}
{\bar g}_{aN}^{2} = \frac{1}{2}(g_{an}^{2} + g_{ap}^{2}) + \frac{1}{3} g_{ap}g_{an} \,\ .
\end{equation}
In the limit of vanishing pion mass, the energy-loss rate per unit mass is given by
\begin{equation}
\varepsilon^{\pi}_{a} =
{\bar g}_{aN}^{2} 2.6 
\times 10^{37} \,\ \textrm{erg} \,
 \textrm{g}^{-1} \textrm{s}^{-1} 
\left(\frac{T}{30~\textrm{MeV}}\right)^6 \left(\frac{Y_p}{0.3}\right)
\left(\frac{z_\pi}{0.4} \right) \,\, ,
\end{equation}
for typical conditions in a SN core. The four-particles interaction vertex shown in Fig.~\ref{fig:processes}-(i), when integrating out the heavy mediator, is also known as contact interaction. Its impact on the axion emissivity is significant and it was discussed in Refs.~\cite{Carena:1988kr,Choi:2021ign}. Recently, Ref.~\cite{Ho:2022oaw} showed also the importance of the vertex with the $\Delta$ resonance, Fig.~\ref{fig:processes}-(i), leading to a sizable increase in the axion emissivity. Therefore, none of these interactions can be neglected in an accurate evaluation of the axion emission from SNe (see Refs.~\cite{Carenza:2023lci,Lella:2022uwi} for more details).

\emph{Axions from nuclear reactions and de-excitations:}
Axions can also be produced in nuclear processes, such as nuclear fusion and transitions. These axions typically carry an energy of the order of $\sim$~MeV, the typical nuclear energy scale. However, as we shall see these energies may be also considerably lower (in the keV region), in some nuclear transitions. A well studied nuclear reaction is $p~d\to$ $^{3}{\rm He}~a$, which produces a monochromatic 5.5~MeV axion flux. Solar axions from this reaction have been experimentally searched, as we shall discuss in Sec.~\ref{sec:Sun}. A list of relevant nuclear reactions can be found in Ref.~\cite{Massarczyk:2021dje} (particularly interesting are Table II and III therein). 

Among the most studied nuclear processes producing axions are thermal excitation and subsequent de-excitation of the nuclei of stable isotopes. These processes have been studied especially in the Sun. The solar core temperature, $T_c\sim 1~$keV, is considerably below the typical MeV range of nuclear energies. As we shall see in Sec.~\ref{sec:Sun}, however, there are low lying nuclear excitations of stable isotopes with a significant abundance inside the sun  that can be thermally excited (see, e.g., Ref.~\cite{Asplund:2009fu}).

The axion rate from these processes can be expressed in terms of the photon emission~\cite{Avignone:1988bv}
\begin{align}
\label{eq:Gamma_Ratio_general}
\frac{\Gamma_a}{\Gamma_\gamma}=
\left( \frac{k_a}{k_\gamma} \right)^3
\frac{1}{2\pi\alpha_{\rm em}}\frac{1}{1+\delta^2}
\left[ \frac{\beta \, g_{aN}^{0} + g_{aN}^{3}} 
{\left( \mu_0-\frac12 \right) \beta + \mu_3 -\eta} \right]^2,
\end{align}
where $g_{aN}^{0} =(g_{ap}+g_{an})/2$ and $g_{aN}^{3} =(g_{ap}-g_{an})/2$ are the iso-scalar and iso-vector couplings, respectively, $k_a,~k_\gamma$ are the axion and photon momenta, $\mu_0$ and $\mu_3$ are the iso-scalar and iso-vector nuclear magnetic moments (expressed in nuclear magnetons), $\delta$ is the E2/M1 mixing ratio for the nuclear transition under consideration, and $\beta$, $\eta$ are constants dependent on the nuclear structure~\cite{Avignone:1988bv}. 

Given these premises, the axion emission rate per unit mass of solar matter is given by
\begin{equation}
\label{eq:axionspersolarmatter}
\mathcal{N}_a=\mathcal{N}\,\omega_1(T)\, \frac{1}{\tau_0}\frac{1}{1+\alpha} \frac{\Gamma_a}{\Gamma_\gamma},
\end{equation}
where $\mathcal{N}$ is the isotope number density per stellar mass, 
\begin{align}
\label{eq:occupation_number}
\omega_1=\frac{(2J_1+1)e^{-E^*/T}}{(2J_0+1)+(2J_1+1)e^{-E^*/T}},
\end{align}
the thermal occupation number of the relevant excited state (usually, the first), $\tau_0$ the lifetime of the excited state, and $\alpha$ the internal conversion coefficient. The parameters in Eq.~\eqref{eq:occupation_number} are reported in Tab.~\ref{tab:isotopes} for some typical examples. These transitions produce very narrow  lines, whose width gets contribution from the natural line width and from the Doppler broadening. We shall discuss more about searches for these lines in Sec.~\ref{sec:Sun} in the context of axions from the Sun.

\begin{table}
			\caption{Isotopes with a nuclear M1 transition and $E^*< 20$~keV.  $J_0$ and $J_1$ represent the total angular momenta of the ground and excited state respectively, $\tau$ is the total lifetime of the state, and $\alpha$ is called internal conversion coefficient. Data taken from Ref.~\cite{DiLuzio:2021qct}.
   } 
	\renewcommand{\arraystretch}{1.5}
	\centering
		\begin{tabular}{l||c|c|c|c|c}
			& $^\textbf{57}$\textbf{Fe}	& $^\textbf{83}$\textbf{Kr}	& $^\textbf{169}$\textbf{Tm} &  $^\textbf{187}$\textbf{Os}	& $^\textbf{201}$\textbf{Hg}  \\
			\hline
			\hline
			$E^*$ [keV] & 14.4 & 9.4 & 8.4 & 9.7 & 1.6\\
		    $J_0$ & 1/2 & 9/2 & 1/2 & 1/2 & 3/2 \\
		    $J_1$ & 3/2 & 7/2 & 3/2 & 3/2 & 1/2 \\
		    $\tau_0$ [ns] & 141 & 212 & 5.9 & 3.4 & 144\\
		    $\alpha$ & 8.56 & 17.09 & 285 & 264 & 47000 \\
      \hline
		\end{tabular}

		\label{tab:isotopes}
\end{table}
\hspace{-1cm}

\subsubsection{Axion-nucleon electric dipole portal}
\label{sec:dipole}

As we discussed in Sec.~\ref{sec:axionintro}, the nEDM coupling [third term at r.h.s  of Eq.~\eqref{eq:Laint2}] is the most fundamental axion coupling, deriving directly from the axion-gluon coupling. The nEDM interaction is particularly important for axion dark matter searches. Indeed, the oscillating axion dark matter field would imprint the same oscillations into the EDM of protons and neutrons~\cite{Graham:2013gfa}. The detection of such a feature is the ambitious goal of the Cosmic Axion Spin Precession ExpeRiment (CASPERe) experiment~\cite{Budker:2013hfa,JacksonKimball:2017elr}. Oscillating electric-dipole moments of atoms and molecules can also be generated by the fundamental axion interactions with fermions and gluons~\cite{Stadnik:2013raa,Flambaum:2019ejc,Roussy:2020ily}.

Furthermore, the nEDM axion coupling can lead to axion production in the core of a SN. The two processes which contribute to this rate are the Compton scattering~\cite{Graham:2013gfa,Lucente:2022vuo}, $N~\gamma \to N~a$, and the nucleon bremsstrahlung process, $N~N \to N~N~a$~\cite{Lucente:2022vuo} [see Fig.~\ref{fig:processes}-(j/k)]. Of these, the Compton scattering gives the largest contribution while the bremsstrahlung is suppressed by more than one order of magnitude~\cite{Lucente:2022vuo}. Then, focusing on the Compton process, the emissivity per unit mass can be written as
\begin{equation}
\begin{split}
    \varepsilon_a = \frac{30\,g_d^2}{\pi^3}\,\frac{T^6}{m_N} \approx 10^{36} {\rm erg}~{\rm g}^{-1}{\rm s}^{-1}  
    \left(\frac{g_d}{{\rm GeV}^{-2}}\right)^2\,\left(\frac{T}{30~{\rm MeV}}\right)^6\,\left(\frac{938~{\rm MeV}}{m_N}\right)\,. 
\label{eq:qacompt}
\end{split}
\end{equation}


\section{The Sun and other nearby   stars}
\label{sec:Sun}


The Sun is a main sequence star with mass $M = 1.989\times 10^{33}~{\rm g}\equiv 1\,{\rm M}_{\odot}$. This value defines the solar mass unit $M_\odot$. Its core temperature is $T_c\simeq1.5\times 10^7$~K, corresponding to about 1.3 keV, while the core density is $\rho_c\sim 10^2\,{\rm g~cm}^{-3}$. The solar radius is $R_{\odot}= 6.9598 \times 10^5$~km and the solar luminosity $L_{\odot}= 3.8418 \times 10^{33}$~erg\,${\rm s}^{-1}$~ \cite{Serenelli:2009yc}. The chemical composition of the core consists mostly of ionized $^{1}$H and $^{4}$He  (see Ref.~\cite{SolarElementalAbundances}). The Sun produces nuclear energy by hydrogen fusion into helium that proceeds through the pp chains,
\begin{equation}
2e^-+4{\rm p} \rightarrow {}^4{\rm He}  + 2 \nu_e  \,\ ,
\label{eq:Hburn}
\end{equation}
which releases an energy of  26.73~MeV. Therefore, for every ${}^{4}{\rm He}$ produced, two protons convert to neutrons. Hence, each ${}^{4}{\rm He}$ generated in the pp-chain is associated with the production of two electron neutrinos, $\nu_e$, which escape the solar core unimpeded. The search for these solar neutrinos has been carried out over decades starting from 1968~\cite{Davis:1968cp} (see, e.g. Ref.~\cite{Antonelli:2012qu} for a review), and has been rewarded with the 2002 Nobel Prize to R. Davis~\cite{Davis:2002fb} and M. Koshiba for their detection. The pp solar neutrino fluxes have been  recently measured by the  Borexino collaboration~\cite{BOREXINO:2018ohr} and by the PandaX-4T dark matter collaboration~\cite{Lu:2024ilt}.

For decades, the Sun has served as a laboratory for setting limits on the properties of neutrinos, starting from the pioneering work of J. Bernstein and collaborators, who studied the effects of neutrino electromagnetic form factors in the Sun~\cite{Bernstein:1963qh}. The history of solar neutrino measurements is tightly connected with the discovery of flavor conversions. Notably, the deficit of measured solar neutrino fluxes compared to Standard Solar Model (SSM) predictions has provided some of the most compelling evidence for neutrino oscillations (see Ref.~\cite{Antonelli:2012qu} for a review). This discovery led to the 2015 Nobel Prize awarded to A. McDonald for the evidence of solar neutrino oscillations, through the Sudbury Neutrino Observatory (SNO) experiment~\cite{SNO:2001kpb}.

In more recent years, the Sun has become an important laboratory for exotic physics. This progress was facilitated by a deeper and deeper understanding of the Sun in the past decades, particularly thanks to advancements in helioseismology and neutrino observations. The importance of the Sun in the study of light and weakly interacting particles, such as the axion, relies in large part (though, not only) on its potential to thermally produce these particles. Furthermore, being very close to the Earth, the Sun provides one of the most efficient WISP sources, widely used in several terrestrial experiments. The case of axions, which is the focus of our review, will be discussed extensively below. Other important cases include dark photons~\cite{OShea:2023gqn,Redondo:2008aa,Redondo:2013lna,Redondo:2015iea,Caputo:2021eaa}, and chameleons~\cite{Brax:2010xq,Brax:2011wp,Vagnozzi:2021quy,OShea:2024jjw}. It is important to remark that the solar potential as laboratory for exotic processes is not limited to light particles (understood as particle with mass not much larger than the solar core temperature). As we shall see, nuclear reactions or nuclear de-excitation processes occurring in the Sun can also efficiently produce particle of mass considerably larger than its temperature. Finally, the Sun has also been used to test dark matter Weakly Interacting Massive Particles (WIMPs). In this case, particles in the dark matter halo would be trapped in the Sun core and the further annihilation can lead to an observable flux of high-energy photons or neutrinos~\cite{Cirelli:2005gh}.

Other MS stars would produce axions in much the same way as the Sun does and are, in this sense, other stellar laboratories for such particles. Obviously, due to their much larger distances, they are not quite as useful, especially as \emph{axion factories}, as the axion flux decreases quadratically with the distance. However, as we shall see, the flux from all MS stars is not that small and may possibly lead to some interesting phenomenology. Moreover, other stars with properties different from the Sun could provide independent and, in some cases, more efficient axion factories than the Sun itself. The case of supergiant stars, whose core temperature may be considerably larger than that of the Sun, could provide a possibly large source of axions, which could be observed if they convert into photons in the Galactic magnetic field. 

In this Section we describe the physics potential of the Sun and of other nearby stars in probing axions. In Sec.~\ref{sec:SSM} we briefly recall the main features of the SSM. In Sec.~\ref{sec:AxionproductionintheSun} we describe the axion production processes in the Sun associated with the couplings with photons, electrons and nucleons. In Sec.~\ref{sec: bounds} we present the current constraints on axions from solar physics. In Sec.~\ref{sec:helioscope} we discuss the status and the future perspectives of direct solar axion searches through helioscopes. In Sec.~\ref{sec:otherdec} we  revise solar axion detection in other classes of experiments, namely underground dark matter detectors (Sec.~\ref{sec:axiondm}), underground neutrino detectors (Sec.~\ref{sec:neutrin}) and X-ray satellites (Sec.~\ref{sec:Xrays}). In Sec.~\ref{sec:axionnearby} we present the axion fluxes from nearby stars. In particular, we characterize the diffuse axion flux from Galactic MS stars (Sec.~\ref{sec:axionfromstar}) and the potential of X-ray observations in probing axions from nearby supergiant stars and pre-SNe (Sec.~\ref{sec:preSN}).

\subsection{Standard Solar Model}
\label{sec:SSM}

Solar models describe the Sun's evolutionary phases from its origin, 4.6 billion years ago, to the present moment. These models give predictions about contemporary Sun features: composition, temperature, pressure, sound-speed profiles, and neutrino fluxes. The SSM extends the general theory of MS stellar evolution to the specific case of the Sun, by means of accurate measurements which are possible only for this star. The SSM is based on the following fundamental assumptions:
\begin{itemize}
\item Energy transport mechanisms involve radiation and convection. In the inner part of the Sun, constituting 98~\% of mass or about 72~\% of radius, energy transport occurs radiatively. In the outer envelope, where the temperature gradient is steeper, energy transport is dominated by convection. The region where energy is generated through thermonuclear reactions is deep within the radiative zone. The position of the boundary between the radiative interior and the convective zone is dependent on opacity, a factor that also impacts various internal features such as the sound-speed profile. The SSM represents convection using the mixing length theory (see Sec.~\ref{sec_mixing}). 

\item As discussed in the Introduction, the Sun produces its energy by fusing protons into ${}^4$He [see Eq.~\eqref{eq:Hburn}] via the pp chain (99~\%) and CN I cycle reactions (see, e.g., Ref.~\cite{Vitagliano:2019yzm}).

\item The model is constrained to reproduce today's solar radius, mass, and luminosity.  An essential premise of the SSM posits that the proto-Sun experienced an extensively convective phase, resulting in a uniformly composed Sun until the onset of MS burning. The initial composition by mass is conventionally categorized into hydrogen ($X_{\rm ini}$), helium ($Y_{\rm ini}$), and everything else (the metals, denoted $Z_{\rm ini}$), with $X_{\rm ini}+Y_{\rm ini}+Z_{\rm ini}=1$. The relative proportions of metals are derived from a combination of meteoritic and solar photospheric data. The absolute abundance $Z_{\rm ini}$ can be taken from the modern Sun's surface abundance $Z_{\rm S}$, after corrections for the effects of diffusion over 4.6 Gyr of solar evolution. Finally, $Y_{\rm ini}/X_{\rm ini}$ is adjusted along with $\alpha_{\rm MLT}$, a parameter describing solar mixing, until the model reproduces the modern Sun's luminosity and radius. The resulting ${}^4$He/H mass fraction ratio is typically $0.27\pm0.01$, contrasting with the Big-Bang value of $0.23\pm0.01$, showing that the Sun was formed from previously processed material. The metallicity  $Z_{\odot}$ is a fundamental diagnostic of the evolutionary history of the Sun. Up to 1998, the state-of-the-art was given by the spectroscopic measurements of Anders and Grevesse (AG89)~\cite{Anders:1989zg} and Grevesse and Sauval (GS98)~\cite{Grevesse:1998bj}. These  presented metallicities $Z_{\odot} = 0.0202$ and $Z_{\odot} = 0.0170$, respectively. Moreover, heavy element mixtures provided by AG89 and GS98 also yielded good agreement with inferences from helioseismology. However, after 1998 new metallicity determinations, lower than the previous ones, produced a tension between models of the seismic sound-speed profile and depth of the convective zone. In particular, the sets of abundances known as AGS05~\cite{Asplund:2004eu} and AGSS09~\cite{Asplund:2009fu} report a metallicity of $Z_{\odot} = 0.0122$ and $Z_{\odot} = 0.0133$, respectively. This tension between models and helioseismolgy became known as  the  ``solar metallicity problem'' (see also Ref.~\cite{Sokolov:2019cbs} for exotic physics implications for this problem). Finally, in 2022 a new photospheric abundance determination~\cite{Magg:2022rxb} reestablished the agreement with helioseismology.

\end{itemize}

\subsection{Axion production in the Sun}
\label{sec:AxionproductionintheSun}

\subsubsection{Axion-photon coupling}
\emph{Primakoff production in electric field of electrons and ions--} The solar flux of light axion coupled to photons is produced mostly by the Primakoff process. In the case of more massive axions ($m_a\gtrsim 10$~keV), the photon coalescence process may become dominant. A detailed discussion of these processes is found in Sec.~\ref{sec:photon}.

Strictly speaking, Primakoff production refers to the generation of axions from thermal photons scattering off an external electric or magnetic field. In the case of the Sun, the term \emph{Primakoff axions} typically refers only to axions produced by the electric field of electrons and ions, as the existence of a large magnetic field in the solar core remains uncertain. In this review, we will adopt this standard convention.

\begin{figure}[t]
    \centering
        \includegraphics[width=0.95\textwidth]{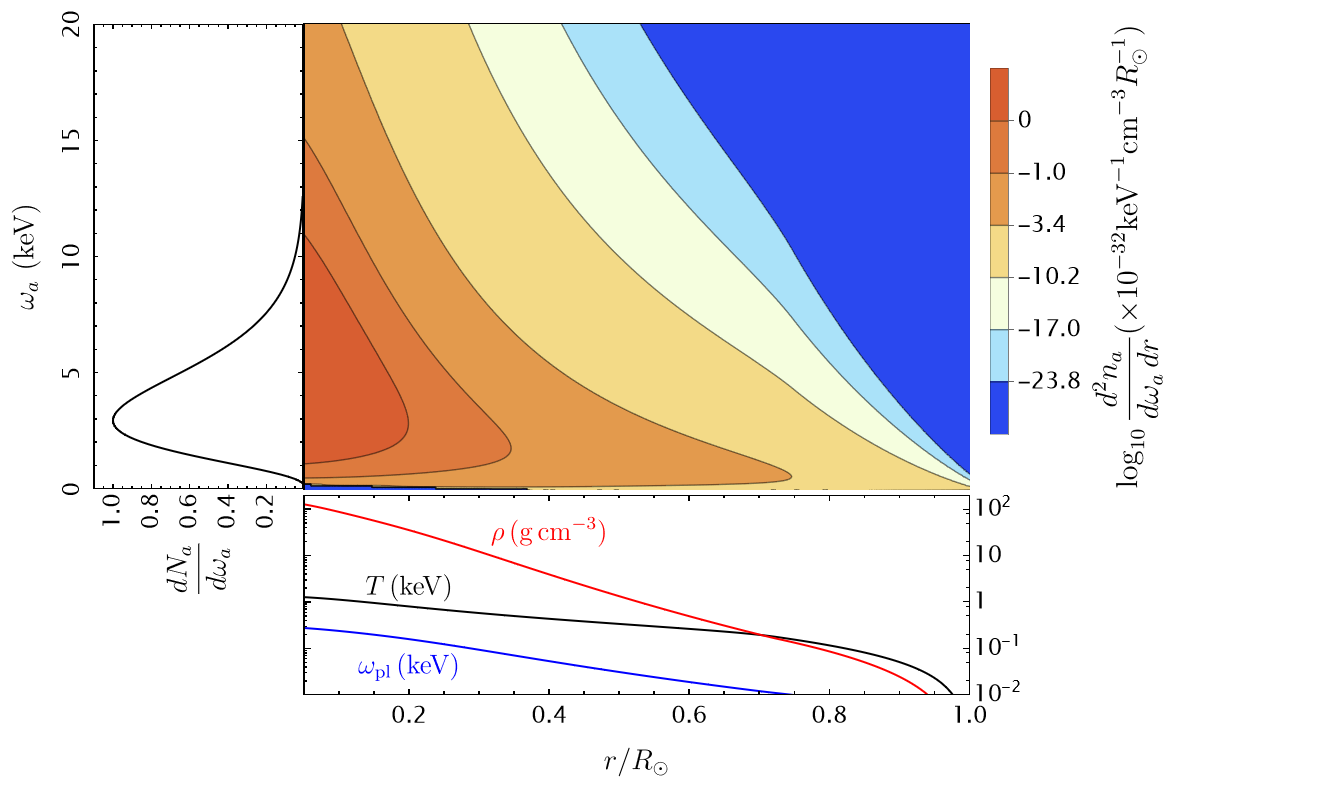}
        \caption{
      Axion emissivity by Primakoff process per unit density for the AGSS09 solar model. We considered massless axions and $g_{a\gamma}=10^{-11}~{\rm GeV}^{-1}$. On the left panel we show the axion emissivity per unit energy integrated over the solar model, in arbitrary units.
        In the lower panel we show    temperature (black), density (red) and plasma frequency (blue) profiles of the given model.
        }
    \label{fig:diagrams}
\end{figure}

The total Primakoff axion number flux at the Earth (average solar distance $D_\odot=1.49 \times 10^8$~km from the Earth) is~\cite{CAST:2007jps}
\begin{equation}
  \Phi_{\rm a}=\frac{1}{4\pi D_{\odot}^2} \int_0^{R_\odot}
  \!\!{\rm d}r\,4\pi\,r^2\int_{\omega_{\rm pl}}^\infty \!\!{\rm d}E\,
  \frac{4 \pi k^2}{(2\pi)^3}\,
  \frac{{\rm d}k}{{\rm d}E}\,2f_{\rm B}\, \Gamma_{\gamma\to{\rm
  a}}\;,
  \label{eq:sunaxionflux}
\end{equation}
where the Primakoff rate $\Gamma_{\gamma\to{\rm a}}$ is given in Eq.~\eqref{generalrate}, $f_{\rm B}= ({\rm e}^{E/T}-1)^{-1}$ is the Bose-Einstein distribution of the thermal photon bath in the solar plasma and $r$ is the solar radial variable, integrating between the origin and the solar radius $R_{\odot}$. The solar plasma frequency $\omega_{\rm pl}$ is never larger than about 0.3~keV. In order to perform the integrations in Eq.~\eqref{eq:sunaxionflux} one has to refer to solar models for which detailed data are publicly available, see e.g. Refs.~\cite{Serenelli:2009yc,Serenelli:2009ww}, from which we show the radial profile of temperature, density and plasma frequency for the AGSS09 model in Fig.~\ref{fig:diagrams}, together with the axion production per unit volume and energy as a function of the solar radius and energy, for the AGSS09 model. From the figure it results that most axions from Primakoff fluxes are generated inside the solar core ($r/R_{\odot} < 0.25$) and almost the entire flux is emitted within a radius of about $r/ R_{\odot} = 0.5$.

\begin{figure}[ht]
    \centering
    \includegraphics[width=0.6\textwidth]{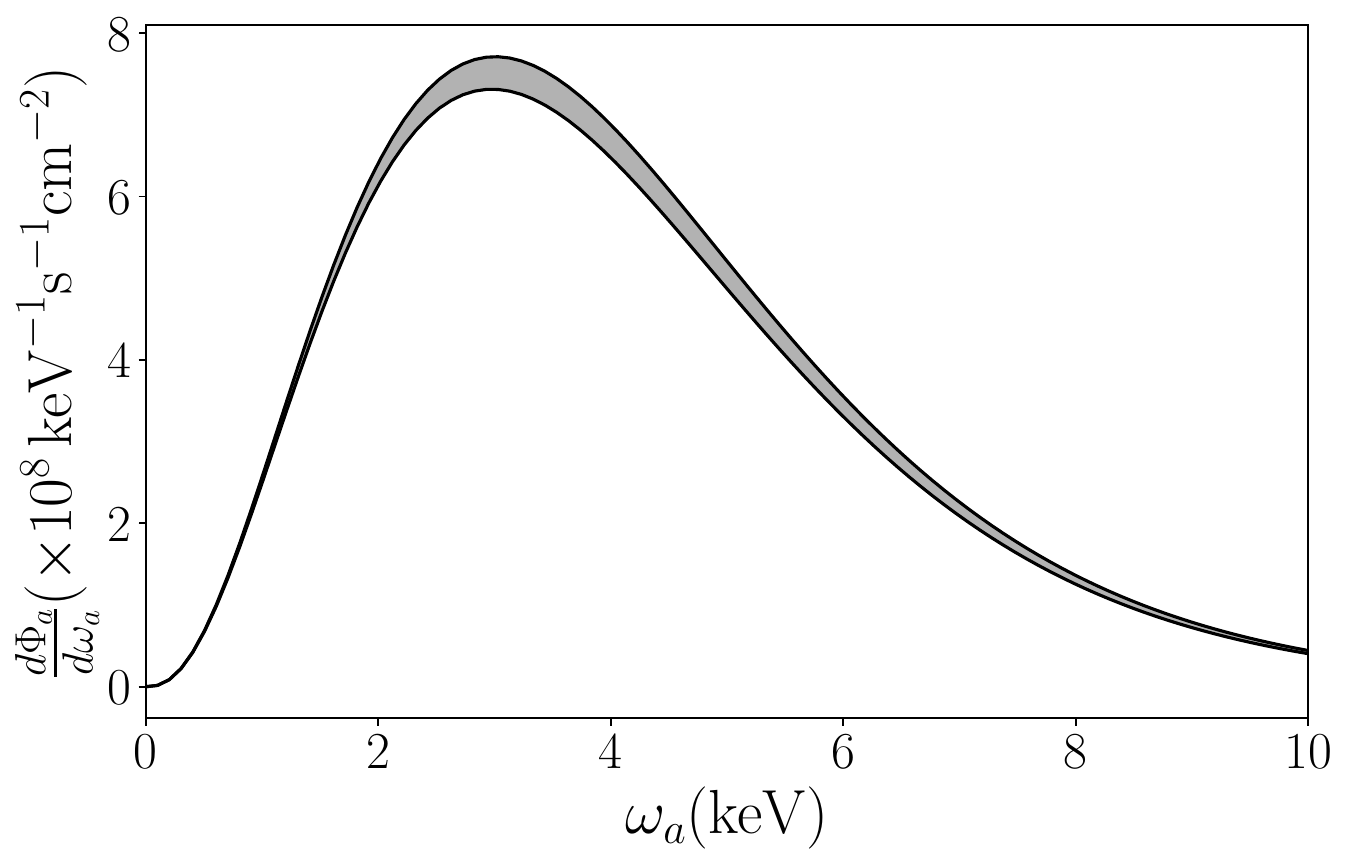}
    \caption{Axion flux from Primakoff conversion for massless axions with $g_{a\gamma} = 10^{-11}{\textrm{GeV}^{-1}}$. The shaded area shows the uncertainties of the fit in Eq.~\eqref{eq:fitprim}.} 
    \label{fig:fitprim}

        \includegraphics[width=0.6\textwidth]{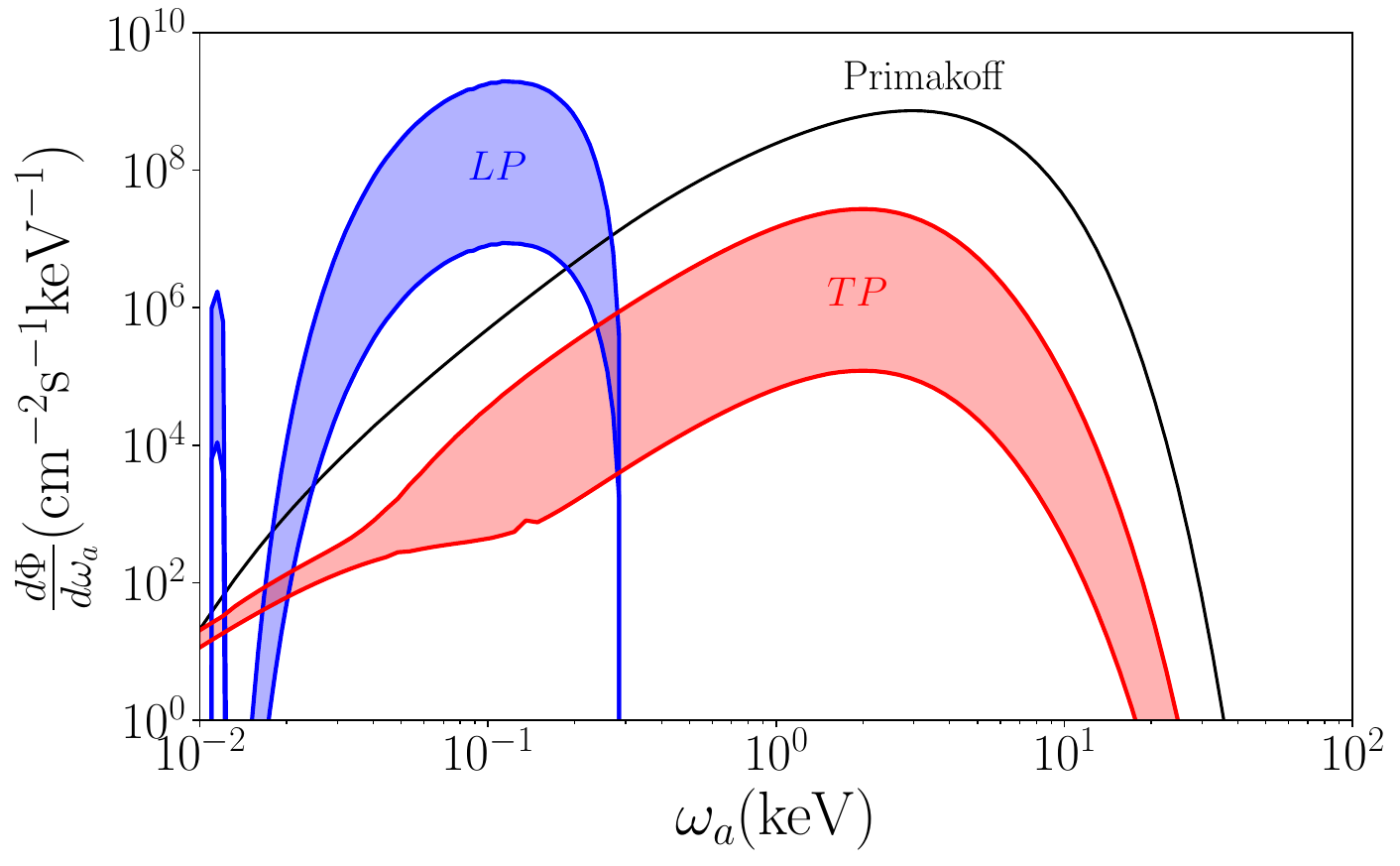}
    \caption{Axion flux from longitudinal~(LP) and transverse~(TP) plasmon interactions for $g_{a\gamma} = 10^{-11}{\textrm{GeV}^{-1}}$ and negligible axion mass. The colored regions show the LP and TP flux between the minimal and maximal magnetic field reference values, respectively. The low-energy bump in the LP spectrum is due to the weaker magnetic field in the convective region.
    The  black line shows the Primakoff flux for comparison, without its uncertainty band that cannot be appreciated on this scale.
    }
    \label{fig:plasmons} 
\end{figure}

Below, we provide an accurate fit of the Primakoff solar axion flux at Earth obtained after integrating over a solar model (a different but analogous fit can be found in Ref.~\cite{CAST:2007jps})
\begin{equation}
     \frac{d\Phi_{a}}{dE_{a}}=C_{0}\left(\frac{g_{a\gamma}}{g_{\rm ref}}\right)^{2}\left(\frac{E}{E_{0}}\right)^{\beta}e^{-(1+\beta)\frac{E}{E_{0}}}\,,
     \label{eq:fitprim}
\end{equation}
where the axion flux parameters are shown in Tab.~\ref{tab:parametersfit}, obtained by considering the screening scale in the non-degenerate limit.\footnote{A simple code to reproduce this calculation can be found at \href{https://github.com/pcarenza95/SolarAxionCode}{https://github.com/pcarenza95/SolarAxionCode}, being a simplified version of Refs.~\cite{Hoof:2021mld,Wu:2024fsf}. This code is also used to produce Fig.~\ref{fig:plasmons}. }
\begin{table}[]
    \caption{Summary of the fitting parameters to be used in Eq.~\eqref{eq:fitprim} to reproduce the axion emission from the Sun via Primakoff (coupling to photons $g_{a\gamma}$), Bremsstrahlung and Compton (coupling to electrons $g_{ae}$). The uncertainty on the fitting parameters includes the most recent solar models~\cite{Magg:2022rxb}. 
    }
    \centering
\begin{tabular}{|c|c|c|c|c|}
\hline
&$g_{ax}$& $C_{0}({\rm keV}^{-1}~{\rm s}^{-1}~{\rm cm}^{-2})$& $E_{0}({\rm keV})$   & $\beta$ \\
\hline 
{\rm Primakoff,} $x=\gamma$ &$10^{-12}~{\rm GeV}^{-1}$&$(2.19\pm0.08)\times10^{8}$&$4.17\pm0.02$&$2.531\pm0.008$\\
{\rm Bremsstrahlung,} $x=e$ &$10^{-12}$&$(3.847\pm0.007)\times10^{11}$&$1.63\pm0.01$&$0.8063\pm0.0003$\\
{\rm Compton,} $x=e$ &$10^{-12}$&$(8.8\pm0.1)\times10^{11}$&$5.10\pm0.03$&$2.979\pm0.001$\\
\hline
\end{tabular}

    \label{tab:parametersfit}
\end{table}

Given this flux, the resulting integrated axion flux and luminosity are
\begin{equation}
    \begin{split}
        \Phi_{a}&=(3.6\pm0.1)\times10^{7}\left(\frac{g_{a\gamma}}{10^{-12}~{\rm GeV}^{-1}}\right)^{2}~{\rm s}^{-1}~{\rm cm}^{-2}\,,\\
         L_{a}&=(1.79\pm0.06)\times10^{-7}\left(\frac{g_{a\gamma}}{10^{-12}~{\rm GeV}^{-1}}\right)^{2}~L_{\rm \odot}\, .\\
    \end{split}
\end{equation} 
As shown in Fig.~\ref{fig:fitprim}, where we plot the Primakoff flux with its associated uncertainty, the axion spectrum peaks around $3$~keV.

\emph{Plasmon conversions in the solar magnetic field--}
Even though the SSM~\cite{Bahcall:2004pz,Serenelli:2009yc} assumes the Sun to be a quasi-static environment, seismic solar models have been developed that include large-scale magnetic fields in various regions of the solar interior~\cite{TurckChieze:2001ye,Couvidat:2003ba}. According to these models, the magnetic field can be found in three different regions in the Sun, the \textit{radiative} zone ($r \lesssim 0.7 \; R_\odot$), the \textit{tachocline} ($ r \sim 0.7 \; R_\odot$), and the exterior (\textit{convective}) zone ($r \gtrsim 0.9 \; R_\odot$), and have intensities~\cite{Hoof:2021mld} $B_\text{rad} \in [200 \,{\rm T},\,3000 \,{\rm T}]$, $B_\text{tach} \in [4 \,{\rm T},\,50 \,{\rm T}]$, and $B_\text{conv} \in [3 \,{\rm T},\,4 \,{\rm T}]$.

The presence of these magnetic fields may trigger conversions of the thermal photons into axions, creating an additional axion flux besides the one produced by the Primakoff conversions~\cite{Caputo:2020quz} (see Sec.~\ref{sec:photon}). Incidentally, in this case LP can also contribute to the axion production and, in fact, would produce a quite interesting spectrum, possibly searchable in future helioscope experiments~\cite{OHare:2020wum}. Recently, the problem of the production of an axion flux from solar magnetic fields has been reconsidered in a series of publications. In particular, Refs.~\cite{Caputo:2020quz,OHare:2020wum} considered the production through conversion of LP while Ref.~\cite{Guarini:2020hps} reconsidered the case of TP. In both cases, the flux can be fairly large (though subject to severe uncertainties, especially on the size of the solar magnetic field), and may surpass the electric Primakoff production in certain region of the axion parameter space (see Fig.~\ref{fig:plasmons}). Progress in this respect was made in Ref.~\cite{Hoof:2021mld}, which showed that the adoption of the Rosseland average for the opacity has a relevant effect on the expected axion spectrum, especially for the transverse case.

\subsubsection{Axion-electron coupling}
\label{sec:Axion-electron coupling}

\begin{figure}[t]
	\centering
	\includegraphics[width=5in]{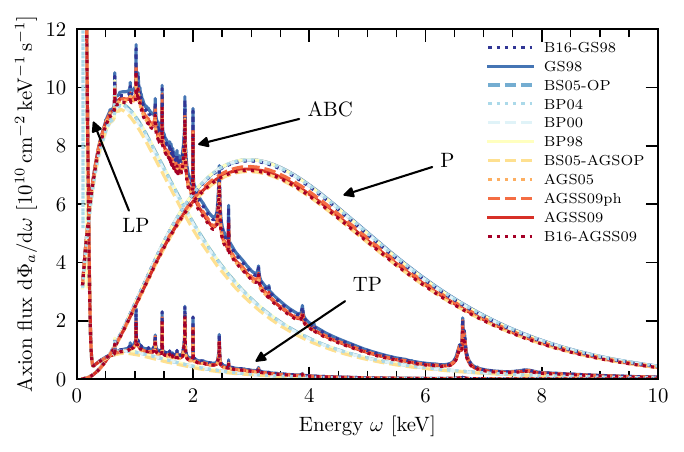}
	\caption{Overview of solar axion fluxes for different solar models.  There are  shown the ABC fluxes for $g_{ae} = 10^{-12}$ and the Primakoff~(P) as well as combined LP and TP fluxes for $g_{a\gamma} =  10^{-10} \,\ {\textrm{GeV}^{-1}}$. (Figure taken from \cite{Hoof:2021mld} with permission). \label{fig:solarabc}
}
\end{figure}

\emph{ABC processes--} If axions couple to electrons at tree level, which is the case of non-hadronic axion models such as the DFSZ~\cite{Zhitnitsky:1980tq,Dine:1981rt}, they can be produced in the Sun through Atomic transitions, Bremsstrahlung and Compton processes, collectively known as ABC processes (see Sec.~\ref{sec: electroncoupl}). In some allowed regions of the parameter space, the ABC processes can dominate over the Primakoff production. From  Fig.~\ref{fig:solarabc}, it results that as in the Primakoff case, most of the ABC axion fluxes are produced in the Sun core. Notably, bremsstrahlung on hydrogen and helium nuclei dominates the emission of low-energy axions, axio-recombination of metals (mostly O, Ne, Si, S and Fe) contributes sizably at intermediate energies and Compton takes over at higher energies. The contribution of axio-deexcitation is dominated by Lyman transitions (mostly Ly$-\alpha$) and is significant only in the case of iron which dominates the axion flux around $\sim 6.5$~keV. In Tab.~\ref{tab:parametersfit} we provide the fitting parameters for the bremsstrahlung and Compton solar axion fluxes at Earth, including the relative uncertainties. For these processes, the resulting integrated axion flux and luminosity are
\begin{equation}
    \begin{split}
        \Phi_{a,B}&=(2.01 \pm 0.01)\times10^{11}\left(\frac{g_{ae}}{10^{-12}}\right)^{2}~{\rm s}^{-1}~{\rm cm}^{-2}\,,\\
         L_{a,B}&=( 3.81\pm0.05 )\times10^{-4}\left(\frac{g_{ae}}{10^{-12}}\right)^{2}~L_{\rm \odot}\,,\\          \Phi_{a,C}&=( 1.08\pm 0.02)\times10^{11}\left(\frac{g_{ae}}{10^{-12}}\right)^{2}~{\rm s}^{-1}~{\rm cm}^{-2}\,,\\
         L_{a,C}&=( 6.4\pm 0.1)\times10^{-4}\left(\frac{g_{ae}}{10^{-12}}\right)^{2}~L_{\rm \odot}\,.\\
    \end{split}
\end{equation}
The bremsstrahlung flux includes both electron-ions and electron-electron contributions. The most accurate evaluation has been presented in Ref.~\cite{Hoof:2021mld}. These fluxes are shown in Fig.~\ref{fig:solarabc}. In this Figure it was included also the emission of axions in the electron capture by an ion, dubbed ``axio-recombination'' (free-bound transition) and atomic ``axio-deexcitation'' (bound-bound transition), which can lead   to an increase the total flux by a factor of ${\mathcal O}$(1).

\subsubsection{Axion-nucleon coupling}
\label{sec:solnucl}
\label{sec:Axion-nucleon_coupling}

The axion-nucleon coupling would also originate a solar axion flux~\cite{Raffelt:1982dr}. In these cases, however, thermal production mechanisms such as the nuclear bremsstrahlung process, relevant in PNS (see Sec.~\ref{sec:SN_NS}), are inhibited because of the low temperature of the Sun. However, as discussed in Sec.~\ref{ref:nucleoncoupling} there are two noteworthy mechanisms for the nuclear production of axions in the Sun: nuclear fusion and thermal excitation and subsequent de-excitation of the nuclei of stable isotopes.  Axions produced in nuclear fusion and nuclear de-excitation processes typically carry an energy from $\sim 10$ keV  to  $\sim$\,MeV. A recent comprehensive analysis of the axion nuclear processes relevant in the Sun can be found in Ref.~\cite{Massarczyk:2021dje}. Here, we summarize briefly the solar nuclear processes producing axions which have been considered so far.  

For what follows, it is useful to remember that the effective nucleon coupling is~\cite{DiLuzio:2021qct} (see also  Eq.~\eqref{eq:Gamma_Ratio_general})
\begin{align}
\label{eq:}
g_{aN}^{\rm eff}=\beta \, g_{aN}^{0} + g_{aN}^{3}
= \left(\frac{\beta+1}{2}\right)g_{ap}+ \left(\frac{\beta-1}{2}\right)g_{an} \,\ .
\end{align}

\emph{pp-chain:}
Most of the energy in the Sun is produced through the pp chain [see Eq.~\eqref{eq:Hburn}]. Namely,
\begin{equation}
\begin{aligned}
p~p \rightarrow d &~e^{+}~\nu_e \\
 \downarrow & \\
p~d & \rightarrow{ }^3 {\rm He}~\gamma(5.49 \,\ {\rm MeV})\,.
\end{aligned}
\end{equation}
According to the SSM, in the first step of the pp-chain $99.7 \%$ of deuterium is produced after the fusion of two protons, $p~p \rightarrow d~e^{+}~\nu_e$, and the remaining $0.3 \%$ via the $p~p~e^{-} \rightarrow d~\nu_e$ process. Practically, every single deuterium produced in such a way ends up capturing a proton, undergoing the reaction $p~d \rightarrow{ }^3 {\rm He}~\gamma$ on a time scale of ${\rm O}(1 {\rm ~s})$. Though this is not the only deuterium reaction allowed in the Sun, the low relative abundance of deuterium with respect to protons makes reactions such as $d~d \rightarrow p~ { }^3 {\rm H}$ or $d~d \rightarrow n~{ }^3{\rm He}$ extremely unlikely. Consequently, for all practical purposes the SSM predicts one neutrino and one photon for each deuterium nucleus produced in the first stage of the $p p$ chain, $\Phi_{\gamma p p}=\Phi_{\nu p p}$.

Concerning the axion production, the second step of the reaction, $p~d \rightarrow{ }^3 {\rm He}~\gamma(5.5 \,\ {\rm MeV})$, is particularly relevant as occasionally an axion can substitute for the photon. The resulting axion flux on Earth is~\cite{Borexino:2012guz}
\begin{align}
\label{eq:}
\phi=3.23\times 10^{10}\, \left( g_{aN}^{3} \right)^2 {\rm cm^{-2} s^{-1}} \,\ .
\end{align}
This flux has been studied by  the neutrino experiments Borexino, Jiangmen Underground Neutrino Observatory (JUNO), SNO and by the CERN Axion Solar Telescope (CAST).

\emph{Lithium Decay:}
Consists in the decay of the excited Lithium $^7{\rm Li}^* \to\, ^7{\rm Li}~a\,({\rm 0.478\, \,\ MeV})$ following from $^7$Be electron capture ($^7{\rm Be}~e^- \to\, ^7{\rm Li}^* ~\nu_e$)~\cite{Krcmar:2001si,CAST:2009klq}
\begin{equation}
\begin{aligned}
{ }^7 {\rm Be}~e \rightarrow { }^7 &{\rm Li}^*~ \nu_e \\
& \downarrow \\
{ }^7 &{\rm Li}^* \rightarrow{ }^7 {\rm Li}~\gamma(477.6 \,\  {\rm keV})
\end{aligned}
\end{equation}
with the axion substituting for the photon. In this case $\beta =1$, $\eta =0.5$ (pure proton coupling). The flux is 
\begin{align}
\label{eq:}
\phi=5.23\times 10^{8}\, \left( g_{aN}^{0}+g_{aN}^{3} \right)^2 {\rm cm^{-2} s^{-1}} \,\ .
\end{align}
It was studied by the CAST~\cite{CAST:2009klq} and Borexino~\cite{Borexino:2008wiu} collaborations. A direct detection in laboratory through the inverse resonant process $a~{ }^7 {\rm Li} \rightarrow{ }^7 {\rm Li}^* \rightarrow{ }^7 {\rm Li}~\gamma(477.6~ {\rm keV})$ was proposed in Ref.~\cite{Krcmar:2001si}. The most restrictive limit from this channel are found from resonance excitation of the ${ }^7 {\rm Li}$ nuclei in LiF crystals~\cite{Belli:2012zz}.

\emph{Decay of $^{57}{\rm Fe}$:}
The $^{57}{\rm Fe}$ iron isotope (relative abundance 2.2\%) has a low threshold excited state, $14.4$~keV, which decay through a M1 transition
\begin{align}
\label{eq:}
 ^{57}{\rm Fe}^* \to\, ^{57}{\rm Fe}~\gamma\,({\rm 14.4\,keV}) \,\ .
\end{align}
As the transition is magnetic, an axion may substitute the photon. Notice that the matrix elements for the this transition have recently been revised in Ref.~\cite{Avignone:2017ylv}, and show a $\sim 30\%$ increase in the axion emission rate (see also Ref.~\cite{DiLuzio:2021qct}). The updated axion flux on Earth is 
\begin{align}
\label{eq:}
\phi=5.06 \times 10^{23}\ (g_{aN}^{{\rm eff}})^2 \ \rm{cm}^{-2}\rm{s}^{-1} \,\ ,
\end{align}
where $g_{aN}^{{\rm eff}}=0.16\, g_{ap} +1.16\, g_{an}$. 
The flux is considerably large, compared to other nuclear production processes, due in part to the relative abundance of this isotope in the Sun and to the low threshold (see Tab.~\ref{tab:isotopes}).

\emph{Decay of $^{83}{\rm Kr}$:}
This is an M1 transition process, similar to the decay of $^{57}{\rm Fe}$
\begin{align}
\label{eq:}
^{83}{\rm Kr}^* \to\, ^{83}{\rm Kr} ~ a\,({\rm 9.4\,keV})
\end{align}
where the axion is emitted in place of a photon. $^{83}{\rm Kr}$ has an even lower excitation threshold than $^{57}{\rm Fe}$, at 9.4~keV. However, the $^{83}{\rm Kr}$ abundance in the Sun is considerably lower and this leads to a much smaller flux (Cf. Tab~\ref{tab:isotopes}). The nuclear coupling in this case is characterized by $\beta=-1$ (essentially, a coupling only to neutrons). Additionally, $\eta=0.5$. According to Ref.~\cite{DiLuzio:2021qct}, the flux is very small, see also Tab.~\ref{tab:isotopes}. Experimentally, this transition can be searched through the resonance absorption reaction $a~{ }^{83} \mathrm{Kr} \rightarrow{ }^{83} \mathrm{Kr} * \rightarrow{ }^{83} \mathrm{Kr}~\gamma, e(9.4$ $\mathrm{keV})$, using a proportional gas chamber filled with krypton and placed in a low-background~\cite{Gavrilyuk:2015aea,Akhmatov:2018kjv}.

\emph{Thulium Decay:}
The most recent case studied is the M1 decay of thulium 169 from an excited state, at 8.4 keV~\cite{Derbin:2023yrn}: 
\begin{align}
\label{eq:}
^{169}{\rm Tm}^* \to\, ^{169}{\rm Tm}~a\,({\rm 8.4 \,keV}) \,\ .
\end{align}
Though the flux is small, the advantage of this process is that it is a nearly pure proton $\mathrm{M} 1$ transition with $\beta\simeq 1$. 

\subsection{Solar bounds on axions}

\label{sec: bounds}

Given its low density and, more importantly, its proximity, the Sun provides a good environment to test the axion-photon coupling. The solar structure is very well know,  revealed by helioseismology and solar neutrinos, and well determined by modern solar models, such as~\cite{Vinyoles:2016djt}. In general, as we shall see, the bounds derived from the solar properties are not as strong as the ones derived from other stars, e.g. HB stars in GCs (see Sec.~\ref{sec:GC}). Nevertheless, the fact that we know the Sun better than any other star presents many advantages and, in general, represents a platform to test new methods. For example, the Sun is the only star for which we have direct detection of neutrinos from nuclear processes and this may serve as a tool to constraint new physics. 

\begin{figure}[t]
	\centering 
\includegraphics[width=15.0cm]{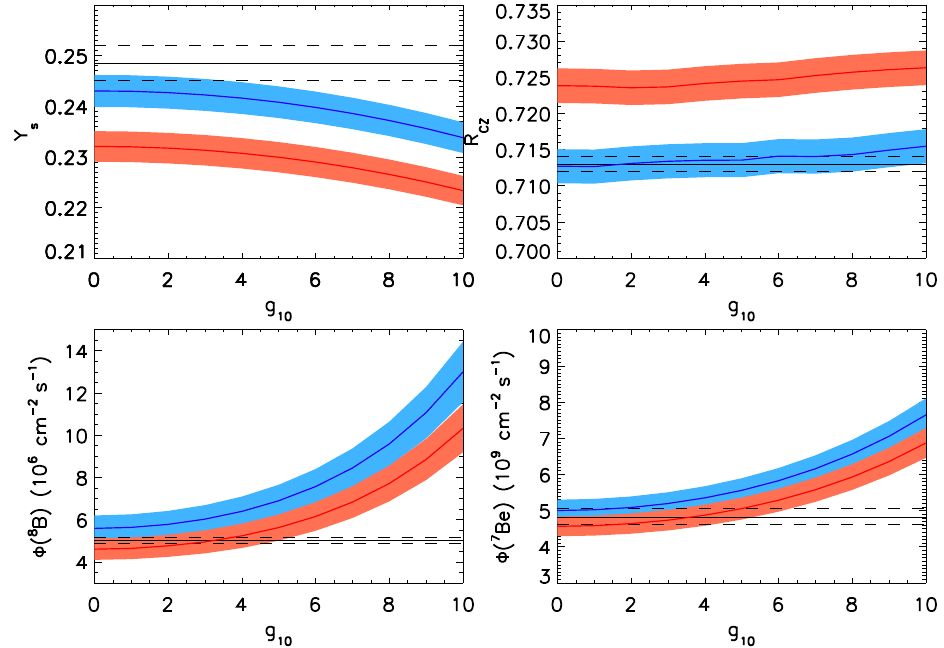}
        \caption{Evolution of the model parameters:
        surface helium abundance  $Y_S$, the convective radius  $R_{CZ}$, and solar neutrino fluxes $\rm{\Phi(^7Be)}$, $\rm{\Phi(^8B)}$ as a function of the axion-photon coupling constant $g_{10} \equiv g_{a\gamma}/10^{-10} {\rm GeV}^{-1}$. The red lines correspond to AGSS09 abundances and the blue ones to the GS98 abundances. Black lines represent the observational value with the errors and the shaded zones show model errors. (Figure taken from Ref.~\cite{Vinyoles:2015aba} with permission).  }\label{fig:obsmodelax}
\end{figure} 

Solar constraints on axions have been generally obtained by applying limits to variations of either neutrino fluxes~\cite{Schlattl:1998fz,Gondolo:2008dd} or the sound speed profile derived from helioseismology~\cite{Schlattl:1998fz}. An upper limit for $g_{a\gamma}$ was derived in Ref.~\cite{Schlattl:1998fz} considering the deviation that axions can impart to the solar model sound speed at a given depth in the Sun, corresponding to $L_{a}\lesssim 0.20\,L_\odot$. Furthermore, the energy loss by solar axion emission requires enhanced nuclear burning and thus a somewhat increased temperature in the Sun, producing an increase of the solar $^8$B and $^7$Be neutrino fluxes~\cite{Schlattl:1998fz}, affecting their measurement in the SNO experiment~\cite{Gondolo:2008dd}. Recently, in Ref.~\cite{Vinyoles:2015aba} it was performed a global analysis of the different observables described above in a systematic approach aimed at combining different sources of data accounting in detail for observation and theoretical errors. Fig.~\ref{fig:obsmodelax} shows the dependence of the solar neutrino fluxes $\rm{\Phi(^8B)}$ and $\rm{\Phi(^7Be)}$ and of the convective envelope properties $Y_S$ and $R_{CZ}$ on $g_{a\gamma}$. Red lines correspond to solar models implementing AGSS09 composition and blue ones to GS98. The shaded zones depict the $1\sigma$ theoretical errors. The black lines shows the experimental values with $1\sigma$ errors (see Ref.~\cite{Vinyoles:2015aba} for details). Combining the different observables, it was found a bound $g_{a\gamma} < 4.1  \times 10^{-10} {\rm GeV^{-1}}$  at a 3-$\sigma$ CL.

In Ref.~\cite{Vinyoles:2015aba}, it was observed that this global constraint corresponds to an axion contribution to the solar luminosity of only 3\%, resulting in a much more restrictive limit compared to the one derived from  $L_{a}\lesssim 0.20\,L_\odot$. The argument that an anomalous energy loss cannot exceed 3\% of the standard flux can be applied also to other axion emission processes, leading to some of  the bounds shown in Tab.~\ref{tab:solar_bounds}. There, besides the bounds on the axion-photon coupling discussed above, we consider the bound on the axion-electron coupling, from the ABC axion flux, and the bound on axions from the de-excitation of $^{57}$Fe~\cite{DiLuzio:2021qct}.

\begin{table}[]
\centering
\caption{
Summary of the solar bounds on the axion couplings. When available, we report the C.L. The references refer to the latest analysis.
($^\dag$):  For higher masses, the bound is Boltzmann suppressed;
($^\ddag$): At higher masses the bounds weaken because of loss of coherence in the axion-photon conversion.}
\label{tab:solar_bounds}
\renewcommand{\arraystretch}{1.8}
\begin{tabular}{|l|l|l|c|}
\hline
\textbf{Bound} &
  \textbf{Analysis} &
  \textbf{Notes} &
  \textbf{Ref.} \\ \hline
$g_{a\gamma}\leq 10\times 10^{-10}\,{\rm GeV^{-1}}$ &
  Helioseismology &
  \multirow{4}{*}{\begin{tabular}[c]{@{}l@{}}Cooling bounds; \\ $m_a\lesssim$ a few keV $^\dag$\end{tabular}} &
  \cite{Schlattl:1998fz} \\ \cline{1-2} \cline{4-4} 
$g_{a\gamma}\leq 7\times 10^{-10}\,{\rm GeV^{-1}}$ &
  \begin{tabular}[c]{@{}l@{}}$^8$B neutrino flux, compared \\ with SNO experimental data\end{tabular} &
   &
  \cite{Gondolo:2008dd} \\ \cline{1-2} \cline{4-4} 
$g_{a\gamma}\leq 4.1\times 10^{-10}\,{\rm GeV^{-1}}$ &
  \begin{tabular}[c]{@{}l@{}}Global analysis of the Sun, \\ with anomalous loss \\ $\leq$ 3\% of standard loss\end{tabular} &
   &
  \cite{Vinyoles:2015aba} \\ \cline{1-2} \cline{4-4} 
$g_{ae}\leq 6.6\cdot 10^{-12}$ &
  \begin{tabular}[c]{@{}l@{}}ABC axion production \\ with anomalous \\ loss $\leq$ 3\% \\ of standard loss\end{tabular} &
   &
  This work \\ \hline
$0.16 g_{a p}+1.16 g_{a n} \leq 1.9 \times 10^{-6}$ &
  \begin{tabular}[c]{@{}l@{}}Axion production from \\ $^{57}$Fe decay. \\ Assumes anomalous loss \\ $\leq$ 3\% of standard loss\end{tabular} &
  $m_a\leq$ 14.4 keV &
  \cite{DiLuzio:2021qct} \\ \hline
$g_{a\gamma}\leq 5.7 \times 10^{-11} \mathrm{GeV}^{-1}$ &
  CAST analysis (2024) &
  \begin{tabular}[c]{@{}l@{}}95\% C.L. \\ $m_a\leq$ 0.06 keV $^\ddag$ \end{tabular} &
  \cite{CAST:2024eil} \\ \hline
$g_{a\gamma}\leq 6.9 \times 10^{-12} \mathrm{GeV}^{-1}$ &
  NuSTAR analysis (2024) &
  \begin{tabular}[c]{@{}l@{}}95\% C.L. \\ $m_a \lesssim 2 \times 10^{-7} \mathrm{eV}$ $^\ddag$ \end{tabular} &
  \cite{Ruz:2024gkl} \\ \hline
$\frac12 |g_{ap}-g_{an}|\leq 2 \times 10^{-5}$ &
  \begin{tabular}[c]{@{}l@{}}Analysis of SNO data,\\ production $p~d \rightarrow{ }^3 \mathrm{He}~a(5.5 \mathrm{MeV})$\\ detection $a~d \rightarrow p ~n$\end{tabular} &
  95\% C.L. &
  \cite{Bhusal:2020bvx} \\ \hline
$|0.19 g_{a p}+2.19 g_{a n}|\leq 6.0 \times 10^{-6}$ &
  \multirow{4}{*}{\begin{tabular}[c]{@{}l@{}}Experimental analysis, \\production $\mathrm{X}^* \rightarrow \mathrm{X}\,a$\\ detection $a\,\mathrm{X} \rightarrow \mathrm{X}^* \rightarrow \mathrm{X}\,\gamma$\end{tabular}} &
  X=$^{57}$Fe; 95\% C.L. &
  \cite{Derbin:2011zz} \\ \cline{1-1} \cline{3-4} 
$|g_{a n}|\leq 8.4 \times 10^{-7}$ &
   &
  X=$^{83}$Kr; 95\% C.L. &
  \cite{Akhmatov:2018kjv} \\ \cline{1-1} \cline{3-4} 
$|g_{a p}|\leq 8.9 \times 10^{-6}$ &
   &
  X=$^{169}$Tm; 90\% C.L. &
  \cite{Derbin:2023yrn} \\ \cline{1-1} \cline{3-4} 
$|g_{a p}|\leq 5.7 \times 10^{-4}$ &
   &
  X=$^{7}$Li; 90\% C.L. & \cite{Belli:2012zz}

   \\ \hline
\end{tabular}
\end{table}

\subsection{Search of solar  axions with helioscopes}
\label{sec:helioscope}

\subsubsection{Basics and state-of-the-art}

The leading technique to detect solar axions is the axion \emph{helioscope}, one of the oldest concepts used to search for axions. The first idea of an axion helioscope was proposed in a seminal work by Pierre Sikivie~\cite{Sikivie:1983ip}, and a practical design for a detector was proposed in Ref.~\cite{vanBibber:1988ge}. The idea was that axion fluxes produced in the solar core, as shown in Fig.~\ref{fig:solarabc}, reconvert into X-ray photons in strong laboratory magnetic fields. The expected number of photons $\mathcal{N}_{\gamma}$ from axion conversion in a given detector is obtained by integrating the product of the differential axion flux with the conversion probability and the detection efficiency over the total range of energies 
\begin{equation}\label{ngamma}
	\mathcal{N}_{\gamma} =  \int_{\omega_{0}} ^{\omega_{f}}{\rm d}\omega\, \Big( \frac{{\rm d}\Phi_a}{{\rm d}\omega}\Big)_{\rm total}\, \mathcal{P}_{a\rightarrow \gamma}\, \epsilon \, S\, t \,\ ,
\end{equation}
where $S$ is the detection area perpendicular to the flux of axions, $t$ is the exposure time, and $\epsilon$ the detection efficiency. The axion-photon conversion probability in a transverse homogeneous magnetic field $B$ over distance $L$ is~\cite{Raffelt:1987im} [see Eq.~\eqref{a16}]
\begin{equation}\label{prob}
	P_{a \rightarrow \gamma} = \left(\frac{g_{a\gamma}B L}{2}\right)^2{\rm sinc}^2\left(\frac{q L}{2}\right)\,,
\end{equation}
where sinc$\,x=(\sin x)/x$ and $q$ is the momentum transfer provided by the magnetic field. For relativistic axions ($\omega_a\gg m_a$), this is $q\simeq (m^2-{\omega^2_{\rm pl}})/2\omega$~\cite{vanBibber:1988ge}. Assuming ${\omega_{\rm pl}}=0$ (conversion in vacuum), a coherent $a$-$\gamma$ conversion along the full magnetic length happens when
\begin{align}
\label{eq:}
\frac{m_a^2\,L}{4\,\omega}
=0.013 \left(\frac{L}{10\,{\rm m}}\right)
\left(\frac{m_a}{{\rm meV}}\right)^2
\left(\frac{\omega}{{\rm keV}}\right)^{-1}\ll 1\,,
\end{align}
i.e.\ when the momentum transfer is smaller
than about $1/L$. For a sufficiently large mass, the conversion is not coherent and the probability gets suppressed by a factor $\sim(4\omega/m_a^2 L)^2$. The loss of coherence can be compensated by adding a buffer gas which would guarantee a finite index of refraction ${\omega^2_{\rm pl}}\neq0$~\cite{vanBibber:1988ge}. Historically, solar axion searches focused primarily on the Primakoff production and thus placed bounds on $g_{a\gamma}$. Remarkably, in this case the expected number of photons $\mathcal{N}_{\gamma}$ scales as  $g_{a\gamma}^4$. The first-generation helioscope was a Brookhaven experiment with a length of $L=1.8$~m and a magnetic field of $B=2.2$T, which collected data for only a few hours~\cite{Lazarus:1992ry}. In the vacuum case, it set a $3\sigma$ bound of $g_{a\gamma} < 3.6 \times 10^{-9}$~GeV$^{-1}$ (at 95\% C.L.), for $m_a < 0.03$~eV, while with two gas pressure settings, it provided the bound $g_{a\gamma} < 7.7 \times 10^{-9}$~GeV$^{-1}$ (at 95\% C.L.) for $0.03 < m_a < 0.11$~eV. The second generation experiment was represented by the Tokyo Helioscope (SUMICO) with $L=2.3$~m and a magnet with $B=4$~T. This setup resulted in an improved upper limit $g_{a\gamma} < 6 \times 10^{-10}$~GeV$^{-1}$  (at 95 \% C.L.) in the mass range up to 0.03~eV~\cite{Moriyama:1998kd}.

\begin{figure*}
	\includegraphics[width=1.0\textwidth]{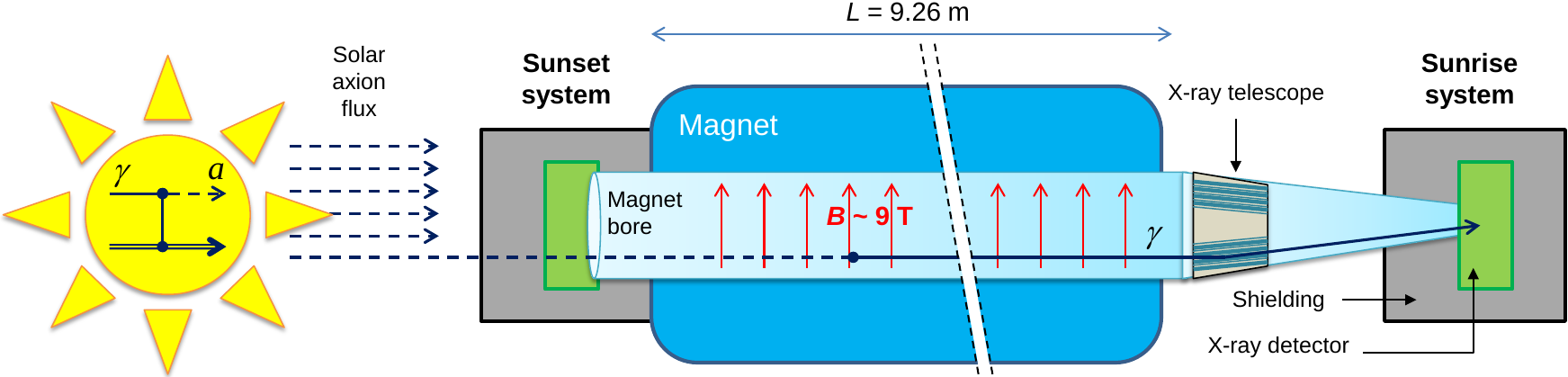}
	\caption{Sketch of the CAST helioscope at CERN to search for solar axions. These	hypothetical low-mass bosons are produced in the Sun by Primakoff scattering on charged particles and converted back to X-rays in the $B$-field of an LHC test magnet. The two straight conversion pipes have a cross section of 14.5~cm$^2$ each. The magnet can move by $\pm8^\circ$ vertically and $\pm40^\circ$ horizontally, enough to follow the Sun for about 1.5~h at dawn and dusk with opposite ends. Separate detection systems can search for axions at Sunrise and	Sunset, respectively. The Sunrise system is equipped with an X-ray telescope to focus the signal on a small detector area, strongly increasing signal-to-noise. (Figure taken from Ref.~\cite{CAST:2017uph} with  permission). 
	}\label{fig:cast}
\end{figure*}
\begin{figure}[t]
 \centering\includegraphics[width=0.8\textwidth]{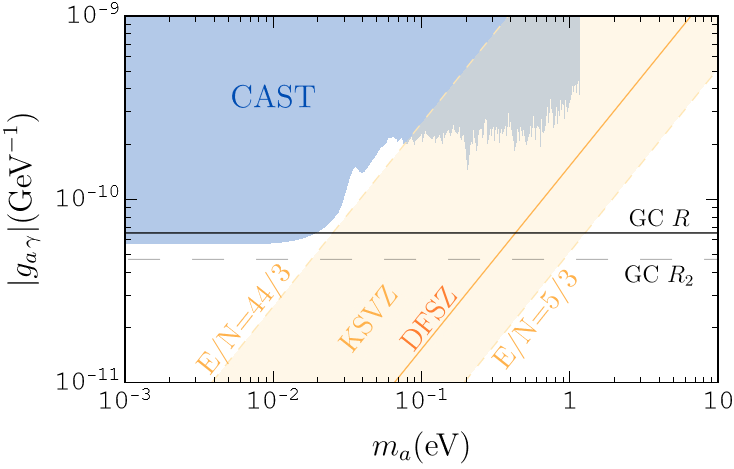}
	\caption{ 
 CAST excluded region at 95\%~C.L. (blue region) in the $m_a$--$g_{a\gamma}$--plane, from Ref.~\cite{CAST:2024eil} (low-mass region) and Ref.~\cite{CAST:2017uph} (high-mass region). Horizontal lines represent the astrophysical bounds from GCs observations from the $R$ (solid line) and the $R_2$ (dashed line) parameters  (see Sec.~\ref{sec:Rparam}). {\em Diagonal orange band:\/} Region expected for the benchmark KSVZ axion models with $g_{a\gamma}=(E/N-1.92)\,\alpha/(2\pi f_a)$ and $5/3\leq E/N\leq 44/3$~\cite{DiLuzio:2016sbl}. The line corresponding to the DFSZ axion model, $E/N=8/3$, is also shown [see Tab.~\ref{tab:QCD_axion_couplings}]. (Data for bounds taken from the repository~\cite{AxionLimits}).}\label{fig:limits}
\end{figure}

Finally the third generation experiment is represented by the CERN Axion Solar Telescope (CAST) with $L=10$ m and $B=9$ T (see Fig.~\ref{fig:cast}). Since 2003, CAST has explored the $m_a$--$g_{a\gamma}$ parameter space. In the regime of coherent conversion the axion-photon conversion probability can be expressed as ~\cite{Sikivie:1983ip,Zioutas:2004hi,Andriamonje:2007ew}:
\begin{equation}
     P_{{\rm a}\to\gamma} =2\times10^{-21}\left(\frac{g_{a\gamma}}{10^{-12}~{\rm GeV}^{-1}}\right)^{2}\left(\frac{L}{10~\rm m}\right)^2
 \left(\frac{B}{9.0~\rm T}\right)^2\,,\\
    \label{eq:helio_conversion_prob}
\end{equation}
giving a differential flux of photons of  
\begin{eqnarray}\label{eq5}
 \frac{{\rm d}\Phi_{\rm a}}{{\rm d}E}\,P_{{\rm a}\to\gamma} =0.1~{\rm cm^{-2}~day^{-1}~keV^{-1}}\left(\frac{g_{a\gamma}}{10^{-10}~{\rm GeV}^{-1}}\right)^{4}\left(\frac{E}{\rm keV}\right)^{2.531}e^{-\frac{E}{1.181~{\rm keV}}}\left(\frac{L}{10~\rm m}\right)^2
 \left(\frac{B}{9.0~\rm T}\right)^2\,.
\end{eqnarray}
The CAST experiment during almost twenty years has explored different ranges of the the axion mass. Specifically, the low-mass part $m_a < $0.02~eV corresponds to the first phase 2003--2004 using evacuated magnet bores~\cite{Zioutas:2004hi,Andriamonje:2007ew}. For the setup of the experiment, coherence is lost for $m_a\gtrsim 0.02~{\rm eV}$, explaining the loss of sensitivity for larger $m_a$. The higher-mass range was explored  by filling the conversion pipes with $^4$He~\cite{Arik:2008mq, Arik:2015cjv} and $^3$He~\cite{Arik:2011rx,Arik:2013nya} at variable pressure settings to provide photons with a refractive mass and in this way match the $a$ and $\gamma$ momenta. The sensitivity is smaller because at each pressure setting, data were typically taken for a few hours only. Despite this limitation, CAST has reached realistic QCD axion models and has superseded previous solar axion searches using the helioscope~\cite{Moriyama:1998kd} and Bragg scattering technique~\cite{Paschos:1993yf,Bernabei:2001ny}. Notably, the strongest bound with this technique, $g_{a\gamma} < 1.45 \times 10^{-9}$~GeV$^{-1}$, has been placed by the MAJORANA experiment~\cite{Majorana:2022bse}. The CAST data were also interpreted in terms of other  axion production channels in the Sun~\cite{Andriamonje:2009ar,Andriamonje:2009dx,Barth:2013sma}. Moreover, CAST constraints on other low-mass bosons include chameleons~\cite{Anastassopoulos:2015yda} and hidden photons~\cite{Redondo:2015iea}. During this long experimental program, CAST has used a variety of detection systems at both magnet ends, including a multiwire time projection chamber~\cite{Autiero:2007uf}, several Micromegas detectors~\cite{Abbon:2007ug}, a low-noise charged coupled device attached to a spare X-ray telescope from the ABRIXAS X-ray mission~\cite{Kuster:2007ue}, a $\gamma$-ray calorimeter~\cite{Andriamonje:2009ar}, and a silicon drift detector~\cite{Anastassopoulos:2015yda}. In the latest data taking campaign (2013--2015), CAST has returned to evacuated pipes, improving the sensitivity to solar axions of about a factor $\sim$3 in signal-to-noise ratio over a decade ago, thanks to the development of novel detection systems, notably new Micromegas detectors with lower background levels, as well as a new X-ray telescope built specifically for axion searches. In this phase CAST placed the best bound on axion-photon coupling, notably $	g_{a\gamma}< 6.6 \times 10^{-11} \,\ \textrm{GeV}^{-1}$, comparable with the bound on energy-loss from HB stars in GCs (see Sec.~\ref{sec:GC}). More recently, in 2024 CAST presented a new analysis based on an extended  run with a Xe-based Micromegas detector~\cite{CAST:2024eil}, improving the previous bound down to
\begin{equation}
g_{a\gamma}< 5.7 \times 10^{-11} \,\ \textrm{GeV}^{-1} \,\ ,
\end{equation}
the most restrictive experimental limit to date. The envelope of the best CAST results described above  are shown by the blue  band in Fig.~\ref{fig:limits}, compared with the astrophysical bounds from HB stars in GCs (see Sec.~\ref{sec:Rparam}). Remarkably, it has been shown that the CAST bound can be relaxed in models where the axion-photon coupling is suppressed in matter due to environmental effects (see, e.g., Refs.~\cite{Jaeckel:2006xm,Pallathadka:2020vwu}) or in string-inspired models where multiple ultralight axions 
are present (see, e.g., Refs.~\cite{Chadha-Day:2021uyt,Chadha-Day:2023wub}).

CAST has also constrained combinations of axion couplings. 
In particular, in case of DFSZ axions assuming the ABC emission processes, in Ref.~\cite{Barth:2013sma} it was derived a bound (at 95 \% CL) on 
\begin{equation}
	g_{a \gamma} \times g_{ae} < 8.1 \times 10^{-23} \,\ \textrm{GeV}^{-1} \,\ ,
\end{equation}
for $m_a \lesssim 10$~meV.

\subsubsection{Helioscope searches for non-thermal axions}
\label{sec:Helioscope_searches__non-thermal_axions}

Helioscopes can, in principle, also be equipped with a $\gamma$-ray detector to look for  axions produced in the $p ~ d\to$ $^{3}{\rm He} ~ a$\,(5.5\,MeV) reaction (see Sec.~\ref{sec:solnucl}), which provides one of the most intense axion fluxes from nuclear reactions. For example, the CAST helioscope installed a $\gamma$-ray calorimeter for some time to gain sensitivity to these high-energy axions~\cite{CAST:2009klq}. At this regard, Ref.~\cite{CAST:2009klq} estimated an axion flux of the order of $10^{10}  (g^{3}_{aN})^{2} /({\rm cm^2s})$ for the reaction mentioned above. This is quite small for reasonable values of the axion-nucleon coupling. Notice that the axion-nucleon coupling can induce axion absorption after the axiodissociation of nuclei $a\,~\,Z \rightarrow Z_1\,~\,Z_2$. Axions with energy 5.5 MeV can dissociate $^{17}$O, $^{13}$C and $^2$H. It has been shown that couplings  $g_{3aN}\lesssim 10^{-3}$ are required for axions not to be trapped inside the Sun~\cite{Raffelt:1982dr}. 

Another -- perhaps more promising -- direction to search for axions coupled to nucleons is to look for low lying nuclear excitations of stable isotopes with a significant abundance inside the Sun and that can be thermally excited. As discussed in Sec.~\ref{sec:Axion-nucleon_coupling}, the largest flux in this case is expected from the decay of $^{57}$Fe, which may produce an axion line at 14.4~keV~\cite{Moriyama:1995bz,Krcmar:1998xn}. This line has been searched by CAST~\cite{CAST:2009jdc}, which found the bound 
\begin{equation}
	|g_{\rm{a}\gamma}\times g_{\rm{aN}}^{\rm eff}|<1.36\times 10^{-16} \,\ \textrm{GeV}^{-1} \,\ ,
\end{equation}
for $m_{\rm{a}}<0.03$~eV at the 95\% confidence level, where $g_{a N}^{\mathrm{eff}}=0.16 g_{a p}+1.16 g_{a n}$, as discussed in Sec.~\ref{sec:Axion-nucleon_coupling}. The potential of the next generation of axion helioscopes, for various configurations of the optics and detectors, was studied in Ref.~\cite{DiLuzio:2021qct}.\footnote{Notice that the CAST analysis~\cite{CAST:2009jdc} and the sensitivity study for Baby-IAXO use slightly different matrix elements for the $^{57}$Fe transition. Specifically, the last study used the updated matrix elements estimated in~\cite{Avignone:2017ylv}. } Additional experimental results on the detection of solar axions from nuclear transitions are presented in Tab.~\ref{tab:solar_bounds}.

\subsubsection{Next-generation helioscopes}

\begin{figure}[!t]
\begin{center}
\includegraphics[height=10cm] {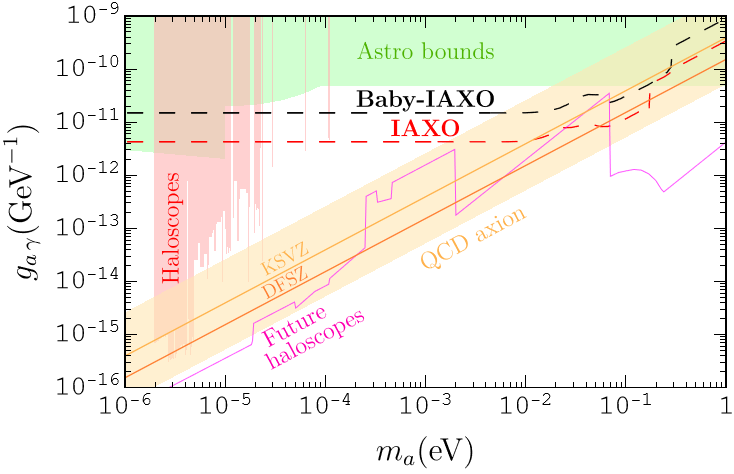}
    \caption{Sensitivity prospects of Baby-IAXO and IAXO experiments in the overall context of other experimental and observational bounds. (Data for bounds taken from the repository~\cite{AxionLimits}).  }
    \label{fig:excluion_iaxo}
\end{center}
\end{figure}

There is currently an intense experimental activitiy to define the detection technologies suitable for the proposed much larger next-generation axion helioscope International Axion Observatory (IAXO)~\cite{Armengaud:2014gea}. IAXO is  currently at the design state, and it aims at improving the CAST signal-to-noise by more than a factor of $10^4$. This corresponds to more than an order of magnitude sensitivity in $g_{a\gamma}$~\cite{IAXO:2019mpb} (see Fig.~\ref{fig:excluion_iaxo}). IAXO will use a X-ray focusing optics and will have as a central component a superconducting magnet with a multibore configuration, to produce an intense magnetic field over a large volume. IAXO is expected to probe a large fraction of unexplored axion parameter space, including the region corresponding to QCD axion models in the mass range $m_a \sim 1 \,\ \textrm{meV}$ -- 1 eV. As an intermediate step towards IAXO, the collaboration aims at building Baby-IAXO, which would allow to obtain results at an intermediate level between CAST current best limit and the future IAXO.  The expected sensitivity of Baby-IAXO and IAXO in the overall context of other experimental and observational bounds is shown in Fig.~\ref{fig:excluion_iaxo}.

The detection of solar axions with IAXO would allow the use of these particles as a new probe of the Sun, complementary with neutrinos. Indeed, as shown in Ref.~\cite{Jaeckel:2019xpa}, the solar axion spectrum would give information about the solar interior and might help solving the conflict between high and low metallicity solar models. Furthermore, it might allow the reconstruction of the solar temperature profile~\cite{Hoof:2023jol}. Of course, for this purpose an accurate determination of the uncertainties in the solar axion flux is mandatory~\cite{Hoof:2021mld}.  

Finally, Ref.~\cite{OHare:2020wum} proposed that the detection of low-energy axion flux produced by LP conversions in the solar magnetic fields (see Fig.~\ref{fig:plasmons}) may constitute a new probe to track for the $B$-field. Remarkably, if the detector technology eventually installed in IAXO has an energy resolution better than 200~eV, then solar axions could probe the magnetic field in the core of the Sun. For energy resolutions better than 10~eV, IAXO could access the inner 70 \% of the Sun and begin to constrain the field at the tachocline.

\subsection{Search of solar axions with other detectors}
\label{sec:otherdec}

\subsubsection{Search of solar axions in dark matter detectors}
\label{sec:axiondm}

Axions could also interact with matter via their coupling with electrons or with nucleons. For example, solar axions could produce visible signals in ionization detectors by virtue of the axioelectric effect~\cite{Ljubicic:2004gt,Derbin:2011gg,Derbin:2011zz,Derbin:2012yk,Bellini:2012kz}, through the $g_{ae}$ coupling. Large liquid Xe detectors, aiming at the detection of dark matter  WIMPs, like XMASS~\cite{Abe:2012ut}, XENON~\cite{Aprile:2014eoa}, PANDAX-II~\cite{Fu:2017lfc}, LUX~\cite{Akerib:2017uem}, XENON 1T~\cite{XENON:2020rca}, XENON nT~\cite{XENON:2022ltv} have all performed this search as a byproduct of their experiments. The latter has produced the most competitive result to date, setting an upper bound $g_{ae} < 2.0\times 10^{-12}$ (90\% C.L.). However this value is still considerably larger than the limit from astrophysics, presented in Sec.~\ref{sec:white_dwarfs}.

Even assuming no electron coupling, the solar Primakoff flux can still trigger a signal in a dark matter detector, through keV-photons from inverse Primakoff production~\cite{Gao:2020wer,Dent:2020jhf,Abe:2020mcs}. In this context, the XENON nT~\cite{XENON:2022ltv} experiment has placed a bound $g_{a \gamma} < 4.8\times 10^{-10}$~GeV$^{-1}$. More exotic scenarios have also been considered. For example, in Ref.~\cite{DiLella:2000dn} it was explored the physics potential of a dark matter underground detector for observing axionic Kaluza-Klein excitations coming from the Sun within the context of higher-dimensional theories of low-scale quantum gravity. In these theories, the heavier Kaluza-Klein axions are relatively short-lived and may be detected by a coincidental triggering of their two-photon decay mode. The bound placed is $g_{a \gamma} < 2.5\times 10^{-11}$~GeV$^{-1}$ and $g_{a \gamma} < 1.2\times 10^{-11}$~GeV$^{-1}$ for theories with 2 and 3 large compact dimensions, respectively.

\subsubsection{Search of solar axions in underground neutrino detectors}
\label{sec:neutrin}

\begin{figure}[!t]
\centering
\includegraphics[width=0.7\textwidth]{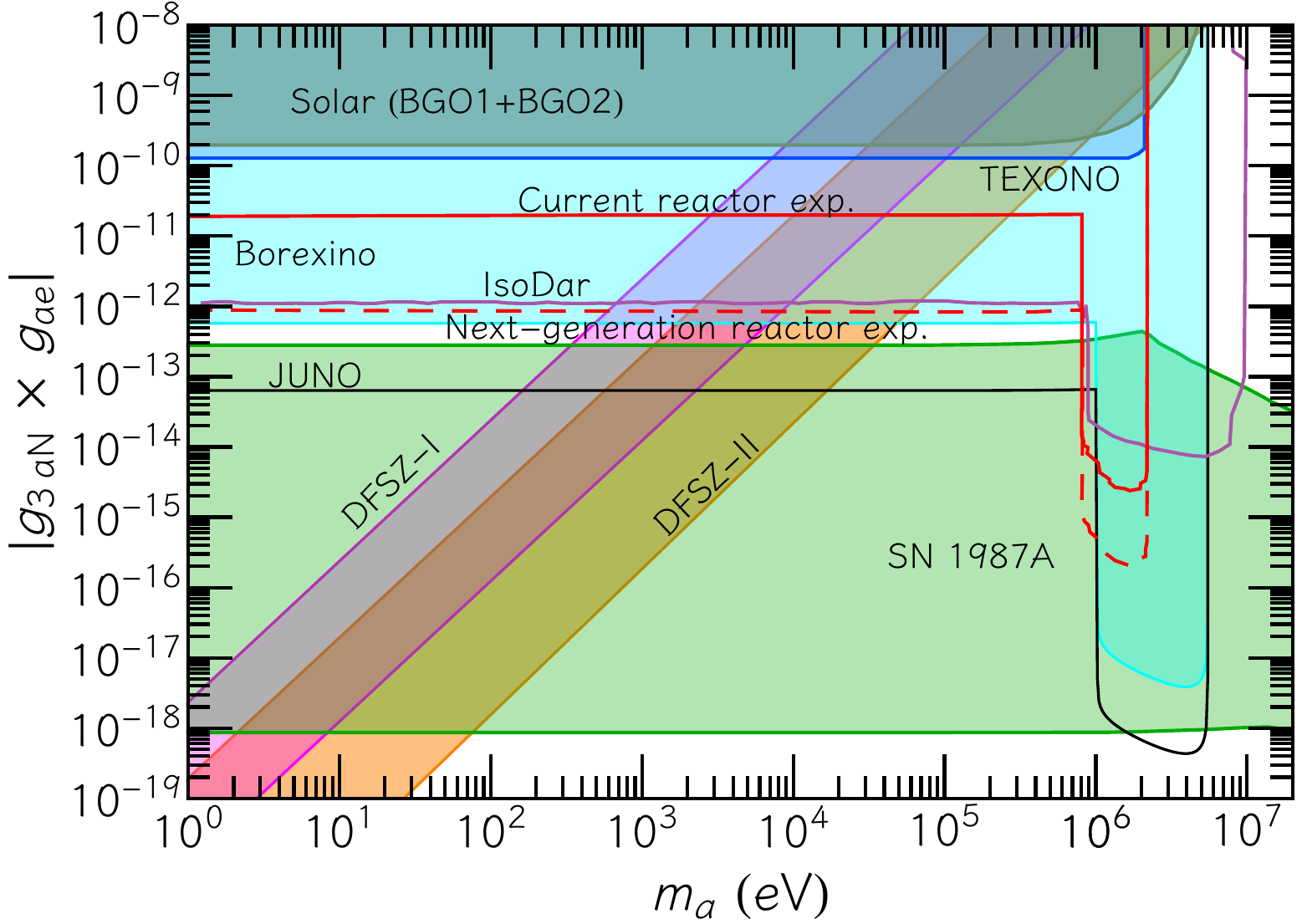}
\caption{\footnotesize  Exclusion region plot in the ($|g_{3aN} \times g_{ae}|$, $ m_a $) plane at 90\% C.L. 
The solid black line represents the JUNO sensitivity.
(Figure taken from Ref.~\cite{Lucente:2022esm} with permission) 
}
\label{fig:gAegA3NVaryMa}
\end{figure}

The axion flux from the $p ~ d\to$ $^{3}{\rm He} ~ a$\,(5.5\,MeV) reaction, which provides one of the most intense axion fluxes from nuclear reactions, has been experimentally searched for by the underground neutrino experiment Borexino~\cite{Borexino:2012guz}. More recently, it was shown that the neutrino experiment JUNO would have the potential to probe regions unexplored by Borexino (or other terrestrial experiments) and thus to push the axion bound further~\cite{Lucente:2022esm} (see Fig.~\ref{fig:gAegA3NVaryMa}). The same flux has also been scrutinized in the SNO data~\cite{Bhusal:2020bvx} for traces of these 5.5~MeV axions, providing a bound $| g_{aN}^{3}|<2 \times 10^{-5}(95 \% \text { C.L.})$.

\subsubsection{Search of   axions in X-ray  telescopes}
\label{sec:Xrays}
In a seminal paper~\cite{Carlson:1995xf}, it was proposed to search axions produced in the Sun core by Primakoff process, looking for their conversions into X-rays in the presence of the Sun's external magnetic dipole field and sunspot-related magnetic fields. Using measurements by such as the Soft X-ray telescope (SXT) on the Yohkoh satellite, it was placed a bound $g_{a \gamma} \lesssim 1.0\times 10^{-10}$~GeV$^{-1}$  for $m_a \lesssim 7 \times 10^{-6}$~eV. A comparison of sensitivity of different X-rays experiments was presented in Ref.~\cite{Hudson:2012ee}. Recently, the solar axion flux has been studied by a dedicated observation with the Nuclear Spectroscopic Telescope Array (NuSTAR)~\cite{NuSTAR:2013yza,Grefenstette:2016esh,Ruz:2024gkl}.

In Ref.~\cite{Davoudiasl:2005nh,Davoudiasl:2008fy} it was proposed to search for solar axions by looking for their conversions into X-rays in the Earth's magnetosphere with a low-Earth-orbit X-ray detector. It was estimated the sensitivity down to $g_{a \gamma} \lesssim 1.0\times 10^{-11}$~GeV$^{-1}$  for $m_a \lesssim  10^{-4}$~eV. Ref.~\cite{DeRocco:2022jyq} pointed out that a small fraction axions with keV masses, produced by Primakoff process, with speeds below the escape velocity, would accumulate over time, creating a basin of slow-moving axions trapped on bound orbits. These axions could decay to two photons, yielding an observable signature. Limits  on the solar basin of axions were set using recent quiescent solar observations made by the NuSTAR X-ray telescope. 
However, in Ref.~\cite{Langhoff:2022bij} it was remarked that strong constraints can be placed on axions produced in the early universe, decaying into photons, severely limiting the parameter space allowed for the solar basin phenomenology.

\subsection{Axions from nearby stars}
\label{sec:axionnearby}

\subsubsection{Axions from other Galactic stars}
\label{sec:axionfromstar}

\begin{figure}
   \centering
  \includegraphics[width=0.8\linewidth]{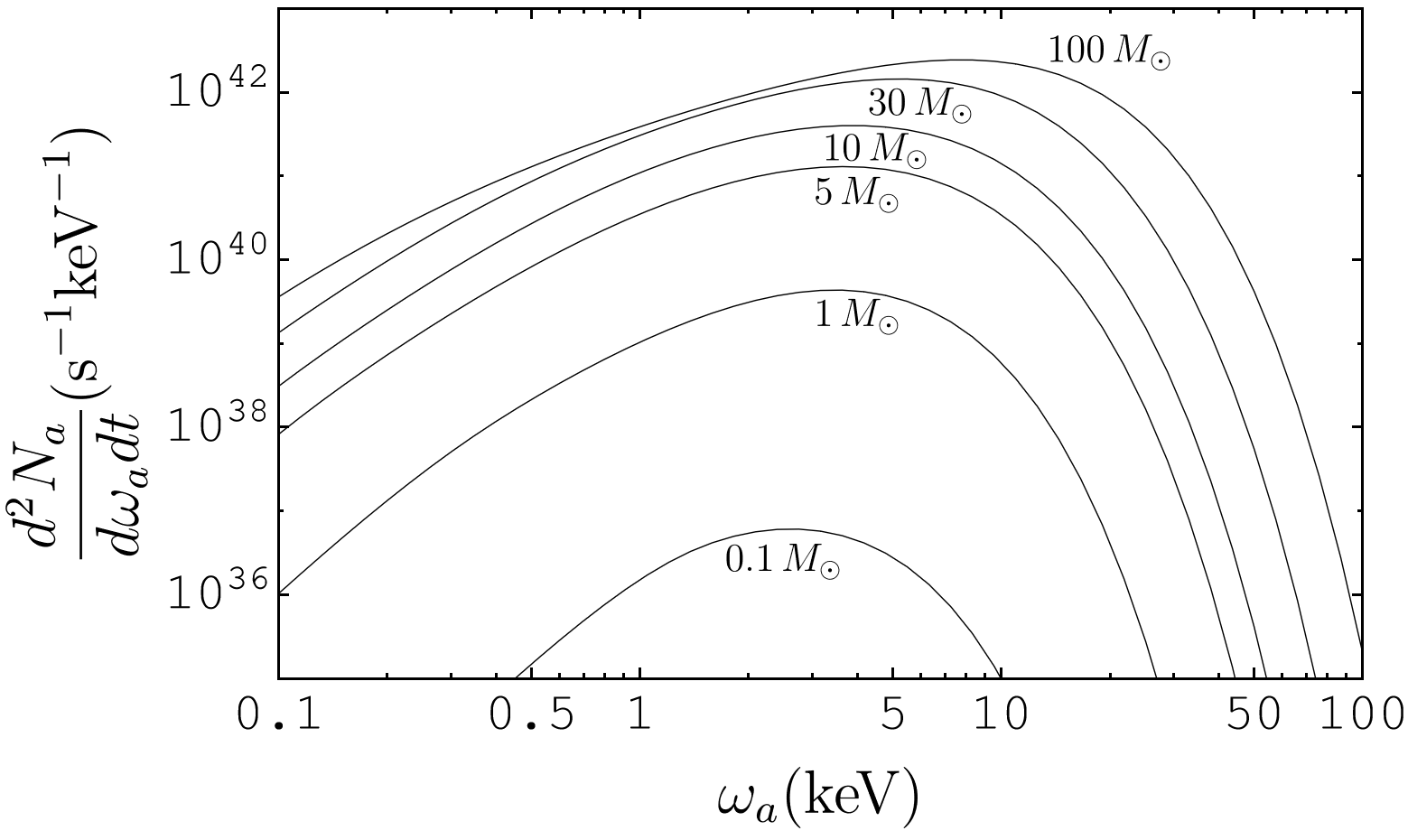}
  \caption{Number of axions  per unit energy and time produced at the source. Here we show single MS stars of different masses $M$, for axions with $g_{a\gamma}=0.65\times10^{-10}{\rm GeV}^{-1}$ and  $m_a\ll {\rm keV}$.  }
  \label{fig:dLadEa}
\end{figure}

Similarly to the Sun other MS stars can produce axions through the $g_{a\gamma}$ coupling~\cite{Lopes:2021mgy,Nguyen:2023czp}. A recent work~\cite{Nguyen:2023czp} characterized the axion flux for 34 representative MS stars, with masses in the range $ 0.1$--$100~M_{\odot}$ (see Fig.~\ref{fig:dLadEa}) and assuming axion masses in the range $0.1-50$~keV.  The study found that the axion emission from the solar-mass star is close to that from the Sun. Instead, the axion production from photon coalescence (see Sec.~\ref{sec:photon}) tends to dominate over that from the Primakoff process for massive stars and sufficiently high axion masses.


\begin{table}[t!]
    \centering
     \caption{Fitting parameters for Eq.~\eqref{eq:fitprim}, where $M$ is in units of solar masses ${\rm M}_{\odot}$ and these fits are valid in the range $1-100~{\rm M}_{\odot}$. The code to obtain these results can be found at \href{https://github.com/pcarenza95/MainSequence-Axion}{https://github.com/pcarenza95/MainSequence-Axion}.}
    \setlength{\tabcolsep}{3pt} 
    \resizebox{\textwidth}{!}{
    \begin{tabular}{|c|c|c|c|c|}
    \hline
    & $g_{a x}$ & $C_{0}(\times10^{40}\text{ keV}^{-1}\text{ s}^{-1})$ & $E_{0}(\text{keV})$ & $\beta$ \\
    \hline
    \text{Prim.} $a=\gamma$ & $10^{-12}\text{ GeV}^{-1}$ & $\begin{cases}
            -0.140+0.053\,M^{-0.347}e^{M^{0.379}} & \text{for } M<10 M_{\odot} \\
            -0.014+0.011\,M^{1.081} & \text{for } M\geq 10 \, M_{\odot} \\
        \end{cases}$ & $3.70+1.13\,M^{0.355}$ & $1.23+3.63\,e^{-M^{0.29}}$ \\
    \hline
    \text{Brem.} $a=e$ &$10^{-12}$& $55.21+1.62\times10^{4}M^{-0.65}$ & $0.06+1.80\,M^{0.23}$ & $\begin{cases}
          0.57 + 0.18 e^{-M^{1.09}} & \text{for } M\le 10 \, M_{\odot} \\
          0.48+0.05 \,M^{0.19} & \text{for } M\geq 10 \, M_{\odot} \\
        \end{cases}$ \\
        \hline
    \text{Compt.}  $a=e$ & $10^{-12}$ & $0.14+1.01\,M^{1.49}$ & $0.025+6.014\,M^{0.225}$ & $2.99-0.56\,e^{-M^{0.09}}$ \\
    \hline      
    \end{tabular}
    }
   
    \label{tab:fitMS}
\end{table}

\begin{figure}[t!]
\centering
\includegraphics[width=0.7\columnwidth]{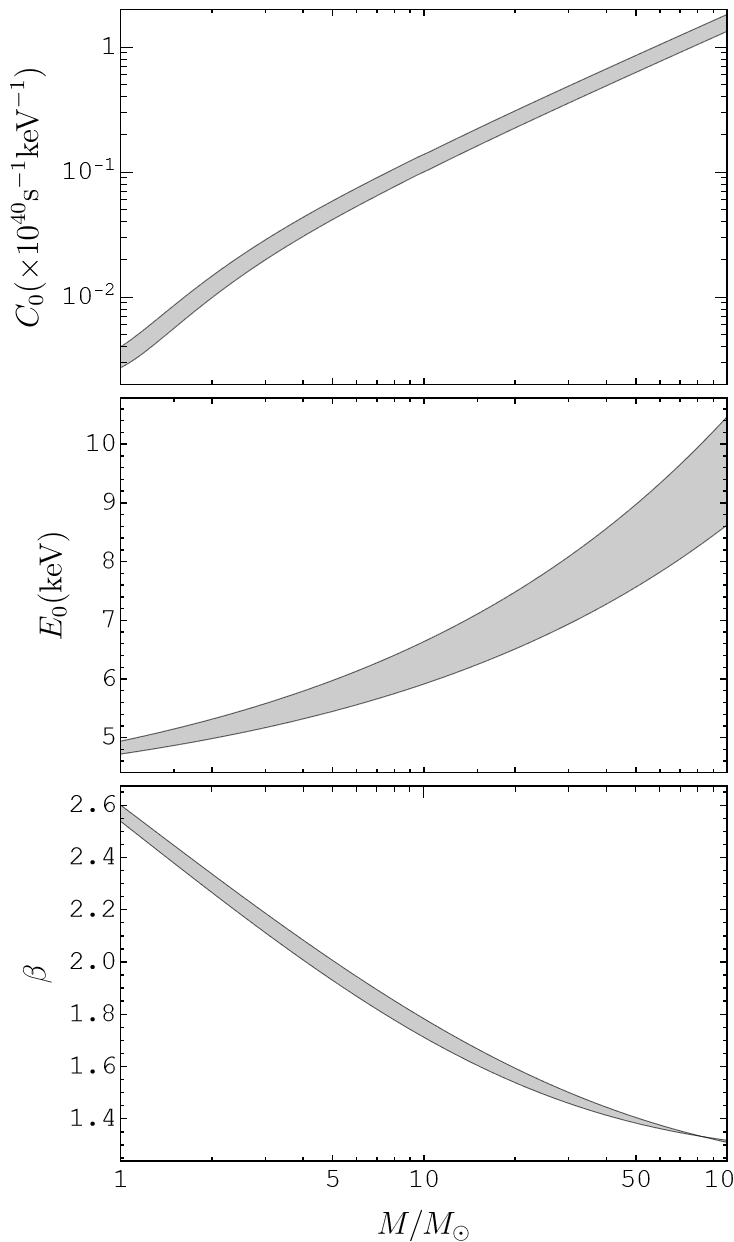}
\caption{Fitting parameters for the Primakoff axion flux from a MS star, as a function of the stellar mass, in the case $m_{a}\ll {\rm keV}$ and $g_{a\gamma}=10^{-12}~{\rm GeV}^{-1}$. Stellar profiles taken from Ref.~\cite{Nguyen:2023czp}. The gray band denotes the uncertainty on the represented quantity. }
\label{fig:MSandSun}
\end{figure}

The axion production from a MS star via Primakoff, Compton and bremsstrahlung would lead to a quasi-thermal axion spectrum (at the source) that can be parameterized as in Eq.~\eqref{eq:fitprim} where the parameters, as function of the stellar mass, are given in Tab.~\ref{tab:fitMS}. In Fig.~\ref{fig:MSandSun} we show the behavior of these parameters, for the Primakoff case, and their uncertainties.

The direct detection of such axion flux from other Galactic stars with an helioscope is not competitive with respect to the solar axion searches due to the distance suppression. Nevertheless, the cumulative flux from all the stars in the Universe would lead to a diffuse axion background. The fluxes of this stellar axion background and its decay photons are subdominant to but can in principle be disentangled from those expected from the Sun and the early universe based on their different spectral and spatial profiles. Ref.~\cite{Lopes:2021mgy} proposed that if MS stars have a strong magnetic field in their stellar envelope, axion-photon conversions for axions with masses in the range $m_{a}\in[10^{-5};10^{-7}]$~eV would lead to a characteristic X-ray spectrum, observable with NuSTAR.

\subsubsection{Supergiant stars and pre-supernovae}
\label{sec:preSN}

\begin{figure}[t]
\centering
\includegraphics[width=10.cm]{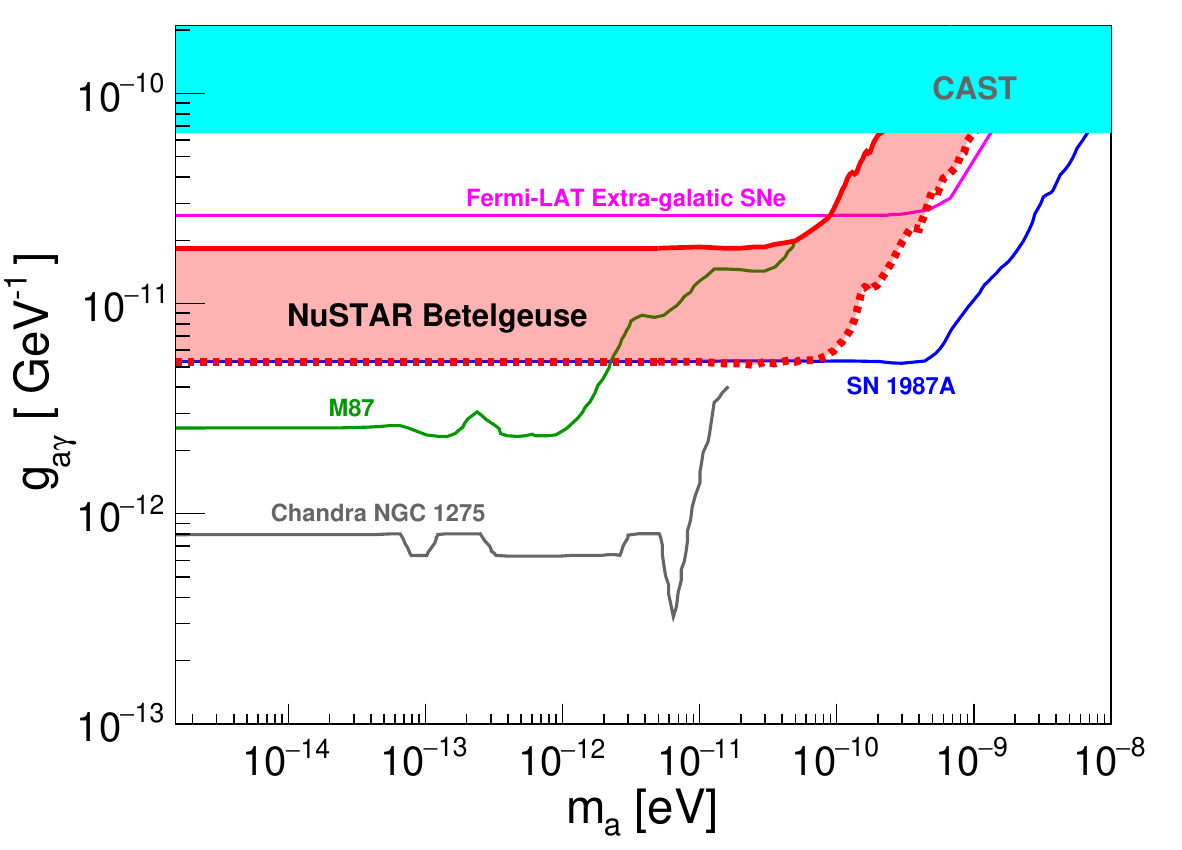}
\caption{Comparison of 95\% C.L. upper limits of $g_{a\gamma}$ from NuSTAR Betelgeuse observation (red band). This band is obtained by assuming the most conservative and most optimistic combination of stellar model and Galactic magnetic field along the line-of-sight. SN 1987A~\cite{Payez:2014xsa}, Fermi-LAT extra-galactic SNe~\cite{Meyer:2020vzy}, M87~\cite{Marsh:2017yvc}, and Chandra NGC 1275~\cite{Reynolds:2019uqt} bounds are also shown. (Figure taken from Ref.~\cite{Xiao:2020pra} with permission). 
}
\label{fig:upperlimts}
\end{figure}
More promising is the search for axions from nearby supergiant stars. Although these stars are considerably less numerous than MS stars, their high core temperatures and the steep dependence of the axion production rate on temperature make them compelling sources of stellar axions. In the radius $d\lesssim 1$~kpc from the Earth there are $\sim 20$ supergiants with masses ranging from $10-30~ {\rm M}_{\odot}$~\cite{Mukhopadhyay:2020ubs}. An example studied quantitatively is the red supergiant star Betelgeuse ($\alpha$ Orionis), spectral type M2Iab at a distance $d \simeq 197$~pc, proposed as axion target in a seminal paper by E.~Carlson~\cite{Carlson:1995wa}. Betelgeuse, whose luminosity, effective temperature and metallicity are, respectively, $\log L/L_\odot=5.10\pm0.22$~\cite{Bertre:2012bh}, $T_{\rm eff}=3641\pm53\,$K~\cite{Perrin:2004ce}, and $[\mathrm{Fe/H}]=+0.1\pm0.2$~\cite{1984ApJ...284..223L}, constraining the initial mass in the range 18-22~${\rm M}_\odot$, in agreement with previous determinations~\cite{Meynet:2013pqa, 2016ApJ...819....7D}. 

Betelgeuse is an ideal candidate for axion searches, as it  has a hot core, and thus is potentially a copious producer of axions via Primakoff,  Bremsstrahlung and Compton process. The axion production rate for the Compton and Bremsstrahlung processes was discussed in Refs.~\cite{Raffelt:1996wa,Carenza:2021osu} while the Primakoff production was presented in Ref.~\cite{Xiao:2020pra}. For all the production mechanisms, the axion source spectrum can be very well approximated by~\cite{Andriamonje:2007ew}
\begin{equation}
\begin{split}
\frac{d\dot{N}_{a}}{dE} &= \frac{10^{42}}{\textrm{keV}~\textrm{s}}\Bigg[ \alpha^{B}g_{13}^{2}\left(\frac{E}{E_{0}^{B}}\right)^{\beta^{B}}e^{-(\beta^{B}+1)E/E_{0}^{B}} + \\
&\quad\quad\quad +\alpha^{C}g_{13}^{2}\left(\frac{E}{E_{0}^{C}}\right)^{\beta^{C}}e^{-(\beta^{C}+1)E/E_{0}^{C}}+
\alpha^{P}g_{11}^{2}\left(\frac{E}{E_{0}^{P}}\right)^{\beta^{P}}e^{-(\beta^{P}+1)E/E_{0}^{P}}\Bigg] \,\ ,  
\label{eq:alp_product}
\end{split}
\end{equation}
where $g_{11}=g_{a\gamma}/10^{-11}\,{\rm GeV}^{-1}$, $g_{13}=g_{ae}/10^{-13}$, $\alpha^{B/C/P}$ is the normalization, $E_0^{B/C/P}$ is the average energy, and $\beta^{B/C/P}$ is the spectral index for Bremsstrahlung, Compton and Primakoff processes, respectively. The values of $\alpha$, $E_0$ and $\beta$ depend on quantities characterizing the core of the star, mostly temperature and density. In order to model the structure of Betelgeuse as a 20~${\rm M}_\odot$ star with solar composition, with profiles computed by using the FuNS code (see Ref.~\cite{Straniero:2019dtm} for a detailed description of this code and the adopted input physics). 

The general strategy to analyze the possible axion emission from supergiant stars is to search for their conversions in the Galactic magnetic field. This would produce a photon signal peaked in the hard X-ray range ($E_\gamma > 10~\rm{keV}$), with its average energy depending on the generally unknown stellar evolutionary stage. In the specific case of Betelgeuse, Ref.~\cite{Xiao:2020pra} used the data of a dedicated 50~ks observation by the NuSTAR satellite to place a 95\% C.L. upper limit on the axion-photon coupling $g_{a\gamma} < (0.5 - 1.8) \times 10^{-11}$~GeV$^{-1}$ for axion masses $m_{a} < (5.5 - 3.5) \times 10^{-11}$~eV, assuming only Primakoff production (see Fig.~\ref{fig:upperlimts}).
Enlarging the production channels to include Bremsstrahlung, Compton and Primakoff, Ref.~\cite{Xiao:2022rxk} derived the constrain $g_{a\gamma} \times g_{ae}< (0.4-2.8)\times10^{-24}$ \textrm{GeV}$^{-1}$ for masses ${m_{a}\leq(3.5-5.5)\times10^{-11}}$~eV. 

In general, the axion flux increases rapidly with the age of the star, as it does its average energy, reflecting the rise in its core temperature. Ref.~\cite{Mori:2021krp} found that in the latest pre-SN stages of Betelgeuse, when the star undergoes O and Si burning, the axion flux may be peaked at MeV energies, producing photons possibly detectable with gamma ray detectors after converting in the Galactic magnetic field, if $m_a\lesssim 1$~neV. Failure to detect a gamma-ray signal from a SN progenitor with next-generation gamma-ray telescopes just after pre-SN neutrino alerts would lead to an independent constraint on axion  parameters as stringent as the SN 1987A limit (see Sec.~\ref{sec:axionlike}). This basic idea can be extended to other, not necessarily single, stellar objects. For example, in Ref.~\cite{Dessert:2020lil} the authors focused on the Quintuplet and Westerlund 1 super star clusters, which host large numbers of hot, young stars including Wolf-Rayet stars; these stars produce axions efficiently through the axion-photon coupling, and back-converting in the  Galactic magnetic field may produce an X-ray flux. Using  NuSTAR data in the $10-80$~keV energy range they constrain the axion-photon coupling $g_{a\gamma} \lesssim 3.6 \times 10^{-12}$~GeV$^{-1}$ for masses $m_a \lesssim 5 \times 10^{-11}$~eV at 95\% CL. The idea was extended even further to the analysis of entire galaxies in Ref.~\cite{Ning:2024eky}, in which the authors analyzed the flux of axions emitted by the galaxies M82 and M87. The first of these galaxies provides a very competitive bound for axion masses up to neV or so. M82 is known for its starburst activity, meaning it has a very high rate of star formation compared to typical galaxies. This results in a large number of hot, massive stars, which are prime candidates for axion production through the Primakoff process, due its steep dependence on the temperature. Furthermore, the central region of M82 hosts strong magnetic fields due to the intense star formation activity, providing an ideal environment for axion-photon conversion.


\section{Globular Clusters}\label{sec:GC}
GCs are the building blocks of any kind of galaxy. They are found in spirals (such as the Milky Way or M31), ellipticals (M87) as well as in Dwarf Spheroidals or irregular galaxies (e.g., the Magellanic Clouds). Hundreds of GCs populate the Milky Way halo and bulge. They are old ($\sim 11-13$~Gyr) and  contain a large number of stars bound by the reciprocal gravitational interaction. Typically a GC contains in between $10^5$ and $10^7$ stars within a radius of a few parsecs. Most of their stars are nearly coeval, even though there exists a growing amount of observational evidences showing that they host multiple stellar populations characterized by different chemical compositions. 

Because of the rather large number of almost coeval stars, GCs are natural laboratories of fundamental physics and have been widely exploited to investigate new physics. During the more advanced phases of their evolution, GC stars develop core temperatures high enough to efficiently produce weakly interacting particles with masses up to a few keV.  In particular, the energy loss induced by the possible production of axions in evolved GC stars may leave characteristic imprints in several observable properties of these stellar systems. In this Section, the potential of GC stars as probes of new physics is illustrated with some examples. In particular, in Sec.~\ref{sec:tip} we present the bounds on axion-electron coupling $g_{ae}$ as derived from measurements of the luminosity of stars at the tip of the RGB, while in Sec.~\ref{sec:Rparam} we illustrate the methods to constrain  the axion-photon coupling $g_{a\gamma}$, based on the observed numbers of stars in different evolutionary phases (RGB, HB and AGB). 

\subsection{The tip of the red giant branch}\label{sec:tip}
The RGB of a GC is populated by stars with quite similar masses. The precise value of the mass depends on the cluster age and composition. Assuming an age in the range $11-13$~Gyr, the mass of GC RGB stars is $\sim 0.8-0.9$~M$_\odot$. The internal structure of these stars is characterized by distinct layers and conditions, especially as they approach the brightest luminosity (hereinafter RGB tip). As illustrated in Sec.~\ref{sec:stelev_H_He}, the core consists mainly of helium, which has been produced through nuclear fusion of hydrogen in earlier stages of the star's life. Surrounding the helium core is a shell where hydrogen burns into helium through the CNO cycle. The He left behind by the H burning makes the core mass progressively bigger. Hence, as the core mass increases, both core density and temperature rise up. Noteworthy, the shell H-burning provides the energy that sustains the losses by radiation at the stellar surface. Hence, as the core mass increases, the shell becomes hotter, thinner and more efficient, so that the stellar luminosity increases. In practice, it exists a well-known relation between the core mass and the stellar luminosity for RGB stars~\cite{1970AcA....20...47P}. In addition, owing to the intense energy flux generated by the H burning shell, the outer envelope expands and a deep convective instability sets on. If the core contraction is a source of gravitational energy, energy sink processes are also in action near the center. The combination of energy sources (gravitational contraction and nuclear fusion)  and sinks (plasma neutrinos), coupled to an efficient heat conduction (by degenerate electrons), shapes the internal temperature profile (see Fig.~\ref{fig_RGB_1}). Then, when the core mass is $\sim 0.5$~M$_\odot$ and the maximum temperature attains $\sim 100$~MK, the fusion of He into C, as triggered by the triple-$\alpha$ reaction, starts. Owing to the degenerate equation of state, a thermonuclear runaway takes place (He-flash, see Fig.~\ref{fig_RGB_flash}). 

\begin{figure}[t]
\centering
\vspace{0.cm}
\includegraphics[width=0.6\columnwidth]{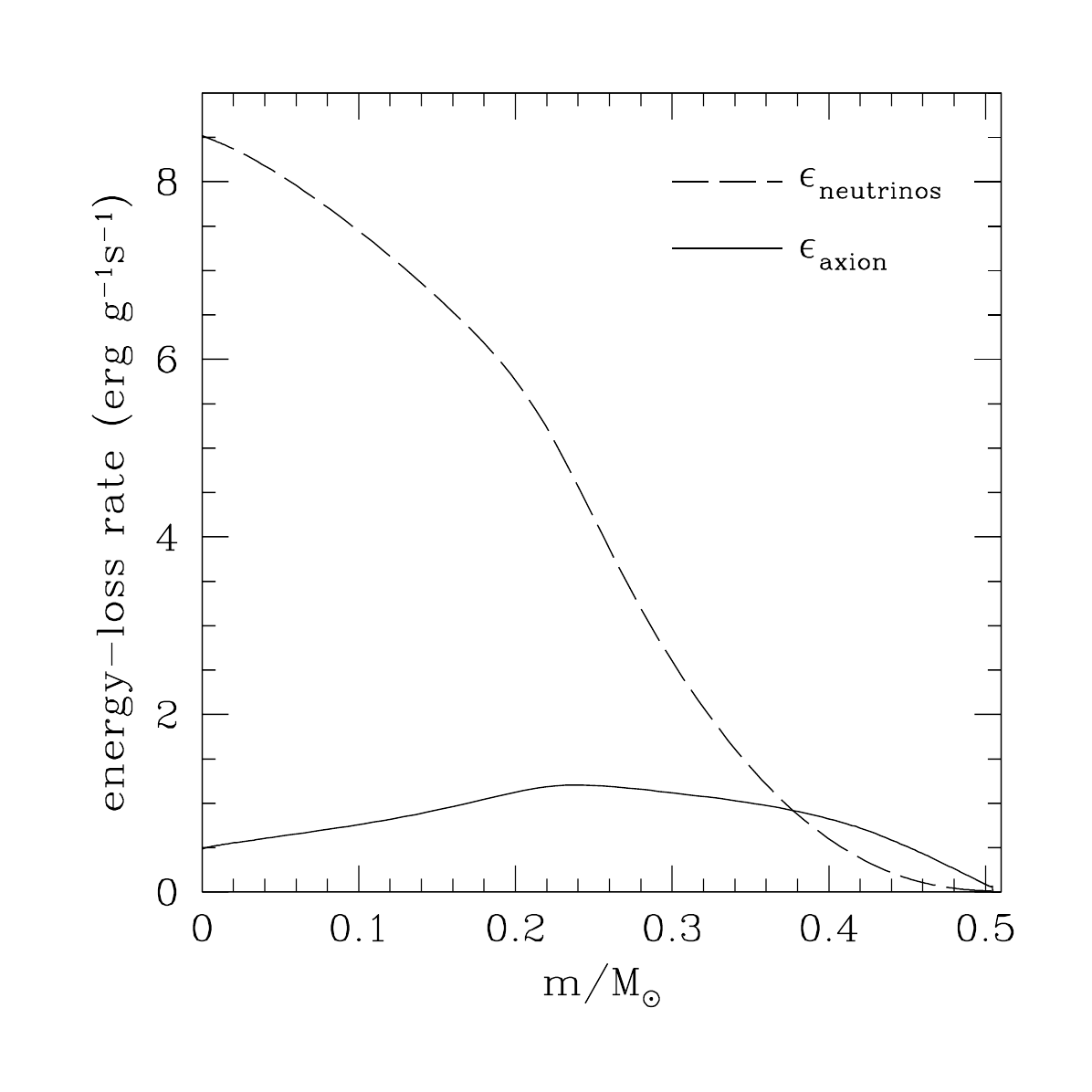}
\caption{Energy-loss rates for axions (Bremsstrahlung) and neutrinos (plasmon decay) within the core of a RGB model close to the RGB tip ($M = 0.82~M_\odot$, $Y = 0.25$, $Z = 0.001$, and $g_{ae}=10^{-13}$). }
\label{fig:loss_flash}
\end{figure}

\begin{figure}[h]
\centering
\vspace{0.cm}
\includegraphics[width=0.6\columnwidth]{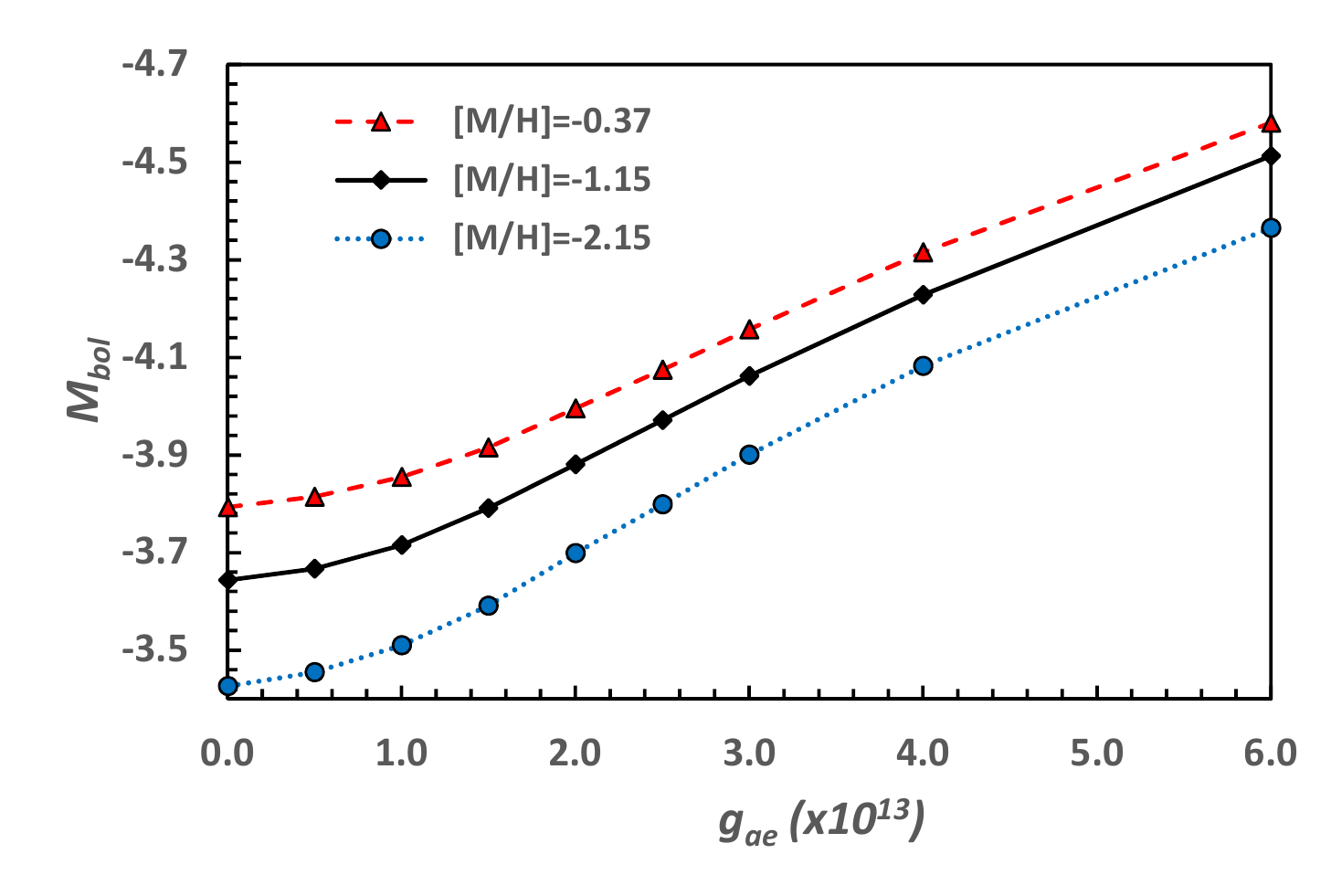}
\caption{The bolometric magnitude ($M_{\rm bol}=4.75-2.5\log(L/L_\odot)$ versus the axion-electron coupling strength. The three lines refer to different stellar metallicities, i.e., $[M/H]=\log(Z/X)-\log(Z/X)_\odot$, where $X$ is the hydrogen mass fraction and $Z$ is the sum of the mass fractions of all the elements with atomic number $\geq 6$. (Figure adapted from Ref.~\cite{Straniero:2020iyi}).} 
\label{fig:gc_mbol}
\end{figure}

If axions are produced in the core of RGB stars, an additional energy loss would contribute to the shaping of the core temperature profile. At a density of $10^5$--$10^6$~g/cm$^3$, as in the core of a bright RGB star, the plasma frequency becomes of the order of the axion energy, a condition for which the Primakoff process is largely suppressed (see, e.g., Refs.~\cite{Raffelt:1987yu,Straniero:2020iyi}). Similarly, due to the Pauli blocking, the Compton scattering on degenerate electrons is unlikely. Instead axions eventually produced by Bremsstrahlung  may provide a non-negligible contribution to the energy-loss~\cite{Raffelt:1994ry,Viaux:2013lha,Straniero:2020iyi}. Hence, If Bremsstrahlung axions are actually produced in RGB stars, a higher core mass would be required to attain the conditions for the He ignition and, according to the core mass-luminosity relation, the RGB tip should appear brighter than expected in case of standard energy-loss. In this framework, the RGB-tip luminosity may be used to constrain the strength of the axion-electron coupling $g_{ae}$. In Fig.~\ref{fig:loss_flash}, the energy loss due to Bremsstrahlung axions is compared to the one of plasma neutrinos, in the core of a star close to the RGB tip, while in Fig.~\ref{fig:gc_mbol}, we show the expected variation of the bolometric magnitude [$M_{\rm bol}=4.75-2.5\log(L/L_\odot)$] of the RGB tip versus the axion-electron coupling strength.  Note that the axion production rate scales as $g_{ae}^2$ and $T^4$. Unlike neutrinos, whose production is heavily concentrated near the center, the axion production peaks at the off-center temperature maximum.

\begin{figure}[h]
\centering
\vspace{0.cm}
\includegraphics[width=0.8\columnwidth]{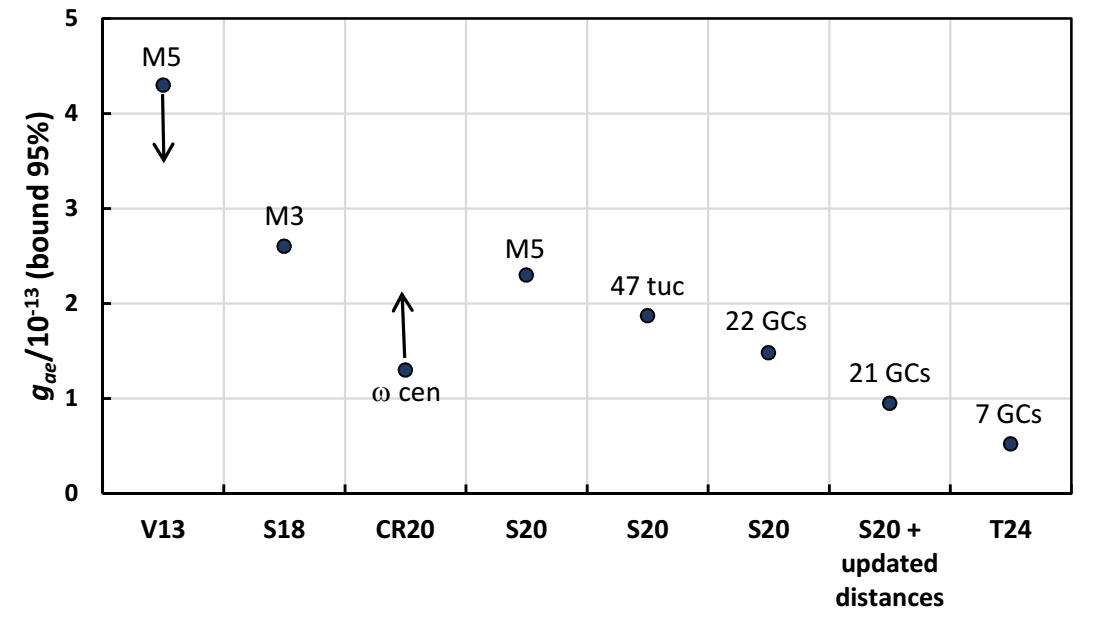}
\caption{10 years of $g_{ae}$ bounds from observations of RGB-tip luminosities of galactic GCs: V13 \cite{Viaux:2013lha}, S18 \cite{Straniero:2018fbv}, CR20 \cite{Capozzi:2020cbu}, S20 \cite{Straniero:2020iyi}, T24 \cite{troitsky:2024stellarevolutionaxionlikeparticles}. The S20(+updated distances) has been obtained by adopting the distances reported in \cite{2021MNRAS.505.5957B}. The two arrows represent the corrections required  to account for the electron screening (V13) and the $\omega$ Cen distance (CR20) issues (see text for details).
}
\label{fig:gc_tip_teorici}
\end{figure}

The first systematic application of this method was presented by Ref.~\cite{Viaux:2013lha}. On the base of an accurate photometric study of NGC 5904 (also M5), the authors derived the I band magnitude $M_I$ of the brightest RGB stars, finding that the observed RGB tip was definitely brighter than predicted by a {\it standard} stellar model. After a detailed analysis of the various uncertainties, they reported an upper bound of g$_{ae}<4.3\times 10^{-13}$ (95\% C.L.). More recently, other studies were extended to other clusters and the difference between observed and predicted RGB-tip luminosity was found substantially smaller than that claimed by Ref.~\cite{Viaux:2013lha} (see Fig.~\ref{fig:gc_tip_teorici}). In particular Ref.~\cite{Straniero:2018fbv}, based on near-infrared photometry of NGC 5272 (also M3), did not found a clear evidence of a brighter than predicted RGB tip and, after considering the major uncertainties, obtained a new and more stringent bound, namely $g_{ae}<2.57\times 10^{-13}$. Such a result was mainly due to a difference in the predicted luminosities of the RGB tip models. Indeed, those obtained by Ref.~\cite{Viaux:2013lha} are about 0.09 mag fainter than those computed by Ref.~\cite{Straniero:2018fbv} [or $\delta\log L/L_\odot=0.037$]. As noted in~\cite{2017A&A...606A..33S}, a large part of this discrepancy is likely due to the evaluation of the screening enhancement factors that multiply the nuclear fusion rates among bare nuclei and the triple-$\alpha$ one, in particular. Indeed, Ref.~\cite{Viaux:2013lha} adopted the weak-screening prescriptions derived by Ref.~\cite{1954AuJPh...7..373S}, where the theoretical framework was derived for the calculation of the  screening potential in the two extreme cases of {\it weak} and {\it strong} regime of the stellar plasma. In the weak-screening regime, the coupling of charged particles (electrons and nuclei) is weak or, more precisely, the Coulomb interaction energy is much smaller than the thermal energy $k\,T$. On the other hand, the strong regime refers to the case of strong coupling, i.e., Coulomb interaction energy much greater than the thermal energy. The weak regime condition is usually fulfilled in the core of H-burning stars with mass $\sim 0.8$~M$_\odot$. However, at the higher density of the core of the brightest RGB stars ($\rho \sim 10^6$~g/cm$^3$), the interaction energy is comparable to the thermal energy and the weak screening prescription by Ref.~\cite{1954AuJPh...7..373S} largely overestimates the screening factors. As shown in Ref.~\cite{1973ApJ...181..439D} (see also Ref.~\cite{1973ApJ...181..457G}), an intermediate regime is more  appropriate to describe the plasma conditions in RGB stars close to the the He ignition. 
 
 Later on, Ref.~\cite{Capozzi:2020cbu} brought this bound down to $g_{ae}<1.3\times 10^{-13}$. The authors used the same RGB stellar models as in Ref.~\cite{Viaux:2013lha}, but applying a shift to the predicted $I$ magnitude of the RGB tip to account for the electron-screening issue. 
This bound was obtained from the RGB tip of the cluster $\omega$ Cen (or NGC 5139), whose distance was derived from  Ref.~\cite{2019MNRAS.482.5138B} by fitting the velocity dispersion profile, after the second data release of the Global Astrometric Interferometer for Astrophysics (Gaia-DR2), with N-body models of globular clusters. As later noted by Ref.~\cite{2021MNRAS.505.5957B}, this kinematic distance is about 200 pc shorter than that derived from the parallax and other methods, such as MS fitting or RR Lyrae period-luminosity relation. This discrepancy is likely due to the flattening of $\omega$ Cen ($e>0.17$~\cite{Mackey:2004kx}). In practice, in the kinematic method, the distance is varied until the observed velocity dispersion profile best matches that of spherically-symmetric N-body models. Therefore, in case of a not fully relaxed cluster dynamics,  the resulting kinematic distance underestimates the true cluster distance. Summarizing,  Ref.~\cite{Capozzi:2020cbu} adopted the kinematic distance to $\omega$ Cen previously reported by Ref.~\cite{2019MNRAS.482.5138B}, i.e., $5.24\pm0.05$ kpc, while Ref.~\cite{2021MNRAS.505.5957B} corrected this distance in $5.426\pm0.047$. However, Ref.~\cite{Soltis:2020gpl}, based on the parallax from Gaia early data release 3, got a distance of $5.24\pm0.11$, which is similar to that adopted by Ref.~\cite{Capozzi:2020cbu}, but with a doubled  error. In any case, the $g_{ae}$ bound should be relaxed. 
Noteworthy, Ref.~\cite{Capozzi:2020cbu} also derive a new bound $g_{ae}<1.6\times 10^{-13}$ from the RGB tip of NGC 4258, a spiral galaxy in the Canes Venatici constellation. Owing to the presence of an active galactic nucleus (AGN), NGC 4258 is also classified as a Seyfert 2 galaxy with a characteristic H2O maser emission from the accretion disk around the central black hole. Such a maser source allows an accurate distance determination for this galacy Ref.~\cite{Reid:2019ApJ886}. 
However, unlike GCs, a galaxy hosts many generations of stars, characterized by different ages and chemical compositions. In practice, the superposition of the giant stars  belonging to different stellar generations implies a spread in both the $I$ magnitude and the $V-I$ color of the stars close to the RGB tip. Thus, to account for these variations, a correction is applied to the observed $I$ magnitude of the brightest RGB stars, as derived from an empirical  $I$-$(V-I)$ relation (see, e.g., Ref.~\cite{Jang:2017ApJ835}). In this way, a unique absolute $I$ magnitude ($M^{TRGB}_I$) is provided at the same reference color, i.e., $(V-I)^{TRGB}=1.8$ mag. The choice of this reference color is not arbitrary. Indeed, it coincides approximately to the color of the brightest stars in the GC M5, thus allowing the use of the same theoretical models adopted by \cite{Viaux:2013lha}, eventually corrected to account for the electron-screening issue. For this reason, \cite{Capozzi:2020cbu} also applied the same correction to the $M^{TRGB}_I$ of the GC $\omega$ Cen, a procedure that would not have been necessary if they had used stellar models with composition and age appropriate for this stellar cluster.  In any case, the use of this relation introduces a further uncertainty to the derived bound, which is probably small only in the case of a limited spread in the $V-I$ color.   
In this context, a controversial conclusion was reached by   \cite{dennis2023tipredgiantbranch}\footnote{
This paper has been uploaded to the ArXiv on May 4, 2023 and, so far, never accepted for publication in a peer-reviewed journal.}.  Basing on a detailed statistical analysis, they argued that the correlations among the uncertainties of only 3 model parameters, i.e, metallicity, He abundance and age, imply a substantially weaker than previously found bound for the axion-electron coupling. Similarly to \cite{Capozzi:2020cbu}, they compared the observed $I$ magnitude of the RGB tip of one globular cluster ($\omega$ Cen) and two galaxies (NGC 4258 and the Large Magellanic Cloud) with those predicted by a machine learning (ML) code, previously trained with a large set of RGB stellar models. However, they assumed an unrealistic large range of values for the 3 input parameters, ignoring the many spectroscopic and photometric constraints. Moreover, as in 
\cite{Capozzi:2020cbu}, they applied a correction to the observed $I$ magnitude to account for the differences between the observed and the reference $V-I$ colors. As previously noted, this correction is needed in case of a galaxy showing a superposition of multiple stellar populations characterized by different ages and compositions, but it is unnecessary in the case of a globular cluster whose composition and age have been accurately measured. Then, at variance with  \cite{Capozzi:2020cbu}, they also apply the same correction to the theoretical prediction. In practice, the predicted $M^{TRGB}_I$ given by the ML is converted to a prediction at the reference $V-I$  color.   
This correction, when coupled to the errors of the cluster  parameters,  amplify the uncertainty on the predicted $M^{TRGB}_I$. For example, a variation of the metallicity within its error bar, implies a change of the theoretical $V-I$ and, in turn, a different correction to the predicted $I$ magnitude. In this way, the many uncertainties affecting the the theoretical $V-I$ are propagated into the predicted $M^{TRGB}_I$. In addition to the adopted bolometric corrections, the resulting axion-electron coupling bound also depends on the model uncertainties due to the radiative opacity, the surface boundary conditions, the equation of state of partially ionized matter, as well as to the adopted mixing-length parameter and the mass-loss rate. Summarizing,
\cite{dennis2023tipredgiantbranch} demonstrated that the method they adopted is heavily affected by these model uncertainties and, for this reason, it fails to produce a reliable and robust constraint for the axion-electron coupling. Note that the previous bounds obtained by \cite{Viaux:2013lha, Capozzi:2020cbu} are not affected by this theoretical $V-I$ uncertainties, except for those related to the bolometric corrections.    

\begin{figure}[t]
\centering
\vspace{0.cm}
\includegraphics[width=0.8\columnwidth]{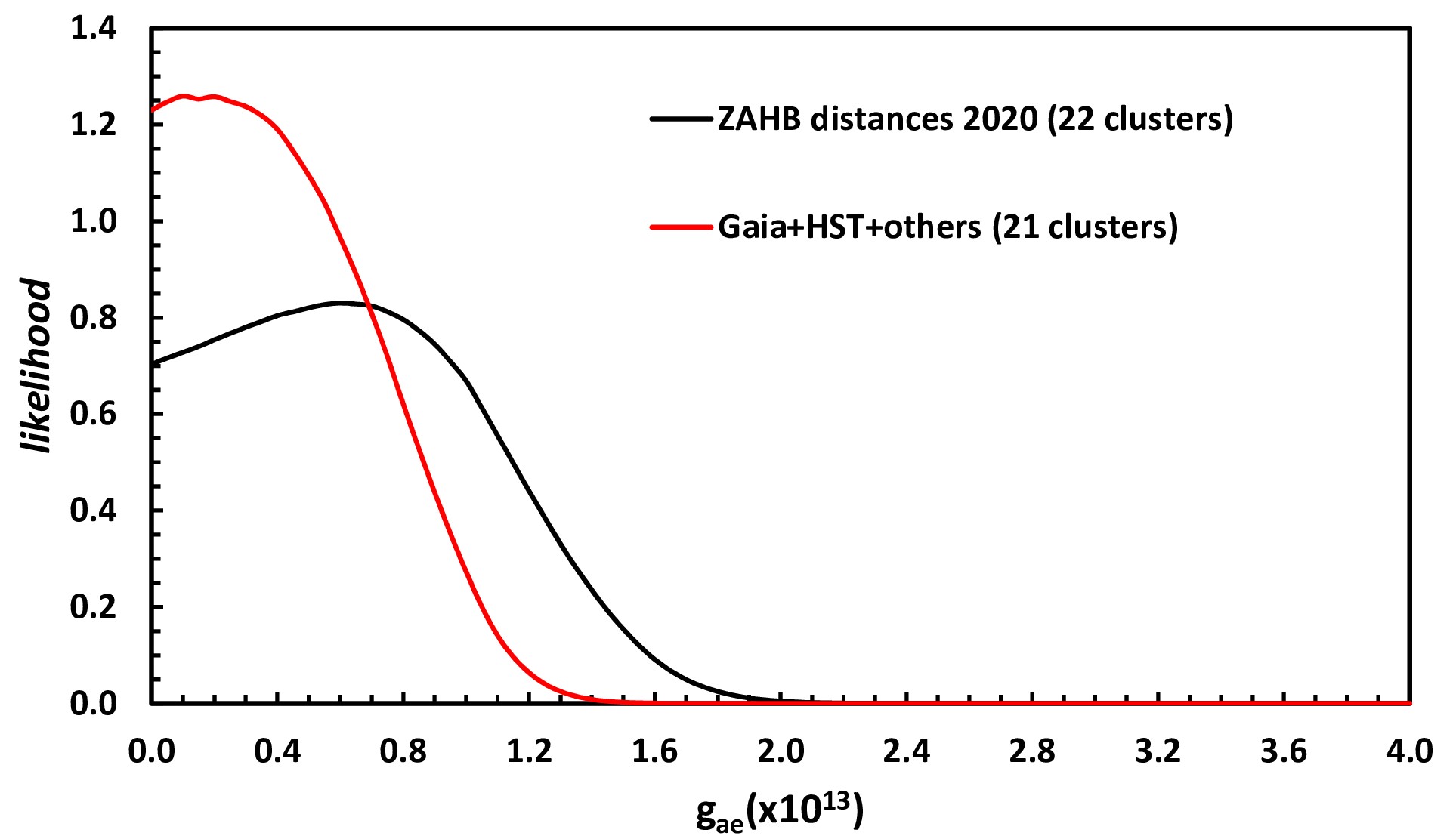}
\caption{The result of the likelihood analysis obtained adopting the ZAHB distance scale~\cite{Straniero:2020iyi} is compared to that obtained with improved distances, as reported in Ref.~\cite{2021MNRAS.505.5957B}. 
}
\label{fig:gc_tip_like}
\end{figure}

A different approach was followed by Ref.~\cite{Straniero:2020iyi} (also see \cite{Straniero:2018fbv}), where 22 GCs have been considered by combining optical and near-infrared data from public archives of ground and space based telescopes. Thanks to these multi-wavelength analyses, they derived the bolometric magnitude of the brightest RGB stars of each cluster, a quantity that can be directly compared to the corresponding luminosity predicted by stellar models accounting for the additional axion cooling. Distances were derived by scaling the ZAHB luminosity to that of 47 Tuc (NGC 104), for which a rather precise parallax was available after Ref.~\cite{2018ApJ...867..132C}. As it is well known, the luminosity of a bright RGB star does not depends on its total mass, but, rather, on its He-core mass \cite{1970AcA....20...47P}. Therefore, the uncertainties affecting the physics of the H-rich envelope, e.g., those related to the mixing-length parameter, the mass-loss rate, the atomic and molecular opacity or the adopted surface boundary conditions, produce negligible effects on the predicted luminosity and, in turn, on the predicted bolometric magnitude. Note that this is a major advantage of this method with respect to the procedure followed in Refs.~\cite{Viaux:2013lha, Capozzi:2020cbu, dennis2023tipredgiantbranch}.  
First, they applied this  method to three GCs, namely,  47 Tuc, M5 and NGC 362, finding  $g_{ae}/10^{13}$ bounds of 1.87, 2.30 and 1.37, respectively (95 \% C.L.). Moreover, they performed a global likelihood analysis combining the RGB-tip measurements of 22 GC, obtaining the more stringent bound to the axion-electron coupling, namely: $g_{ae}<1.48\times 10^{-13}$ (see Fig.~\ref{fig:gc_tip_like}). 
This approach allows to reduce both systematic and statistical errors.
The authors also analyzed the various uncertainties of the stellar models and how they affect the predicted RGB-tip luminosity. Also, comparisons with results obtained by means of other stellar evolution codes were discussed in detail. 

As noted by many authors, among the uncertainties affecting the derived bound, a major contribution to the error budget is given by the distance determination. Also the bolometric corrections (see, e.g., Refs.~\cite{Buzzoni:2010mh,Worthey:2006gj}) used to transform the measured magnitudes into the absolute stellar luminosity (or viceversa) limit the derived constraint to the axion couplings.  Concerning distances, an important improvement has been achieved after the third Gaia data release (Gaia-DR3). A quantification of the impact of improved distances can be obtained by adopting the updated distances reported by Ref.~\cite{2021MNRAS.505.5957B} for 21 of the 22 GCs of the Ref.~\cite{Straniero:2020iyi} sample. In this way an improved bound of  $g_{ae}<0.95\times 10^{-13}$ may be obtained (see Fig.~\ref{fig:gc_tip_teorici} and Fig.~\ref{fig:gc_tip_like}). While we were finishing writing this review, a new analysis based on Gaia-DR3 photometry and astrometry of GCs was uploaded to the ArXiv by Ref.~\cite{troitsky:2024stellarevolutionaxionlikeparticles}\footnote{This paper has been recently accepted for publication on Pis’ma v ZhETF.}. The derived bound, i.e., $g_{ae}<0.58\times 10^{-13}$ (95\% CL), confirms and reinforces previous results.    

More in general, since the discrepancy between the standard theoretical prediction and the observed tip luminosity is very small, if any, the derived bound is essentially determined  by the error budget. For this reason, statistical and systematic errors affecting the measured stellar luminosity should be carefully estimated, as well as all the possible uncertainties affecting the corresponding theoretical predictions. In any case, even if a further reduction of this global uncertainty will be achieved in the future, when  $g_{ae}$ is substantially smaller than $10^{-13}$, the RGB tip becomes practically insensitive to the thermal production of axions, so that other astronomical constraints should be considered to investigate the extremely weak coupling region of the axion parameter space.

On the other hand, one may argue that these bounds are affected by  hidden systematic errors in the assumed standard physics inputs. Among them, the rate of the triple-$\alpha$ reaction plays a fundamental role in determining the RGB-tip luminosity. At $T>0.1$~GK, this reaction  proceeds as a two step process: $\alpha+\alpha \rightarrow ^{8}$Be($0_1^+$) followed by $^{8}$Be$(0_1^+)+ \alpha\rightarrow ^{12}$C($0_2^+$). Therefore, in the intermediate-temperature domain $0.1$-$2$ GK, the total cross section is fully determined by the properties of the two resonant states, $^{8}$Be$(0_1^+)$ and $^{12}$C$(0_2^+)$  (the latter being the famous Hoyle state), whose strengths can be measured in laboratory experiments. Under this assumption, the rate is relatively well known (within 10\%, see Ref.~\cite{ISOLDE:2005rym}). Note that in the RGB tip analysis of Ref.~\cite{Straniero:2020iyi}, a conservative $\pm 20\%$ error has been assumed for the triple-$\alpha$ reaction rate. However, at relatively low temperatures, the thermal energy of $\alpha$ particles is insufficient to populate the narrow resonances, and a direct three-body capture becomes the favored channel for this reaction. Many methods have been developed to calculate the cross section of a three-body process (see, e.g., Ref.~\cite{Nguyen:2011cy,Suno:2015boa}, and references therein). In general, a large discrepancy is found for $T<0.07$~GK, but at the temperature of the He ignition in RGB stars there is a substantial agreement with what is reported in the extant libraries of stellar reaction rates (e.g., Ref.~\cite{Angulo:1999zz}). 

\subsection{Horizontal branch stars: the \texorpdfstring{$R$, $R_1$ and $R_2$}~ parameters}\label{sec:Rparam}
The series of thermonuclear runaways that follows the He ignition in a low-mass star is a quite rapid transient phase, at the end of which the star settles on the ZAHB (see Sec.~\ref{sec:stelev_H_He}). During the HB phase, a quiescent He-burning takes place in the center of the star and an extended convective core develops. 

\begin{figure}[t!]
\centering
\vspace{0.8cm}
\includegraphics[width=0.7\columnwidth]{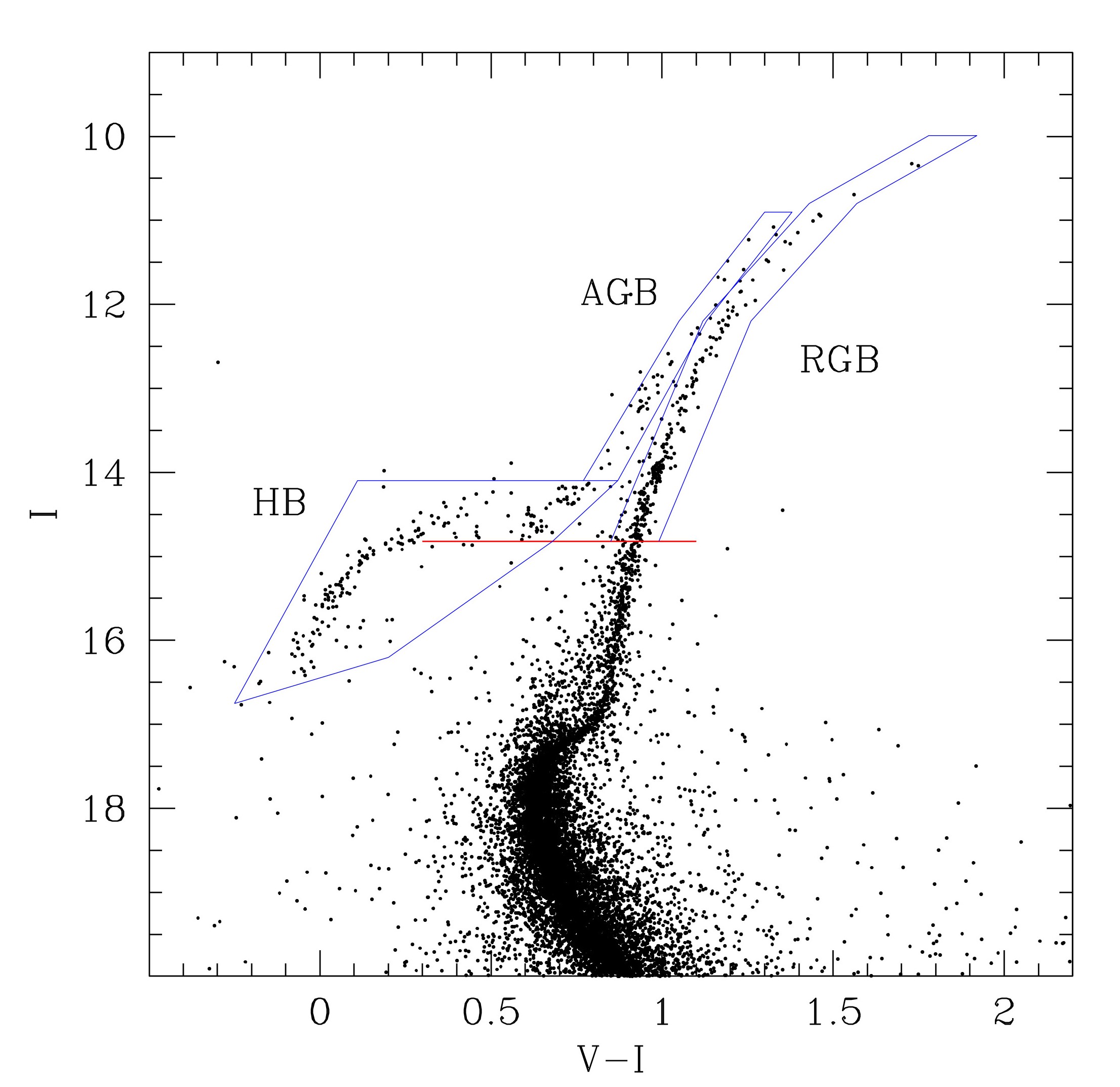}
\caption{ Color-magnitude diagram of NGC 5904~\cite{2000A&AS..145..451R}. The blue polygons delimit the three portions of the CMD to be considered in the  evaluation of the $R$ and the $R_2$ parameters, i.e., $R={N_{\rm HB}}/{N_{\rm RGB}}$ and $R_2={N_{\rm AGB}}/{N_{\rm HB}}$. The red horizontal line represents the ZAHB level.}
\label{fig:gc_cmd}
\end{figure}

The interpretation of the different HB morphologies in the Color-Magnitude Diagram (CMD) of galactic GCs has been a longstanding issue for stellar astrophysics. Firstly, it should be noted that all the HB stars belonging to a GC share a very similar He-rich core mass ($\sim 0.5$~M$_\odot$), so that their ZAHB luminosity is practically constant.  It is for this reason that this phase is called horizontal branch. On the other hand, their effective temperature depends on the amount of mass lost during the previous RGB phase. Note that the RGB mass loss is a rather stochastic process and the resulting spread of the HB masses determines a spread in the HB effective temperature. In practice, the total mass of a HB star is in between 0.1 and 0.3 M$_\odot$ smaller than the initial one.  These characteristics are partially lost when passing from the $L-T_e$ plane to the magnitude-color plane. For instance, if the I magnitude of the coolest HB stars ($\log T_{e} < 3.75$) remains practically constant, hot HB stars, those with smaller masses, appear fainter. This occurrence is evidently due to the bluest spectral energy distribution of the hottest stars. For the same reason, the (V-I) color of hot HB stars saturates.  In practice, these stars form a characteristic blue tail in the observed CMD. As an example, Fig.~\ref{fig:gc_cmd} shows the  I vs (V-I) diagram of the Milky Way GC NGC 5904 (also M5). The almost flat portion of the HB coincides with the instability strip of the RR-Lyra variables, while the blue tail contains stars whose envelope mass was eroded by a more intense RGB mass loss. A substantial reduction of the  envelope mass may also be the consequence of a close encounter, an event more frequent in the crowded core of a GC, or of mass transfer processes occurring in binary systems.          

The variation of the HB color distribution is also influenced by the cluster metallicity, which affects the radiative opacity in the stellar envelope. In this context, also the presence of more than one stellar population, each one characterized by a different initial composition, may affect the observed color spread. Indeed, some clusters showing HB stars particularly faint in optical and near-infrared filters, also present evidences of the presence of He enhanced stellar populations. Since the higher the initial He abundance, the faster the faster  its  evolution, the initial mass of He-enhanced HB stars should be smaller than that of star with ``normal'' He.       

As explained in Sec.~\ref{sec:stelev}, stars evolve because they lose energy. This energy loss is replaced by some internal source of energy, nuclear fusion or gravitational contraction. Hence, the duration of a nuclear driven evolutionary phase depends on the amount of the primary fuel and on the efficiency of the energy generation processes. As illustrated in Sec.~\ref{sec:stelev_H_He}, the HB lifetime is essentially determined by the rates of the two competitive reactions, the triple-$\alpha$ and the $^{12}$C$(\alpha,\gamma)^{16}$O, coupled to the available He fuel and, in turn, the efficiency of the convective and semi-convective processes active in the stellar core. In addition, the H-burning shell, which is active just outside the core, also contributes to the stellar luminosity. 

For a typical HB star of a galactic GC, the core-He burning phase lasts for about 80 Myr, to be compared to the $\sim 13$ Gyr of the previous core-H burning phase. This occurrence, that depends on the smaller amount of energy released by the He fusion, compared to that released by the H fusion, and to the higher luminosity, implies that the number of stars in the HB phase is a small fraction of the total number of stars in a GC. Indeed, the larger the time spent by a star in a particular evolutionary phase, the higher the number of stars observed in this phase. This occurrence also applies to the HB phase. If axions are produced within the stellar core, mainly through the Primakoff process and, in the presence of an axion-electron coupling, also by Compton scattering, they may provide an additional energy loss. In particular, if the axion mass is lower than a few keV, the energy-loss caused by the free streaming of these particles accelerates the He consumption and less stars would be observed on the HB. This occurrence has been widely used to constrain the axion-photon coupling $g_{a\gamma}$.     

A  first rough estimation of the maximum allowed strength of the axion-photon coupling was reported in Ref.~\cite{Raffelt:1996wa}. According to that seminal work, the true HB lifetime was assumed to agree within a 10\% with the theoretical prediction. Hence, a reduction $\leq 10\%$ of the HB lifetime might be obtained if the strength of the axion-photon coupling would be $g_{a\gamma}\leq 10^{-10}$~GeV$^{-1}$. More recently, Ref.~\cite{Ayala:2014pea} suggested to use the so-called $R$ parameter, originally introduced by Ref.~\cite{1969Natur.224.1006I} to estimate the primordial He. In practice, $R$ is the total number of HB stars ($N_{\rm HB}$) normalized to the number of RGB stars brighter than the ZAHB ($N_{\rm RGB}$, see Fig.~\ref{fig:gc_cmd}). Because of the already mentioned relation between the time spent by stars in a given evolutionary phase and the number of stars observed in this phase, the $R$ parameter may be approximated by the ratio of the HB and the RGB lifetimes, namely:
\begin{equation}
   R=\frac{N_{\rm HB}}{N_{\rm RGB}}\sim \frac{\tau_{\rm HB}}{\tau_{\rm RGB}} \,\ .
\end{equation} 
Note that $R$ does not depend on the cluster parameters, such as the distance and the reddening, and it is slightly affected by variations of the cluster age. In addition, it is also mildly dependent on the metallicity of the cluster stars, so that its value may be averaged over a large sample of galactic GCs. However, it depends on the original He mass fraction $Y$. Indeed, the higher the $Y$, the larger the ZAHB luminosity and, in turn, less stars are expected on the brighter then the ZAHB portion of the RGB. In this context, $R$ is considered an excellent estimator of the He content in the pristine gas from which globular clusters formed about $\sim 11-13$ Gyr ago. Nevertheless, Ref.~\cite{Ayala:2014pea} noted that, if the axion-photon coupling is switched on, the HB lifetime is reduced and, in turn, also $R$ becomes smaller. In practice $R$ depends on both $Y$ and $g_{a\gamma}$. Therefore, an independent estimation of $Y$ is needed to estimate $g_{a\gamma}$. Ref.~\cite{Ayala:2014pea} combined a weighted average value of $R=1.39\pm 0.04$, as measured on a sample of 39 GCs with [Fe/H]$<-1.1$, and the weighted average value of $Y=0.2535\pm0.0036$, as measured in extragalactic H II molecular clouds with metallicity in the same range of the 39 galactic GCs \cite{Aver:2013wba}. In this way, it was obtained a quite stringent bound for the axion-photon coupling of $g_{a\gamma}<0.66\times10^{-10}$ GeV$^{-1}$ (95 \% C.L.).  

As a further test, Ref.~\cite{Ayala:2014pea} repeated the analysis under two opposite and extreme assumptions about the original He content, namely: $Y=0.2472\pm0.0003$, corresponding to the primordial He mass fraction from extant Big-Bang-Nucleosynthesis calculations with effective number of neutrinos $N_{\rm eff}=3.046$ and baryon density from the Planck legacy~\cite{Planck:2015fie}, and $Y=0.269$, that is the solar He mass fraction, as obtained with extant standard solar models~\cite{Piersanti:2006zh}. Note that the latter is a very conservative assumption, considering that the Sun is a pop I star, belonging to the galactic disk, and formed about 8 Gyr later than the bulk of the GCs. As a result, it was found  $g_{a\gamma}<0.5 \times 10^{-10}$~GeV$^{-1}$ and $g_{a\gamma}<0.76\times 10^{-10}$~GeV$^{-1}$, for the primordial and the solar He assumptions, respectively.    

\begin{figure}[t!]
\centering
\vspace{0.8cm}
\includegraphics[width=1\columnwidth]{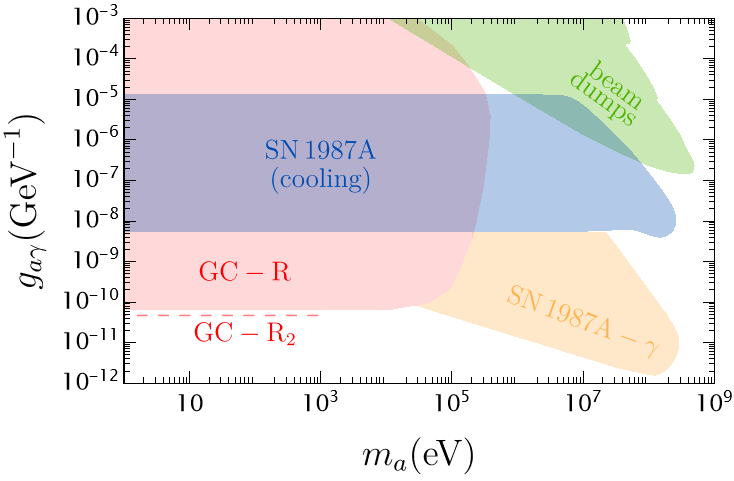}
\caption{Collection of some axion bounds, for axions coupled to photons. The focus is on the HB bounds (in red) coming from the $R$-~\cite{Ayala:2014pea,Carenza:2020zil,Lucente:2022wai} and the $R_{2}$-parameters~\cite{Dolan:2022kul}, with the latter shown in the massless axion case only. For comparison, we report also the SN 1987A cooling bound~\cite{Lucente:2020whw} (blue) and the gamma-ray decay constraint~\cite{Jaeckel:2017tud,Muller:2023vjm,Hoof:2022xbe} (orange) [see Sec.~\ref{sec:SN_NS}]. Moreover, in green we show the laboratory bounds from beam dumps~\cite{CHARM:1985anb,Riordan:1987aw,Dolan:2017osp,Blumlein:1990ay,NA64:2020qwq}. (Data for bounds taken from the repository~\cite{AxionLimits}).}
\label{fig:HBfull}
\end{figure}

As previously stated, this bounds are applicable in case of rather light axions, but what happens in case of massive axions? Given the typical temperature $T \sim 10$~keV in the stellar core of a HB star, one expects the thermal production of particles to be Boltzmann suppressed for $m_a \gtrsim 30$~keV relaxing the previous bound. However, as shown in Ref.~\cite{Carenza:2020zil} for masses $m_a \gtrsim 50$~keV the axion production via coalescence process $\gamma \gamma\to a$ becomes dominant over the Primakoff production (see Fig.~\ref{fig:energyloss}). Furthermore, for sufficiently large values of the axion mass and coupling to photons, one should expect a significant fraction of axions to decay inside the star. These axions do not contribute to the energy loss but rather lead to an efficient energy transfer inside the star. As shown in Ref.~\cite{Lucente:2022wai}, one expects the axion energy deposition to become especially relevant for $m_a \sim 0.4$~MeV and $g_{a\gamma} \gtrsim 10^{-6}$~GeV$^{-1}$. Ref.~\cite{Lucente:2022wai} proposed a  ballistic recipe to cover both the energy-loss and energy-transfer regimes, and presented a dedicated simulation of GC stars including the axion energy transfer. This argument allowed the constraining of the region with $m_a \sim 0.5$~MeV and $g_{a\gamma} \gtrsim 10^{-5}$~GeV$^{-1}$. The HB bound on $g_{a\gamma}$ in function of the axion mass $m_a$ is shown in Fig.~\ref{fig:HBfull}, in comparison with other constraints.

\begin{figure}[t!]
\centering
\vspace{0.8cm}
\includegraphics[width=1\columnwidth]{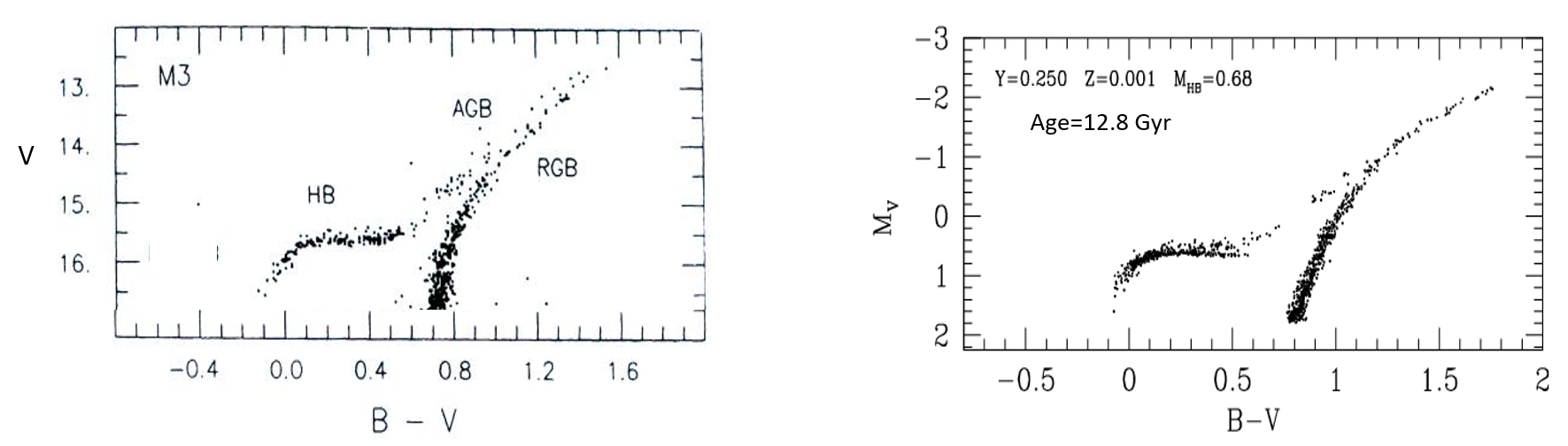}
\caption{A theoretical SCMD is compared to a real (observed) diagram. {\it Left}: the color-magnitude diagram of NGC 5272 (also M3); apparent magnitudes and colors of observed stars are shown. {\it Right}: a SCMD, as obtained for the following cluster parameters: age 12.8 Gyr, helium mass fraction $Y=0.25$, metallicity $Z=0.001$, average HB mass 0.68~M$_\odot$; absolute magnitudes and colors are shown.}
\label{fig:synthetic}
\end{figure}

As already recalled in the case of the RGB tip constraint (see Sec.~\ref{sec:tip}), also in this case is the error budget that determines the quoted bound. In this context, one of the main issues of the $R$ parameter method described above concerns the identification of the ZAHB luminosity level on the CMD (see Fig.~\ref{fig:gc_cmd}). This quantity  determines the faintest RGB stars that contribute to $N_{\rm RGB}$ and a systematic shift of it directly affects the estimated $R$. Measuring the ZAHB level is not an easy task and the result may be ambiguous. Usually, this magnitude is identified by eyes, looking at the fainter HB stars whose effective temperature is $\log T_e \sim 3.7$ (see, e.g., Ref.~\cite{Salaris:2004xd}). Alternatively the mean magnitude of the RR-Lyrae stars  is sometime used (see, e.g., Ref.~\cite{Harris:2010ut}). The first method is evidently limited by the relatively small number of HB stars in the available GC stellar catalogs. On the other hand, the implicit assumption that the RR-Lyrae mean brightens coincides with the ZAHB level has been widely criticized, as, depending on the composition, a star may start the HB evolution within the instability strip or enter it later, during the core-He burning evolution \cite{Lee:1990zzg,1992ApJ...386..663C}. An additional issue concerns the identification of the ZAHB in the theoretical models of HB stars. For instance, in Ref.~\cite{Ayala:2014pea} the theoretical ZAHB on the evolutionary track was identified  at the first appearance of a central convective core. This definition does not necessarily coincides with the observed ZAHB, as derived with one of the two methods previously described. As suggested by Ref.~\cite{Ferraro:1999xa}, Synthetic Color-Magnitude Diagrams (SCMD) can provide a more accurate determination of the ZAHB level, by accounting for both statistical fluctuations due to the limited stellar sample, the statistical errors in both magnitude and color, and the HB evolution. Based on theoretical evolutionary tracks of stars with chemical composition close to those of a real cluster, a synthetic diagram can simulate the observed diagram. An example is illustrated in Fig.~\ref{fig:synthetic}. Note that no corrections to account for the distance and for the absorption along the line of sight have been applied to the observed magnitudes and colors, but they do not affect the derivation of the $R$ parameter. On the other hand, random sampling on the stellar mass function  and on the photometric error band is properly accounted in the SCMD. In practice, by comparing real and synthetic CMD, one may consistently calculate the observed and the predicted $R$. 
This method also allows for a direct evaluation of the global theoretical error, i.e., the combinations of all the uncertainties affecting the theoretical recipes, as those due to the assumed cluster parameters (age, $Y$ and $Z$) and those affecting the physics inputs. As an example, we report the result of an analysis described by Ref.~\cite{Straniero:2015nvc}. In this case, the uncertainties of four of the most relevant inputs affecting RGB and HB stellar models are considered, namely: the He mass fraction ($Y$) and the rates of the three most relevant nuclear reactions, i.e., the $^{14}$N$(p,\gamma)^{15}$O, the $^{4}$He$(2\,^{4}$He$,\gamma)^{12}$C and  the $^{12}$C$(^{4}$He$,\gamma)^{16}$O. Adopted central values and standard deviations are reported in Tab.~\ref{tab_mcR}, where Gaussian error distributions are assumed. 

\begin{figure}[t!]
\centering
\vspace{0.8cm}
\includegraphics[width=1\columnwidth]{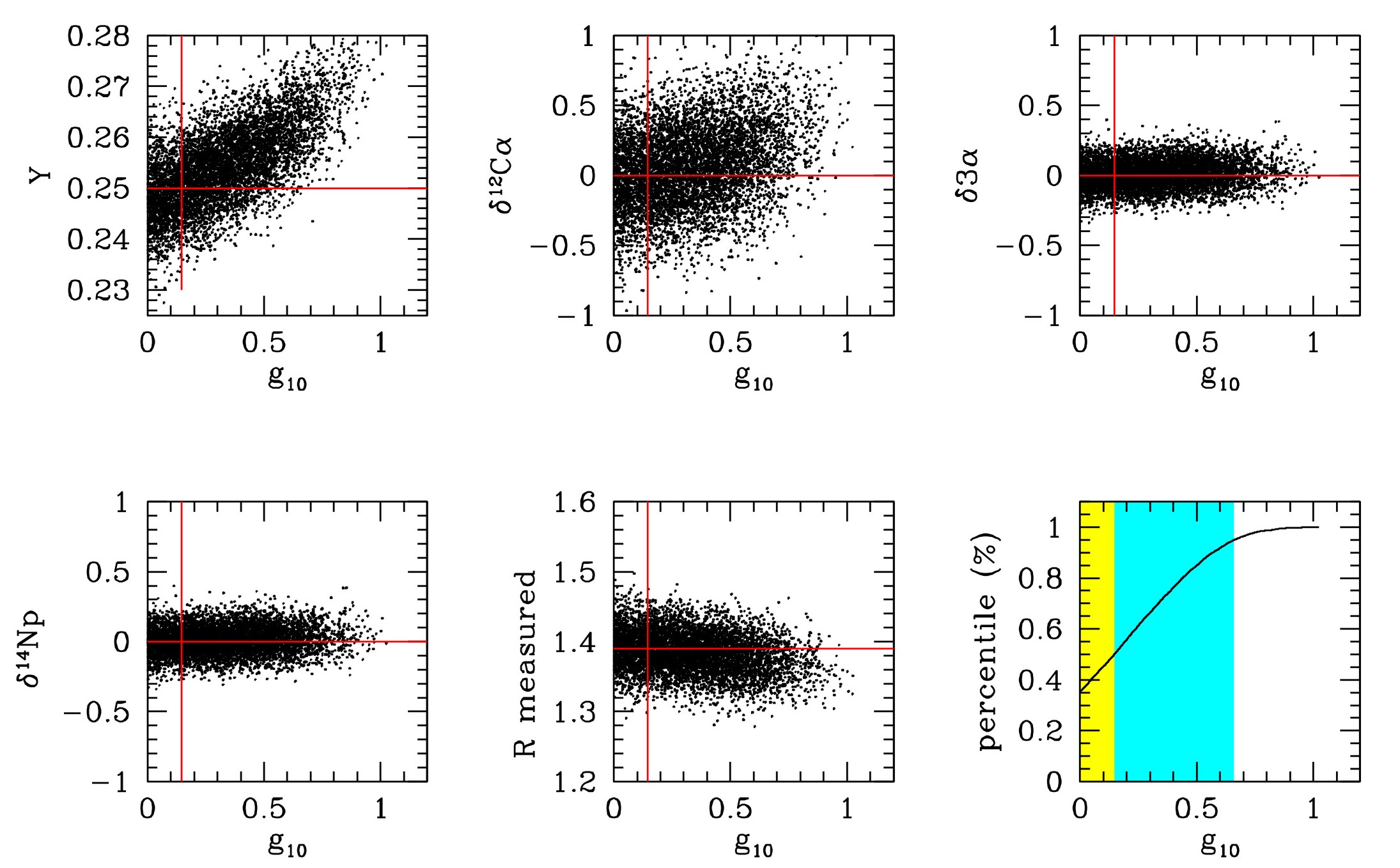}
\caption{Results of the MC analysis of $10^4$ SCMDs, as obtained by Gaussian-distributed samplings of 5 parameters, namely: the He mass fraction ($Y$), the measured $R$ parameter and 3 of the most relevant nuclear reaction rates (see Tab.~\ref{tab_mcR}). The dimensionless  quantity  reported in the horizontal axis of each panel is $g_{10}=g_{a\gamma}/$($10^{-10}$~GeV). The vertical-red lines correspond to the median of the resulting $g_{a\gamma}$ distribution, while the horizontal-red lines show the adopted central values of the 5 parameter distributions (see Tab.~\ref{tab_mcR}). Eventually, the curve in the right-bottom panel shows the percentiles of the resulting $g_{a\gamma}$ probability distribution and the two colored areas delimit the 50th and the 95th percentiles, respectively. }
\label{fig:mcR}
\end{figure}

Then, a simple Monte Carlo (MC) is used to generate series of 4 parameter sets. At each extraction, a SCMD and the corresponding $R$ are calculated. Eventually, also the measured $R=1.39\pm0.03$~\cite{Ayala:2014pea} is sampled according to a Gaussian error distribution, and an appropriate  value of $g_{a\gamma}$ is obtained by requiring that the difference between the predicted and the measured $R$ would be reduced to 0. The results of this MC procedure are shown in Fig.~\ref{fig:mcR}. Since a non-zero $g_{a\gamma}$ is expected to reduce $R$, only $\delta R=R_{\rm{th}}(g_{a\gamma}=0)-R_{\rm{meas}} > 0$ implies a $g_{a\gamma} \neq 0$. Indeed, a negative $\delta R$ would require an energy source in the core rather than a cooling. Therefore, the smaller  the measured $R$ the larger the $\delta R$ and, in turn, a larger $g_{a\gamma}$ is needed to reconcile theory with observations. On the other hand, also an increase of the predicted $R$ leads to stronger coupling. For instance, larger $g_{a\gamma}$ are obtained in case of higher He mass fractions (see left-upper panel in Fig.~\ref{fig:mcR}). In fact, an increase of $Y$ implies a brighter ZAHB  and, in turn, less stars brighter than the ZAHB are counted on the RGB. Similarly, a faster $^{12}$C$(^{4}$He$,\gamma)^{16}$O reaction implies longer HB lifetime and, once again, larger $\delta R$ and $g_{a\gamma}$ (central-upper panel in Fig.~\ref{fig:mcR}). Let us finally stress that this method naturally account for possible correlations among the variations of the various inputs. As a whole, this analysis based on MC error sampling and SCMDs confirms and reinforces the previous bound for the axion-photon coupling, namely $g_{a\gamma}<0.66\times 10^{-10}$~GeV$^{-1}$ (95\% C.L.) for light axions.

\begin{table}[t!]
    \centering    
    \caption{Adopted values and relative uncertainties of the quantities varied in the MC analysis in Fig \ref{fig:mcR}.}
    \setlength{\tabcolsep}{3pt} 
{
\begin{tabular}{ |c| c| }
\hline
  input parameter & central value and STD   \\ 
 \hline
 Y & $0.25\pm0.01$  \\  
 $^{14}$N$(p,\gamma)^{15}$O & as in \cite{2011RvMP...83..195A} $\pm 10\%$   \\  
  $^{4}$He$(2^{4}$He$,\gamma)^{12}$C & as in \cite{1999NuPhA.656....3A} $\pm 10\%$  \\
 $^{12}$C$(^{4}$He$,\gamma)^{16}$O & as in \cite{2001PhRvL..86.3244K} $\pm 30\%$ \\ 
 \hline
\end{tabular}
}

\label{tab_mcR}
\end{table}

\begin{figure}[t!]
\centering
\vspace{0.8cm}
\includegraphics[width=1\columnwidth]{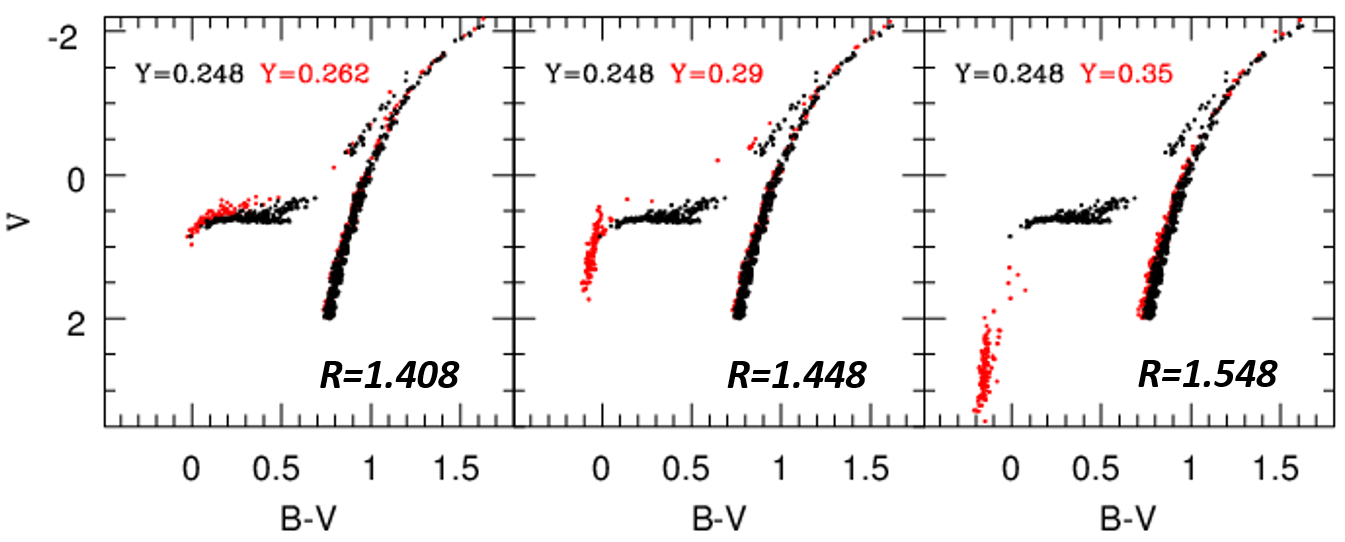}
\caption{SCMDs of GCs harboring two stellar populations, having the same age and metallicity, but different He mass fractions: black dots refer to stars with primordial He abundance ($Y=0.248$), while red dots represent stars with enhanced He. The corresponding values of $R$ are reported in each panel. Note that $R=1.408$ is found in the case of a single stellar population with primordial He.}

\label{fig:synthetic_multipop}
\end{figure}

SCMDs of HB stars may also be used to estimate the impact of multiple stellar populations on the derivation of the $R$ parameter. This is particularly important in case of significant variations of the He content among the different populations. He-enhanced HB stars have smaller mass and, in turn, they are hotter than stars with lower He. In addition, their core mass is higher, so that they are brighter than stars with primordial He. As shown in Fig.~\ref{fig:synthetic_multipop}, both these occurrences affect the HB morphology in the CMD and, in turn, the $R$ parameter. In case of limited variations of the He mass fraction, i.e., $\delta Y<0.03$, He-enhanced stars appear brighter than those with lower He, so that their presence does not affect the determination of the ZAHB level. In this case, the corresponding $R$ parameter is only slightly larger than expected in case of a single stellar population. A non-negligible increase of the $R$ value is instead found in case of larger $\delta Y$. Nevertheless, heavily He-enhanced stars can be easily identified and, eventually, isolated because they are concentrated in a blue tail of the HB and are well separated from the primordial stellar population. Summarizing, the use of the $R$ parameter to constrain the new physics may be less straightforward in case of HB showing extended blue tails. In this case, a special attention should be given to the evaluation of the $R$ parameter, to properly account for the possible presence of multiple stellar populations. 

\begin{figure}[t!]
\centering
\vspace{0.8cm}
\includegraphics[width=1\columnwidth]{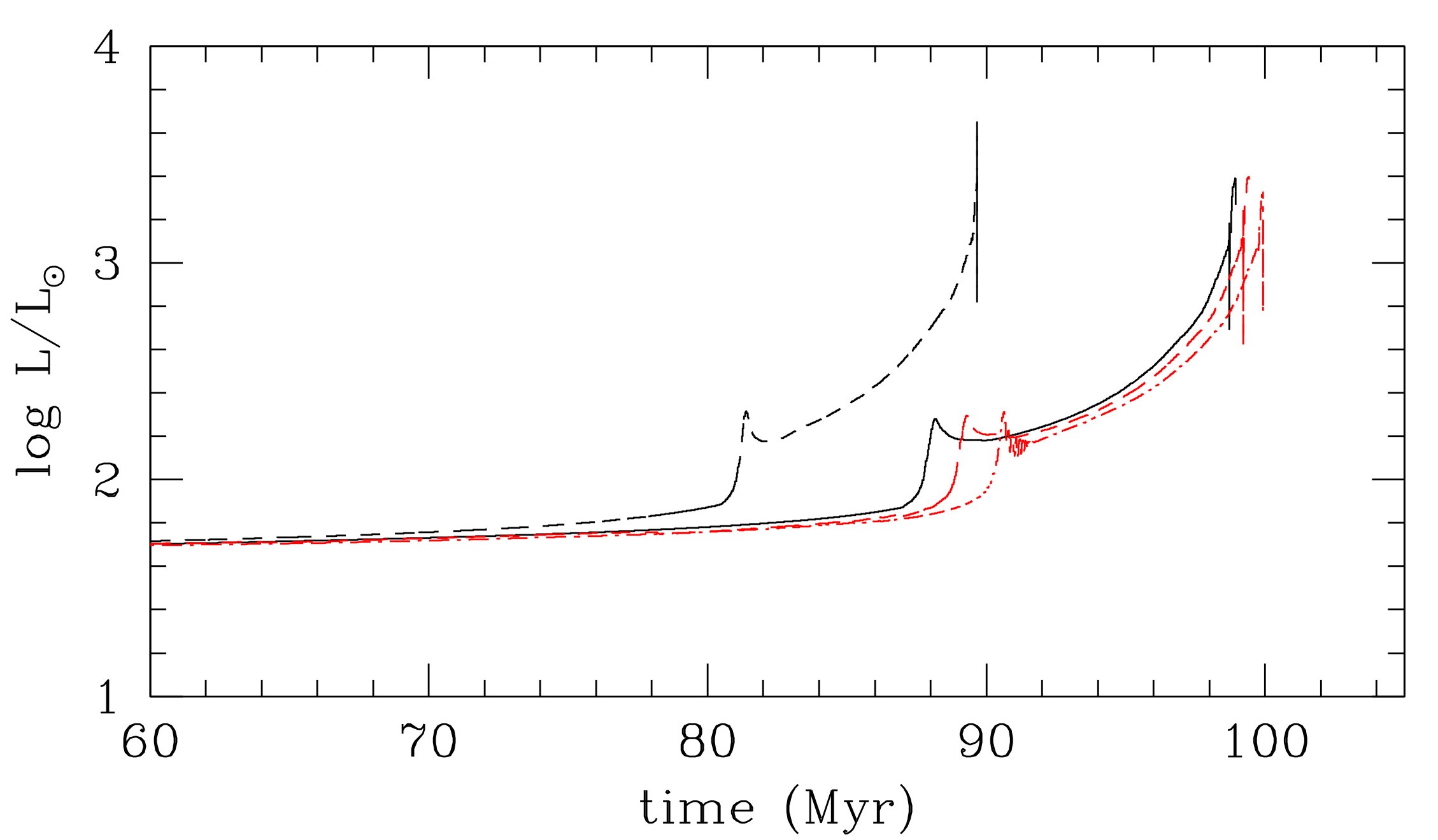}
\caption{Evolution of the stellar luminosity for four stellar models ($M=0.66~M_{\odot}$, $Y=0.248$1, $Z=0.001$) in the transition phase between the HB and the AGB. Three models are computed under different assumptions for the treatment of convection, but without additional energy leaks due to a possible axion production. In particular, the model represented by the black solid line includes overshooting and semiconvection as described in Ref.~\cite{1985ApJ...296..204C,Straniero:2002pf}. This is the same scheme adopted in Ref.~\cite{Ayala:2014pea}. Instead the two red lines refers to models computed by adding an arbitrary overshoot zone on top of the formal border of the convective core (see the text and Eq.~\eqref{eq:overshoot}). In the latter, the parameter $f_{\rm ov}$, which controls the exponential decay of the mean convective velocity in the overshoot zone, is 0.001 and 0.01, for the red-dashed and the red-dot-dashed lines, respectively. For comparison, the track of a model with thermal axion production by Primakoff ($g_{a\gamma}=5\times 10^{-11}$~GeV$^{-1}$), is also reported (black dashed line).}

\label{fig:R2_HB_AGB}
\end{figure}

So far, the uncertainties regarding our knowledge of the nuclear power that counterbalances the radiative energy loss have been considered in some detail, but those affecting the amount of He fuel available in the core of a HB star have been ignored. As described in Sec.~\ref{sec:stelev_H_He}, convection, semiconvection, overshoot all affect the HB lifetime and, in turn, $R$. This problem may introduce a systematic error in the evaluation of the $g_{a\gamma}$. If we are interested to the upper bound of the coupling strength, a conservative approach consists in adopting a convection algorithm that maximizes the HB lifetime.  A quite common approach is to use the classical Schwarzschild  criterion to distinguish between stable and unstable regions against convection, eventually extending the mixing outside the  borders of the convective regions to account for the penetration of convective bubbles or plumes (convective overshoot). In this case, the extension of this overshooting and the mixing efficiency are not predicted by the theory and are regulated by tuning some free parameter. In the case of the core-He burning phase, this arbitrary overshoot prevents the formation of a semi-convective zone above the fully-convective core. A more effective approach has been early proposed by Ref.~\cite{1985ApJ...296..204C} (see also Ref.~\cite{Straniero:2002pf}). In this case, overshooting and semiconvection are simulated by means of an algorithm that searches for the layers where the convective neutrality is established, i.e., $\nabla_{\rm rad}= \nabla_{\rm ad}$. As illustrated in Sec.~\ref{sec:stelev_H_He}, when He is converted into C and O, the boundary of the convective core becomes unstable and even a minimal boundary mixing pushes the instability outward. This overshoot is a natural consequence of the higher opacity of C and O. To account for this phenomenon, the mixing is progressively extended beyond the formal convective border, mesh by mesh,  until a stable mesh is found ($\nabla_{\rm rad}\leq \nabla_{\rm ad}$). The mixing procedure above described will allow thermal neutrality  to be first reached in correspondence of the minimum of $\nabla_{\rm rad}$. So far, the new border of the fully convective core is stable. Further outside, the radiative $T$ gradient still exceeds the adiabatic one and a further mixing, but limited to the region beyond the minimum, is applied, until the $\nabla_{\rm rad}=\nabla_{\rm ad}$ condition is established also above the convective core (semiconvection). According to this procedure, the extension of the convective core overshoot and the chemical composition in the semiconvective region above it are not fixed a priori, e.g., by tuning some free parameter,  but is obtained naturally by means of successive mixings and convective-neutrality checks. With respect to a classical HB model, those that account for  the convective core overshoot induced by the conversion of He into C and O, and an appropriate treatment of semiconvection, result in a substantially longer HB lifetime. Let us finally note that a mechanical overshoot, i.e., an additional sizable penetration of convective bubbles (or plumes) in the region above the convective core, a sort of dynamical overshoot,  cannot be excluded. In extreme cases, it  may result in a even longer HB lifetime.  A definitive test and, perhaps, a more stringent bound, could be achieved by means of stellar structure constrains derived from astroseismology studies of evolved GC stars.  Fig.~\ref{fig:R2_HB_AGB} and Tab.~\ref{tab:HB_AGB} compare models obtained under different assumptions about the treatment of convective. In particular, the black lines refer to models with an implicit treatment of convection and semiconvection~\cite{1985ApJ...296..204C,Straniero:2002pf}, while the red lines refer to models computed by assuming an arbitrary penetration of convective bubbles (or plumes) in the region above the convective core. In the latter, the mixing is calculated by means of an advective scheme and the mean convective velocity is assumed to decay exponentially in the overshoot zone, namely:
\begin{equation}
    v_{\rm ov}= v_0\exp\left(  -\frac{\delta r}{f_{\rm v}H_P}\right) \,\ ,
\label{eq:overshoot}
\end{equation}
where, $\delta r=r-r_0$ is the distance from the external border of the convective core, $v_0$ is the average convective velocity at the external border of the convective core, as computed by means of the classical mixing-length theory, $H_P$ is the pressure scale height and $f_{\rm ov}$ is a free parameter that regulates both the extension of the overshoot zone and the mixing efficiency. 

\begin{table}[t!]
 \caption{Comparison between HB$+$AGB models computed under different assumptions about the input physics. In the first block, there are four models all computed with the treatment of convection and semiconvection described in Ref.~\cite{1985ApJ...296..204C}, while in the second block, there are model computed by assuming a parameterised dynamical penetration of convective bubbles (or  plumes) outside the convective core. The parameter $f_{\rm ov}$ is defined in the text. The second and the third columns report the HB and the AGB lifetimes, respectively, while the parameters $R$, $R_1$ and $R_2$ are listed in the last three columns. The initial parameters, $M=0.66~ M_\odot$, $Z=0.001$ and $Y=0.248$, are the same for all the model. 
    }
    \centering
    \begin{tabular}{|c|c|c|c|c|c|}
\hline
     &	$\tau_{\rm HB}$	&  $\tau_{\rm AGB}$ & $R$ & $R_1$ & $R_2$  \\
     \hline
 \multicolumn{6}{|c|}{models with induced overshoot and semiconvection}\\
 \hline
reference  &    87.52 & 11.17 & 1.41 & 0.18 &  0.13 \\ 
$^{12}$C$(^{4}$He$,\gamma)^{16}$O$+30\%$   &    88.98 & 11.89 & 1.43 & 0.19 & 0.13 \\ 
$^{4}$He$(2\,^{4}$He$,\gamma)^{12}$C$+10\%$  &    88.46 & 11.10  & 1.42 & 0.18 & 0.13 \\ 
$g_{a\gamma}=5\times 10^{-11}$ GeV$^{-1}$  &	80.84  & 8.58  &	1.30	&	0.14 & 0.11	\\
 \hline        
  \multicolumn{6}{|c|}{models with dynamical penetration beyond the convective core}\\
\hline
$\,\ f_{\rm ov}=0.005$  &      88.66 & 10.52 &  1.43 & 0.17 & 0.12 \\   
$f_{\rm ov}=0.01$   &      90.12 & 9.57 &   1.45 & 0.15 & 0.11  \\  
$f_{\rm ov}=0.03$   &      95.03 & 8.22 &   1.53 & 0.13 & 0.09   \\ 
\hline
    \end{tabular}
   
    \label{tab:HB_AGB}
\end{table}

Once the He fuel is almost fully consumed, the convective core disappears and the star rapidly leaves the HB and enters in to the AGB phase. While the newborn C-O core contracts, the stellar luminosity increases and the effective temperature decreases. Later on, a He-burning shell settles just outside the previous semiconvective region. This occurrence causes a temporary stop of the luminosity increase. Fig.~\ref{fig:R2_HB_AGB} shows the evolution of the luminosity in the transition between HB and AGB for stellar models with mass $M=0.66~{\rm M}_\odot$, metallicity  $Z=0.001$ and He mass fraction $Y=0.248$. In all cases, an evident  slowdown of the evolution occurs at $\log L/L_\odot\sim 2.2$, when the He-burning shell becomes fully efficient.  Looking at the CMD shown in Fig.~\ref{fig:synthetic}, a bump of stars in the AGB region (in between $V=14.5$ and $15$ mag) is clearly distinguished. It is the consequence of the longer time spent by these stars at that luminosity.

If the $R$ parameter may be used to constraint the duration of the HB phase, the AGB evolutionary timescale can be probed by means of other two parameters, namely: $R_1=N_{\rm AGB}/N_{\rm RGB}$ and $R_2=N_{\rm AGB}/N_{\rm HB}$. A recent work, Ref.~\cite{Dolan:2022kul}, analyzing a limited number of clusters for which $R_2$ measurements are available from extant photometric studies, reports a quite stringent bound for the axion-photon coupling, namely $g_{a\gamma} \lesssim 0.47 \times 10^{-10}$~GeV$^{-1}$. It has been also proposed that this bound could be improved to $g_{a\gamma} \lesssim 0.34 \times 10^{-10}$~GeV$^{-1}$ once  asteroseismology constraints are taken into account. In principle, an advantage of both $R_1$ and $R_2$ is that the core temperature during the AGB phase is higher than in the HB phase, so that the axion production rate is enhanced and weaker axion-photon couplings could be investigated. However, after the central He exhaustion, electron degeneracy suddenly develops in the  C-O core, so that the axions production by Primakoff and Compton are suppressed. Eventually, in case of axion coupled to electrons, Bremsstrahlung could provide an additional energy sink during the AGB phase (see the seminal work of  Ref.~\cite{1999MNRAS.306L...1D}). In practice, the axion production by Primakoff and Compton is more efficient in the region where the shell-He burning takes place. In contrast, a clear disadvantage of $R_1$ and $R_2$ is that the evolution is much faster after the central-He exhaustion, so that a rather small number of stars are found in the AGB phase, very few out of the AGB bump. This occurrence implies a low significance of the AGB statistical sample in the majority of the photometric studies of galactic GCs. On the theoretical grounds, it should be noted that also the AGB lifetime is influenced by the same uncertainties affecting HB models, such as, in particular,  those related to the treatment of turbulent convection and semiconvection.  In particular, as noted by Ref.~\cite{1985ApJ...296..204C}, the presence of a semiconvective region, where the chemical composition (He/C/O) is modeled by the adopted mixing scheme, affects the leaving from the HB and the following early-AGB phase. The evolutionary tracks in Fig.~\ref{fig:R2_HB_AGB} illustrate some examples. The corresponding HB and AGB lifetimes, as well as the $R$, $R_1$ and $R_2$ parameters estimated by combining these lifetimes with the $\tau_{\rm RGB}$, are reported in  Tab.~\ref{tab:HB_AGB}. In principle, a more stringent constraint could be obtained by combining independent measurements of these three star number ratios. This procedure may alleviate the problem of the degeneracy among the effects caused by the uncertainties of the standard input physics and a non negligible axion-photon coupling.

\section{White dwarfs}
\label{sec:white_dwarfs}

WDs are the late stage of the evolution of low and intermediate mass stars, $M \lesssim 8-9~{\rm M}_\odot$. Their structure is relatively simple: an electron degenerate core containing the bulk of mass, surrounded by a thin non-degenerate or partially degenerate mantle made of helium and hydrogen which is absent in $\sim 20$\% of cases. Because of degeneracy, WDs cannot obtain energy from nuclear reactions and, as a consequence, their evolution is just a process of gravo-thermal evolution, see Ref.~\cite{2010A&ARv..18..471A}  for a review, that can be described as:
\begin{equation}
L + {L_\nu } =  - \int\limits_{{M_{\rm WD}}} {{c_V}\frac{{d{T_C}}}{{dt}}dm - \int\limits_{{M_{\rm WD}}} {T{{\left( {\frac{{\partial P}}{{\partial T}}} \right)}_{V,x}}\frac{{dV}}{{dt}}dm + \left( {{l_s} + {e_s}} \right){{\dot m}_e} + {{\dot g }_z} \pm \left( {{\dot\varepsilon _e}} \right)} } \,,
\label{sec:white_dwarfs_lum}
\end{equation}
where the left hand side contains the classical energy losses by photons and neutrinos, and the first two terms on the right hand side represent the energy release associated to the gravothermal readjustment of the structure. Additionally, ${\dot g }_z$ represents the energy released by the gravitational settling of neutron rich species like $^{22}$Ne, $\dot m$ is the rate at which matter crystallizes and $l_s$ and $e_s$ the associated latent heat and gravitational energy caused by the sedimentation of heavy species and, finally, $\dot\varepsilon_e$ represents any additional source or sink that could exist.

This equation has to be complemented with a relationship between the temperature of the core, $T_C$, and the luminosity. Typically $L \propto T_C^\beta$ with $\beta \approx 2.5-2.7$, although it depends on the nature of the envelope. Solving Eq.~\eqref{sec:white_dwarfs_lum} allows one to obtain $L=L(t_{\rm cool})$ and $T_C=T_C(t_{\rm cool})$ and use WDs as an efficient forensic tool to study the temporal evolution of the Galaxy. Notice, however, that the total lifetime of a WD is $t_{\rm WD}=t_{\rm PS}+t_{\rm cool}$ where $t_{\rm PS}$ is the age of the progenitor and that the real age of a WD is dominated by the uncertainty of the age of the progenitor except in the case of massive WDs where $t_{PS}$ can be much smaller than the cooling time.

In this Section we will show how WDs provide strong limits on the axion-electron coupling $g_{ae}$. The structure of this Section is as follows. In Sec.~\ref{subsec:structure} we will recall the structure and the evolution of WDs.  In Sec.~\ref{sec:drift} we describe axion bounds from the secular drift of the pulsation period. Then, in Sec.~\ref{sec:axwdlf} we present the axion bounds form the WD Luminosity Function (WDLF).

\subsection{Structure and evolution of white dwarfs}
\label{subsec:structure}

\begin{figure}[!t]
\minipage{0.45\textwidth}
  \includegraphics[width=\linewidth]{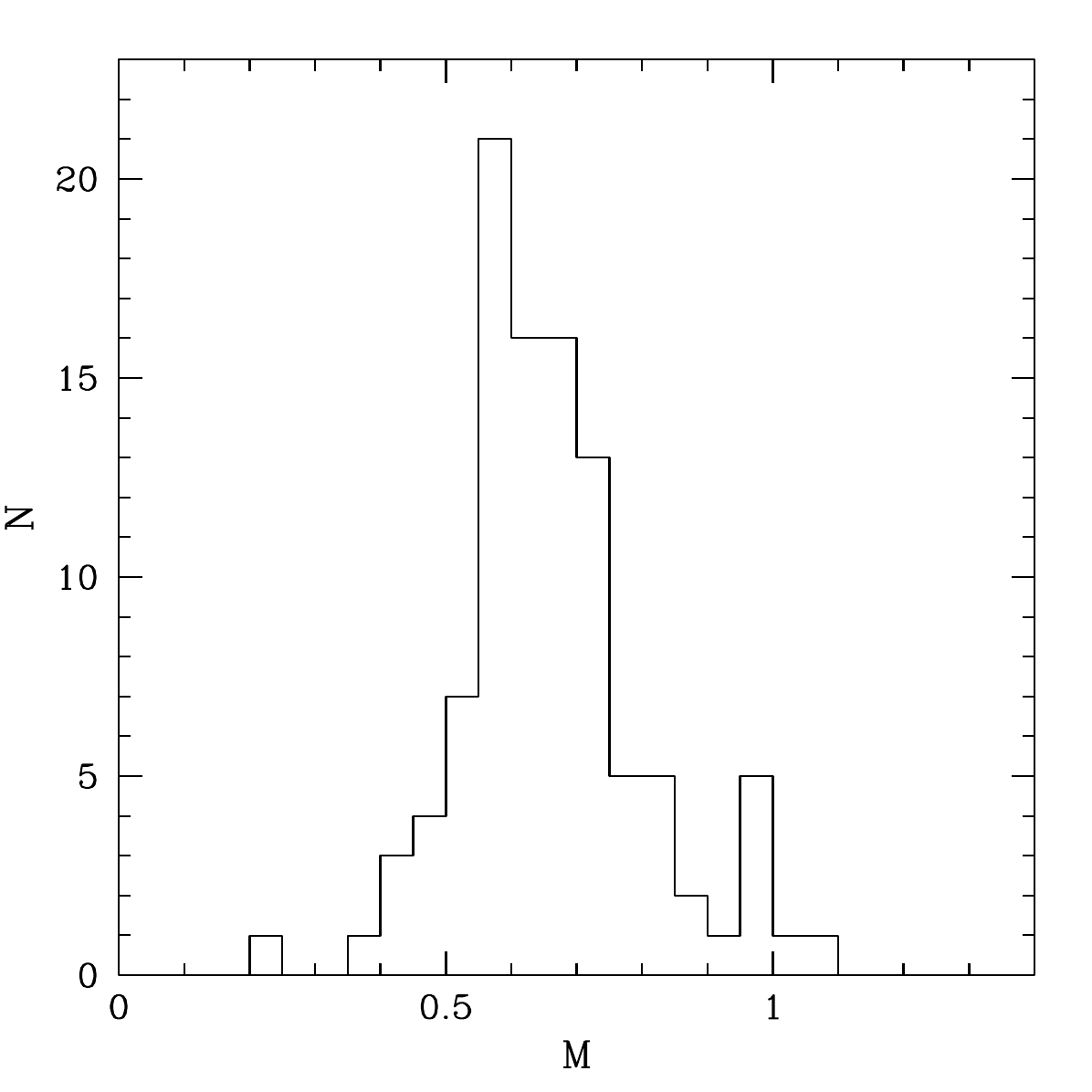}
\endminipage\hfill
\minipage{0.45\textwidth}
  \includegraphics[width=\linewidth]{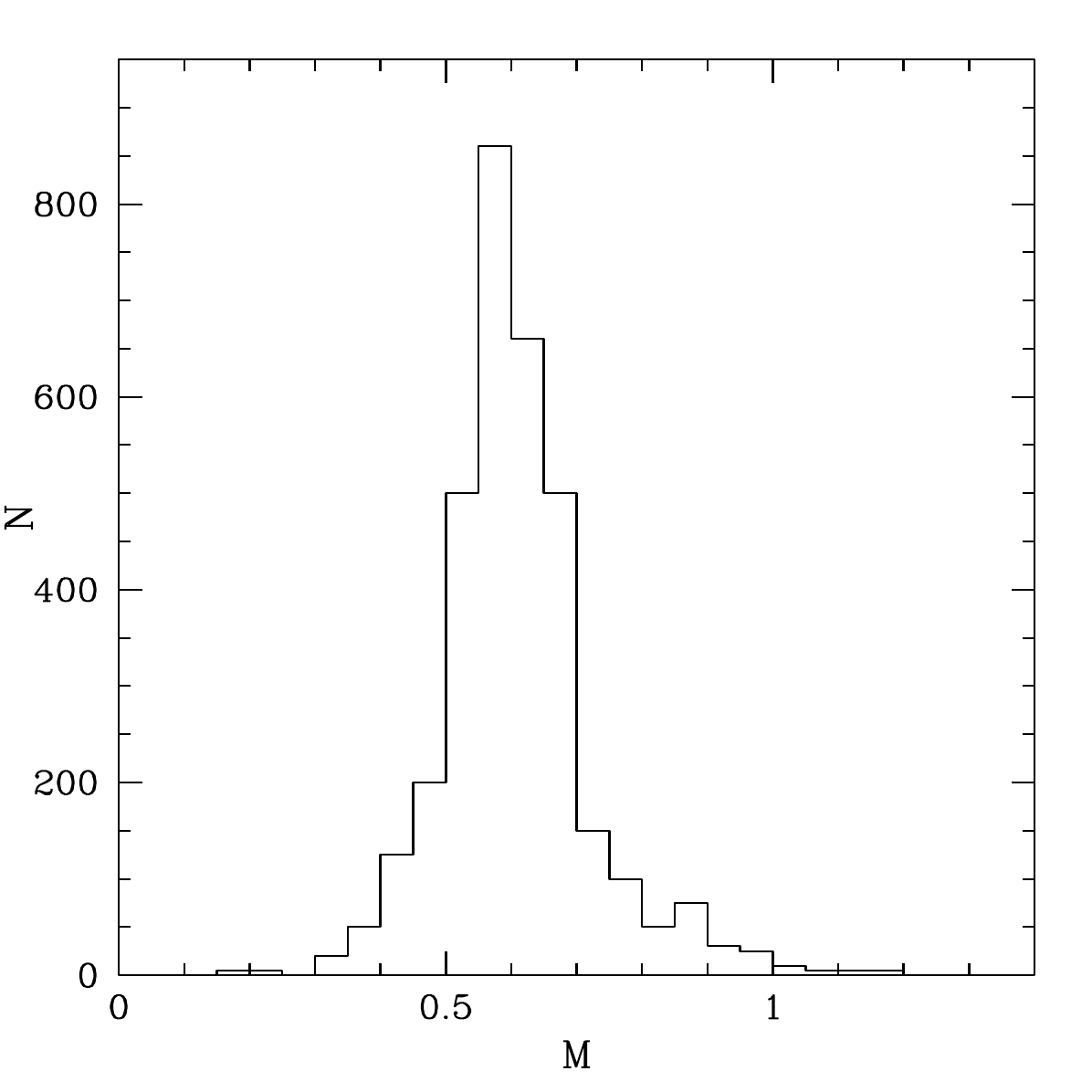}
\endminipage\hfill
\caption{WD mass distribution: on the left a 20 pc sample; on the right, the SDSS DR7 sample. (Data taken from   Ref.~\cite{2016MNRAS.461.2100T} with permission).}\label{wdmd}
\end{figure}

There are three types of WDs according to the chemical composition of their degenerate core. Those with a mass $\gtrsim 1.05~{\rm M}_\odot$ are made of a mixture of oxygen and neon (O/Ne-WDs) and are descendants of stars with a mass in the range of 8--9~${\rm M}_\odot$, these figures being rather uncertain, those with a mass $0.4~{\rm M}_\odot \lesssim M \lesssim 1.05~{\rm M}_\odot$ are made of a mixture of carbon and oxygen (C/O-WDs) and are descendants of stars with a mass in the range $0.8~{\rm M}_\odot\lesssim M \lesssim 8~{\rm M}_\odot$, and those with a mass $\lesssim 0.4$~M$_\odot$ that are made of helium (He-WDs) and are the byproduct of the evolution of interacting binaries since stars with masses smaller than $0.8~M_\odot$ have no time to ignite helium. Fig.~\ref{wdmd} displays the mass distribution of WDs within a distance of 20 and 40~pc (left and right panels, respectively). The distribution is strongly peaked around $\sim 0.6~{\rm M}_\odot$ and progressively declines with the mass. The structures present in the region of $0.7-0.9~{\rm M}_\odot$ are still under discussion and could be a consequence, at least in part, of the merging of double degenerates~\cite{2020ApJ...898...84K}.

One of the persisting problems is the determination of the initial conditions of WD when they are born. These initial conditions are very complex and strongly dependent on the amount of residual hydrogen left at the end of the AGB stage. If it is large enough, $M_H \gtrsim 10^{-4}~{\rm M}_{\odot}$, hydrogen burning via pp-reactions never stops and it can be even dominant at low luminosities. In fact, the overabundance of bright WDs in the GCs M13 and NGC 6752 can be interpreted as a consequence  of a residual thermonuclear burning of hydrogen in the outer layers of low metallicity WDs~\cite{2021NatAs...5.1170C,2022ApJ...934...93C}. Fortunately, asteroseismological observations seem to constrain the mass of hydrogen well below such a critical value.

\begin{figure}[!h]
\begin{center}
\includegraphics[width=11cm, clip=true, trim= 0cm 0cm 0cm 0cm]{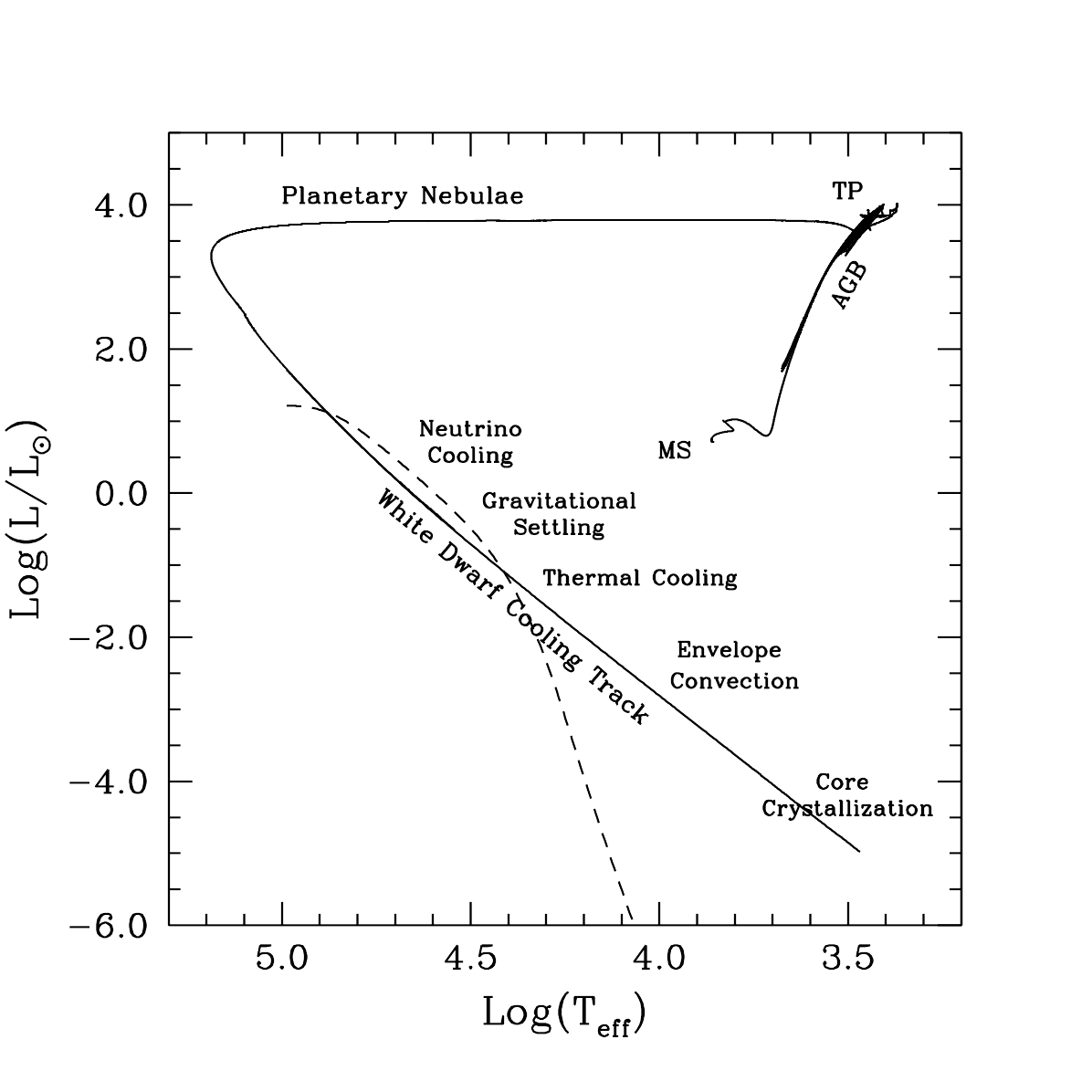}
\caption{Evolution of a star of 1.5~M$_\odot$ in the CMD according to the La Plata models~\cite{2010A&ARv..18..471A}. The dashed line represents the neutrino luminosity. (Figure taken from Ref.~\cite{Isern:2022vdx} with permission).
}
\label{wdhr}
\end{center}
\end{figure}

At very early times, the main contributor to the luminosity of the nascent WD is hydrogen burning via CNO cycle and it is during this epoch that the final thickness of the hydrogen layer is configured. This stage is very short, $\sim 10^4$~yr, nuclear reactions cease abruptly and the luminosity drops from $\log (L/L_\odot) \approx 4$ to $\log (L/L_\odot) \approx 1-2$~\cite{2010A&ARv..18..471A}. From this moment neutrino emission becomes dominant, Fig.~\ref{wdhr}, and forces the different thermal structures to converge to a unique one for models dimmer than $\log (L/L_\odot) \approx -1.5$~\cite{1989ApJ...347..934D}. The comparison between the evolution times obtained with the LPCODE and the BaSTI codes using exactly the same inputs differ at this epoch by an 8\% as a consequence  of the different converged models at the beginning of the cooling sequence~\cite{2013A&A...555A..96S}. This means that hot WDs have to be used with care when testing new physics.

After $10^7 - 10^8$~yr, depending on the mass of the star, neutrino emission rapidly drops at this stage because of its steep dependence on the core temperature, $L_\nu \propto T^7_C$, and photon losses become dominant. At this epoch photon luminosity is controlled by a thick non-degenerate layer with an opacity dominated by hydrogen, if present, and helium that is weakly dependent of the metal content~\cite{1984ApJ...282..615I,1989ApJ...347..934D,2010A&ARv..18..471A}. At the same time, the core of the WD behaves like a fluid  that can be described as a Coulomb plasma not very strongly coupled, $\Gamma <178$~\cite{1994ApJ...434..641S,2010CoPP...50...82P,2021ApJ...913...72J}\footnote{The Coulomb  parameter  $\Gamma =  Z^2e^2/akT$,  where $a=[3/(4\pi n_{\rm  i})]^{1/3}$ is the ion-sphere  radius, $n_{\rm i}$ is the ion number density and  $k$ is the Boltzmann constant, compares the energy of the Coulomb interaction with the thermal energy. In the case of one component plasma the transition liquid-solid occurs for $\Gamma \simeq 178$}. A key ingredient is the electron conductivity at the frontier between moderate and strong degeneracy. Models predict longer cooling times at high luminosities and shorter cooling times at low luminosities~\cite{2022MNRAS.509.5197S} when the Blouin et al. conductivities~\cite{2020ApJ...899...46B} are used instead of those of Cassisi \emph{et al}.~\cite{2007ApJ...661.1094C}.

\begin{figure}[!h]
\center
  \includegraphics[width=0.8\linewidth,clip=true, trim=4cm 0cm 4cm 0cm]{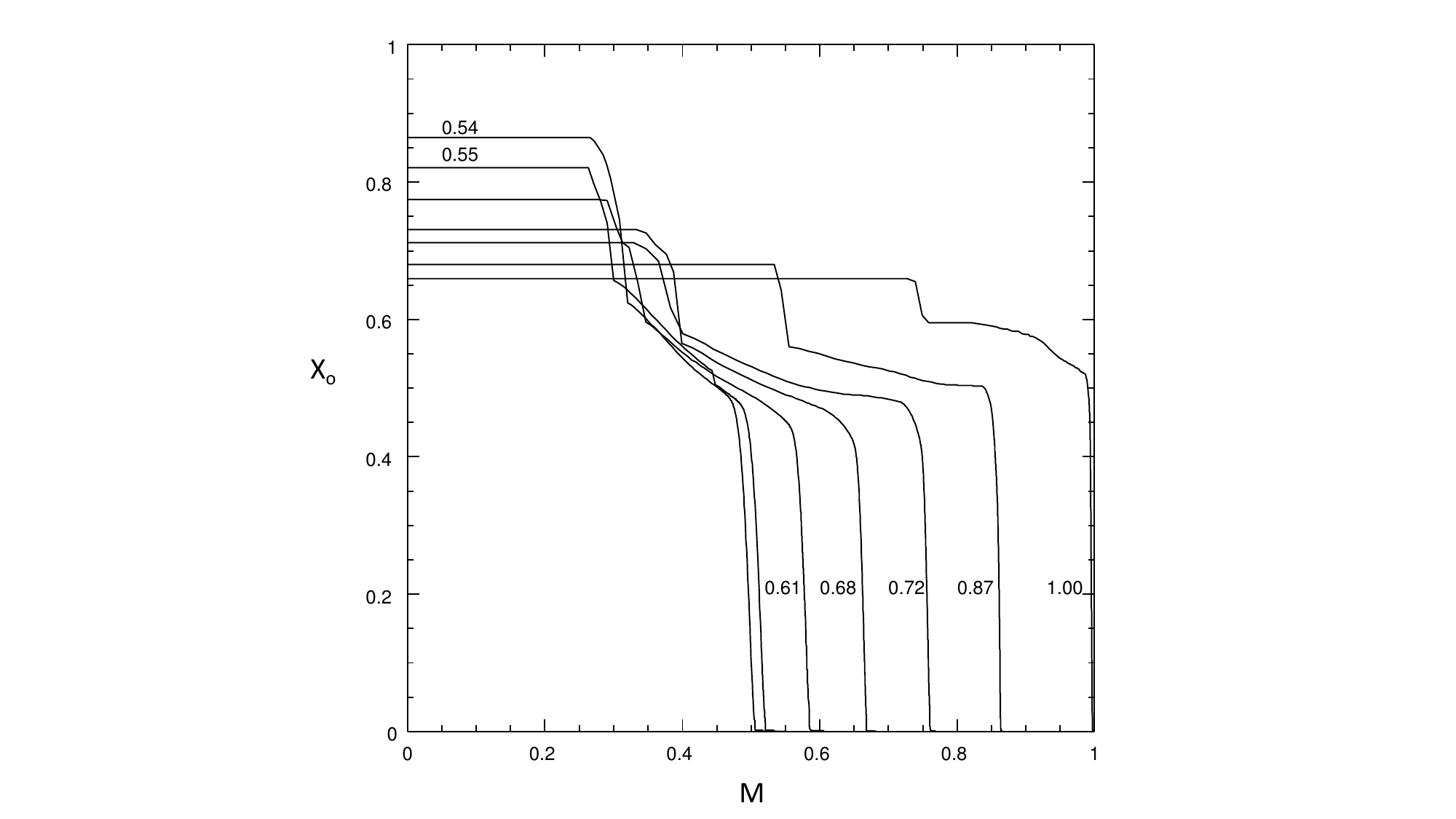}
\caption{Oxygen distribution in the interior of WDs with different masses~\cite{1997ApJ...486..413S}.
}\label{wdprof}
\end{figure}

The dependence of the specific heat on the chemical composition introduces an important source of uncertainty in the cooling rate of WDs. The abundance of oxygen in the center is larger than in the outer layers and this tendency increases when the mass decreases, Fig~\ref{wdprof}. This behavior is controlled by the rate of the reaction $^{12}$C$(\alpha,\gamma)^{16}$O and by the treatment given to semiconvection and overshooting~\cite{1997ApJ...486..413S,2017RSOS....470192S} which are both uncertain. Fortunately, asteroseismological techniques have started to provide direct information about the internal structure and have confirmed the stratified nature of the C/O core~\cite{2017A&A...599A..21D,2017ApJ...851...60R,2018Natur.554...73G}.

Isotopes  with small $Z/A$ ratios, like $^{22}$Ne, can play an important role during the cooling of WDs. This isotope is the result of the $\alpha$-burning of the  $^{14}$N  left at the end of the H-burning stage and its abundance is of the order  of the sum of C, N and O initial abundances of the parent star i.e. $X(^{22}$Ne$)\approx 0.02$ in the case of solar metallicities. Because of this neutron excess  and the high sensitivity of degenerate stellar structures on the electron number profile, $Y_e$, its migration towards the central regions of the WD can represent an important source of gravitational energy despite its low abundance~\cite{1991A&A...241L..29I,1998JPCM...1011263I}.

Mixtures of C/O/Ne are miscible during the liquid phase. However, as a consequence of its neutron excess $^{22}$Ne isotopes experience a downwards force $F=2m_pg$, where $m_p$ is the mass of the proton and $g$ the local acceleration of gravity, that induces its migration towards the central regions of the WD~\cite{1992A&A...257..534B,2001ApJ...549L.219B,2002ApJ...580.1077D}. This mechanism, however, is only efficient in the hot interiors and in the envelope of WDs, and only induces significant delays in massive WDs, see Ref.~\cite{Isern:2022vdx} and references there in for a description of this process.

When the temperature is low enough, the plasma experiences a phase transition and crystallizes into a classical body centered crystal (bcc). This transition provides two additional sources of energy, latent heat and gravitational sedimentation, Eq.~\eqref{sec:white_dwarfs_lum}. The contribution of the latent heat was introduced by Refs.~\cite{1968ApJ...151..227V,1976A&A....51..383S} who found it was of the order of $0.75 k T_S$ per nucleus, where $T_S$ is the temperature of solidification. The total energy released is not very high but since it happens at relatively low luminosities the cooling delay is not negligible.

During the process of solidification of Coulomb plasma mixtures, the chemical composition of solid and liquid phases in equilibrium is not equal. Therefore, if the solid is denser than the liquid it sinks towards the inner regions of the WD while the remaining lighter liquid rehomogenizes via Rayleigh-Taylor instabilities. On the contrary, if the solid is lighter it rises upwards and melts when the temperature of solidification, which depends on the density as $T_S \propto \rho^{1/3}$, becomes equal to that of the approximately isothermal core and rehomogenizes via Rayleigh-Taylor instabilities.

\begin{figure}[t!]
\center
\includegraphics[width=16cm, clip=true, trim= 0cm 10cm 0cm 1cm]{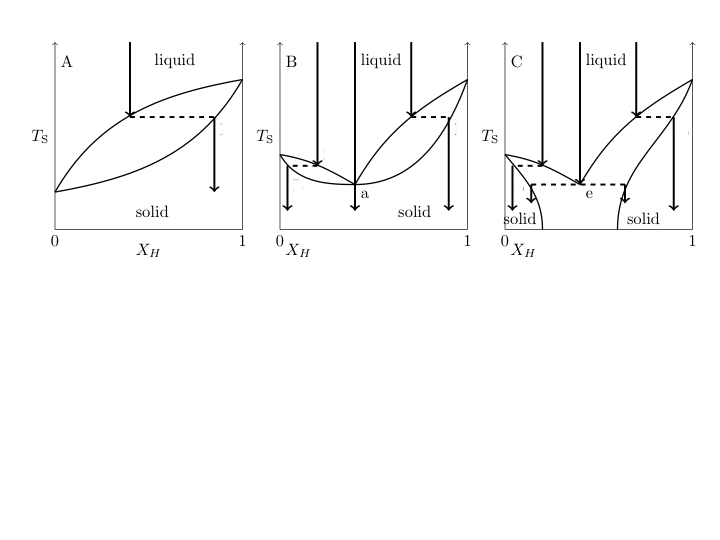}
\caption{From left to right, {\it A} spindle, {\it B} azeotropic and {\it C} eutectic phase diagrams for a binary mixture. Solid lines display the temperatures at which the liquid starts to solidify (liquidus line) and the solid starts to melt (solidus line) for each chemical composition. $T_S$ is the temperature at which the lightest species crystallizes, $x_H$ is the abundance of the heaviest species  by number, and the cooling tracks are represented by arrows. (Figure taken from Ref.~\cite{Isern:2022vdx} with permission).
}\label{wdphdco}
\end{figure}

The first phase diagram for a C/O mixture predicted a eutectic behavior~\cite{1980JPhys..41C..61S}, Fig.~\ref{wdphdco}~C, with a complete separation of both elements at the solid phase. This pioneering calculation was improved by Refs.~\cite{1988A&A...199L..15B,1993A&A...271L..13S} who found a phase diagram of the spindle form (Fig.~\ref{wdphdco}~A), while Refs.~\cite{1988ApJ...334L..17I,1993PhRvE..48.1344O} found an azeotropic behavior (Fig.~\ref{wdphdco}~B) and improved calculations by Refs.~\cite{2010PhRvL.104w1101H,2010PhRvE..81c6107M,2020A&A...640L..11B} confirmed such a behavior. In particular Blouin et al.~\cite{2020ApJ...899...46B} have obtained an azeotropic  abundance of oxygen by number of $x\approx 0.16$. Since the relative abundance of oxygen versus carbon  in WDs is larger than this value, the solid that forms is richer in oxygen than the liquid and settles down while the carbon rich liquid remnant mixes with the upper layers. The outcome of this process is the formation of a solid, oxygen rich core surrounded by a convective mantle, a structure that remembers that of the interior of the Earth~\cite{2017ApJ...836L..28I}.

As already stated, the migration of neutron rich impurities like $^{22}$Ne and $^{56}$Fe towards the central regions can be an important source of gravitational energy. In a first calculation it was found an azeotropic behaviour when the C/O/Ne ternary mixture was represented by a binary mixture in which C and O were substituted by an average nucleus and predicted, under this assumption, the formation of a neon-rich core surrounded by a mantle depleted of neon with a subsequent huge delay in the cooling~\cite{1991A&A...241L..29I}~\footnote{Notice that this effect depends on the initial metallicity of the parent star and that the oldest, cooling time plus age of the parent star, WDs are descendants of low metallicity stars.}. Nowadays, the nature of the phase diagram of the C/O/Ne mixture is an open question, but depending on the chemical composition and mass two scenarios are possible, one in which a neon rich core forms, and another in which a carbon-neon shell without oxygen forms~\cite{2021ApJ...911L...5B}. See Refs.~\cite{Isern:2022vdx,Saumon:2022gtu} for more details. 

When the temperature is small enough and the star is almost completely solid, the specific heat follows the Debye’s law and the WD cools down very quickly. However, since the outermost layers are still far from this regime, their energy content prevents the sudden disappearance of the WD~\cite{1989ApJ...347..934D}. The nature and the evolution of the envelope is crucial since the luminosity not only depends on the properties of the core but also on the properties (mass, chemical composition and structure) of the outer layers. As a consequence of the strong gravitational field the envelope has the tendency of becoming stratified, the lightest elements floating above the heaviest ones. This tendency is counterbalanced by molecular diffusion, convection and other processes trying to restore the chemical homogeneity. The outcome of such physical processes is a rather rich phenomenology as it can be seen in Tab.~\ref{table:wdclass}.

\begin{table}[t!]
\caption{The WD zoo~\cite{1983ApJ...269..253S}.}
\centering
\begin{tabular}{l l} 
 \hline 
DA   & Pure hydrogen layers \\
        & 90,000~K > T$_e$ 6,000~K, below this temperature Balmer lines are not seen \\
DAZ & DA type with metal traces \\
DO  & Spectrum dominated by ionized helium \\
       & 100,000~K > T$_e$ >  45,000~K \\
       & C, N, O, and Si are present in the photosphere \\
       & The coolest are hydrogen poor \\
DB   & Helium dominated atmospheres \\
       & 30,000~K > T$_e$ > 12,000~K \\
       & There is a gap between DO and  Db classes \\
DBZ & DB with metal traces \\
DQ  & Helium dominated atmospheres \\
       & 12,000~K > T$_e$ > 6,000~K \\
       & Carbon abundances in the range of $ 10^{-7} - 10^{-2}$ \\
DQZ& DQ with metal traces \\
DZ   & Only metallic feature (CaII H-K lines) \\
DC   & So cool that the dominant components are not seen \\
       & No lines deeper than the 5\% \\
\hline
\end{tabular}

\label{table:wdclass}
\end{table}

WDs displaying hydrogen in their spectra are called DAs, while those in which hydrogen is absent are called non-DAs which, in turn, are composed by stars with different spectroscopic properties that are classified in order of decreasing temperatures as DO, DB, DC, and DQ. The most common interpretation  is that the DAs have a double layer envelope made of hydrogen ($M_H \lesssim 10^{-4}$~M$_\odot$) and helium ($M_{He} \sim 10^{-2}$~M$_\odot$), while  the non-DAs have just a single He layer or an extremely thin H layer on the top.

The DAs represent $\sim 80$\% of the total but this proportion changes along the cooling sequence for which reason a mechanism able to change this property must exist~\cite{1997ASSL..214..165S}. The problem is that the initial conditions of the formation of WDs are not well known and, for the moment, it is not possible to disentangle which part of this behavior is inherited and which part has an evolutionary origin.  In any case it is possible to adjust the parameters of the AGB progenitors to obtain $\sim 20$\% of WDs completely devoid  of the hydrogen layer~\cite{2010A&ARv..18..471A}. 

It is believed that DAs have their origin on planetary nebula cores that have a mass of hydrogen in the range of $10^{-9} - 10^{-4}$~M$_\odot$. As the star cools down, the outer convective envelope deepens and, depending on the mass, the hydrogen layer is completely engulfed by the He-layer in such a way that the WD becomes a non-DA for which reason the ratio DA/non-DA decreases with the effective temperature.

\begin{figure}[h]
\center
  \includegraphics[width=0.7\linewidth,clip=true,trim=0cm 1cm 0cm 3cm]{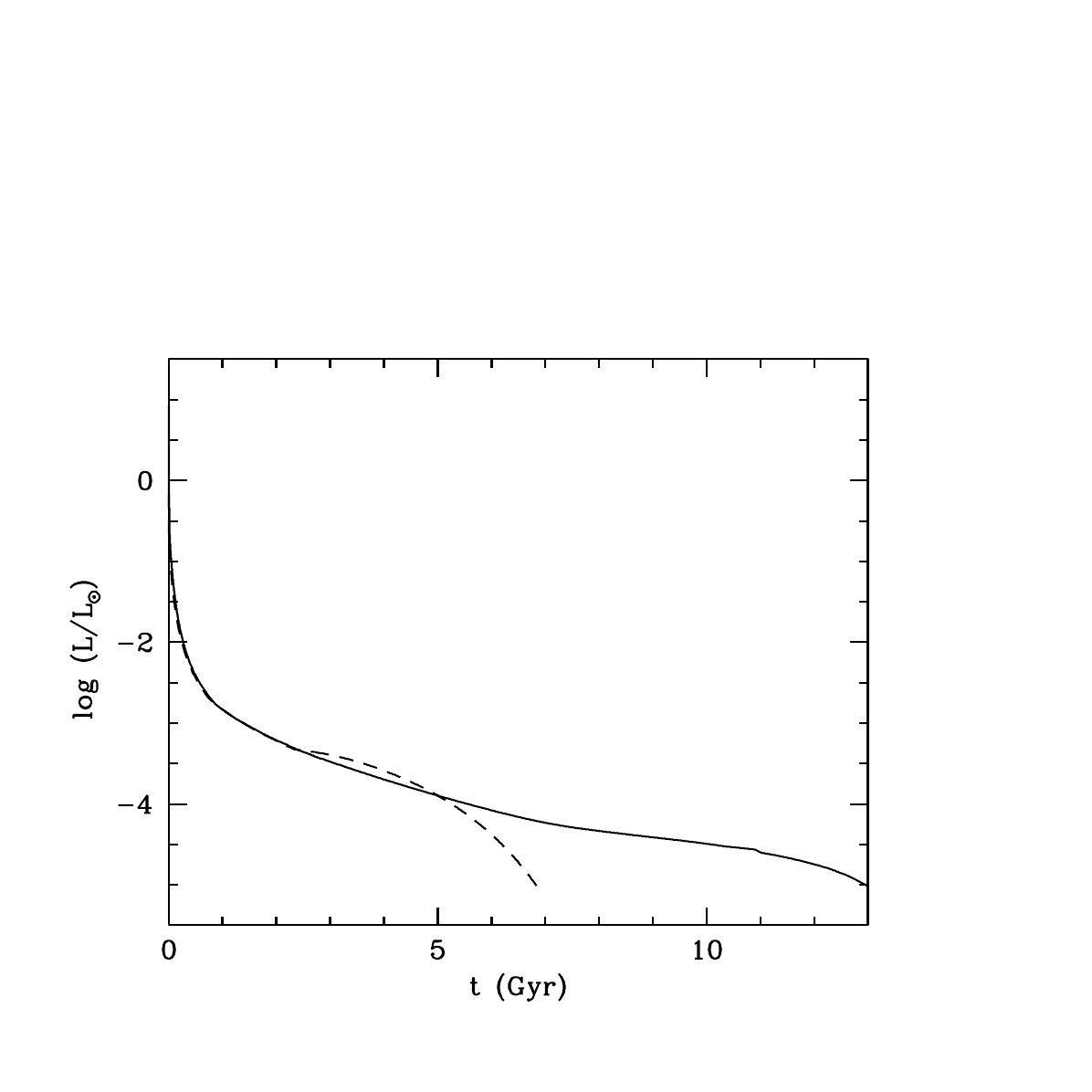}
\caption{Luminosity versus time for a 1.0~M$_\odot$ DA (solid line) and a non-DA (dashed line) WD.
(Data taken from Ref.~\cite{2022MNRAS.509.5197S}).}\label{wdlt}
\end{figure}

The non-DA WDs are thought to be the descendants of He-rich central stars of planetary nebulae. At the beginning they look as PG 1159 stars and later on, as they cool downs, as DOs. During this epoch  the residual hydrogen present in the helium envelope floats up to the surface  and forms a layer that, when the temperature is of the order of  50.000~K, is thick enough to hide the He-layer and the white dwarf becomes a DA. When the temperature goes below 30.000~K, the He-layer becomes convective and the star recovers the non-DA character, now as a DB WD, and as the star continues to cool down it becomes a DC~\footnote{Notice that a fraction of DCs has a DA origin.}. The lack of non-DA WDs  in the temperature range 50.000-30.000~K is known as the DB gap. Gaia data have shown that the WD domain of the CMD contains two branches, A and B, being the stars of the second branch more massive than those in the first branch~\cite{2018A&A...616A..10G}. The A-branch contains $\sim 94$\%  of DA WD while the B-branch contains $\sim 64$\% of non-DAs and only  $\sim 36$\% of DAs~\cite{2023MNRAS.518.5106J}.

Besides the phenomenological differences, the most important property is that DAs cool down more slowly than non-DAs since hydrogen is more opaque than helium, Fig.~\ref{wdlt}. Therefore, the nature of the envelope plays a very important role during the cooling stage dominated by photons. At present there are several properties of WDs that allow them to be used as physical laboratories: {\it i)} the secular drift of the period of pulsation of variable WDs, and {\it ii)} the shape of the luminosity function.

\subsection{Axion bounds from the secular drift of the pulsation period}
\label{sec:drift}
At different stages of their evolution, WDs experience episodes of variability~\cite{2010A&ARv..18..471A,Corsico:2019nmr,Corsico:2020eup,2021RvMP...93a5001A}. The frequency spectrum of the luminosity variations depends on the internal structure of the star and on the nature of the pulsation mode. Radial modes are the simplest ones and maintain the spherical symmetry~\footnote{They are characteristic of Cepheids and RR Lyr stars.} while non-radial modes can be classified as toroidal and spheroidal. According to the restoring force, the spheroidal ones can be divided into g- and f-modes if the restoring force is gravity, and p-modes if the restoring force is the pressure gradient. In general, g-modes are characterized by low oscillation frequencies and horizontal displacements while p-modes have higher frequencies and radial displacements. 

The driving mechanism of pulsations in WD stars is the $\kappa-\gamma$ mechanism that relies on an increase of the opacity as a consequence of partial ionization of the dominant chemical species~\cite{1981A&A...102..375D}, and the 'convective driven' mechanism that acts when the outer convection zone deepens~\cite{1991MNRAS.251..673B}. One of the characteristics of these pulsations  is that the number of detected periods is very small and the inversion techniques cannot be applied. The only way to obtain information about the interior is comparing the properties of the observed periods with those of a computed set of pulsation models. These solution are not  unique and external conditions like surface gravity and effective temperature independently obtained have to be used to break the degeneracy. 

There are at least six classes of pulsating WDs~\cite{Corsico:2019nmr}, but in this work only DAVs, DBVs, and DOVs will be considered. The multifrequency character and the amplitude of the period of pulsation ($10^2-10^3$~s) indicate they are g-mode pulsators, i.e. the driven mechanism is buoyancy. In principle, if there are enough data it is possible to obtain information about the mass, the internal chemical stratification, the rotation rate, the presence of magnetic fields, the cooling timescale and the core composition of the WD. In this sense, the {\it Kepler} mission plus its {\it K2} extension (90 days of uninterrupted observations) have noticeably improved the number of variable WDs that have been intensively studied and it is expected that the data from the missions {\it TESS}, {\it Cheops} and {\it Plato} will qualitatively change our knowledge of these stars.

\begin{table}[t!]
\centering
\caption{ Main characteristics of variable white dwarfs.}
\begin{tabular}{lccccc}
\hline
\hline
\\
Class & $T_{\rm eff} $     & $\log g$ & Amplitude &  Period  &  $\dot P $  \\   
          &                (K)  &  (cgs)      & (mag)         &   (s)       &  (ss$^{-1}$)   \\
\\
\hline
\\
DOV (GW Vir)    & 80,000-180,000  &5.5-7.7 &0.02-0.1 &300-2,600 &  $10^{-10}-10^{-12}$ \\
DBV (V777 Her) & 22,400-32,000   &7.5-8.3 &0.05-0.3 &120-1,080 &  $10^{-12}-10^{-13} $\\
DAV (ZZ Cet)     & 10,400-12,400   &7.5-9.1 &0.01-0.3 &100-1,400 &  $(1-6)\times 10^{-15}$ \\
\\
\hline
\end{tabular}
\label{tab:DV}
\end{table}

As the cooling proceeds, degeneracy increases, the buoyancy decreases, and the star contracts inducing  a secular change in the period of pulsation $P$. The temporal evolution of this period can be estimated to be~\cite{1983Natur.303..781W}
\begin{equation}
\frac{\dot P}{P}= -a\frac{\dot T}{T}+ b\frac{\dot R}{R}\,,
\label{pdot}
\end{equation}
where $a$ and $b$ are positive constants of the order of unity. The first term of the r.h.s. reflects the decrease of the Brunt-V\"ais\"al\"a frequency with the temperature, while the second term reflects the increase of the frequency induced by the residual gravitational contraction. In the case of DAVs and DBVs
this second term is  always negligible for which reason the secular drift of the period of pulsation is always positive and of the order of $10^{-15}$ and $10^{-12} - 10^{-13}$~s~s$^{-1}$   respectively. On the contrary, this second term cannot be neglected in the DOV case and the secular drift can be positive or negative. Tab.~\ref{tab:DV} displays the characteristic parameters of these stars.

\begin{figure}[t]
\center
  \includegraphics[width=0.7\linewidth,clip=true,trim=4.5cm 9.3cm 2cm 10cm]{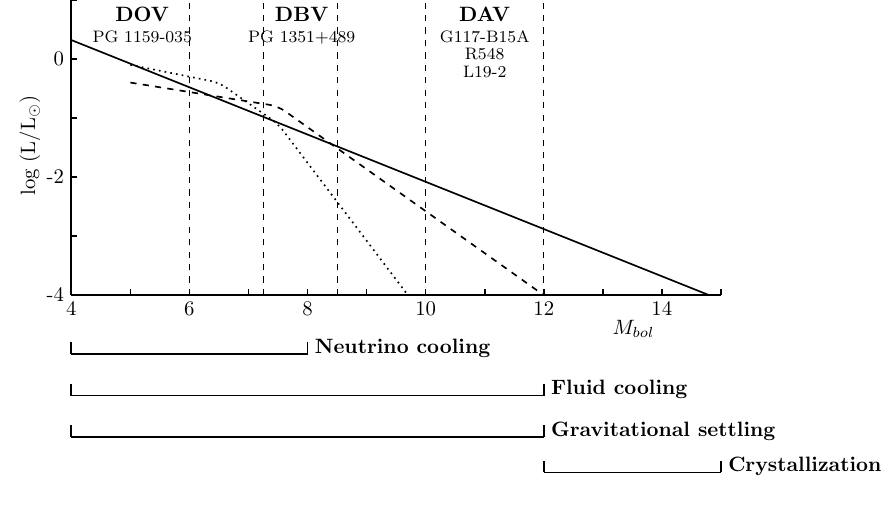}
\caption{Sketch showing where the cooling processes processes are acting during the evolution of white dwarfs: photon, neutrino and axion luminosities (continuous, dotted and dashed lines respectively)as well as the regions occupied by the different sources of energy. The figure also shows the position of the variable white dwarfs.}\label{wdsketch}
\end{figure}

 These secular drifts can be used to test the predicted evolution of white dwarfs and, if the models are reliable enough, to test any physical effect able to change the pulsation period of these stars. In order of magnitude,
\begin{equation}
\frac{{{L_{0}} + {L_x}}}{{{L_{0}}}} \approx \frac{{{{\dot P}_{\rm obs}}}}{{{{\dot P}_{0}}}}
\label{eqise92}
\end{equation}
where $\dot P_{\rm obs}$ is the observed period drift,  $L_0$ and $\dot P_0$ are obtained from standard models and $L_x$ is the extra luminosity necessary to fit the observed period~\cite{Isern:1992gia}.

Notice that if there is a resonance between  the local wavelength of a pulsation mode and the thickness of a layer, like the H or the He envelopes, the mode is trapped and the drift of the period can be substantially modified because the radial term of Eq.~\eqref{pdot} is not negligible since these external layers are still contracting. The information obtained in this way is based on the individual properties of WDs and is independent of the star formation rate. However, the pulsation period and its evolution not only depend on the cooling rate but also on the detailed structure of each individual which, in turn, depends on the poorly known mass and the unknown metallicity of the progenitor. Furthermore, the presence of neutron rich impurities can modify  in a substantial way the seismological properties of WDs (values of the periods as well as spacing and secular drift) as a consequence of the strong dependence of the Brunt-V\"ais\"al\"a (BV) frequency on the number of electrons per nucleon in the case of a degenerate plasma.

As it can be seen in Fig.~\ref{wdsketch}, the cooling rate of DOVs and DBVs is controlled by the photon and neutrino luminosities, while DAVs is only controlled by photon emission as a consequence of the rapid decrease of the neutrino bremsstrahlung emission with the temperature.

DAVs, also known as ZZ Ceti WDs, were the first to be discovered~\cite{1968ApJ...153..151L}, and constitute the most numerous group, including more than $400$ WDs~\cite{2022MNRAS.511.1574R}. Their atmosphere is made of almost pure hydrogen and they are characterized by relatively low effective temperatures and high gravity, see Tab.~\ref{tab:DV}. The secular drift is of the order  of $10^{-15}$~s$^{-1}$ for which reason it has only been possible to measure it in three stars (G117-B15A, R548, and L19-2).

\begin{figure}[h]
\center
  \includegraphics[width=0.6\linewidth,clip=true,trim= 5cm 9cm 5cm 9cm]{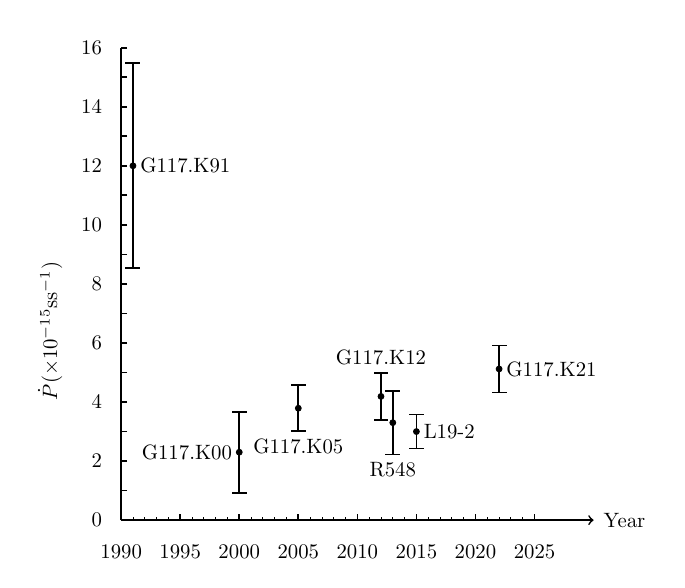}
\caption{Evolution of the period drift measurements of G117-B15A~\cite{1991ApJ...378L..45K,2000ApJ...534L.185K,2005ApJ...634.1311K,2012ASPC..462..322K,2021ApJ...906....7K}, R548~\cite{2013ApJ...771...17M} and L19-2~\cite{2015ASPC..493..199S}. }
\label{pdotdv}
\end{figure}

\begin{figure}[ht]
\center
  \includegraphics[width=0.9\linewidth,clip=true, trim=3cm 8cm 0cm 7cm]{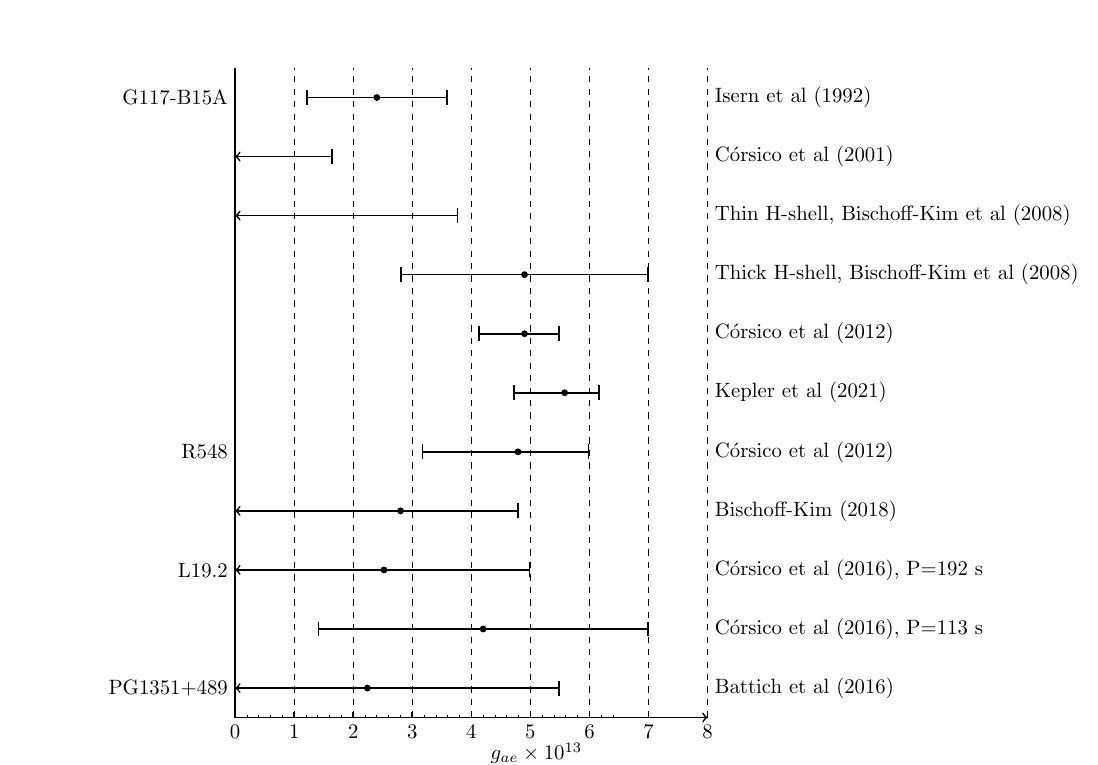}
\caption{Evolution of the axion-electron constant $g_{ae}$ obtained from WD variables. The amplitude of the segments represent the $1\,\sigma$ uncertainty introduced by observations and models. Dots represent the values of $g_{ae}$ that fit the observations if axions alone were responsible of the extra cooling.}
\label{wddv}
\end{figure}

The monitoring of G117-B15A started in 1974 and is still continuing. Fig.~\ref{pdotdv} displays the historical evolution of the drift measurement of the 215~s period mode. The first value that was obtained,  $(12.0 \pm 3.5)\times 10^{-15}$~ss$^{-1}$~\cite{1991ApJ...378L..45K},  was much larger than the one predicted by the existing models at the epoch, $(2-6)\times 10^{-15}$~ss$^{-1}$ ~\cite{1991ASIC..336..153F}. The two conventional ways advanced to solve this discrepancy were rejected. One of them assumed that the envelope was more transparent than previously thought but this would imply an age of the Galactic disk of 4.8 Gyr, a value much smaller than the one necessary to account for the luminosity function~\cite{1991ASIC..336..143K}\footnote{Interestingly enough, asteroseismological techniques have shown  that the mass of H-layer of G117-B15A is two orders of magnitude smaller than  the standard one and that at least a $\sim 10$\% of DAVs have thin hydrogen layers~\cite{2012MNRAS.420.1462R}.}. The other suggested that the average mass number of the core was $<A> \approx 33$ but this was incompatible with the standard evolution of isolated stars~\cite{1991ASIC..336..153F}.

Axions were proposed as a possible solution to this puzzle~\cite{Isern:1992gia}. At the temperatures and densities  typical of hot DAVs and assuming they can interact with electrons, axions are emitted via bremsstrahlung (see Sec.~\ref{sec: electroncoupl}). Since they are bosons, they scale with the temperature as $\epsilon \propto T^4$ while neutrinos, which are fermions, scale as $\epsilon \propto T^7$, for which reason they can still be an important sink of energy at  the luminosity domain of DAVs, Figure~\ref{wdsketch}. Using Eq.~\eqref{eqise92} and a simple model of WDs, Ref.~\cite{Isern:1992gia} showed that axions of DFSZ type with a mass $m_a\cos^2 \beta \approx 8.5 $~meV ($ g_{ae} \approx 2.36\times 10^{-13}$) could cause the drift.

Measuring the secular drift is difficult and the observed values have been evolving with time. In the case of G117-B15A the last value that has been obtained is $(5.12 \pm 0.82) \times 10^{-15}$~ss$^{-1}$~\cite{2021ApJ...906....7K}, as shown in Fig.~\ref{pdotdv}. The codes that have been used to analyse these data are the La Plata Code (LPCODE), which computes model for the ZZ Ceti white dwarf from the appropriate parent star~\cite{2012MNRAS.420.1462R,2022A&A...663A.167A}, and the White Dwarf Evolution Code (WDEC), which parametrizes the chemical structure of the WD until the optimal model is obtained~\cite{2018AJ....155..187B}. As a consequence of the improving of both, theoretical and observational results, the value of $g_{ae}$ necessary to account for the pulsation drift of G117-B15 A via axions has been changing with time (Fig.~\ref{wddv})~\cite{2001NewA....6..197C,2008ApJ...675.1512B,2012MNRAS.424.2792C,2012MNRAS.420.1462R}. 
At present the predicted value is $g_{ae} = (5.66 \pm 0.57)\times 10^{-13}$ ($m_a \cos^2 \beta =20 \pm 2 $~meV for DFSZ axions)~\cite{2021ApJ...906....7K} (Fig.~\ref{wddv}).

The secular drift of R548, also known as ZZ Ceti,  is $(3.3 \pm 1.1) \times 10^{-15}$~ss$^{-1}$~\cite{2013ApJ...771...17M}. The theoretical analysis performed with WDEC provided an upper limit of $g_{ae} \lesssim 4.8 \times 10^{-13}$ ($m_a \cos^2 \beta \lesssim 17 $~meV for DFSZ axions), and $g_{ae} = 2.8 \times 10^{-13}$ ($m_a \cos^2 \beta =10 $~meV for the exact fit~\cite{2018phos.confE..28B}. The analysis with the LPCODE provided a value $g_{ae} =4.79^{+1.2}_{-1.7} \times 10^{-13}$~ \cite{2012JCAP...12..010C}\footnote{Notice that the values of $g_{ae}$ quoted here were obtained from the observations and the code versions at that moment and that both have been changing with time.}.

 L19-2 has been observed from 1976 to 2012 and presents five independent pulsation modes of which two, the 113 and 192 seconds ones, are non-trapped in contrast with the G117-B15A and R548 cases and have the same pulsation drift, $\dot P= (3.0 \pm 0.6) \times 10^{-15}$~ss$^{-1}$~\cite{2015ASPC..493..199S} (Fig.~\ref{pdotdv}). The values of $g_{ae}$ necessary to account for the discrepancies between the observed and the predicted drifts of the LPCODE are $g_{ae} =2.8^{+4.2}_{-1.4} \times 10^{-13}$ ($m_a = 10^{+15}_{-5}$~meV)  for the 113 period, and $g_{ae} \lesssim 5 \times 10^{-13}$ ($m_a \cos^2 \beta \lesssim 18 $~meV) and 
$g_{ae} \approx 2.52 \times 10^{-13}$ ($m_a \cos^2 \beta \approx 9$~meV) for the 192~s one (Fig.~\ref{wddv})~\cite{2016JCAP...07..036C}.

The analysis performed with the WDEC code \cite{2022ApJ...934...34C} suggests that L19-2 is massive, hot, with a core very rich in oxygen and with a relatively thick H-layer. Concerning pulsation periods, the predicted secular drifts are smaller than the observed ones and not entering in contradiction with the existence of axions

A similar analysis has been performed with PG 1351 + 489, a DB variable with a drift of $\dot P= (2.0 \pm 0.69) \times 10^{-13}$~ss$^{-1}$ for a period of 489~s, the one with the largest amplitude~\cite{2011MNRAS.415.1220R}. The theoretical drift is expected to be $\dot P= (0.81 \pm 0.5) \times 10^{-15}$~ss$^{-1}$~\cite{2016JCAP...07..036C}, for which reason the axion-electron coupling constant necessary to fit the observed value is 
$g_{ae} \approx 2.24 \times 10^{-13}$ ($m_a \cos^2 \beta \approx 8$~meV), and an upper limit of $g_{ae} < 5.5 \times 10^{-13}$ ($m_a \cos^2 \beta \approx 8$~meV)~\cite{2016JCAP...08..062B}, a value that is concordant with those obtained  with the DAVs (Fig.~\ref{wddv}). Notice that this WD still has a strong neutrino emission and the result could be influenced by an extra emission if neutrinos have a magnetic moment.

These results are concordant and point out towards the introduction of an extra cooling term in the energy balance of WDs. However, the uncertainties, both observational and theoretical, prevent for the moment to decide if the anomaly has a systematic origin or is caused by a FIPs or anything else~\cite{Isern:2022vdx}.

\subsection{Axion bounds from the luminosity function}
\label{sec:axwdlf}
The luminosity function of a WD ensemble is just their distribution in luminosities or, equivalently, in magnitudes. Traditionally it is presented as the number of WDs of a given absolute magnitude per unit of magnitude interval. In the case of single WDs, the number $N$ of WD that have an absolute magnitude within the range  $M_{\rm abs}\pm 0.5~\Delta M_{\rm abs}$ at the present time, $T_G$, per unit of magnitude interval is given by
\begin{equation}
N\left( M_{\rm abs}\right) = \int\limits_{{M_l}}^{{M_u}} {\Phi \left( M \right)\Psi \left( {{T_G} - {t_{\rm cool}} - {t_{\rm ps}}} \right){\tau _{\rm cool}}dM} \,,
\label{lf}
\end{equation}
where $M$ is the mass of the parent star (for convenience all WDs are labeled with the mass of the ZAMS progenitor to avoid the derivative of the poorly known initial final mass relationship or IFMR that connects the mass of the WD with that of the progenitor), $t_{\rm cool}$ is the cooling time down to the magnitude $M_{\rm abs}$, $\tau_{\rm cool} = dt/dM_{\rm abs}$ is the characteristic cooling time of the WD at this magnitude, $t_{\rm ps}$ is the lifetime of the progenitor of the WD, $T_G$ is the age of the Galaxy or the population under study, and $M_u$ and $M_l$ are the maximum and the minimum mass of the MS stars able to produce a WD, therefore $M_l$ satisfies the condition $T_G=t_{\rm cool}(M_{\rm abs},M_l)+t_{\rm ps}(M_l)$. $\Phi(M)$ is the Initial Mass Function (IFM) and $\Psi(t)$ is the Star Formation Rate (SFR) of the population under consideration. Additionally, hidden, there is the IFMR connecting the properties of the progenitor with those of the WD. In order to compare theory with observations, and since the total density of WDs is not yet  well known, the computed WDLF is usually normalized to a bin with a small error bar, usually $\log L/L_\odot \simeq 3$ or the corresponding magnitude. This equation contains three sets of terms, the observational ones, $N(M_{\rm abs})$, the stellar ones, $t_{\rm cool},\tau_{\rm cool}, t_{\rm PS}, M_{u}, M_{l}$, plus the IFMR, and the galactic ones $\Phi$ and $\Psi$.

\begin{figure}[h]
\center
  \includegraphics[width=1.\linewidth]{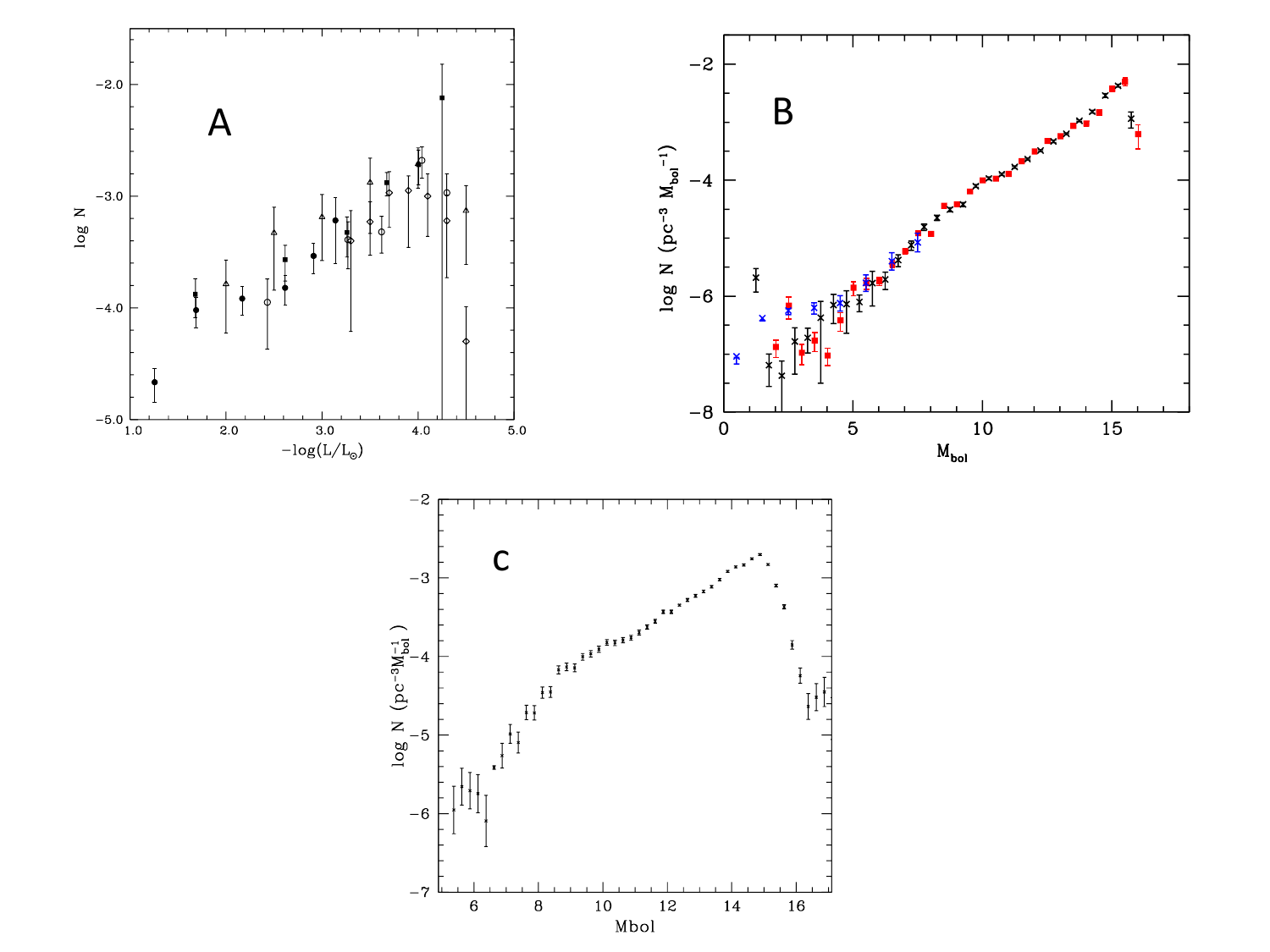}
\caption{{\it A)} Early luminosity functions: Ref.~\cite{1988ApJ...332..891L}, full circles; Ref.~\cite{1992MNRAS.255..521E}, full squares; Ref.~\cite{1996Natur.382..692O}, open triangles; Ref.~\cite{1998ApJ...497..294L}, open diamonds; Ref.~\cite{1999MNRAS.306..736K}, open circles. {\it B)} WDLFs obtained with the SCSS catalogue~\cite{2011MNRAS.417...93R} (black crosses), with the SDSS catalogue~\cite{2006AJ....131..571H} (red squares), and from UV-excesses~\cite{2009A&A...508..339K} (blue crosses) normalized at $M_{\rm bol}\approx12$. {\it C)}  Luminosity function of WDs located within 100 pc of the Sun  obtained from the Gaia Early Data Release 3~\cite{2021A&A...649A...6G}. (Figure taken from Ref.~\cite{Isern:2022vdx} with permission).
}\label{wdlf}
\end{figure}

A practical way to compute the WDLF is from
\begin{equation}
N({M_{\rm abs}},{T_G}) = \frac{1}{\Delta M_{\rm abs}}\int\limits_t {\int\limits_{{M}} {\Phi \left( {{M}} \right)\Psi \left( t \right) d{M}\,dt} }\,,
\label{eq:wdlf1}
\end{equation}
since written in this way it avoids not only he numerical derivative of the IFMR but also that of the cooling time \textbf{necessary} to obtain the characteristic cooling time~\cite{1998ApJ...503..239I}. In deriving this equation it has been assumed that WDs are not destroyed nor created via merging of binaries and that the ensemble is closed, i.e. there is not an important flux of stars in and out of the ensemble under consideration.

The values of $M$ are constrained to the domain that satisfies the conditions,
\begin{eqnarray}
 t + {t_{\rm ps}}[{M,Z(t)}] + {t_{\rm cool}}[{m_{\rm wd}},Z(t),{M_{\rm abs}} - 0.5\Delta {M_{\rm abs}}] = {T_G} \,,\notag  \\
 t + {t_{\rm ps}}[{M,Z(t)}] + {t_{\rm cool}}[{m_{\rm wd}},Z(t),{M_{\rm abs}} + 0.5\Delta {M_{\rm abs}}] = {T_G} \,,
\label{eq:cwd1}
\end{eqnarray}
for each $t$, and $0 \le t\le T_G$. In these equations the dependence on the metallicity, $Z(t)$, of the parent star has been made explicit since it is a source of uncertainty.

The first WDLF was derived by Weidemann~\cite{1968ARA&A...6..351W}. Fig.~\ref{wdlf} (panel A) shows the state of the art by the end of the 90s, before the advent of the large cosmological surveys.  The monotonic growing of the number of stars with the magnitude strongly supported the idea that the evolution of WDs was just a cooling process and, consequently, that the sudden cut-off  was caused by the finite age of the Galaxy~\cite{1987ApJ...315L..77W} but the large dispersion of bins prevented the determination of the slope, and the unique tools to constrain physics were the expected population at each bin and the position of the cut-off~\cite{Raffelt:1985nj}. With the advent of the large cosmological surveys, like the SDSS and the SCSS catalogues, the size of the samples increased to several thousand of stars thus allowing the use of the slope as a tool for testing new physics, Fig.~\ref{wdlf}, panel B. For instance, the change of the slope in this figure shows the change from a cooling dominated by neutrinos to one dominated by  photons. At present, thanks to the accuracy of the photometric and astrometric data provided by the Gaia mission, it has been possible to obtain the WDLF for  a statistically complete 100~pc sample, Fig.~\ref{wdlf} (panel C).

\begin{figure}[!t]
\center
  \includegraphics[width=0.8\linewidth,clip=true,trim=0cm 0cm 0.5cm 1cm]{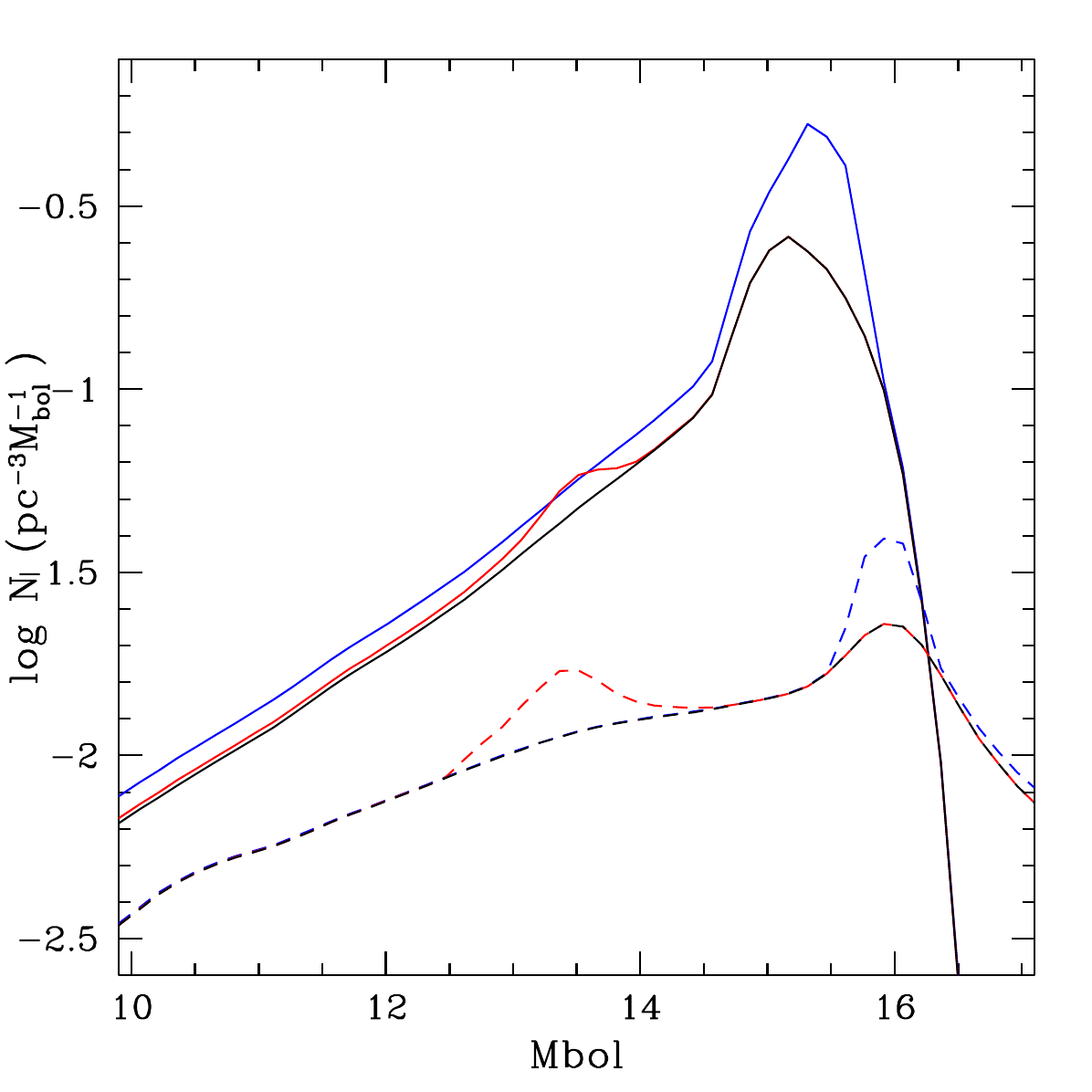}
\caption{WDLFs obtained with a constant SFR (black), with a constant SFR plus a burst 11 Gyr old (blue), and with a constant SFR plus a burst 2 Gyr old (red). Dashed lines represent the luminosity function ($\times 10$) for the same cases but restricted  to stars in the mass range 0.9-1.1~M$_\odot$.}\label{wdlfburst}
\end{figure}

Since the WDLF depends on the characteristic cooling time it can be used to obtain information about the existence of any extra source or sink of energy. This technique is based on the fact that the bright branch of the luminosity function, i.e. bins with $M_{\rm bol} \lesssim 13$ or, equivalently with $\log {L/L_\odot} \gtrsim -3.5$, is dominated by WDs coming from low-mass MS stars. Eq.~\eqref{lf} can be written as
\begin{equation}
{N \propto \left\langle {{\tau _{\rm cool}}} \right\rangle \int\limits_{{M_i}}^{{M_s}} {\Phi \left( m \right)\Psi \left( T_G-t_{\rm ps}-t_{\rm cool}\right)} } \,dm\,.
\label{eq:wdlf2}
\end{equation}
Bright WDs have had no time to cool down, i.e. $t_{\rm cool}$ is small, and, as a consequence of the strong dependence of the MS lifetimes with mass, $M_i$ is almost constant and independent of the luminosities under consideration. As a consequence, the slope essentially depends on the averaged characteristic cooling time and only weakly on the star formation history. This average is dominated by low-mass WDs and, as far as the mass distribution is not strongly perturbed by the adopted SFR or initial mass function, is representative of the intrinsic properties of WDs. Notice that this is not true if recent star formation bursts are present since, in this case, low-mass MS stars have no time to become WDs and $M_i$ becomes dependent on the luminosity under consideration. On the contrary, if the burst is old enough, the slope of the bright region is maintained and differences appear at larger magnitudes. See Fig.~\ref{wdlfburst}, solid lines. Another important property is that in this bright region the slope of the relationship between the luminosity and the core temperature  of DA and non-DA WDs almost coincide and both luminosity functions almost overlap in this luminosity interval after normalization~\cite{2009JPhCS.172a2005I}. 

\begin{figure}[h]
\center
  \includegraphics[width=0.8\linewidth,clip=true,trim=5cm 11cm 5cm 11cm]{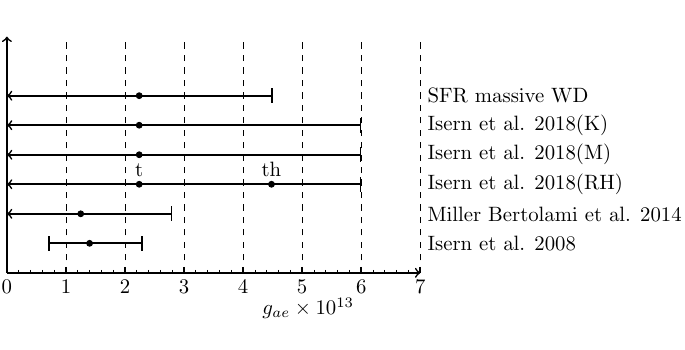}
\caption{Values of the axion-electron coupling constant $g_{ae}$ obtained from different white dwarf luminosity functions. The amplitude of the segments roughly represents the values compatible with the observations taking into account the present uncertainties. Dots represent the values that better fit the observations. See the text for details.}\label{wdlfgae}
\end{figure}

In the case of axions, this technique was used for the first time by Ref.~\cite{2008ApJ...682L.109I} 
using a preliminary form of the luminosity function obtained by \cite{2006AJ....131..571H} displayed in Fig.~\ref{wdlf}-B
obtaining $g_{ae} =(1.4^{+0.9}_{-0.8})\times 10^{-13}$ (Fig.~\ref{wdlfgae}, label Isern et al. 2008). This result was reexamined by Ref.~\cite{2014JCAP...10..069M} using all the luminosity functions represented in Fig.~\ref{wdlf}-B and a self-consistent treatment of the neutrino cooling, concluding that the existence of DFSZ axions with a coupling constant $g_{ae}\sim 1.4\times 10^{-13}$ could improve the agreement between theoretical and observed WDLFs and values $g_{ae} \gtrsim 2.8\times 10^{-13}$ could be excluded (Fig.~\ref{wdlfgae}, label Miller Bertolami et al. 2014).

Given the degeneracy between the stellar and galactic terms of Eq.~\eqref{eq:wdlf2} it is natural to wonder if the shape of the WDLF can be attributed to axions or if it is just a consequence of changes in the SFR like those introduced by recent bursts of star formation (solid red line in Fig.~\ref{wdlfburst}).

There are several ways to disentangle, at least partially, such a degeneracy. One of them is to examine the luminosity function of populations that have different formation histories. If axions really modify the cooling rate of WDs, their imprint will be present in all the independent luminosity functions at roughly the same luminosities~\cite{2018MNRAS.478.2569I}. Effectively:
\begin{itemize}
\item  The discrepancies between the theoretical calculations and the observed luminosity functions of the thin and thick disks and halo~\cite{2011MNRAS.417...93R} decrease if axions are introduced. The best results are obtained for $g_{ae} \approx 2.24\times 10^{-13}$~and~$4.48\times 10^{-13}$ (($m_a \cos^2 \beta \approx 8,\, 16$~meV) for the thin disk and the thick disk and halo respectively  (Fig.~\ref{wdlfgae}, label RH, dots labelled t and th)~\cite{2018MNRAS.478.2569I}. 
\item Munn \emph{et al}.~\cite{2017AJ....153...10M} improved the disk luminosity function assuming a constant scale height above the Galactic plane and recomputed that of the halo. As before, the inclusion of axions improved the concordance  between theory and observations both in the disk and in the halo. The best it was obtained for $g_{ae} \approx 2.24\times 10^{-13}$ (Fig.~\ref{wdlfgae}, label M)~\cite{2018MNRAS.478.2569I}.
\item Kilic \emph{et al}.~\cite{2017ApJ...837..162K} reexamined the Munn \emph{loc.cit.} data assuming different variable scale heights above the Galactic plane. Although this improved the results, the concordance observed and computed luminosity function was better when axions with a $g_{ae} \approx 2.24 \times 10^{-13}$ were included (Fig.~\ref{wdlfgae}, label K)~\cite{2018MNRAS.478.2569I}.
\end{itemize}
Fig.~\ref{wdlfgae} shows that the agreement between observed and theoretical luminosity functions improve if an additional cooling term is included.

Another possibility to break the degeneracy between stellar and galactic properties is provided by massive WDs. Since the lifetime of their progenitors is short compared with the cooling time, their luminosity function closely follows the temporal variation of the SFR, dashed lines of Fig.~\ref{wdlfburst} and in this case it is possible to directly obtain the star formation rate $\Psi$~\cite{2019ApJ...878L..11I}.
The luminosity function for WDs  with a mass in the range of 0.9-1.1~M$_\odot$ and within a distance of 100~pc has been recently obtained by~\cite{2019Natur.565..202T}. This WDLF presents a bump that has been associated to the process of crystallization, but, because of the degeneracy of Eq.~\eqref{eq:wdlf2}, it is always possible to find a SFR able to reproduce the observed WDLF. 

\begin{figure}[h]
\center
  \includegraphics[width=0.55\linewidth,clip=true,trim=0cm 0cm 0cm 0cm]{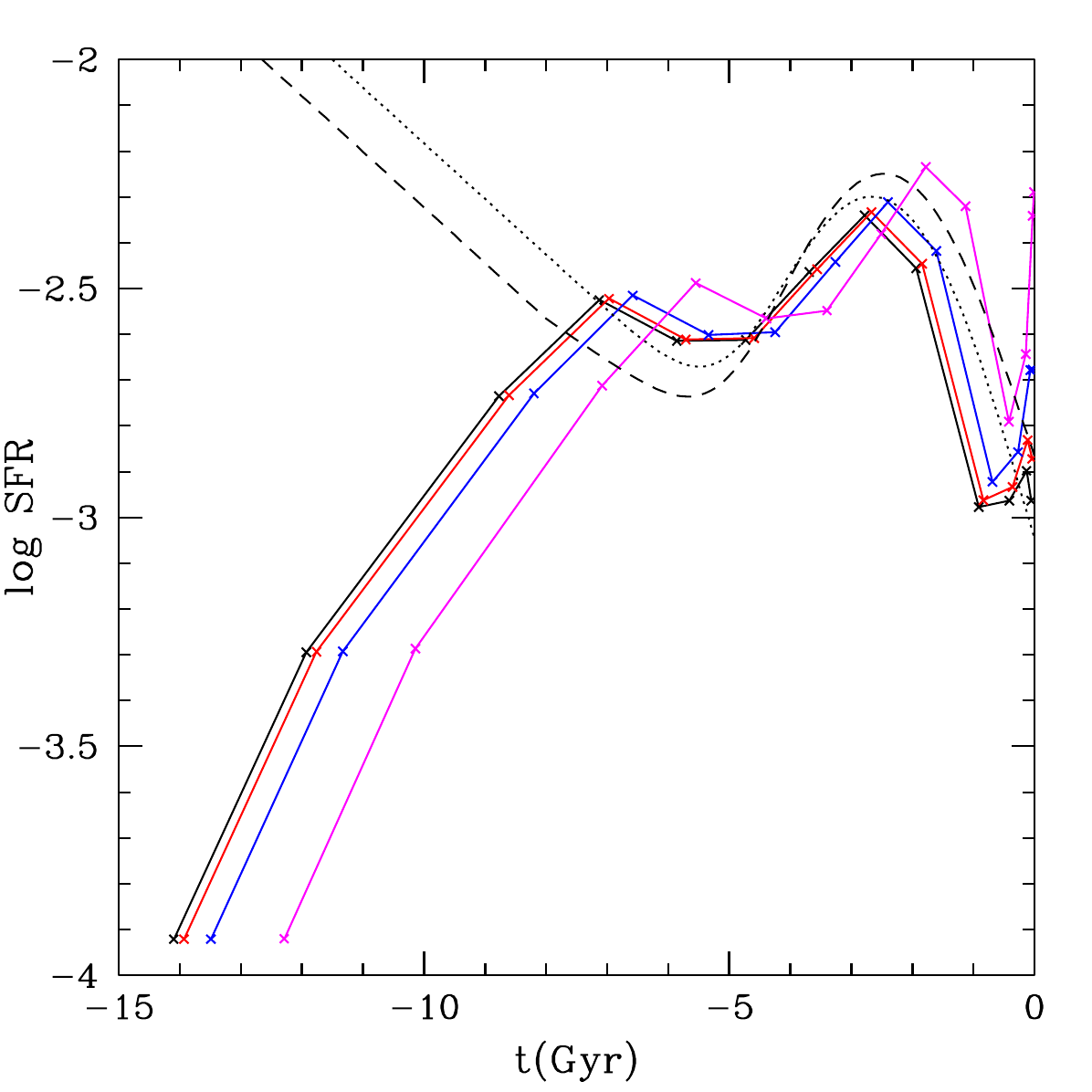}
\caption{SFRs per unit volume obtained assuming solidification and axion cooling with $g_{ae} = 0.0, 1.12, 2.24, and 4.48~\times 10^{-13}$ (black, red, blue and magenta solid lines). Dotted and dashed lines represent the SFRs obtained from MS stars~\cite{2019A&A...624L...1M} divided, respectively, by a constant and an age dependent scale height (see text for details).}\label{wdsfrax}
\end{figure}

The black solid line of Fig.~\ref{wdsfrax} represents the SFR obtained from the luminosity function of massive white dwarfs using the old BaSTI models~\cite{2013A&A...555A..96S} which include crystallization and the associated separation of carbon and oxygen for white dwarfs with solar metallicity. As it can be seen the SFR increases with the age and reaches a maximum around 7-8 Gyr ago and decreases at larger ages. A noticeable feature is the existence of a prominent peak around 2.5 Gyr ago, being the its exact position and height model depending (see Refs.~\cite{2019ApJ...878L..11I,Isern:2019nrg} for details). It is important to notice the the SFR obtained in this way does not takes into account the secular evolution of the sample caused by radial migrations, scale height inflation or merging of double degenerate stars. The peak appearing around 0.2 Gyr is in the limit of applicability of the method and deserves more attention.

The Gaia data release 2 together the Besan\c{c}on Galaxy Model has allowed to obtain the SFR per unit of disk surface for MS stars with $G <12$~\cite{2019A&A...624L...1M}. In order to compare this rate with the SFR per unit volume obtained from white dwarfs it is necessary to introduce the galactic disk scale height. If a constant scale height is adopted, both methods predict a concordant burst of star formation at $\sim 2.5$ Gyrs ago but diverge at large ages, Fig.~\ref{wdsfrax}. This tension may have several origins: a local delay in starting the star formation process, a different behaviour of the outer and inner regions of the disk, a vertical dilution larger than expected caused by an early collision or just an efficient conversion of DA WDs into non-DAs~\cite{2019ApJ...878L..11I}. 

If axions are included, red, blue and magenta solid lines in Fig.~\ref{wdsfrax}, the bump moves towards shorter ages and values $g_{ae} \gtrsim 4\times10^{-13}$ are not compatible with the results obtained with Gaia. 
If a time dependent height scale like that of~\cite{2023MNRAS.522.1643C}, where  $h_{\rm old} > h_{\rm young}$ (dashed line), is adopted, values of $g_{ae} \sim 2.24\times10^{-13}$ are acceptable while values $g_{ae}\gtrsim 4.5\times10^{-13}$ can be rejected\footnote{Notice that these results are very sensitive  to  the evolution of the DA/non-DA ratio and to the contribution of binary mergers.}~\cite{Isern:2019nrg}. 

\begin{figure}[!h]
\center
  \includegraphics[width=0.55\linewidth]{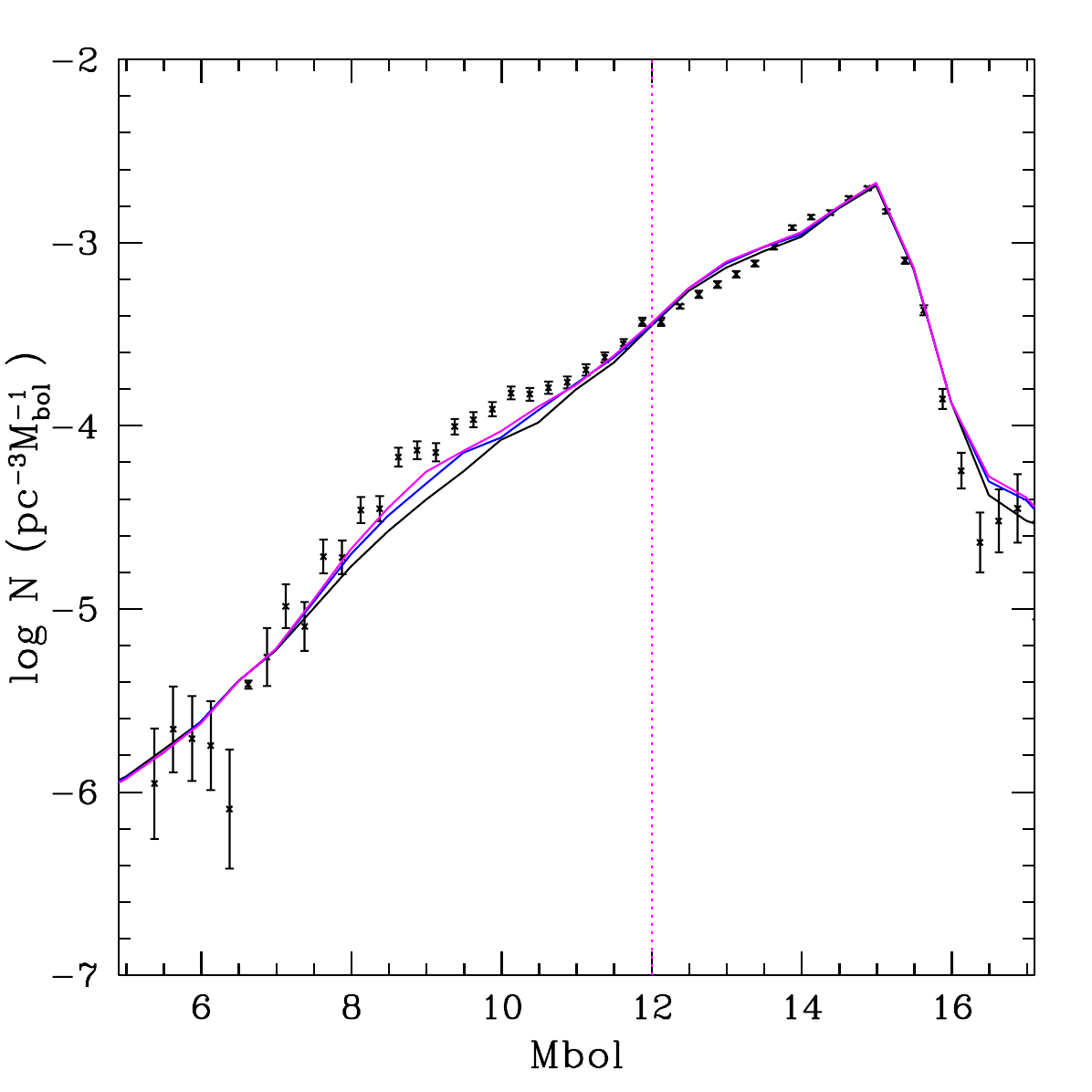}
\caption{Luminosity functions resulting from the SFRs displayed in Fig.~\ref{wdsfrax} normalized to $M_{\rm bol}=12$ as compared whith that of WDs within a distance of 100~pc \cite{2021A&A...649A...6G}.}\label{wdlfsinax}
\end{figure}

Fig.~\ref{wdlfsinax} compares the WDLFs built with the SFRs obtained from massive WDs assuming different axion masses represented in Fig.~\ref{wdsfrax} with the luminosity function of the 100~pc sample \cite{2021A&A...649A...6G}. The deficit of WDs predicted by theoretical models around $M_{\rm bol}=10$ may have different origins, the dependence on the initial models, which are poorly known, the contribution of mergers, or the aforementioned  very recent burst of star formation~\cite{2024gacv.confE...9I}.

The shortage of white dwarfs in the region $8\lesssim M_{\rm bol} \lesssim 13$ of the luminosity functions of disc and halo  points out towards an intrinsic origin, being axions a possibility. Nevertheless, other factors not yet well known could be acting over the white dwarf distribution like migration of stars and inflows/outflows of gas in the solar neighborhood, vertical inflation, merging of stars in binary systems, evolution of the DA/non-DA character and so on. The analysis of the globular cluster 47 Tuc \cite{2015ApJ...809..141H, 2016ApJ...821...27G} seems not to favor the axion option, although the influence of the presence of a black hole in the center, the hydrogen burning in low metallicity WDs, and the presence of several generations of stars may change this perception. If an intrinsic origin is adopted, the problem to be solved is the apparent inconsistency with the results obtained with variable white dwarfs and the bounds set by red giant stars.



\section{Supernovae and Neutron Stars}
\label{sec:SN_NS}

Core-collapse SNe correspond to the terminal phase of massive stars (with mass $M \gtrsim 8$~M$_{\odot}$),  which become unstable at the end of their lives, collapsing and ejecting their outer mantle in a shock-wave driven explosion. During such a process $99~\%$ of the emitted energy [$E \sim {\mathcal O}(10^{53})$~erg] is released by neutrinos and antineutrinos of all the flavors with average energies of 15~MeV, on a timescale of ten seconds. Due to the huge amount of energy released in neutrinos, a SN can be really considered a cosmic neutrino factory. During such an event, neutrinos play the role of astrophysical messengers allowing us to probe both fundamental neutrino properties (like mass, mixing, interactions)  and the SN explosion mechanism (see, e.g. Refs.~\cite{Mirizzi:2015eza,Janka:2006fh,Janka:2012wk,Nakamura:2016kkl,Horiuchi:2018ofe,Capozzi:2022slf} for reviews). Furthermore, SNe are powerful laboratories to study novel weakly interacting light particles (see, e.g., Refs.~\cite{Raffelt:1990yz,Raffelt:1999tx,Raffelt:1996wa}). For typical SN core temperatures, $T \sim {\mathcal O}(30)$~MeV, particles with masses $m_a < 100$~MeV can be thermally produced without being Boltzmann suppressed.

Neutrinos from a stellar gravitational collapse have been observed only once during the explosion of the SN 1987A in the Large Magellanic Cloud~\cite{Kamiokande-II:1987idp,Hirata:1988ad,Bionta:1987qt,IMB:1988suc,Alekseev:1988gp}. Despite the sparseness of the recorded neutrino data, the neutrino observation of the few events from  SN 1987A is considered a milestone achievement of astroparticle physics. Indeed, SN 1987A neutrinos allowed for  a direct confirmation of the SN explosion mechanism paradigm (see, e.g., Refs.~\cite{Loredo:2001rx,Pagliaroli:2008ur,Li:2023ulf,Fiorillo:2023frv}).  Furthermore, significant constraints were placed on many novel particles~\cite{Raffelt:1987yt} including axions~\cite{Brinkmann:1988vi,Burrows:1988ah, Burrows:1990pk,Keil:1996ju,Chang:2018rso,Carenza:2019pxu}, dark photons~\cite{DeRocco:2019njg,Chang:2016ntp}, sterile neutrinos~\cite{Dolgov:2000jw,Mastrototaro:2019vug,Carenza:2023old}, Kaluza-Klein gravitons~\cite{Hannestad:2001jv}, unparticles~\cite{Hannestad:2007ys}, particles excaping into extra dimensions~\cite{Friedland:2007yj}, whose emission  would have affected the observed neutrino burst.

In the specific case of axions, different arguments have been proposed to constrain their emissivity (see Refs.~\cite{Raffelt:1996wa,Caputo:2024oqc} for reviews). If sufficiently weakly coupled, axions stream freely (i.e., without re-interacting with the stellar medium) out of the star, providing a very efficient cooling mechanism. This consideration has allowed to set stringent bounds on the axion-nucleon coupling from the comparison with the observation of the SN 1987A neutrino burst  (see, e.g., Refs.~\cite{Raffelt:1987yt,Turner:1987by,Brinkmann:1988vi,Burrows:1988ah}). This argument, however, cannot be used to constrain arbitrarily large couplings. In fact, for a sufficiently large coupling, axions would be \emph{trapped} in the SN core and thermalize. In this case, axions would be emitted from a surface, the \emph{axion-sphere}, in close analogy with the neutrino case. Even trapped axions may therefore provide an efficient \emph{energy-transfer} mechanism which leads to further constraints~\cite{Raffelt:1987yt,Burrows:1990pk}. This topic was revisited from a modern perspective in recent years, as we will document in this Section. Furthermore, in the case of massive axions (with mass $m_a \gtrsim 10$~MeV) coupled with photons one can have also  the possibility that these particles decay into photons into the star behind the shock-wave, leading to an \emph{energy-deposition} which can be used to obtain other constraints requiring that this effect does not lead to too energetic explosions~\cite{Caputo:2022mah}. Conversely, if the axion-photon coupling is sufficiently small, the decay can happen outside the SN leading to an observable \emph{gamma-ray signal}. A similar signal is expected also in the case of ultra-light axions (with $m_a \lesssim  10^{-10}$~eV)  converting into photons in the Galactic magnetic-field \cite{Grifols:1996id,Brockway:1996yr,Payez:2014xsa,Hoof:2022xbe}. Therefore, one expects a lot of opportunities from a SN explosion to probe axions as we will document in detail in this Section.

In Sec.~\ref{sec:expl} we will recall the main steps leading to  a core-collapse SN explosion. In Sec.~\ref{sec:nusign}  we present the expected features of the SN neutrino burst and we compare them  with the SN 1987A neutrino observation. In Sec.~\ref{sec:novelpart} we discuss how the duration of the  SN 1987A neutrino burst can be used as a tool to constrain FIP properties. In particular, we discuss the \emph{cooling bound} in both the case of free-streaming and trapped particles. In Sec.~\ref{sec:QCD} we focus on  axions coupled with nucleons, discussing in detail the emissivity in nuclear medium and the progresses in the SN mass bound. Then, in Sec.~\ref{sec:axionlike} we consider axions coupled with photons and we discuss the bounds coming from gamma-ray signal and energy-deposition. in Sec.~\ref{sec:miscellanea} we  comment on other axion scenarios recently discussed in the literature. Finally, in Sec.~\ref{sec:neutrstar} the impact of QCD axions on neutron star cooling is also reviewed.

\subsection{SN explosion mechanism}
\label{sec:expl}

As discussed in Sec.~\ref{sec:stellarev}, a star dies when there is no more fuel to be burnt and any star leaves fossils after its death. Low mass stars, as the Sun, eject the mantle and leave a carbon-oxygen WD with a radius comparable to that of the Earth and a central density $\rho\sim 10^{6}{\rm g}~{\rm cm}^{-3}$. Eventually, the WD cools on a timescale of billion years. For masses larger than 8~M$_{\odot}$, the star burns elements until the core is composed by iron, the most stable nucleus, and then it cannot be burnt to produce energy. When the core reaches the Chandrasekhar mass, 1.4~M$_{\odot}$, the electron degeneracy is not enough to contrast gravity and the core collapses on its own. The core grows thanks to the infalling material, reaching the degenerate regime: as the mass rises, the radius shrinks to increase the electron Fermi momentum (and the pressure). When the nuclear density is reached, $\rho\sim 3 \times 10^{14}{\rm g}~{\rm cm}^{-3}$, matter becomes incompressible due to the nucleon repulsion. Then the core bounces back, triggering a shock wave which drives the SN explosion. This explosion ejects oxygen, carbon, magnesium, silicon, calcium, sulphur, radioactive ${\rm \,^{56}{Ni}}$. What remains after such a process is a compact object, i.e. a PNS or sometimes a black-hole. For a detailed review, see, e.g.,  Ref.~\cite{Mirizzi:2015eza}

There is about one core-collapse SN explosion in the Universe per second, and a few per century in the Milky Way~\cite{Rozwadowska:2020nab}. The mean distance of a Galactic SN is expected to be around $10$~kpc, as indicated by the observations of other Galaxies, or in our Galaxy by  the distribution of pulsars and SN remnants~\cite{Mirizzi:2006xx,Adams:2013ana}.

The physics of a SN explosion is a fascinating process where SN models are produced on the basis of large-scale numerical simulations having as input the (mangeto-) hydrodynamics of the stellar plasma, the relativistic gravity, nuclear equations of state, neutrino transport and progenitor conditions (see, e.g., Ref.~\cite{Janka:2006fh,Janka:2012wk,Burrows:2020qrp} for reviews). Despite the detailed features of the SN explosion models depend on the different assumptions on these inputs, one can generically picture different
phases present in all the simulations as shown in Fig.~\ref{fig:cc}:
\begin{itemize}
\item \emph{Initial phase of the collapse} (post-bounce time $t_{\rm pb} \sim 0$): The collapse of the core starts when the temperature is high enough to dissociate iron nuclei by means of photodisintegration $\gamma ~ \,^{56}{\rm Fe}\rightarrow 13\alpha ~ 4n$. At the beginning of the collapse, the electron degeneracy is reduced due to the electron capture $ep\rightarrow n\nu_{e}$ and neutrinos freely escape. The collapse is accelerated by the lower degeneracy pressure, causing heavier nuclei to fall in the inner star, where their $\beta$ decay subtracts energy from the core. This is the first neutronization phase.	
\item \emph{Neutrino trapping} ($t_{\rm pb}\sim 0.1$~s): When the nuclear matter in the core reaches a density $\rho \sim10^{12}{\rm g}~{\rm cm}^{-3}$, neutrinos start to  become trapped	by coherent scatterings on heavy nuclei~\cite{Bethe:1990mw}: the collapse is faster than the neutrino diffusion time. In this phase the lepton number per baryon $Y_{L}$, is nearly conserved causing the formation of a degenerate $\nu_{e}$ sea. The inner part of the core, the ``homologous core'', collapses at a subsonic velocity because of the incompressibility of nuclear matter~\cite{1980ApJ...238..991G}. In the meanwhile the outer part shrinks at supersonic velocity.
	
\begin{figure}[h!]
	\centering
	\vspace{0.cm}

 \includegraphics[width=0.85\columnwidth]{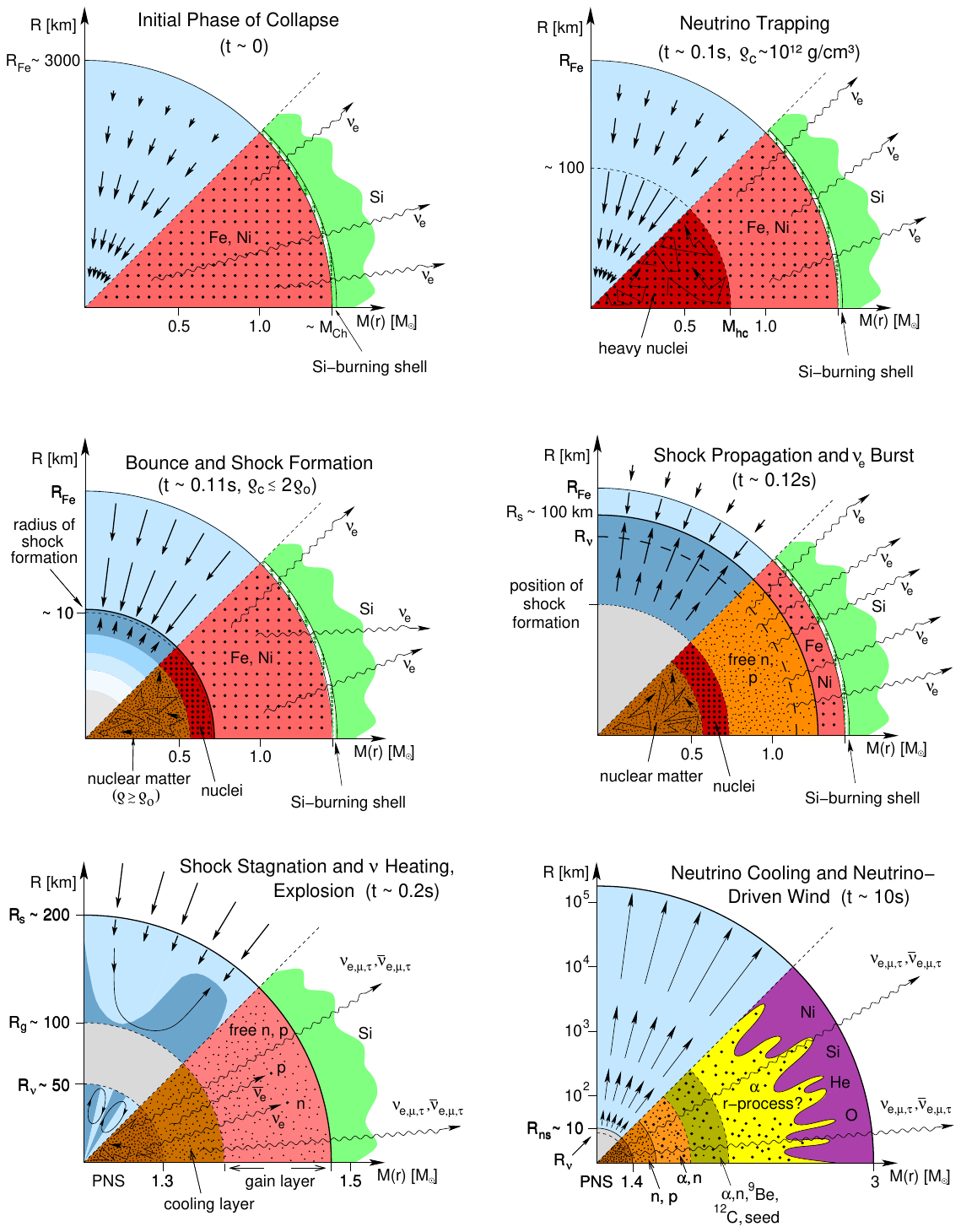}
	\caption{Scheme of the six phases of a core collapse SN. (Figure taken from Ref.~\cite{Janka:2006fh} with permission). 
}
	\label{fig:cc}
\end{figure}

\item \emph{Bounce and shock formation} ($t_{\rm pb}\sim 0.11$~s): The core collapse decelerates when the density reaches $\rho \sim 3 \times 10^{14}{\rm g}~{\rm cm}^{-3}$, because the EoS of nuclear matter stiffens. The outer layers continue to fall onto the inner core with supersonic velocity and a shock wave starts to propagate from to the outer core. This bounce reverses the collapse into an explosion.
	\item \emph{Shock propagation and $\nu_{e}$ burst} ($t_{\rm pb}\sim 0.12$~s): From simulations of core-collapse SNe it is evident that the shock wave loses its energy  during its propagation due to  the heavy nuclei dissociation. Therefore the prompt explosion mechanism in which shock-wave would propagate outward in one step, the so-called prompt scenario proposed in Ref.~\cite{Colgate:1960zz} (see also Ref.~\cite{Bethe:1990mw} and references therein),  is not a viable SN explosion mechanism. The dissociation of heavy nuclei increases the number of free protons, which capture electrons to produce neutrons and electron neutrinos. It leads to further neutronization of matter. In this phase $10^{51}$~erg of energy are released in the \emph {prompt electron neutrino burst}.
\item \emph{Shock stagnation and $\nu$ heating} ($t_{\rm pb}\sim 0.2$~s): The shock wave dissipates its energy, dissociating heavy nuclei in its propagation and eventually stalls at a distance ${\mathcal O}(100)$~km from the SN core. Meanwhile infalling matter is accreting on the SN core.  If nothing revitalizes it, the shock-wave  eventually recollapses leading to a black-hole. At this regard, neutrinos play a crucial role. Neutrinos are trapped in a SN core and are emitted from the last scattering surface, the so-called ``neutrinosphere'' (with a radius $R_{\nu}\sim {\mathcal O} (100)$~km). A small  fraction of these neutrinos can be reabsorbed, depositing  energy behind  the stalled shock wave by $n\nu_{e}\rightarrow pe^{-}$ and $p\bar{\nu}_{e}\rightarrow ne^{+}$ reactions. Thanks to these processes the shock wave is revitalized, resuming its motion as proposed in the \emph{delayed} neutrino-driven SN explosion mechanism~\cite{Bethe:1984ux,Wilson:1986ha}.	
\item \emph{Neutrino cooling and neutrino driven wind} ($t_{\rm pb}\sim 10$~s):  After $0.5-1$~s thanks to the neutrino	heating, the shock wave that has resumed its motion eventually ejects the outer layers of the star (``neutrino driven wind'').
The PNS cools down by emission of (anti)neutrinos of all flavors, and most of its energy is released in the first $5-10$~s.  In this phase a SN can be roughly seen as a black-body radiating neutrinos~\cite{Burrows:1988ba}. 
\end{itemize}

The scenario depicted above is called the \emph{neutrino-driven} mechanism, since essentially an explosion is associated with the shock-wave reheating by neutrinos. The main challenge of SN simulations is to realize self-consistently  this  complicated dynamics~\cite{Cardall:2005zy,Janka:2012sb}. Indeed, one of the main problems faced by SN simulations is related to the fact that often the shock-wave cannot be self-consistently revitalized by the neutrino-energy deposition. This is a typical problem often encountered in one-dimensional spherically symmetric SN models (except for low-mass electron capture SNe, with masses in the range $8-10$~$M_{\odot}$~\cite{Fischer:2009af,Hudepohl:2009tyy}). In this situation an explosion can be triggered artificially enhancing the neutrino energy deposition behind the shock-wave. However, it is expected that in multi-dimensional SN simulations~\cite{Bollig:2020phc,Vartanyan:2018iah,Fryer:2002zw,Kifonidis:2005yj,Scheck:2006rw} the explosion would be naturally achieved. Indeed, the onset of the explosion would be supported by matter  convective motions helped by hydrodynamical instabilities~\cite{Herant:1994dd,Janka:1995bx} such as the Standing Accretion Shock Instability (SASI)~\cite{Blondin:2002sm} which enhances the neutrino-energy deposition. At this regard, recent two-dimensional SN models are successfully exploding (see, e.g., Ref.~\cite{Just:2018djz} for an inspection of different models). The  forefront of the research has reached the development of exploding three-dimensional SN simulations by different groups (see, e.g., Refs.~\cite{Glas:2018oyz,Burrows:2019zce,Takiwaki:2016qgc,Bollig:2020phc,Vartanyan:2018iah}). Despite these significant advances,   a proof of the robustness of this neutrino heating scenario is still lacking. Alternative scenarios have been also proposed involving magnetohydrodynamics~\cite{Akiyama:2002xn}, acoustics~\cite{Burrows:2005dv} or phase-transition mechanisms~\cite{Fischer:2010wp}. However, for definitiveness in the following we will refer only to the neutrino-heating picture described above. Indeed, only in this framework the SN axion emission has been characterized. Furthermore, since in order to constrain axions we will be interested in the SN cooling phase up to $t \sim 10-20$~s after core-bounce, we will consider only 1D models which are currently the only ones extending till such late times (see, e.g., Ref.~\cite{Fiorillo:2023frv}).

\subsection{The neutrino signal}
\label{sec:nusign}

A SN model predicts different observables that can be used to probe the explosion model, notably the neutrino signal, gravitational waves, the explosion energies and remnant masses, the optical lightcurves, the output of nucleosynthesis and the explosion asymmetries. Therefore, in order to get the most of information one has to be prepared to a multi-messenger approach (see Ref.~\cite{Nakamura:2016kkl}). 

At this regards, neutrinos are among the most important and studied messengers from a SN explosion. Indeed, the $99\%$ of the entire binding energy of the PNS is carried out by neutrinos of all flavors in neutrino burst which lasts $\sim10$~s. This carried energy can  be estimated by the binding energy of the compact star that formed after collapse~\cite{Burrows:1990ts}
\begin{equation}
	E_{\rm B}=\frac{3}{5}\frac{G M^{2}}{R}\simeq 3 \times10^{53}~{\rm erg} \left(\frac{M}{1.4 \,\ M_{\odot}}\right)^{2} \left(\frac{10 \,\ \textrm{km}}{R}  \right)   \,\ .
\end{equation}
Then, the typical neutrino  luminosity is $L_{\nu}\sim 10^{52}{\rm erg}~{\rm s}^{-1}$ (assuming equipartition among the  six (anti)neutrino species) and it varies depending on the SN explosion phase. Neutrinos are trapped in the SN core and emitted from the ``neutrino-sphere''. Then, from  the virial theorem (see Sec.~\ref{intro}) one can roughly estimate the neutrino mean energy as $2\langle E_{\rm kin}\rangle\simeq GM m_{N}/R$. Taking as neutrino-sphere radius at late times  $R\sim15$~km, one gets $\langle E_{\rm kin}\rangle\simeq 25$~MeV. 

The duration of the neutrino burst is related to the diffusion time $t_{\rm diff}$ inside the SN core, that can be crudely estimated as  the product of the time duration between successive scatterings and the number of steps,
\begin{equation}
	t_{\rm diff}= \frac{\lambda}{c} \frac{R^2}{\lambda^2} \,\ ,
\end{equation}
where $\lambda$ is the neutrino mfp. A typical neutral current neutrino-nucleon cross section is $\sigma\simeq 10^{-40}~{\rm cm}^{2}(E_{\nu}/100~{\rm MeV})^{2}$ where $E_{\nu}$ is the neutrino energy. At the nuclear density, the number density of nucleons is $n_{B}\sim 10^{38}~{\rm cm}^{-3}$, corresponding to a mfp $\lambda = 1/ n_B \sigma  \simeq 30~{\rm cm}$. Then, the diffusion time-scale is of $t_{\rm diff} \sim {\mathcal O}(1)$~s.

\begin{figure}[t]
	\centering
	\includegraphics[width=0.99\columnwidth]{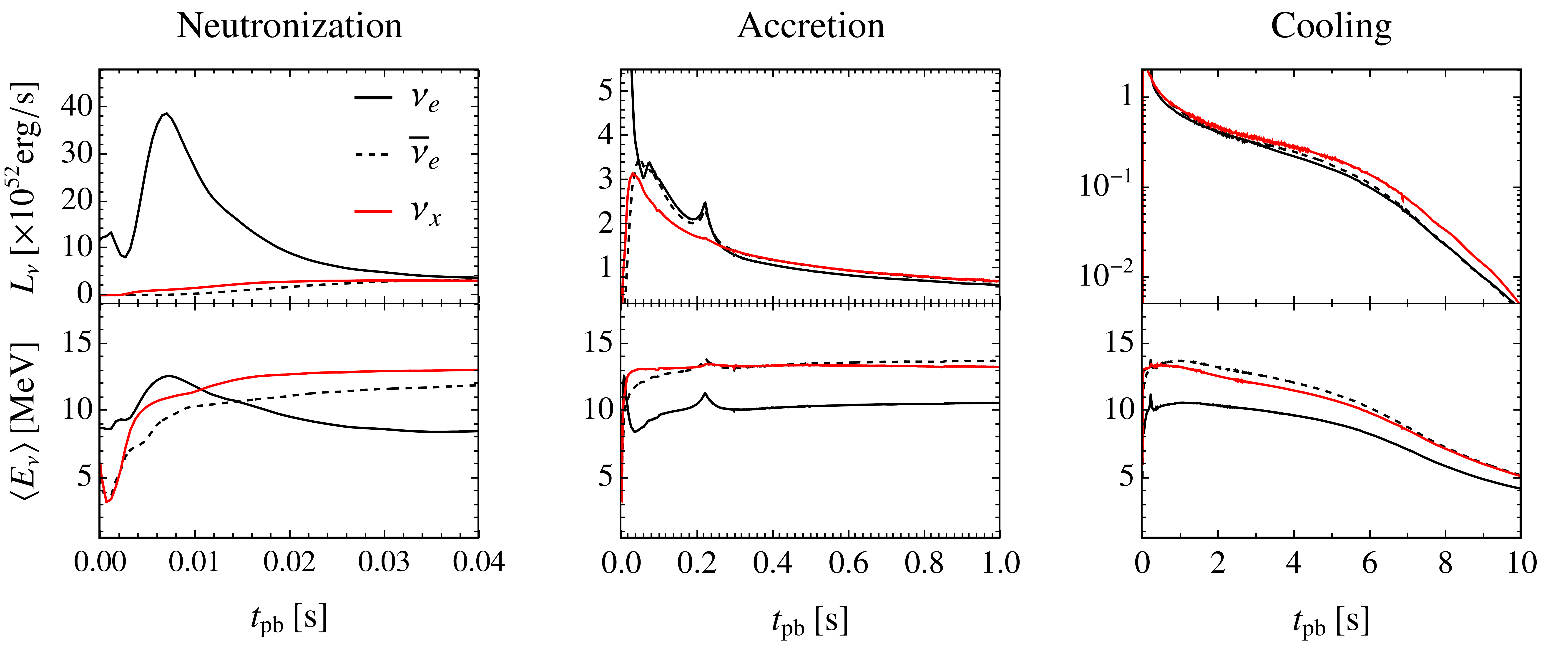}
	\caption{Time evolution of neutrino luminosities and mean energies for a $18.8~{\rm M}_{\odot}$ progenitor mass SN. The three phases of neutrino emission are separated as prompt neutrino burst (left), accretion phase (center) and cooling phase (right). Data taken from Ref.~\cite{SNarchive}. (Courtesy of Alessandro Lella).
}
	\label{fig:Emean}
\end{figure}

These back-by-the-envelope estimations~\cite{Cardall:2007dy} can be compared with the features of the neutrino signal emerging from SN models. From definitiveness in Fig.~\ref{fig:Emean} we show the evolution of the neutrino burst based on 1D spherical symmetric {\tt GARCHING} group's SN model SFHo-s18.8 provided in Ref.~\cite{SNarchive} and based on the neutrino-hydrodynamics code {\tt PROMETHEUS-VERTEX}~\cite{Rampp:2002bq}. The simulation employs the SFHo EoS~\cite{Hempel:2009mc,Steiner:2012rk} and is started from a stellar progenitor with mass $18.8~{\rm M}_\odot$~\cite{Sukhbold:2017cnt}, leading to NS with baryonic mass $1.35~{\rm M}_\odot$.

Notably, the neutrino emission is divided into three main phases:
\begin{itemize}
\item \emph{Prompt neutrino burst} (few tens of ms after bounce, left panel): {As explained above,} the stellar collapse is triggered by the reduction of the electron gas pressure, as electrons are captured by protons bound in heavy nuclei. The electron capture produces only $\nu_{e}$, giving a strong enhancement in the $\nu_{e}$ luminosity and mean energy. This process continues until matter becomes opaque to neutrinos. In only $20$~ms, about $10^{51}{\rm erg}$ of energy are released as electron neutrinos with energy $\langle E_{\nu_{e}}\rangle\sim10-12$~MeV. This phase is largely independent on the progenitor mass and EoS~\cite{Kachelriess:2004ds}.
\item\emph{Accretion phase} (few tens to several hundred ms, middle panel): When the core density reaches the nuclear one, the short-range nuclear repulsion balances gravity, stopping the collapse and producing a shock wave, which stalls around $100-200$~km. Matter is infalling on the SN core, and  mostly  electron (anti)neutrinos are produced by charged-current processes on this accreting matter.  In this phase a visible hierarchy in the neutrino energies is observable, namely $\langle E_{\nu_{e}}\rangle\lesssim\langle E_{\bar{\nu}_{e}}\rangle\lesssim\langle E_{\nu_{x}}\rangle$, where $\nu_{x}$ is any of $\nu_{\mu,\tau}$ or $\bar{\nu}_{\mu,\tau}$. Indeed, neutrinos are emitted from the neutrino-sphere, whose temperature fixes the neutrino mean energies.  Different average energies reflect different interactions with matter. For instance, electron (anti)neutrinos interact with matter via charged interactions $\bar{\nu}_{e}p\rightarrow n e^{+}$ and $\nu_{e}n\rightarrow pe^{-}$. However, muonic and tauonic neutrinos typically  do not have enough energy to produce respectively muons and tauons  in SN core (see, however, Refs.~\cite{Bollig:2017lki,Fischer:2020vie} for recent simulations including the effects of muons creation), then their neutral current interactions are weaker than electron neutrinos and they decouple at higher temperature. The slight difference between electron neutrinos and antineutrinos relies in the fact that since there are more neutrons than protons in the SN core, antineutrinos decouple at higher temperature.
\item \emph{Cooling phase} (up to 10–20~s, right panel): Starting from $0.5-1$~s the PNS enters the cooling phase, where the remaining stellar energy is emitted as neutrinos of all flavors with typical luminosities $L_{\nu}\sim 10^{52}~{\rm erg}~{\rm s}^{-1}$. One observes an approximate luminosity equipartition between all species and only a mild $\langle E_{\bar\nu_e} \rangle / \langle E_{\nu_x} \rangle $ hierarchy.	The decrease of mean energies reflects the gradual cooling of the PNS. 
\end{itemize}

These three phases  offer different physics opportunity. The neutronization and accretion phase,  during which  flavor differences among neutrino species are sizable, offer the best time windows to probe flavor conversion effects (see, e.g., Refs.~\cite{Mirizzi:2015eza,Tamborra:2020cul,Capozzi:2022slf,Volpe:2023met,Dasgupta:2023fdr}). Conversely, during the cooling phase flavor conversion effects are presumably small. However, as we will see in the following, the neutrino cooling time is an important tool to diagnose the emission of other novel particles.

\subsubsection{SN 1987A neutrinos}

\begin{figure}[t!]
	\centering
	\vspace{0.cm}
	\includegraphics[width=0.6\columnwidth]{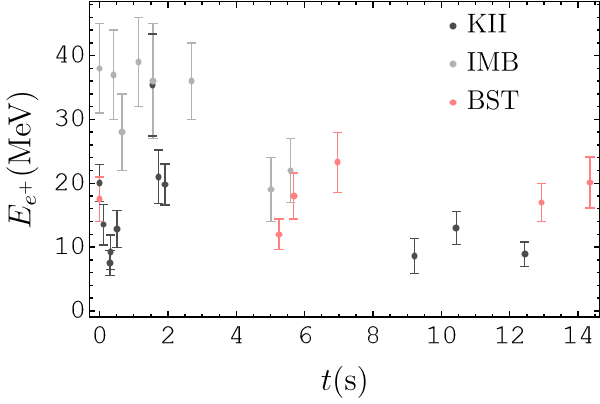}
	\caption{SN 1987A neutrino events at KII, IMB and BST. The energies refer to secondary positrons.
}
	\label{fig:KIIevents}
\end{figure}

The observation of SN 1987A in the Large Magellanic Cloud, a small satellite galaxy of the Milky Way at a distance of $51.4$~kpc, marked a milestone event in astroparticle physics, offering the first and thus far only detection of neutrinos from a stellar gravitational collapse. The SN 1987A progenitor was Sanduleak $-69^{\circ} 202$ a blue supergiant with a mass of $\sim20 {\rm M}_{\odot}$. Different detectors measured the neutrino burst from SN 1987A within a few seconds about $7:35:40$ UT (universal time) on $23^{\rm th}$ February 1987 and the subsequent optical signal was seen at $10:38$ UT. The most important observations were performed by the Kamiokande II (KII)~\cite{Kamiokande-II:1987idp,Hirata:1988ad},  Irvine-Michigan-Brookhaven (IMB)~\cite{Bionta:1987qt,IMB:1988suc} water Cherenkov detectors, originally projected to search for proton decay (see Fig.~\ref{fig:KIIevents}). A few events were detected by the Baksan Scintillator Telescope (BST)~\cite{Alekseev:1987ej}~\footnote{A presumably  spurious observation was also reported  from the Mont Blanc Liquid Scintillator Detector (LSD)~\cite{Dadykin:1987ek}.}. The three detectors revealed positrons produced by the inverse-beta decay (IBD) ${\bar\nu}_e ~ p \to n ~ e^+$  of electron antineutrinos on target protons. In particular KII collected 11 events, IMB 8 events and BST 5 events. This small sample of SN 1987A events was studied in details in several papers, see, e.g., Refs.~\cite{Jegerlehner:1996kx,Loredo:2001rx,Pagliaroli:2008ur,Yuksel:2007mn,Li:2023ulf,Fiorillo:2023frv}. These studies suggest that SN 1987A neutrinos can be taken  as a confirmation of the salient features of our physical comprehension of the core-collapse supernova phenomenon and of the associated neutrino emission~\cite{Raffelt:1990yu}. This observation allowed one to put strong constraints on exotic neutrino properties (e.g., decays, neutrino charge, etc...) that would have altered the SN neutrino emission (see, e.g., Ref.~\cite{Raffelt:1990yu} for a comprehensive review). Most importantly, the total energy of ${\bar \nu}_e$ and the inferred cooling timescale of a few seconds of the PNS put severe limits on non-standard cooling mechanisms associated with new particles emitted from the SN core~\cite{Raffelt:1987yt}. Unfortunately, the poor statistics of the SN 1987A events limited our ability to determine both  SN and neutrino parameters, although some hints could be obtained. For example, the signal of SN 1987A was also analyzed in the light of neutrino oscillations (see, e.g., Refs.~\cite{Jegerlehner:1996kx,Kachelriess:2000fe,Minakata:1988cn,Lunardini:2004bj, Li:2023ulf, Fiorillo:2023frv}).

\subsection{Constraining FIP emission}
\label{sec:novelpart}

\subsubsection{Free-streaming and trapping regimes}

The neutrino cooling time offers a unique opportunity to constraint FIPs such as axions~\cite{Brinkmann:1988vi,Burrows:1988ah, Burrows:1990pk,Keil:1996ju,Chang:2018rso,Carenza:2019pxu}, dark photons~\cite{DeRocco:2019njg,Chang:2016ntp}, sterile neutrinos~\cite{Dolgov:2000jw,Mastrototaro:2019vug,Carenza:2023old}, Kaluza-Klein gravitons~\cite{Hannestad:2001jv} and unparticles~\cite{Hannestad:2007ys}. 

The basic argument was presented in a seminal paper, Ref.~\cite{Raffelt:1987yt}. Let us assume a generic FIP,  $X$, coupled with matter through a coupling constant $g_X$. In the weak coupling limit (i.e., small $g_X$) $X$ escapes the SN core unimpeded, providing an efficient energy loss channel. This may lead to a sizable reduction of the neutrino burst duration, in tension with observations. Typically, the efficient production of exotic particles needs, not only high temperatures, but also high densities and the cooling phase is the only SN explosion phase featuring both of these requirements. A shorter neutrino burst corresponds to fewer events from the observed SN 1987A, ultimately the predicted neutrino events would be incompatible with the measurements of IMB, KII and BST. A constraint based on this observable is called \emph{energy-loss} or \emph{cooling bound}~\cite{Raffelt:1987yt,Raffelt:1990yz,Raffelt:1996wa}. 

The observation of the neutrino events from SN 1987A is compatible with a binding energy of $1-4\times10^{53}~{\rm erg}$  carried out by neutrinos in a $\sim10$ s burst, assuming equipartition among the different neutrino species. As estimated in Refs.~\cite{Ellis:1987pk,Raffelt:1987yt}, at most $E\sim2\times10^{53}~{\rm erg}$ in $10$~s might be stolen by exotic particles, resulting in a luminosity $L\lesssim 2\times10^{52}{\rm erg}~{\rm s}^{-1}$. Assuming a PNS mass of $\sim 1 \,\ M_{\odot}=2\times10^{33}~{\rm g}$, the energy-loss criterion is conveniently formulated as a constraint on the exotic luminosity per unit mass~\cite{Raffelt:1987yt,Raffelt:1990yz,Raffelt:1996wa,Raffelt:2006cw}
\begin{equation}
	\varepsilon_X \lesssim 10^{19}~{\rm erg}~{\rm g}^{-1}{\rm s}^{-1}\,,
	\label{eq:cbound}
\end{equation}
calculated for typical SN conditions $T=30~{\rm MeV}$, $\rho=3\times10^{14}~{\rm g}~{\rm cm}^{-3}$ and $Y_{e}=0.3$. This criterion was calibrated on different SN simulations available at the epoch of SN 1987A including exotic particles (axions in this case~\cite{Mayle:1987as,Burrows:1988ah,Burrows:1990pk}), confirming that the neutrino burst duration is halved if the time-integrated exotic losses are $E_{X}\sim 2\times10^{53}~{\rm erg}$. We remark that the criterion in  Eq.~\eqref{eq:cbound}  has never been checked with current state-of-the-art SN simulations. Furthermore, the recent analysis of SN 1987A data, conducted in Ref.~\cite{Fiorillo:2023frv} using state-of-the-art 1D SN simulations—including NS convection and nucleon correlations in the neutrino opacities—suggests shorter PNS cooling times of 5-9 seconds. This finding conflicts with the late event clusters observed in KII and BST after 8-9 seconds, which are also challenging to attribute to background noise. Thus, there appears to be a potential tension between SN models and late-time SN 1987A data, which may call into question the robustness of SN 1987A constraints on exotic particles. However, in the seminal paper Ref.~\cite{Raffelt:1987yt}, which considered only the IMB events to derive a cooling bound, the effect was found to be limited to a factor of 2. Therefore, we should not expect significant quantitative changes when applying the simplified criterion in Eq.~\eqref{eq:cbound}. In any case, the interpretation of late-time events from SN 1987A warrants further investigation.

As the coupling increases, the exotic particles  become so strongly coupled with matter, that they enter the \emph{trapping regime}, analogously to neutrinos. Details on this regime can be found in Refs.~\cite{Raffelt:1996wa,Caputo:2021rux}. We will follow these references closely in our presentation. Trapped particles can lead to a radiative energy-transfer from the deepest core regions to the surface of the PNS. In this way they compete with the energy transfer by neutrinos and by convection. Furthermore, when they enter the free-streaming regime after their decoupling, they can carry-out energy from the core as well as deposit energy behind the shock-front. Finally, a sufficiently   strong coupling may lead to detectable events in large underground neutrino detectors. 

In the trapping limit, the $X$ particles emerge from their last-scattering surface, the ``X-sphere'' with radius $R_X$. Assuming that they are emitted as a black-body radiation, one can calculate their luminosity applying the Stefan-Boltzmann (SB) law
\begin{equation}
	L_X= 4 \pi R_X^2 \frac{\pi^2}{120} T^4(R_X) \,\ ,
	\label{eq:black}
\end{equation}
where $T(R_X)$ is the temperature at radius $R_X$ and we assumed that $X$ is a boson. This radius is calculated using the optical depth against the exotic particle absorption, defined as~\cite{Burrows:1990pk}
\begin{equation}
	\tau(r)=\int_{r}^{\infty} d r^\prime \Gamma_X(r^\prime) \,,
	\label{eq:tau}
\end{equation}
where the absorption rate $ \Gamma= 1/\lambda_X$, being $\lambda_X$ the mfp of the $X$ particle. The X-sphere  radius $R_{X}$ is defined such that $\tau(r_{X})=2/3$. As remarked in Ref.~\cite{Caputo:2021rux}  from the Boltzmann collisional equation, if the medium is in thermal equilibrium, the detailed balance condition implies that in Eq.~\eqref{eq:tau} one has to consider the \emph{reduced} absorption rate, defined as 
\begin{equation}
	\Gamma_X \equiv \Gamma_A^\ast= \Gamma_A (1-e^{-\omega/T}) \,\ ,  
	\label{eq:abs}
\end{equation}
where $\Gamma_A$ is the absorption rate. Also in this case a constraint on $X$ emissivity  can be obtained  by requiring $L_X \lesssim L_\nu$.

\subsubsection{Modified luminosity criterion}

In the recent literature on SN bounds on novel particles~\cite{Chang:2016ntp,Chang:2018rso} it was proposed a ``modified luminosity criterion'' proving a unifying recipe covering both the free-streaming and trapping regimes described above. This recipe assumes that  the $X$ emission comes from an extended volume in the outer regions of the star, instead of a well-defined surface. Then one can evaluate the $X$ emissivity on a SN unperturbed model as 
\begin{equation}
\frac{d L_X}{d \omega}= \int_{0}^{\infty} dr 4 \pi r^2 \frac{4 \pi \omega^3}{(2 \pi)^3} \Gamma_E(\omega, r) \langle e^{-\tau(\omega,r)} \rangle \,\ ,
\label{eq:modlum}
\end{equation}
where $\omega$ takes into account gravitational redshift effects and the emissivity rate is related to the reduced absorption rate of Eq.~\eqref{eq:abs} by
\begin{equation}
\Gamma_E= \frac{\Gamma_X}{e^{\omega/T}-1} \,\ , 
\end{equation}
and the optical depth $\tau(\omega,r)$ is based on the reduced absorption rate. The directional average of the absorption factor is
\begin{equation}
\langle e^{-\tau(\omega,r)} \rangle = \frac{1}{2} \int_{-1}^{+1} d\mu e^{-\int_{0}^{+\infty} ds \Gamma(\omega, \sqrt{r^2 + s^2+2 rs\mu})} \,\ ,
\end{equation}
where $\mu= \cos \beta$ and $\beta$ is the angle between the outward going radial direction and a given ray of propagation along which $ds$ is integrated.

In Ref.~\cite{Caputo:2021rux} it was performed a comparison of bosons emissions in a SN core from Primakoff process with the SB  approach and  with the modified luminosity of Eq.~\eqref{eq:modlum} recipe integrating over an unperturbed SN model. The obtained emissivity was similar in the two cases, so that one does not expect large changes when presenting bounds. Notably, the problem of  radiative transfer by new feebly-interacting bosons was studied in Ref.~\cite{Caputo:2022rca} starting from first principles.

\begin{figure}[t!]
\centering
\vspace{0.cm}
\includegraphics[width=0.7\columnwidth]{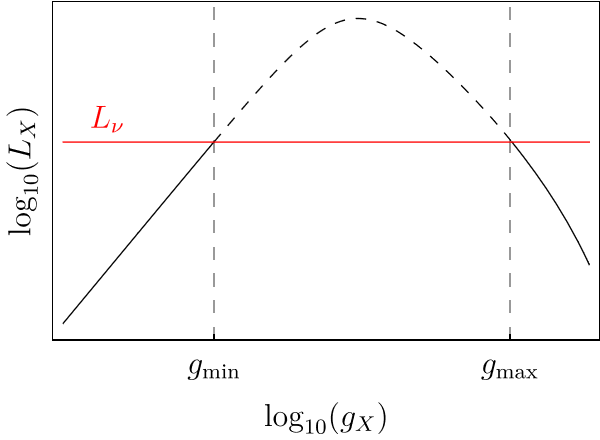}
\caption{Schematic behaviour of the exotic particle luminosity $L_{X}$ as function of the coupling $g_{X}$. The horizontal line indicates when $L_{X}$ exceeds the neutrino luminosity, shown by the dashed line.}
\label{fig:Lavsg}
\end{figure}

\subsubsection{Getting constraints}

Let us summarize the previous results showing an example of the impact of  a FIP $X$ interacting with nuclear matter, electrons or photons with a coupling $g_{X}$,  produced in a SN core. In  Fig.~\ref{fig:Lavsg}, inspired by Ref.~\cite{Raffelt:1987yt}, we compare the luminosity carried out by the exotic species $L_X$ with the neutrino luminosity $L_\nu$. For a sufficiently small coupling, $g_X < g_{\rm min}$, the exotic energy-loss is negligible compared to the neutrino one and there are no observable consequences. The $X$ particles are in the free-streaming regime:  they are emitted from the entire volume of the PNS and escape freely because of the weakness of their interactions. In contrast, the trapped neutrinos are emitted from the surface of the neutrinosphere. As $g_{X}$ increases, the luminosity $L_{X}$ grows. When the coupling exceeds $g_{\rm min}$, the luminosity $L_{X}$ becomes comparable or larger than the neutrino one, thus the SN evolution is affected in an observable way. The effect would be the reduction of the SN 1987A neutrino burst duration, since the energy budget available to neutrinos is reduced. 

Clearly, as $g_{X}$ grows the mfp decreases and the radius $R_{X}$ moves outside from the core where the temperature $T_{X}$ reduces. Thus, the blackbody luminosity in Eq.~\eqref{eq:black} lowers and less energy is subtracted from the PNS, as shown in Fig.~\ref{fig:Lavsg}. Eventually, for couplings larger than $g_{\rm max}$, the exotic luminosity becomes lower than the neutrino one and, again, the impact on the SN evolution is small. In this regime, FIPs contribute to the stellar energy transfer and their impact is maximized when the mfp is comparable with the radius of the PNS. Thus, the non-standard energy transfer becomes efficient on the same scales of the standard one determined by neutrinos. The luminosity bound in Eq.~\eqref{eq:cbound} is usually applied in this regime, but there is no analytic criterion to constrain the diffusion properties. In conclusion, an excessive reduction of the neutrino burst is expected for particles with a coupling in the range $g_{\rm min}\lesssim g_{X}\lesssim g_{\rm max}$.

\begin{figure}[t!]
\centering
\includegraphics[width=0.8\columnwidth]{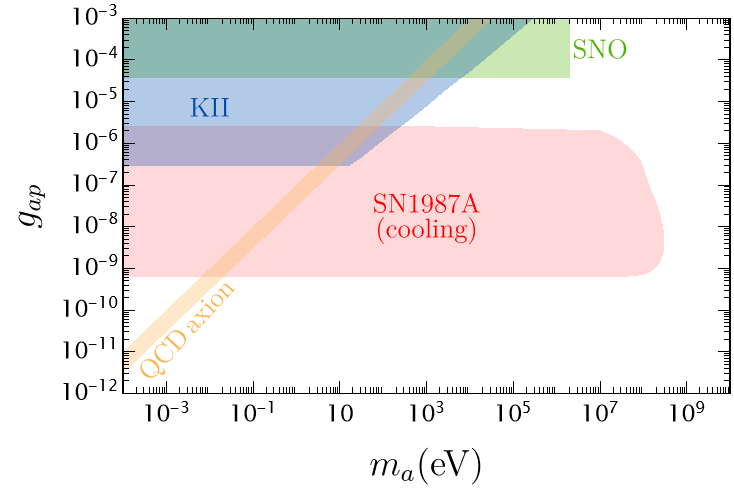}
\caption{Summary plot of the bounds in the $g_{ap}$ vs $m_a$ plane together with the QCD axion band (in orange). For definiteness, we assume here that axions interact only with protons ($g_{an}=0$). The red region labeled ``SN1987A (cooling)'' (light magenta) is ruled out by the cooling argument~\cite{Lella:2023bfb}. The blue one, by the non-observation of extra events inside the KII experiment in coincidence with SN 1987A~\cite{Engel:1990zd,Lella:2023bfb,Carenza:2023wsm}.  For reference, we show also the region excluded by the search for dissociation of deuterons induced by solar axions in the SNO data~\cite{Bhusal:2020bvx} (in green) (see also Sec.~\ref{sec:neutrin}). (Data for bounds taken from the repository~\cite{AxionLimits}).} 
\label{fig:CoolingBound}
\end{figure}

\subsection{Axion-nucleon coupling }
\label{sec:QCD}

One can apply the general argument, described above, to the specific case of axions. On a general ground, given the typical SN core temperature a few tens of MeV, we may expect these bounds to extend up to masses ${\mathcal O}(100)$ MeV, if axions are thermally produced. This is indeed the case, as shown explicitly in Fig.~\ref{fig:CoolingBound} where, in the red region we report the most current cooling limits on the axion-nucleon coupling~\footnote{For simplicity, the figure refers to the case of axions coupled only to protons, as is the case of the KSVZ axion model.}. As the figure shows, the cooling criterions allows us to exclude \mbox{$6\times10^{-10}\lesssim g_{ap}\lesssim 2.5\times10^{-6}$} for \mbox{$m_a\lesssim 10~{\rm MeV}$} and \mbox{$7\times10^{-10}\lesssim g_{ap}\lesssim 3.5\times10^{-7}$} for \mbox{$m_a\sim\mathcal{O}(100)~{\rm MeV}$}. These limits are discussed in detail below. 

We begin with the case of free-streaming axions. As discussed in Sec.~\ref{ref:nucleoncoupling}, for a long time the main channel for axion emissivity in a nuclear medium was identified with the nucleon-nucleon NN bremsstrahlung, i.e. $N~ N\rightarrow N ~ N ~ a$~\cite{Brinkmann:1988vi,Brinkmann:1988vi,Burrows:1988ah, Burrows:1990pk,Keil:1996ju}. However, it was later realized that the Compton pionic process, $\pi^{-} ~ p \to n ~ a$, already discussed in early references~\cite{Turner:1991ax,Raffelt:1993ix, Keil:1996ju}, was likely comparable to the nuclear bremsstrahlung, and might even dominate~\cite{Carenza:2020cis}. In fact, it was estimated, that neglecting the Compton pionic processes,  the cooling bound in the free-streaming regime would be weakened by a factor $\sim 2$~\cite{Carenza:2020cis}.

A reliable calculation of the axion emission rate through the NN bremsstrahlung and pionic processes presents considerable difficulties. In the bremsstrahlung case, the main challenge is the appropriate description of the nuclear-nuclear interactions. In the original literature, these were described through the OPE potential~\cite{Raffelt:1987yt,Turner:1987by} (see  Sec.~\ref{ref:nucleoncoupling}). Criticisms to this approach, however, came quickly in the following years, as it was realized that the OPE approximation  would lead to an overproduction of axions.  In particular, Refs.~\cite{Raffelt:1991pw,Janka:1995ir,Sigl:1995ac} pointed out the important role of nucleon spin fluctuations caused by nucleon self-interactions, in reducing the axion emissivity by NN process (see also Sec.~\ref{ref:nucleoncoupling}). Different corrections to the NN bremsstrahlung as the finite pion mass in the propagator, corrections for more realistic NN interactions and the multiple nucleon scattering effects have been applied with different levels of accuracy in Refs.~\cite{Raffelt:2006cw,Hanhart:2000ae,Chang:2018rso,Carenza:2019pxu} (see also Ref.~\cite{Carenza:2023lci}).

In the case of the pionic processes, the most challenging issue consists in an appropriate estimation of the abundance of negative pions in the nuclear medium. The thermal abundance of pions, estimated from their bare mass $m_{\pi^{ \pm}}=139.6 \,\ \mathrm{MeV}$, and assuming a SN core temperature of 30 MeV and density $\sim 3\times 10^{14}\,{\rm g\,cm^{-3}}$, is quite small, only $Y_\pi=2 \times 10^{-4}$ pions per nucleon (see, e.g., Ref.~\cite{Caputo:2024oqc}). This value is, however, quite underestimated since pions participate in the equilibrium between nucleons and thus their chemical potential should be taken into account. Furthermore, a post-ideal gas approximation, which can be carried out through the virial expansion, presented additional contributions to the abundance of negative pions, namely~\cite{Fore:2019wib}  
\begin{equation} 
n_{\pi^-} = \int \frac{dk}{2\pi^2} k^2 \exp{ (-\beta ( \sqrt{k^2+m^2_\pi}-\hat{\mu}))} + n^{\rm int}_{\pi^-} \,,
\label{eq:pion_number}
\end{equation} 
where 
\begin{equation}
n^{\rm int}_ {\pi^-} = \sum_{N=n,p} z_N z_{\pi^-} b_2^{N \pi^-}\,,  
\end{equation}
is the contribution due to pion-nucleon interactions, where for a given species $i$ the fugacity $z_i=\exp{\beta(\mu_i-m_i)}$ with $\mu_i$ and $m_i$ are the chemical potential and the mass of the particle $i$, and $\beta=1/T$, and $b_2^{N \pi^-}$ is the second virial coefficient for $N$ $\pi^-$ system. A substantial advancement in this direction came with the publication of a seminal paper by S. Reddy and B. Fore~\cite{Fore:2019wib}, who showed that the pion abundance in the nuclear matter typical of the SN core can be as large as a few percent of the nucleon abundance. Such a large fraction of negative pions would imply the dominance of the pionic process in the axion production rate from the SN core~\cite{Carenza:2020cis,Fischer:2021jfm}.	  

The most recent SN axion bound, which includes all the relevant corrections to the OPE prescription for the NN process~\cite{Lella:2022uwi} and which accounts also for $N\pi$ production processes was presented in Ref.~\cite{Lella:2023bfb}.   The bound in the free-streaming regime corresponds to the lower edge of the red region in Fig.~\ref{fig:CoolingBound}.

Let us now consider the trapping regime. 
As discussed above, this corresponds to the case in which axions are reabsorbed in the SN medium and emitted from an axion-sphere, similarly to the neutrino case. In this case, the impact of the pionic processes is substantially more marginal, limiting the effects of the uncertainties associated with the pion abundance. Nevertheless, they can lead to peculiar features in the axion spectrum, that may lead observational signatures  (see Ref.~\cite{Lella:2024hfk}). The limit on the cooling associated with trapped axions is shown in Fig.~\ref{fig:CoolingBound}, and corresponds to the upper edge of the red region. A stronger bound on trapped axions can be derived from the non-observation of axion-related events in KII at the time of the SN 1987A. In fact, strongly-coupled axions emitted during a SN explosion would lead to a detectable signal in large water Cherenkov neutrino detectors, as  proposed in the  seminal paper by Engel \emph{et al.}~\cite{Engel:1990zd}, in which the authors proposed to look for axion-induced excitation of oxygen nuclei with the subsequent emission of a photon to relax the system {$a ~ {^{16}O}\rightarrow {^{16}O}^* \rightarrow {^{16}O} ~ \gamma$}. A revised calculation of the cross-section for this process, using state-of-the-art nuclear models, is presented in Ref.~\cite{Carenza:2023wsm}. This argument  rules out the blue region in Fig.~\ref{fig:CoolingBound} (KII). For reference, in Fig.~\ref{fig:CoolingBound} we also show in yellow the QCD axion band~\footnote{In the specific case of DFSZ, the width of the band may be slightly shifted because of radiative corrections. See Ref.~\cite{DiLuzio:2023tqe}.}. These results show that SN bounds strongly constrain the parameter space available for axions coupled to nucleons, excluding values of the axion-proton coupling $g_{ap}\gtrsim6\times10^{-10}$ for axion masses $m_a\lesssim1~{\rm MeV}$. Assuming axions coupling with both protons and neutrons, the  SN arguments place a bound on the axion masses that varies in the range~\cite{Lella:2022uwi}  
\begin{equation}
    g_{an}^2 + 1.10 g_{ap}^2 - 0.26 g_{an} g_{ap} < 7.8\times10^{-19}\,,
\end{equation}
\begin{equation}
    m_a\in[8,19]~{\rm meV}\,,
\end{equation}
depending on the presence of pions in the SN core. 
%
\begin{figure}[t!] \includegraphics[width=0.8\columnwidth]{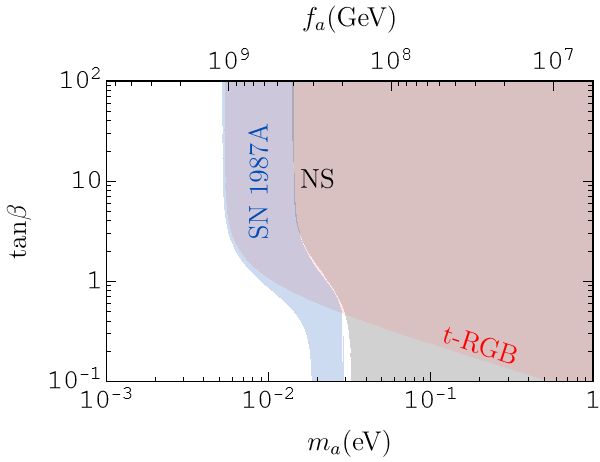}
    \caption{Summary plot of the bounds in the $\tan\beta$ {\emph vs} $m_a$ plane for the DFSZ axion model. The blue region displays the uncertainties on the SN bound related to the presence of pions in the core~\cite{Lella:2023bfb}. The whole region on the right of the blue band is excluded by the SN argument, even though it is not shown in the plot for clarity. The red area shows the region of the parameter space ruled out by the RGB tip bound~\cite{Capozzi:2020cbu}, while the grey  regions refer to the bound placed by isolated NS cooling~\cite{Buschmann:2021juv}. (Figure adapted from Ref.~\cite{Lella:2023bfb} with permission).
   }
    \label{fig:ma_vs_tanbeta}
\end{figure}
%
These results, within the  present uncertainties, are comparable with the bound placed in Ref.~\cite{Buschmann:2021juv}, which ruled out KSVZ axion masses ${m_a\gtrsim16~{\rm meV}}$ from isolated NS cooling (see Sec.~\ref{sec:neutrstar}). We highlight that in the KSVZ axion model also the axion-photon coupling $g_{a\gamma}$ is switched on. However, the SN bounds are remarkably stronger than constraints HB stars in GCs~\cite{Ayala:2014pea,Carenza:2020zil} (see Sec.~\ref{sec:Rparam}), excluding $g_{a\gamma}\gtrsim0.66\times10^{-10}~{\rm GeV}^{-1}$, which corresponds to $m_a\gtrsim440~{\rm meV}$~\cite{ParticleDataGroup:2022pth}.

The constraints for the DFSZ axion model are shown in Fig.~\ref{fig:ma_vs_tanbeta}, in the plane $m_a$ vs  $\tan\beta$. In particular, SN arguments exclude $m_a\gtrsim5~{\rm meV}$ for $\tan\beta\gtrsim5$ and $m_a\gtrsim18~{\rm meV}$ for $\tan\beta\lesssim1$. In the absence of pions this bound is relaxed to $m_a\gtrsim14~{\rm meV}$ for $\tan\beta\gtrsim5$ and $m_a\gtrsim28~{\rm meV}$ for $\tan\beta\lesssim1$ (see the dashed blue line). Moreover, since DFSZ axions couple also to leptons, it is necessary to take into account the RGB tip bound introduced in Refs.~\cite{Straniero:2020iyi,Capozzi:2020cbu} {(see Sec.~\ref{sec:white_dwarfs})}. It is remarkable to notice that, in presence of pions, SN 1987A bounds are comparable to the RGB constraint, as shown by the red area in Fig.~\ref{fig:ma_vs_tanbeta}, representing the bound from Ref.~\cite{Capozzi:2020cbu}.

In general, in the case of canonical QCD axion models, the SN arguments rule out QCD axion masses $m_a \gtrsim \mathcal{O}(10)~{\rm meV}$. This is in contrast with the original literature on the SN 1987A bound, which reported the existence of a window around a QCD axion mass $m_a \sim {\mathcal O}(1)$ eV, classicaly dubbed the ``hadronic axion window''~\cite{Chang:1993gm}. Finally, this bound is stronger than the reach of current and future cosmological experiments, which would probe axion masses $m_a\gtrsim150~{\rm meV}$~\cite{DEramo:2022nvb}.  Thus, it is unlikely that future cosmological probes would find signatures of the QCD axion mass as hot dark matter. Nevertheless, lighter axion can still play the role of dark radiation (see, e.g., Ref.~\cite{DEramo:2021psx,Ferreira:2018vjj,Ferreira:2020bpb}). Recently, in Ref.~\cite{Cavan-Piton:2024ayu} it has been performed a  quantitative study of the role of SN axion emission from hadronic matter beyond the first generation--in particular strange matter, obtaining a bound on  the axial axion-strange-strange coupling, as well as on the axion-down-strange counterpart. In particular, this  latter coupling  can be as small as $10^{-2}$ for $f_a \sim 10^{9}$~GeV.
 
It is worth mentioning certain \textit{nucleophobic} axion models (e.g.,~\cite{DiLuzio:2017ogq,DiLuzio:2023tqe}) that could potentially evade the SN constraint through a carefully engineered suppression of the axion-nucleon coupling. Recently, Ref.~\cite{DiLuzio:2024vzg} demonstrated that the nucleophobia condition is preserved up to supersaturation densities. Moreover, the analysis showed that, although nucleophobia cannot be enforced throughout the entire star and is instead confined to a narrow shell, this condition is still sufficient to relax the SN constraint. In general, however, the full extent of the density impact on the axion emission rate from SN and NS is still under investigation (see Ref.~\cite{Springmann:2024mjp} and references therein)\footnote{It is worth noting that a recent study~\cite{Springmann:2024ret}, still unpublished at the time of writing, identified a new SN production channel applicable to all QCD axion models, including those specifically engineered to suppress the axion-nucleon coupling. In this case, the relaxation of the SN axion bound flux would be diminished, due to the additional axion production.}. 

To conclude, we mention a radical criticism on the SN axion bound, presented in Ref.~\cite{Bar:2019ifz}. The work questioned the validity of the core-collapse SN explosion mechanism for SN 1987A, proposing that it was rather a thermonuclear explosion~\cite{Blum:2016afe}. After the collapse, an accretion disk forms around the PNS. This phase lasts $1-3$~s after the bounce, until the PNS accumulates $\sim2-3 \,\  M_{\odot}$ and collapses into a black hole, giving to a fast decrease of the neutrino luminosity. Matter heats up in the accretion disk around the black hole, restarting the neutrino emission and triggering the explosion. The axion emission in this scenario is negligible due to the low density of the accretion disk ($\rho \sim 10^{9}~{\rm g}~{\rm cm}^{-3}$) compared to the nuclear density encountered in a traditional PNS. However, this alternative explosion mechanism should face astrophysical observations in agreement with the standard SN explosion mechanism, as the recent observation of a NS candidate after the SN 1987A explosion~\cite{Page:2020gsx,Fransson:2024csf}.

\subsection{Axion-photon coupling}
\label{sec:axionlike}

In the case of axions coupled with photons one can extend the  SN 1987A bounds  with other arguments related to the possible observation of axion induced  gamma-ray signal.

Assuming that these particles are coupled only with photons, the main production channel for $m_a < 30$~MeV is  the Primakoff process (see Sec.~\ref{sec:photon} and Fig.~\ref{fig:HBfull}). One can already set a \emph{cooling bound} on such particles by evaluating the SN axion luminosity and  constraining it according to  the criterion in Eq.~\eqref{eq:cbound}. This results in the exclusion of the range $7 \times 10^{-8}$~GeV$^{-1}<g_{a\gamma}< 2 \times 10^{-6}$ ~GeV$^{-1}$, where  the lower value is associated with the free-streaming case while the upper one  corresponds to the trapping regime. These limits on the axion-photon coupling are not competitive with respect to the one that one can place using GC observations (see Sec.~\ref{sec:GC}).

\subsubsection{Ultralight axions}

\begin{figure}[!h]
\centering
\includegraphics[width = .9\linewidth]{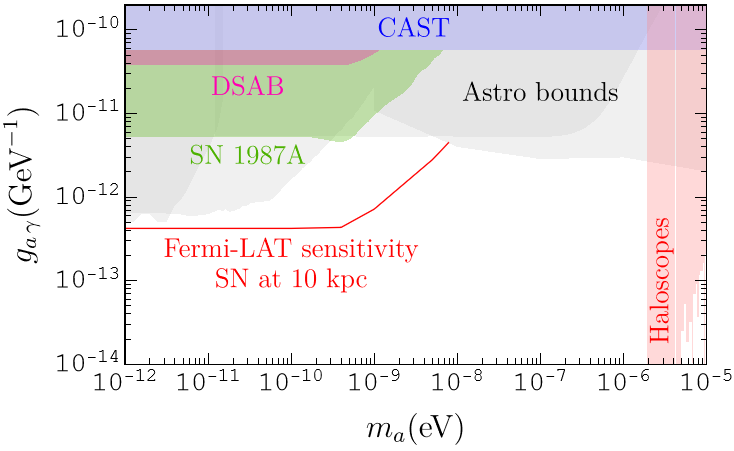}
\caption{\label{fig:SNlimits}Comparison of the expected \emph{Fermi}-LAT sensitivity for axion-photon conversions from a Galactic SN exploding at a distance $d=10$~kpc  (red line) with overall astrophysical and experimental limits. In particular, we show the SN 1987A gamma-ray bound~\cite{Hoof:2022xbe} (green region) and the {\it Fermi}-LAT bound on the Diffuse Supernova Axion background (DSAB)~\cite{Calore:2021hhn} (pink region). (Data for bounds taken from the repository~\cite{AxionLimits}).}
\end{figure}

As mentioned above, in the case of axions coupled with photons, the main production channel for $m_a \lesssim 30$~MeV is given by the Primakoff process. The  axion fluence from the Primakoff process is given, with excellent precision, by the analytical expression~\cite{Payez:2014xsa}
\begin{equation}
\begin{split}
\frac{d^{2}N_a}{dE\,dt} &= C \left(\frac{g_{a\gamma}}{10^{-11}\textrm{GeV}^{-1}}\right)^{2}
\left(\frac{E}{E_0}\right)^\beta \exp\left( -\frac{(\beta + 1) E}{E_0}\right) \,,
\end{split}
\label{eq:time-int-spec}
\end{equation}
where
\begin{equation}
\begin{split}
C&=2.18\times10^{52}~{\rm MeV}^{-1}~{\rm s}^{-1} \left(\frac{t}{\rm s}\right)^{0.89} e^{-\frac{t}{1.38~{\rm s}}}\,,\\
E_{0}&=126.0~{\rm MeV} \left(\frac{t}{\rm s}\right)^{0.51} e^{-\frac{t}{3.0~{\rm s}}}\,,\\
\beta&=2.54\left[1+\left(\frac{t}{7.62~{\rm s}}\right)^{5.01}\right]\,.
\end{split}
\end{equation}
This fit was calculated for the 1D spherical symmetric {\tt GARCHING} group's SN model SFHo-s18.8 provided in Ref.~\cite{SNarchive}~\footnote{Courtesy of Alessandro Lella}. The spectrum described in Eq.~\eqref{eq:time-int-spec} is a typical quasi-thermal spectrum, with mean energy $E_0$ and index $\beta$ ($\beta=2$ corresponds to a perfectly thermal spectrum of ultrarelativistic particles).

This flux could lead to additional bounds on axions through their conversion into photons in the Galactic magnetic field. For ultralight axions ($m_a \ll 10^{-8}$~eV), the flux from SN 1987A would have produced a gamma-ray signal (with $E \sim {\mathcal O}(100)$~MeV) in coincidence with the neutrino burst.  The non-observation of such a signal  in the Gamma-Ray Spectrometer (GRS) of the SMM (Solar Maximum Mission) in coincidence the neutrino signal from SN 1987A provides a strong bound on axions coupled to photons~\cite{Grifols:1996id,Brockway:1996yr}. For $m_a < 4 \times 10^{-10}$~eV one finds  $g_{a\gamma} < 5.3 \times 10^{-12}$ GeV$^{-1}$~\cite{Payez:2014xsa} (see Fig.~\ref{fig:SNlimits}). Temporal  information of the expected signal may allow to improve the limit on the axion-photon coupling by a factor of 1.4~\cite{Hoof:2022xbe}. After SN 1987A, no Galactic explosion occurred. A study considered possible axion signals from extra-galactic SNe, observed by gamma-ray satellite \emph{Fermi} Large Area Telescope ({\it Fermi}-LAT). Under the assumption that at least one SN was contained within the LAT field of view, one finds the bound $g_{a\gamma} < 2.6 \times 10^{-11}$~GeV$^{-1}$ for $m_a < 3 \times 10^{-10}$~eV~\cite{Meyer:2020vzy}. Future perspectives for improvement of the SN 1987A bound have been also investigated. In particular, it has been shown that for a SN exploding in our Galaxy, the {\it Fermi}-LAT would be able to explore the photon-axion coupling down to $g_{a\gamma} \simeq 4 \times 10^{-13}$~GeV$^{-1}$ for an axion mass $m_a \lesssim 10^{-9}$~eV~\cite{Calore:2023srn,Meyer:2016wrm} (see Fig.~\ref{fig:SNlimits}). The sensitivity in reconstructing the axion-photon coupling is within a factor of two, mostly due to the uncertainty in the Galactic magnetic field~\cite{Calore:2023srn}.  The case of axions with both nucleon and photon couplings has also been explored. In this scenario, enhanced axion production occurs in the PNS core via nucleon coupling, and the flux is converted into gamma-rays in the Galactic magnetic field through $g_{a\gamma}$. 
Notably, the gamma-ray spectral shape offers a unique probe of the pion abundance in the PNS~\cite{Lella:2024hfk}. 
Recently, this latter scenario has been extended considering axion-photon conversions in the magnetic fields of the external SN layers. This mechanism would allow one to probe a larger range of the axion mass, though large uncertainties are associated to the $B$-field in SNe~\cite{Manzari:2024jns}.

\subsubsection{Massive axions}

\begin{figure}[!t]
\centering
\vspace{0.cm}
\includegraphics[width=0.8\columnwidth]{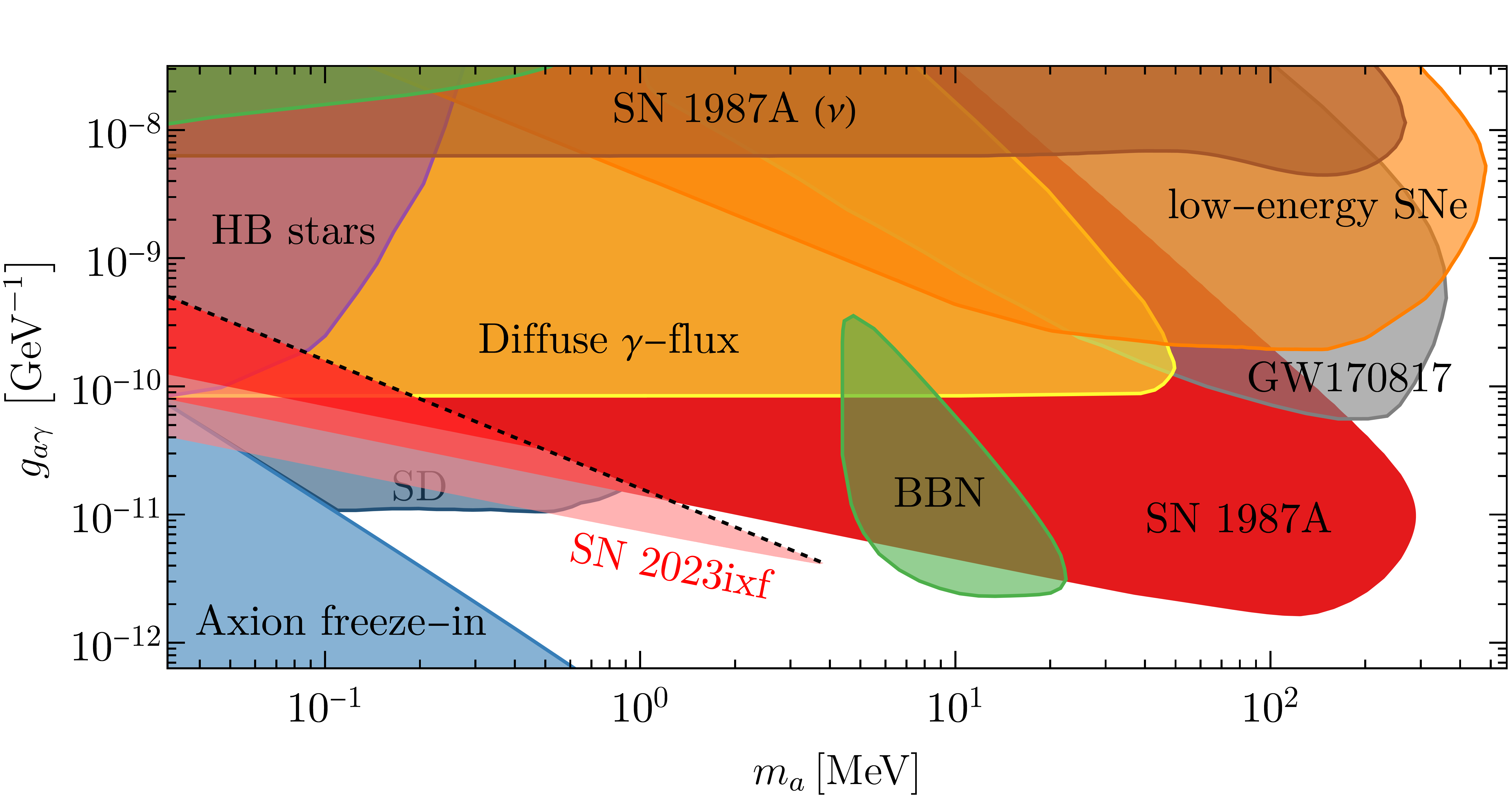}
\caption{Constraints on the axion-photon coupling $g_{a\gamma}$ as a function of mass $m_a$ in the high mass regime. The gray region labeled `SN 1987A ($\nu$)' is the cooling bound on axions coupled to photons~\cite{Lucente:2020whw}. The red exclusion region labeled `SN 1987A' derives from gamma-ray decays of axions produced in SN 1987A and measured by SMM~\cite{Muller:2023vjm}. The red regions labeled `SN 2023ixf'~\cite{Muller:2023pip} derives from similar arguments applied to SN 2023ixf and {\it Fermi}-LAT observations (magenta band).  The other, semi-transparent regions are constraints from the energy axions could deposit in low-energy SNe (orange) and the diffuse gamma-ray flux due to decays of axions produced in all past SNe (yellow), both from Ref.~\cite{Caputo:2022mah}; changes in the evolution of horizontal branch stars (purple) \cite{Ayala:2014pea,Carenza:2020zil,Lucente:2022wai}; the non-observation of X-rays after the multi-messenger observation GW170817 of a neutron star merger~\cite{Diamond:2023cto,Dev:2023hax} (gray); the irreducible cosmic axion density from freeze-in production (blue) \cite{Langhoff:2022bij}; as well as from the dissociation of light elements during BBN (green)~\cite{Depta:2020wmr}, the most conservative bound.  Figure adapted from Ref.~\cite{Muller:2023pip}. (Courtesy of Eike Ravensburg).
}
\label{fig:SNALPGagmassive}
\end{figure}

For Massive axions with a mass $m_a \gtrsim 1$ keV , the axion-photon coupling $g_{a\gamma}$ can be constrained by a series of arguments, obtaining the bounds in Fig.~\ref{fig:SNALPGagmassive}, described in this Section. For $m_a \gtrsim 50$~MeV, the axion production in SNe is dominated by the so-called ``photon coalescence'' or ``inverse decay process''~\cite{DiLella:2000dn}, where two photons can annihilate producing an axion (see Sec.~\ref{sec:photon}).

Furthermore,  axions with masses $m_a \sim {\mathcal O}(0.1-100)$~MeV,  might decay into gamma-rays before reaching Earth. The decay rate into two photons is~\cite{Raffelt:2006rj,Carenza:2019vzg}
\begin{equation}
\Gamma_{a\rightarrow \gamma\gamma}= \frac{g_{a\gamma}^2 m_a^3}{64 \pi} \,,
\label{eq:decayr}
\end{equation}
which grows rapidly with the axion mass. Concretely, Eq.~\eqref{eq:decayr} results in a decay length for the axions given by~\cite{Jaeckel:2017tud}
\begin{eqnarray}
l_{\rm a} &=& \frac{\gamma v}{\Gamma_{a\gamma\gamma}} = 
\frac{E_a}{m_a}\sqrt{1-\frac{m_a^2}{E_a^2}}  \frac{64 \pi}{g_{a\gamma}^2 m_a^3} \nonumber \\
&\simeq & 4 \times 10^{-3} \,\ \textrm{ly} \left(\frac{E_a}{100 \,\ \textrm{MeV}} \right) \left(\frac{10 \,\ \textrm{MeV}}{m_a} \right)^4 
\left(\frac{10^{-10} \,\ \textrm{GeV}^{-1}}{g_{a\gamma}} \right)^2 \,\ .
\end{eqnarray}
Axions with a mass above a few keV have a sizable probability of decaying into photons before arriving to Earth, contributing to the gamma-ray background.  The photon flux from decaying axions from a SN is~\cite{Calore:2020tjw}
\begin{equation}
\frac{d N_\gamma(E_\gamma)}{d E_\gamma} =2 \times [1-\exp(-d/l_{\rm a})] \frac{dN_a(E_a)}{dE_a} \;;
\label{eq:decay}
\end{equation}
where $E_a= 2 E_\gamma$, $dN_{a}/E_{a}$ is the produced axion flux and $l_{\rm a}=\gamma v/\Gamma_{a\rightarrow \gamma\gamma}$ is the ALP mfp. In this range of parameters the decay length is sufficiently large to assume that all the ALPs decay outside the SN photosphere, $l_{\rm a} \gg 3\times10^{12}$~cm~\cite{Kazanas:2014mca}. Also in this case a stringent bound is obtained from the lack of a signal in SMM, precisely $g_{a\gamma} \lesssim2 \times 10^{-12}$ GeV$^{-1}$ at $m_{a}\sim 100$~MeV~\cite{Jaeckel:2017tud,Muller:2023vjm} (red band in Fig.~\ref{fig:SNALPGagmassive}) (see Ref.~\cite{Caputo:2021kcv} for a similar study on massive axions produced in a SN magnetic field and decaying on their route to Earth). Recently, Ref.~\cite{Diamond:2023scc} pointed out that for axion masses of a few tens of MeV and  photon coupling of a few $10^{-10}$~GeV$^{-1}$, the decay photons may not escape and can instead form a fireball, a plasma shell with $T \sim 1$~MeV. Thus, the previous argument seems  not to exclude axions with these parameters. However, another argument can be made to exclude that region. In the fireball scenario, the photon energy is reduced to the sub-MeV range, and in the case of SN 1987A, such photons should have been detected by the Pioneer Venus Orbiter but were not observed.

Not surprisingly, a future galactic SN would allow the exploration of a much larger parameter space. A recent quantitative study of the {\it Fermi}-LAT sensitivity in the case of a future nearby SN has shown that, in the case of a non-observation, the current bound on the axion-photon coupling $g_{a\gamma}$ might be strengthened by about an order of magnitude~\cite{Muller:2023vjm}. In particular, using 
the {\it Fermi}-LAT observations of SN 2023ixf, a Type II supernova in the nearby Pinwheel Galaxy, Messier 101 (M101), the SN 1987A  constraints on the axion-photon coupling can be improved, under optimistic assumptions, by up to a factor of $\sim 2~$\cite{Muller:2023pip} (magenta band in Fig.~\ref{fig:SNALPGagmassive}).

It is also interesting to notice that  heavy axions with masses $m_a \gtrsim 100$~MeV and photon coupling $g_{a\gamma} \gtrsim 4 \times10^{-9}$~GeV$^{-1}$ would decay into photons depositing energy  within the progenitor star  surrounding the collapsing core. This energy deposition contributes to the SN explosion energy, often taken to be 1--2~B, where $1~{\rm B~(bethe)}=10^{51}~{\rm erg}$. Thus, it  provides a ``calorimetric'' constraint on radiative decays that can also include the $e^+e^-$ channel. This idea was originally advanced by Falk and Schramm a decade before SN~1987A~\cite{Falk:1978kf}, recently rediscovered~\cite{Sung:2019xie}, and applied to muonphilic bosons~\cite{Caputo:2021rux} and generic $e^+e^-$ decays~\cite{Calore:2021lih}. Moreover, it was speculated that such effects could power SN  explosions~\cite{Schramm:1981mk,Rembiasz:2018lok,Mori:2021pcv} or gamma-ray bursts~\cite{Berezhiani:1999qh,Diamond:2021ekg}. The basics idea is that  the PNS binding energy released in a core collapse is $2-4\times10^{53}~{\rm erg}$, whereas a typical explosion energy is some $10^{51}~{\rm erg}$, so less than 1\% of the total energy release shows up in the explosion (see also Ref.~\cite{Sung:2019xie} for a recent application of this criterion to the dark photon case).
In order to obtain stronger bounds, in Ref.~\cite{Caputo:2022mah} it was considered a SN population with particularly low explosion energies (the Type II-P SNe) as the most sensitive calorimeters to constrain this possibility. Their low energies limit the energy deposition from particle decays to less than about 0.1~B. In order not to exceed this explosion energy one should exclude axion-photon couplings $g_{a\gamma}$ in the $10^{-10}$--$10^{-8}~{\rm GeV}^{-1}$ range for $m_a \gtrsim 1$~MeV (orange region in Fig.~\ref{fig:SNALPGagmassive}). This bound excludes the possibility that axion energy-deposition behind the shock wave can contribute significantly to shock wave reheating, as proposed in Refs.~\cite{Mori:2021pcv,Mori:2023mjw}

\subsubsection{Diffuse SN axion background}

\begin{figure}[t!]
\centering
\includegraphics[width=0.6\columnwidth]{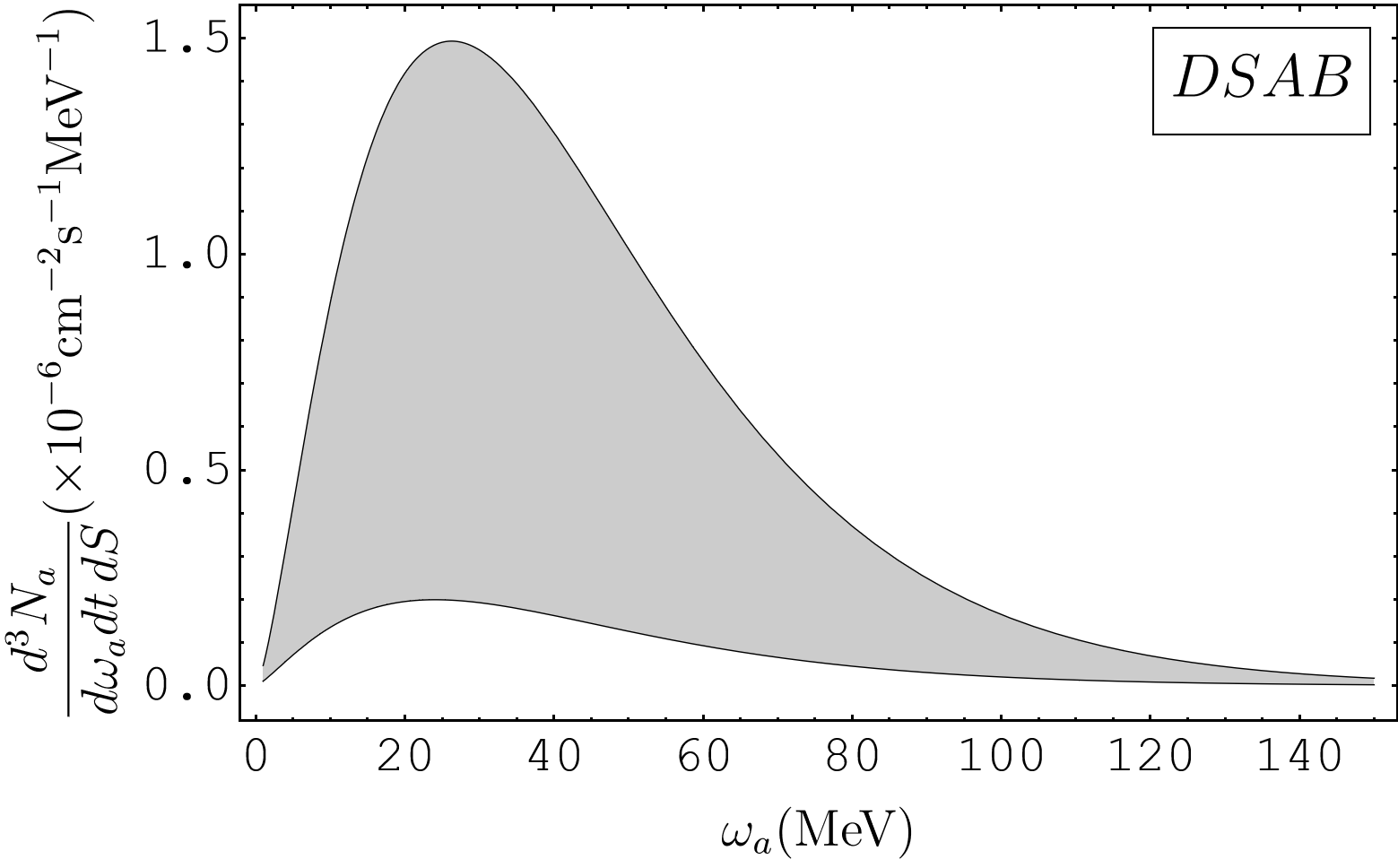}
\caption{DSAB fluxes range of variability (gray band) for  $g_{a\gamma} = 10^{-11}$ GeV$^{-1}$. (Data taken from Ref.~\cite{Calore:2021hhn}).
	 }\label{fig:dsalprange} 
\end{figure}

Let us finally look at the contribution of all pasts SN events to the axion background. This is often dubbed the Diffuse Supernova Axion Background (DSAB)~\cite{Calore:2020tjw}. The non-observation the DSAB allows us to set  limits on the axion-photon coupling. Here, we refer to che case of generic axions coupled only with photons~\cite{Calore:2020tjw}. The QCD axion case was considered in Ref.~\cite{Raffelt:2011ft}. The axion flux from all past SNe can be expressed as
\begin{equation}
\frac{d \phi_a (E_a)}{d E_a} =\int_0^{\infty} (1+z) \frac{dN_a(E_a(1+z))}{dE_a}[R_{SN}(z)] \bigg[ \bigg|\frac{dt}{dz} \bigg| dz \bigg]\,,
\label{eq:diffuse}
\end{equation}
where $z$ is the redshift, $R_{SN}(z)$ is the SN explosion rate~\cite{Priya:2017bmm}, $ |{dt}/{dz} |^{-1}= H_0(1+z)[\Omega_\Lambda+\Omega_M(1+z)^3]^{1/2}$ with the cosmological parameters $H_0= 67.4$ km s$^{-1}$ Mpc$^{-1}$, $\Omega_M=0.3$, $\Omega_\Lambda=0.7$~\cite{Aghanim:2018eyx}. Ref.~\cite{Calore:2021hhn} performed an improved calculation of the DSAB flux a set of core-collapse SN models with different progenitor masses, as well as the effects of failed  SNe -- which yield the formation of black holes. Uncertainties in the SFR have been also included. The result of the DSAB flux with its uncertainty band is shown in Fig.~\ref{fig:dsalprange}.

From this flux, Ref.~\cite{Calore:2021hhn} calculated the diffuse gamma-ray flux produced by conversions in the Milky-Way. This latter flux has been  used  via a template-based analysis  that utilizes 12 years of {\it Fermi}-LAT data, finding  an upper limit on $g_{a\gamma} \lesssim 3.76\times10^{-11}\;\mathrm{GeV}^{-1}$ at 95$\%$ C.L. for $m_a \ll 10^{-11}$ eV (see Fig.~\ref{fig:SNlimits}).

In case of massive axions, if $l_{\rm a} \lesssim 1/H_0$ a large fraction of axions from cosmological distances decay before reaching Earth, producing a diffuse gamma-ray flux. In this way, comparing this diffuse flux with the one measured by COMPTEL and {\it Fermi}-LAT in the range $E\in [1:100]$~MeV, one  excludes the yellow-band in Fig.~\ref{fig:SNALPGagmassive}  (see also Ref.~\cite{Calore:2020tjw}).

\subsection{Beyond the minimal scenario}
\label{sec:miscellanea}

Apart from the case of axions coupled with  nucleons and  photons, the SN bound has been studied for different types of axions with a variety of couplings. We give here a brief overview of different cases explored in the literature. Refs.~\cite{Lucente:2021hbp,Ferreira:2022xlw} studied the case of axions coupled with electrons, which would be produced in a SN via electron-proton bremsstrahlung and electron-positron fusion, possibly including loop effects. In this case, the SN 1987A cooling bound would  exclude currently unprobed regions down to $g_{ae}\sim 2.5\times 10^{-10}$ at $m_a\sim 120$~MeV. In Ref.~\cite{Caputo:2021rux} (see also Ref.~\cite{Bollig:2020xdr}) it was considered the case of muonphilic bosons  both scalars $\phi$ or pseudoscalars $a$ that couple to muons through the Yukawa operators (see Ref.~\cite{Alda:2024cxn} for light particles coupled to tau leptons)
\begin{equation}\label{eq:coupling}
\mathcal{L}_\phi \supset - g_{\phi}\phi \bar{\mu}\mu
\quad\hbox{and}\quad
\mathcal{L}_a \supset -i g_{a} a \bar{\mu}\gamma_5\mu\,.
\end{equation}
If bosons escape freely from the SN core the main constraints originate from SN~1987A gamma rays and the diffuse cosmic gamma-ray background. The latter allows to exclude $g_a \gtrsim 0.9\times10^{-10}$ and $g_\phi \gtrsim 0.4\times10^{-10}$, for $m_{a,\phi} \gtrsim 100$~keV. In the strong coupling regime, in order to avoid too much energy  energy  dumped into the surrounding progenitor-star matter, one should exclude $g_{a}\gtrsim 2\times10^{-3}$ and $g_{\phi} \gtrsim 4\times10^{-3}$. 

Non-minimal axion models with different combinations of couplings have also been considered. In Ref.~\cite{Calore:2021klc} it was investigated the potential of SNe  to constrain axions coupled simultaneously to nucleons and electrons. Axions coupled to nucleons can be efficiently produced in the SN core via NN bremsstrahlung and, for a wide range of parameters, leave the SN unhindered, producing a large axion flux (see also Sec.~\ref{sec:QCD}). For masses exceeding 1 MeV, these axions would decay into electron-positron pairs, generating a positron flux. In the case of Galactic SNe, the annihilation of the created positrons with the electrons present in the Galaxy would contribute to the 511 keV annihilation line. Using the SPI (SPectrometer on INTEGRAL) observation of this line, allows us to exclude a wide range of the axion-electron coupling, $10^{-19} \lesssim g_{ae} \lesssim 10^{-11}$, for $g_{ap}\sim 10^{-9}$ below the cooling bound. Additionally, axions from extra-galactic SNe decaying into electron-positron pairs would yield a contribution to the cosmic $X$-ray background. In this case, one can constrain the axion-electron coupling down to $g_{ae} \sim 10^{-20}$. See also Refs.~\cite{DelaTorreLuque:2023huu,DelaTorreLuque:2023nhh,DelaTorreLuque:2024zsr} for recent developments on this topic.

If axions couple simultaneously with nucleons and photons one can strengthen the SN 1987A bound on ultralight ALPs from non-observations of gamma-rays in GRS down to  $g_{a\gamma} <3.4 \times 10^{-15}$~GeV$^{-1}$ for $g_{ap}\sim 10^{-9}$~\cite{Calore:2020tjw}. In this case, the DSAB bound on the diffuse gamma-ray flux lead to $g_{a\gamma} \lesssim 6 \times 10^{-13}$~GeV$^{-1}$~\cite{Calore:2020tjw}. If axions are heavier than $\sim$ keV, the decay into photons becomes significant, leading again to a diffuse gamma-ray flux. Allowing for a (maximal) coupling to nucleons, the limit improves to the level of $g_{a\gamma} \lesssim 10^{-19}$~GeV$^{-1}$ for  $m_a \sim 20$~MeV~\cite{Calore:2020tjw}, which represents the strongest constraint to date. 

We also remark that in Ref.~\cite{Lella:2024gqc} it was realized that axions coupled to nucleons will have an induced photon coupling. As a matter of fact, the nucleon coupling arises from a fundamental interaction with quarks and/or gluons. A coupling to photons can be induced by a fermion loop or the non-perturbative axion-pion mixing. Therefore, it is possible to obtain constraints on MeV-scale axions decaying into photons and produced via nuclear processes, using the arguments described above. A combination of these bounds allows one to rule out axion-nucleon couplings down to  $g_{aN} \gtrsim 10^{-11}-10^{-10} {\rm GeV}^{-1}$ for $m_a \gtrsim1$~MeV.

In Ref.~\cite{Lucente:2022vuo} (see also Ref.~\cite{Graham:2013gfa}) were considered axions coupled to nucleons and photons only through the nucleon electric-dipole moment  portal  (see Sec.~\ref{sec:dipole}). 
As discussed before, this coupling is a model-independent feature of QCD axions, which solve the strong CP problem, and might arise as well in more general axion-like particle setups.\footnote{More recently, two unrelated model-independent SN axion bounds have been proposed in Ref.~\cite{Chakraborty:2024tyx} and in Ref.~\cite{Springmann:2024ret}. 
These bounds are subdominant for most QCD axion models, which typically also couple to fermions. However, while the coupling to fermions can, in principle, be suppressed, the bounds derived in the aforementioned references—similar to those derived from the nEDM—apply universally to all QCD axion models.}
The SN 1987A bound allows the exclusion of the coupling to EDM in the range   $6.7\times 10^{-9}{\rm GeV}^{-2}\lesssim g_{d} \lesssim  7.7\times 10^{-6}{\rm GeV}^{-2}$.
Intriguingly, there is potential to detect the axion flux through a distinctive gamma-ray signal generated in a large underground neutrino detector, such as the proposed Hyper-Kamiokande, even for couplings below the current lower bound.

\subsection{Neutron stars cooling}
\label{sec:neutrstar}

Besides SNe, NS are also powerful laboratories to constrain the axion-nucleon coupling. Young isolated NS have their temperature evolution driven by neutrino emission from birth, in a SN explosion. As we have seen, the initial  neutrino cooling has time scale of seconds  (refer to Refs.~\cite{Potekhin:2015qsa,Yakovlev:2004iq,Page:2005fq} for reviews). The thermal evolution of NS also provides valuable information about temperature-dependent properties, including transport coefficients, crust solidification, and internal pulsar heating mechanisms. Accurate theoretical cooling calculations should include factors such as the nuclear EoS, stellar mass, magnetic field, superfluidity, meson condensates, and potential existence of color-superconducting quark matter. Therefore, cooling simulations, when matched with observations in soft X-ray, extreme UV, UV, and optical wavelengths of the thermal photon flux emitted from neutron star surfaces offer vital insights into the properties of super-dense hadronic matter and the structure of NS.

NS cooling is described by the energy conservation equation for the star which, in its Newtonian formulation, can be summarized as follows~\cite{Page:2005fq}
\begin{equation}
\frac{dE_\mathrm{th}}{dt} = C_\mathrm{v} \frac{dT}{dt}
                   = -L_\nu - L_\gamma + H \, ,
\label{equ:energy-conservation}
\end{equation}
where $E_\mathrm{th}$ is the thermal energy content of the star, $T$ its internal temperature, and $C_\mathrm{v}$ its total specific heat. The energy sinks are the total neutrino luminosity $L_\nu$, and the surface photon luminosity $L_\gamma$.  The source term $H$ includes all possible ``heating mechanisms'' which, for instance, convert magnetic or rotational energy into heat. The impact of axion emission has been studied also in relation to the late-time cooling of NS, considering as main emissivity process
$n ~ n \to n ~ n ~ a$~\cite{Hanhart:2000ae}.

\begin{figure}[t!]
\centering
\vspace{0.cm}
\includegraphics[width=0.6\columnwidth]{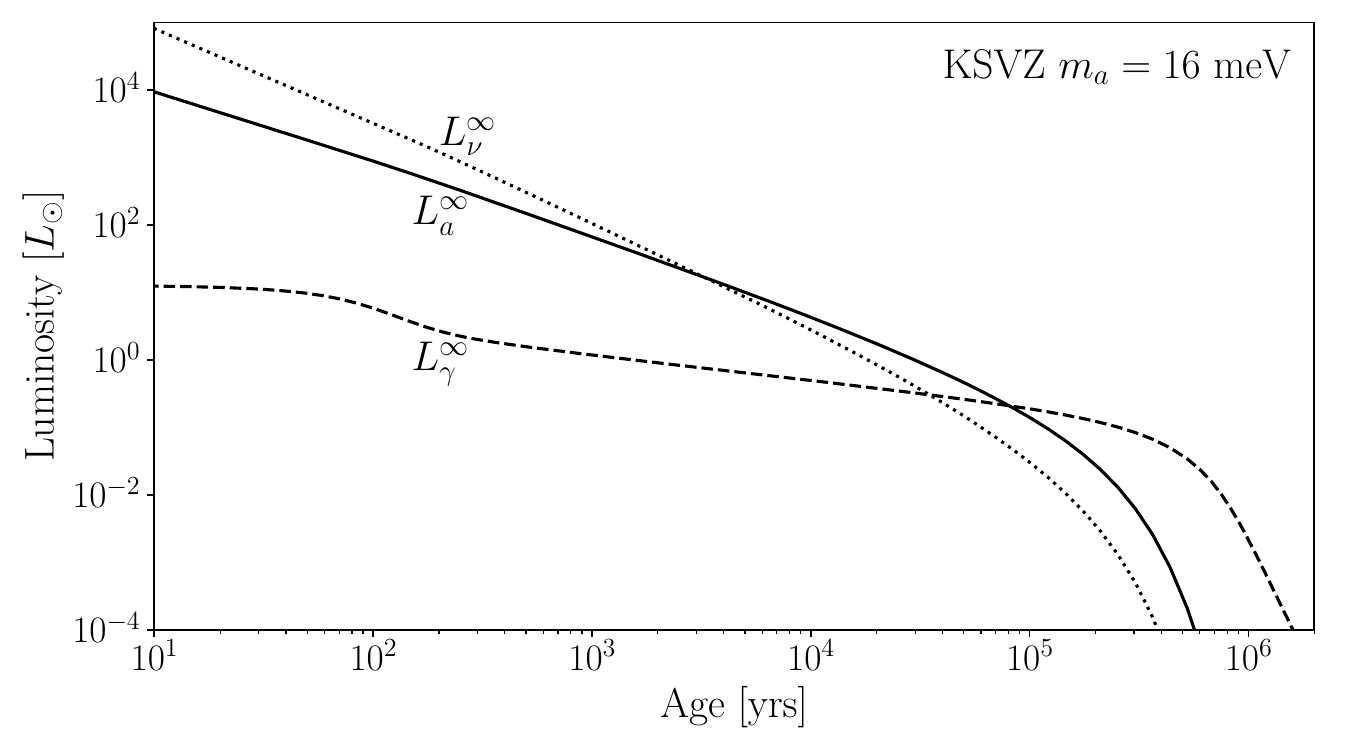}
\caption{The luminosity production from neutrinos, axions, and surface radiation for an example NS with the KSVZ axion at $m_a = 16$ meV (Figure taken from 
Ref.~\cite{Buschmann:2021juv} with permission). }\label{fig:NS}
\end{figure}

After the initial non-isothermal phase the models settle into an equilibrium state which is characterized by an isothermal core and gradient-featuring envelope. Then, integrating the energy balance equation over the interior of the NS leads to the cooling equation
\begin{equation}
\label{eq:cooling}
    L_\gamma^\infty = -C_v \dfrac{dT^\infty}{dt} - L_\nu^\infty - L_a^\infty + H \,,
\end{equation}
where $L_\gamma^\infty = 4\pi R_{*,\infty}^2 (T_s^\infty)^4$ is the photon luminosity, and $t$ is time (see Fig.~\ref{fig:NS}).

The analysis of NS cooling offers a set of advantages with respect to the PNS in the study of new physics. Firstly, the cooling of NS can be observed for several years. Furthermore, there are several NS that can be currently studied. On the other hand, SN events  are rapid and  rare. Practically, however, there are many theoretical uncertainties in the microphysics which complicate the extraction of a reliable axion flux. For example, previous studies of the young NS CAS A had shown an anomalous fast cooling, which was interpreted as evidence for axions coupled to nucleons with $g_{an}\approx 4\times 10^{-10}$~\cite{Leinson:2014ioa}. However,  it was later shown that this apparent anomaly might be due in large part to a systematic evolution of the energy calibration of the detector over time~\cite{Posselt:2018xaf}. Furthermore, as it was soon realized by the same author of Ref.~\cite{Leinson:2014ioa} and by others, the data can also be explained assuming a neutron triplet superfluid transition occurring at the present time, $t \sim 320$ years, in addition to a proton superconductivity operating at $t\ll$ 320 years. Under these assumptions, it was possible to fit the available data well, leaving little room for additional axion cooling and leading to the bound~\cite{Hamaguchi:2018oqw}
\begin{equation}
\label{eq:Hamaguchi}
    g_{ap}^2+1.6 g_{an}^2\leq 1.1\times 10^{-18}\,.
\end{equation} 
In any case,  the new-physics hypothesis was not completely ruled out, though the existence of a possible standard model solution weakened substantially the axion case.

A stronger bound, though only on the axion-neutron coupling, 
\begin{equation}
g_{a n}^2<7.7 \times 10^{-20}~~ \text { at } 90 \% \,\  \text {C.L.}    
\end{equation}
was inferred from observations of the NS in HESS J1731-347~\cite{Beznogov:2018fda}. NS HESS J1731-347 is a much older star than CAS A. The bound was substantially relaxed in Ref.~\cite{Sedrakian:2018kdm}, which found, 
\begin{equation}
    g_{an}\lesssim (2.5-3.2) \times 10^{-9}\,.
\end{equation}

It is important to notice that the impact of the axion emission from NS depends on the star age. So far young NS, in particular Cas A, with age $\sim$300 yrs, have been the most widely studied in relation to axion physics~\cite{Page:2010aw,Shternin:2010qi,Leinson:2014cja,Leinson:2014ioa,Sedrakian:2015krq,Hamaguchi:2018oqw,Leinson:2021ety}. Indeed, for a young star the thermal axion emission is very high as it steeply depends on the core temperature. However, the axion luminosity may be surpassed by the neutrino contribution, which has an even steeper dependence on temperature. In an old NS, on the other hand, the neutrino and axion emissions are suppressed and the cooling is dominated by photons. There is, evidently, a preferred age where the effects of the axion emission rate can be maximized. A systematic study of these effects was presented in Ref.~\cite{Buschmann:2021juv}, which found the ideal age to study axions to be $\approx 10^{5}$ yr. The authors make the case that four of the nearby isolated Magnificent Seven NSs are prime candidates for axion cooling studies  because they are coeval, with ages of a few hundred thousand years, known from kinematic considerations, and they have well-measured surface luminosities. The results of this analysis are shown in Fig.~\ref{fig:NS}.   

Older  NSs, with ages $\sim 10^5$--$10^6$ yrs, have also been studied. In particular, Refs.~\cite{Dessert:2019dos,Buschmann:2019pfp} considered the expected X-ray signal deriving from the conversion of the NS axion flux in the magnetic fields surrounding the NS (see also Refs.~\cite{Fortin:2018ehg,Fortin:2021sst}). Furthermore, the analysis in Ref.~\cite{Buschmann:2019pfp} found that the observed excess of hard $X$-ray emission in the $2 - 8$~keV energy range from the nearby Magnificent Seven isolated NSs could be explained by this conversion of NS axions, if axions have mass $m_a \lesssim 2 \times 10^{-5}$ eV and $g_{a\gamma} \times g_{an} \in (2 \times 10^{-21}, 10^{-18})$ GeV$^{-1}$ at 95\% C.L.. Some attempts to present this as QCD axions were also considered in the literature~\cite{Darme:2020gyx}. It is finally worth mentioning the case of NS merger~\cite{Fiorillo:2022piv}, which however finds less restrictive bounds with respect to the ones obtained from SNe and NS.


\section{Discussions and Conclusions}
\label{sec:conclusions}

Forty years after their introduction as a solution to the strong CP problem, axions are currently experiencing a renaissance. Notably, axions are among the most promising dark matter candidates, which motivates much of the effort to search for these particles~\cite{Adams:2022pbo}. Furthermore, beyond their original motivation, the theoretical framework for axions has been significantly strengthened by the realization that, in string theory compactification scenarios, QCD axions and several light axion-like particles naturally emerge-- a phenomenon known as the string axiverse~\cite{Cicoli:2023opf}. These developments have stimulated a vibrant experimental program worldwide, aiming to discover axions using various sources and experimental approaches~\cite{Irastorza:2018dyq}. In this regard, over the coming decades, planned experiments are expected to reach the sensitivity needed to explore much of the previously inaccessible axion parameter space.

\emph{In this context, what is the role of astrophysics?} As we have documented in this review, stellar astrophysics provides two types of arguments to probe axions: \emph{(a)} the \emph{direct} detection of stellar axions, and \emph{(b)} the \emph{indirect} probing of axions by monitoring the discrepancies their production induces in various stellar observables.

Regarding direct detection, we have shown that axions have been searched for from the Sun and other nearby stars (see Sec.~\ref{sec:Sun}) and from supernovae  (see Sec.~\ref{sec:SN_NS}). 
In particular, the Sun, our closest star, offers the possibility of directly searching for axions with dedicated helioscope experiments, such as CAST, which has set the best experimental limit on the axion-photon coupling over a broad mass range. Next-generation experiments, like the planned IAXO, are expected to significantly improve the current sensitivity to solar axions.
Moreover, axions from stars can be studied using X-ray or gamma-ray telescopes, aiming to detect photon fluxes from stellar axion conversions in cosmic magnetic fields. Nearby supergiants present intriguing targets for X-ray telescopes, while Galactic SNe are predicted to produce axion-induced gamma-ray bursts. Notably, a gamma-ray signal coincident with SN 1987A has been extensively studied, allowing constraints to be placed on a significant portion of the axion parameter space. It is expected that a gamma-ray experiment like \emph{Fermi}-LAT would greatly enhance current sensitivity in the event of a Galactic SN within its field of view.

\emph{What will we learn in the case of a direct detection of an axion signal from the stars?}
Apart from cornering the axion mass and coupling with photons, axions will constitute a new probe to learn about the inner structure of stars, complementary with the information that one can get from photons and neutrinos.

Indirect probes of axions from stellar evolution include a suppression of the number of helium burning stars in Globular Clusters (see Sec.~\ref{sec:GC}), the impact on the white dwarf luminosity function (see Sec.~\ref{sec:white_dwarfs}), the shortening of a SN neutrino burst and the impact on  neutron star light-curves (see Sec.~\ref{sec:SN_NS}). While the axion direct detection is mostly sensitive to the axion-photon coupling, energy-loss argument may allow one to probe different couplings depending on the stellar system. In particular, helium-burning stars are sensitive to the axion-photon coupling, while red giants and white dwarfs allow probing of the axion-electron coupling. Finally, (proto)-neutron stars are sensitive to the axion-nucleon coupling, which is challenging to explore experimentally.

\emph{What are the prospects for improving the energy-loss argument?}
We feel quite optimistic in this respect. An unparalleled progress in the asteroseismology of white dwarfs is expected in the coming years, thanks to space missions such as TESS, and in globular clusters due to the data releases from the astrometric satellite Gaia. Regarding the supernova  and neutron star  environments, improving the cooling bounds will require better characterization of axion emissivity processes in dense nuclear matter. In recent years, several studies have been conducted addressing, for example, corrections beyond the one-pion approximation and the inclusion of pionic processes. However, further research is necessary to achieve a coherent understanding. Moreover, the SN 1987A bound is based on the observation of a sparse neutrino signal. New state-of-the-art 3D supernova neutrino simulations~\cite{Bollig:2020phc,Vartanyan:2018iah}, including axions, are essential to assess the true impact of axion energy loss on the neutrino signal and to interpret the high-statistics neutrino data from the next Galactic supernova.

\emph{Why is it important to rely on indirect axion signatures from stellar evolution when direct experimental searches are possible?}
We believe that these types of indirect arguments remain important and complementary to direct searches. As we noted in the Introduction, astrophysical arguments have already played a key role in guiding laboratory searches, such as in the case of the PVLAS claim of an axion-like particle~\cite{PVLAS:2005sku}, which was quickly shown to be incompatible with the physical understanding of our Sun. This example demonstrates that every experimental measurement and every astrophysical argument carries its own systematic uncertainties and its own recognized or unrecognized loopholes. Therefore, to conclusively detect axions, it is essential to use as many independent interaction channels and as many different approaches as possible.

In conclusion, with this review, we have presented the state-of-the-art in stellar bounds and opportunities to study axions. We hope this work will be useful to the community in guiding future theoretical and experimental studies on axions and will stimulate new ideas and further investigations to improve the arguments discussed here.

\setcounter{secnumdepth}{0}

\section{Acknowledgments}
\label{sec:acknowledgments}

We warmly thank Alessandro Lella and Giuseppe Lucente for the many discussions during the development of this work, and Georg Raffelt for several relevant comments and suggestions. 
This article is based upon work from COST Action COSMIC WISPers CA21106, supported by COST (European Cooperation in Science and Technology).

The work of PC is supported by the European Research Council under Grant No.~742104 and by the Swedish Research Council (VR) under grants  2018-03641, 2019-02337 and 2022-04283.

The work of AM  was partially supported by the research grant number 2022E2J4RK ``PANTHEON: Perspectives in Astroparticle and
Neutrino THEory with Old and New messengers" under the program PRIN 2022 funded by the Italian Ministero dell’Universit\`a e della Ricerca (MUR). 
This work is (partially) supported
by ICSC – Centro Nazionale di Ricerca in High Performance Computing.

JI acknowledges grants Spanish program Unidad
de Excelencia María de Maeztu CEX2020-001058-M, PID2023-149918NB-I00, and 2021-SGR-1526 (Generalitat de Catalunya).

MG acknowledges support from the Spanish Agencia Estatal de Investigación under grant PID2019-108122GB-C31, funded by MCIN/AEI/10.13039/501100011033, and from the “European Union NextGenerationEU/PRTR” (Planes complementarios, Programa de Astrofísica y Física de Altas Energías). He also acknowledges support from grant PGC2022-126078NB-C21, “Aún más allá de los modelos estándar,” funded by MCIN/AEI/10.13039/501100011033 and “ERDF A way of making Europe.” Additionally, MG acknowledges funding from the European Union’s Horizon 2020 research and innovation programme under the European Research Council (ERC) grant agreement ERC-2017-AdG788781 (IAXO+). 

OS acknowledges funding from the large program BRAVOSUN of the Italian National Institute of Astrophysics.

\newpage

\section{Appendix A: Units Convention}
\label{sec:Appendix_Units}

For this review, we adopted the \textit{natural system of units},
{unless otherwise specified}.
These units may result somewhat confusing for those with a background specifically in astrophysics, where normally the CGS (cm, grams, meter) system is quite common. 
However, these are the usually adopted units in high energy physics as well as in astroparticle physics, and used in the standard reference, including in the classic book by G. Raffelt~\cite{Raffelt:1996wa}.
We refer to Appendix A of this book for more details on this system of units.
Even more details can be found in the text by Jackson~\cite{Jackson:1998nia}, which is a standard reference in classical electrodynamics.

In these units, the speed of light $c$, the reduced Planck's constant $\hbar$, and the Boltzmann's constant $k$ are dimensionless and equal to unity.
Therefore, ${\rm length}^{-1}$, ${\rm time}^{-1}$, mass, energy, and temperature can all be measured in the same unit. 

For the electric charge $e$ we use the definition $\alpha=e^2 / 4 \pi$, where $\alpha \approx$ $1 / 137$ is the fine-structure constant, which is dimensionless and has the same value in all systems of units.
This choice corresponds to the \textit{rationalized system of (natural) units}. 
The energy density of an electromagnetic field is then $\frac{1}{2}\left(E^2+B^2\right)$. 
In the older literature and some texts on electromagnetism, unrationalized units are used where $\alpha=e^2$ and the energy density is $\left(E^2+B^2\right) / 8 \pi$.

\newpage

\section{Appendix B: Tables of notations and of acronyms}
\label{sec:Appendix_tables}


\begin{table}[h!]
\centering
\label{tab:Acronyms}
\scalebox{0.95}{
\begin{tabular}{|l|l|}
\hline
\textbf{Acronym} & \textbf{Meaning}                                                       \\ \hline
ABC            & Atomic, Bremsstrahlung and Compton processes                             \\ \hline
AGB              & Asymptotic Giant Branch                                                \\ \hline
ALP              & Axion Like Particle                                                    \\ \hline
BBN              &Big-Bang-Nucleosynthesis                                                \\ \hline
BST              & Baksan Scintillator Telescope                                          \\ \hline
CAST             & CERN Axion Solar Telescope                                             \\ \hline
CMD              & Color Magnitude Diagram                                                \\ \hline
DFSZ              & Dine-Fischler-Srednicki-Zhitnitsky                                              \\ \hline
DMSAB            & Diffuse Main Sequence axion Background                                 \\ \hline
DSAB             & Diffuse Supernova axion Background                                     \\ \hline
EoS              & Equation of State                                                                   \\ \hline
Fermi-LAT        & Fermi Large Area Telescope                                             \\ \hline
FIP              & Feebly Interacting Particle                                            \\ \hline
FuNS & Full Network Stellar evolution
 \\ \hline
Gaia             &   Global Astrometric Interferometer for Astrophysics                  \\ \hline
GC              & Globular Cluster                                                 \\ \hline 
HB               & Horizontal Branch                                                       \\ \hline
KII              & Kamiokande-II detector                                                     \\ \hline
IAXO             & International AXion Observatory                                         \\ \hline
IBD              & Inverse Beta Decay                                                      \\ \hline
IFMR              & Initial-Final Mass Relation                                                      \\ \hline
IMB              & Irvine-Michigan-Brookhaven (IMB) water Cherenkov                        \\ \hline
IMF   & Initial Mass Function
\\ \hline
KSVZ   & Kim-Shifman-Vainshtein-Zakharov
\\ \hline
LPM               & Landau-Pomeranchuk-Migdal                                                    \\ \hline
mfp    & mean free path \\
\hline
MS & Main Sequence
 \\ \hline
 (n)EDM &  (neutron) Electric Dipole Moment 
 \\ \hline
NS               & Neutron Star                                                            \\ \hline
NuSTAR           & Nuclear Spectroscopic Telescope Array                                   \\ \hline
OPE              & One Pion Exchange                                                       \\ \hline
PNS              & Proto-Neutron Star                                                     \\ \hline
RG             & Red Giant                                                        \\ \hline
RGB              & Red Giant Branch                                                        \\ \hline
SCMD             & Synthetic color-magnitude diagrams                                   \\ \hline   
SFR  & Star Formation Rate
\\ \hline
SM               & Standard Model (of particle physics)                                    \\ \hline
SSM & Standard Solar Model
 \\ \hline
SMM & Solar Maximum Mission
 \\ \hline
SN               & Supernova                                                               \\ \hline
TP/LP               & Transverse Photon/Longitudinal Plasmon                                                       \\ \hline
WD               & White Dwarf                                                             \\ \hline
WDLF               & White Dwarf    Luminosity Function                                                         \\ \hline
WIMP             & Weakly Interacting Massive Particle                                     \\ \hline
WISP             & Weakly Interactive Slim Particle (Light FIP, $m \lesssim \mathrm{eV}$ ) \\ 
\hline
ZAHB&Zero Age Horizontal Branch\\\hline
ZAMS&Zero Age Main Sequence\\\hline
\end{tabular}
}
\end{table}

\newpage
\newpage

\bibliography{references_final}

\bibliographystyle{elsarticle-num}
\end{document}